\documentclass[12pt]{article}
\usepackage{amsmath,amssymb,amsfonts,color,graphicx,cite,color,feynarts,soul}
\input paperdef

\graphicspath{{figs/}}

\evensidemargin \oddsidemargin
\allowdisplaybreaks

\newcommand{\trans}{T}

\begin{document}
\thispagestyle{empty}

\def\thefootnote{\fnsymbol{footnote}}

\begin{flushright}
CERN--PH--TH/2011--291\\
FR--PHENO--2011--021\\
KA--TP--34--2011 
\end{flushright}

\vspace{0.5cm}

\begin{center}

{\large\sc {\bf Heavy Scalar Top Quark Decays in the Complex MSSM:}}

\vspace{0.4cm}

{\large\sc {\bf A Full One-Loop Analysis}}

\vspace{1cm}

{\sc
T.~Fritzsche$^{1}$%
\footnote{Email: Thomas.Fritzsche@de.bosch.com}%
, S.~Heinemeyer$^{2}$%
\footnote{Email: Sven.Heinemeyer@cern.ch}%
, H.~Rzehak$^{3}$%
\footnote{Email: heidi.rzehak@cern.ch}%
\footnote{On leave from Albert-Ludwigs-Universit\"at Freiburg, 
  Physikalisches Institut, D--79104 Freiburg, Germany
}%
~and, C.~Schappacher$^{4}$%
\footnote{Email: cs@particle.uni-karlsruhe.de}%
}

\vspace*{.7cm}

{\sl
$^1$Max-Planck-Institut f\"ur Physik (Werner-Heisenberg-Institut),
F\"ohringer Ring 6, \\
D--80805 M\"unchen, Germany
\footnote{Present address: Robert Bosch GmbH, Corporate Sector Research 
and Advance Engineering, 
D-70839 Gerlingen-Schillerh\"ohe, Germany}

\vspace*{0.1cm}

$^2$Instituto de F\'isica de Cantabria (CSIC-UC), Santander,  Spain

\vspace*{0.1cm}

$^3$PH-TH, CERN, CH--1211 Gen\`eve 23, Switzerland

\vspace*{0.1cm}

$^4$Institut f\"ur Theoretische Physik,
Karlsruhe Institute of Technology, \\
D--76128 Karlsruhe, Germany

}

\end{center}

\vspace*{0.1cm}

\begin{abstract}
\noindent
We evaluate all two-body decay modes of the heavy scalar top quark in
the Minimal Supersymmetric Standard Model with complex parameters (cMSSM) 
and no generation mixing. The evaluation is based on a full
one-loop calculation of all decay channels, also including hard QED and
QCD radiation. The renormalization of the complex parameters is
described in detail. 
The dependence of the heavy 
scalar top quark decay  on the relevant cMSSM parameters is
analyzed numerically, including also the decay to Higgs bosons 
and  another scalar quark or to a top quark and the lightest neutralino.
We find sizable contributions to many partial decay widths and  
branching ratios. They are roughly of \order{10\%} of the tree-level 
results, but can go up to $30\%$ or higher. 
These contributions are important for the correct interpretation
of scalar top quark decays at the LHC and, if kinematically allowed, at 
the ILC.
The evaluation of the branching ratios of the heavy scalar top quark
will be  implemented into the Fortran code {\tt FeynHiggs}.
\end{abstract}

\def\thefootnote{\arabic{footnote}}
\setcounter{page}{0}
\setcounter{footnote}{0}

\newpage

\newcommand{\decayhn}{\Stopz \to \Stope h_n}
\newcommand{\decayh}{\Stopz \to \Stope h_1}
\newcommand{\decayH}{\Stopz \to \Stope h_2}
\newcommand{\decayA}{\Stopz \to \Stope h_3}
\newcommand{\decayZ}{\Stopz \to \Stope Z}
\newcommand{\decaygl}{\Stopz \to t \gl}
\newcommand{\decayNe}{\Stopz \to t \neu1}
\newcommand{\decayNz}{\Stopz \to t \neu2}
\newcommand{\decayNd}{\Stopz \to t \neu3}
\newcommand{\decayNv}{\Stopz \to t \neu4}
\newcommand{\decayNk}{\Stopz \to t \neu{k}}
\newcommand{\decayCe}{\Stopz \to b \cha1}
\newcommand{\decayCz}{\Stopz \to b \cha2}
\newcommand{\decayCj}{\Stopz \to b \chap{j}}
\newcommand{\decayCpe}{\Stopz \to b \chap1}
\newcommand{\decayCpz}{\Stopz \to b \chap2}
\newcommand{\decayCpj}{\Stopz \to b \chap{j}}
\newcommand{\decaySbeH}{\Stopz \to \Sbote H^+}
\newcommand{\decaySbzH}{\Stopz \to \Sbotz H^+}                     
\newcommand{\decaySbiH}{\Stopz \to \Sboti H^+}                     
\newcommand{\decaySbeW}{\Stopz \to \Sbote W^+}                     
\newcommand{\decaySbzW}{\Stopz \to \Sbotz W^+}
\newcommand{\decaySbiW}{\Stopz \to \Sboti W^+}
\newcommand{\decayxy}{\Stopz \to {\rm xy}}


\section{Introduction}

One of the most important tasks at the LHC is to search for physics
effects beyond the Standard Model (SM), where the Minimal Supersymmetric
Standard Model (MSSM)~\cite{mssm} is one of the leading candidates. 
Supersymmetry (SUSY) predicts two scalar partners for all SM fermions as well
as fermionic partners to all SM bosons.
Another important task is investigating the mechanism of electroweak
symmetry breaking. 
The most frequently investigated
models are the Higgs mechanism within the SM and within the MSSM.
Contrary to the case of the SM, in the MSSM 
two Higgs doublets are required.
This results in five physical Higgs bosons instead of the single Higgs
boson in the SM; three neutral Higgs bosons, $h_n$ ($n = 1,2,3$), 
and two charged Higgs bosons, $H^\pm$. 

If SUSY is realized in nature and the scalar quarks and/or the gluino
are in the kinematic reach of the LHC, it is expected that these
strongly interacting particles are copiously produced. This includes the
production of scalar top quark pairs or the production of two gluinos 
with the subsequent (possible) decay to a scalar top  quark and a top 
quark.  
An interesting production channel of Higgs bosons at the LHC is 
the decay of the heavy scalar top  quark to the lighter scalar top 
(scalar bottom)  quark and a neutral (charged) Higgs boson, see, for
instance, \citeres{atlas,cms} and references therein. 
At the ILC (or any other future $e^+e^-$ collider such as CLIC) 
a precision determination of the properties of the observed particles is
expected~\cite{teslatdr,ilc}. (For combined LHC/ILC analyses and further
prospects see \citere{lhcilc}.) 
Thus, if kinematically accessible, Higgs production via scalar top quark
decays could offer important information about the stop and Higgs
sector of the MSSM. 

In order to yield a sufficient accuracy, one-loop corrections to
the various scalar top quark decay modes have to be considered.
We take into account all two-body decay modes of the heavy scalar top
quark, $\Stopz$, in the MSSM with complex parameters (cMSSM),
but we neglect flavor violation effects and resulting decay channels 
that rather play a role for the decay of the light scalar top quark, 
$\Stope$, in special regions of the MSSM parameter 
space~\cite{FVstop1decay}. 
More specifically, we calculate the full one-loop corrections to the 
partial decay widths%
\footnote{
  It should be noted that the purely loop induced decay channels 
  $\Stopz \to \Stope \ga/g$ have been neglected because they yield exactly 
  zero; see \refse{sec:calc} for further details.
}
\begin{align}
\label{ststphi}
&\Ga(\Stopz \to \Stope h_n) \qquad (n = 1,2,3)~, \\
\label{ststZ}
&\Ga(\Stopz \to \Stope Z)~, \\
\label{sttneu}
&\Ga(\Stopz \to t \neu{k}) \qquad (k = 1,2,3, 4)~, \\
\label{sttgl}
&\Ga(\Stopz \to t \gl)~, \\
\label{stsbH}
&\Ga(\Stopz \to \Sboti H^+) \qquad (i = 1,2)~, \\
\label{stsbW}
&\Ga(\Stopz \to \Sboti W^+) \qquad (i = 1,2)~, \\
\label{stbcha}
&\Ga(\Stopz \to b \chap{j}) \qquad (j = 1,2)~,
\end{align}
where $\neu{k}$ denotes the neutralinos, $\gl$ the gluino, $\cha{j}$
the charginos, $t$ and $b$ the top and bottom quark and $Z$ and
$W^{\pm}$ the SM gauge bosons.
The total decay width is defined as the sum of the partial decay
widths (\ref{ststphi}) to (\ref{stbcha}), where for a given parameter
point several channels may be kinematically forbidden. 

As explained above, 
we are especially interested in the branching ratios (BR) of the decays of 
the $\Stopz$ to a Higgs boson and another squark, \refeqs{ststphi} and 
(\ref{stsbH}), as part of an evaluation of a Higgs production cross section.
This can be an interesting production channel at the LHC 
(see, for instance, \citere{stopstophiggs-LHC}, where 
$pp \to \aStope \Stope h$ is analyzed, or \citere{Higgsincascades}, where
searches for $\cp$-odd Higgs bosons in top squark decays are discussed).  
However, in order to reach a high accuracy, {\em all} two-body decay 
channels should be evaluated at one-loop.
On the other hand, because we are interested in two-body modes 
(involving Higgs bosons), it is not necessary to investigate three- or 
four-body decay modes as these only play a significant role once the 
two-body decay modes are kinematically forbidden, and thus the relevant 
BR are zero.

We also concentrate on the decays of $\Stopz$ and do not
investigate $\aStopz$ decays. 
In the presence of complex phases this
would lead to somewhat different results. However, such an analysis of
$\cp$-violating effects is beyond the scope of this paper.

Higher-order contributions to scalar fermion decays have been evaluated in
various analyses over the last decade. However, they were in most cases
restricted to one specific channel. In many cases only parts of a
one-loop calculation has been performed, and no higher-order corrections
in the cMSSM are available so far. 
More specifically, the available literature comprises the following. 
First, \order{\als} corrections to partial decay widths of various squark 
decay channels in the MSSM with real parameters (rMSSM) were derived: 
to the decay of a squark to a quark and a chargino or neutralino in 
\citere{squark_q_chi_als}, to the decay of a squark to a quark and a 
gluino in \citeres{squark_q_gl_als,stop_top_gl_als}, to the decay of a 
squark to a squark and a SM gauge boson in \citere{squark_q_V_als}, 
and to the neutral Higgs boson radiation of a scalar top or bottom quark 
in \citere{stopsbot_phi_als}.
(Those \order{\als} corrections have been implemented into the code
{\tt SDECAY}~\cite{sdecay}.)
A tree-level analysis on several $\Stop$ and $\Sbot$ decay modes was
presented in \citere{stopsbot_tree}. In a second step, Yukawa corrections
to the partial decay widths of a scalar quark were evaluated in
\citeres{sbot_top_cha_alt,sbot_stop_Hpm_altb}. 
Finally, full one-loop contributions were derived, for the decay
of a squark to a quark and a chargino or neutralino in 
\citere{squark_q_chi_full}, and for the decay of a scalar fermion 
to a scalar fermion and a gauge boson in \citere{ sferm_f_V_full}. 
One-loop corrections to scalar quark decays in the rMSSM, derived
in a pure \DRbar\ scheme (see below) have been made available in the
program package {\tt SFOLD}~\cite{sfold}.
Also the partial decay width of a $\cp$-even and a $\cp$-odd Higgs boson to 
scalar quarks at the one-loop level is available; see \citeres{hsqsq} and
\cite{A_sq_sq_full}, respectively. A more recent evaluation can be
found in \citere{hsqsq-new}.
Tree-level analyses for the decay of a $\Stop$ or a $\Sbot$ in the cMSSM 
have been published in \citeres{stopsbot_cMSSM,Bartl:2003pd}, and a 
one-loop calculation of a scalar top quark decaying to a bottom quark 
and a chargino in the cMSSM is presented in \citere{Eberl:2009xe},
where an LHC specific analysis can be found in \citere{Kittel:2011sq}.
Finally, results in an effective Lagrangian approach can be found 
in \citere{squark_q_chi_cMSSM_effL}.

Several methods have been discussed in the literature to
extract the complex parameters of the model from experimental
measurements. Branching ratios at a linear collider were analyzed at the
tree-level in \citere{Bartl:2003pd}. 
Triple products of decaying scalar top or bottom  quarks have been
examined in \citere{Bartl:2004jr} (without specifying the production
modes) and in \citeres{Ellis:2008hq,Deppisch:2009nj,Deppisch:2010nc} 
at the LHC. 
Rate asymmetries for decaying top squarks are analyzed in
\citere{Eberl:2009xe}, again without specifying the production modes, 
and especially for the LHC in \citere{Kittel:2011sq}.
Depending on the realized cMSSM parameter space and on some further
assumptions on the LHC performance, it seems to be possible to obtain
limits on, e.g., the phases of $M_1$, $\At$ and $\Ab$ at the LHC. No
corresponding  analysis, to our knowledge, of the phase of $M_3$ has
been performed so far.

In this paper we present for the first time a full one-loop calculation 
for all two-body decay channels of the heavier scalar top in the cMSSM 
(with no generation mixing), taking into account soft and hard QED and 
QCD radiation.
In \refse{sec:renorm} we review the renormalization of  all relevant 
sectors of the cMSSM. Details about the calculation can be found in 
\refse{sec:calc}, and the numerical results for all decay channels are 
presented in \refse{sec:numeval}. The conclusions can be found in 
\refse{sec:conclusions}.
The results will be  implemented into the Fortran code 
{\tt FeynHiggs}~\cite{feynhiggs,mhiggslong,mhiggsAEC,mhcMSSMlong}.


\section{The complex MSSM and its renormalization}
\label{sec:renorm}

All the channels (\ref{ststphi}) -- (\ref{stbcha}) are calculated at the
one-loop level, including hard QED and QCD radiation. This requires the
simultaneous renormalization of several sectors of the cMSSM, including
the colored sector with top and bottom quarks and their scalar partners
as well as the gluon and the gluino, the Higgs and gauge boson sector with 
all the Higgs bosons as well as the $Z$ and the $W$~boson and the
chargino/neutralino sector. 
In the following subsections we briefly review these sectors and their 
renormalization. 
To our knowledge, it is the first time that such a complete 
renormalization of the cMSSM has been performed.


\subsection{The colored sector of the cMSSM}
\label{sec:color}

The colored sector of the cMSSM can be divided into a quark/squark part,
a gluino part and a gluon part. The quark/squark part contains the soft 
SUSY-breaking mass parameters $M_{\sq_L}$, $M_{\tilde{q}_R}$, the trilinear 
couplings $A_q$, the quark masses $m_q$ as well as the quark and the 
squark fields%
\footnote{
  It should be noted that for the renormalization of the quark/squark 
  sector we focus on the third generation -- which is the relevant part 
  for our calculation -- but, in principle, it can be generalized to the 
  other generations.
}
$q$ and $\sq$, while the gluino part comprises the soft SUSY-breaking 
gaugino mass parameter $M_3$ and the gluino field $\gl$. 
From the gluon part, only the renormalization of the strong coupling 
constant $\als$ is needed for our calculation.


\subsubsection{The top and bottom quark/squark sector}
\label{sec:topbottom}

The part of the Fourier transformed Lagrangian that is bilinear in the
quark and the squark fields with $q = \{t, b\}$ and 
$\sq = \{\Stop,\Sbot\}$ can be written as 
\begin{align}
\cL_{q/\sq}^{\text{bil.}} &= 
\begin{pmatrix}{{\sq}_{L}}^{\dagger}, 
               {{\sq}_{R}}^{\dagger} \end{pmatrix}
(p^2 \id - 
\matr{M}_{\sq})\begin{pmatrix}{\sq}_{L}\\
                                  {\sq}_{R}\end{pmatrix} +
\bar{q} (\pslash - m_q) \OM q + \bar{q} (\pslash - m_q) \OP q~,
\end{align}
where $\om_{\pm} = \edz(\id \pm \gamma_5)$ are the right- and left-handed 
projectors, respectively. $m_q$ with $q = \{t,b\}$ is the 
$\{$top, bottom$\}$ quark mass and the stop and sbottom mass matrices, 
$\matr{M}_{\tilde{t}}$ and $\matr{M}_{\tilde{b}}$, are given by 
\begin{align}\label{Sfermionmassenmatrix}
\matr{M}_{\sq} &= \begin{pmatrix} 
M_{\sq_L}^2 + m_q^2 + M_Z^2 c_{2 \beta} (I_q^3 - Q_q \sw^2) & 
 m_q \Xq^* \\[.2em]
 m_q \Xq &
 M_{\sq_R}^2 + m_q^2 +M_Z^2 c_{2 \beta} Q_q \sw^2
\end{pmatrix}
\end{align}
with
\begin{align}\label{kappa}
\Xq &= \Aq - \mu^*\kappa~, \qquad \kappa = \{\cot\beta, \tan\beta\} 
        \quad {\rm for} \quad q = \{t, b\}, 
        \quad c_{2 \be} \equiv \CZb~.
\end{align}
The soft SUSY-breaking mass parameter $M_{\sq_L}$ is equal for all
members of an $SU(2)_L$ doublet, while the soft SUSY-breaking mass
parameter $M_{\sq_R}$ can be different for scalar top and scalar bottom 
type quarks.
$Q_q$ and $I_q^3$ denote the charge and isospin of $q$.
$A_q$ is the trilinear soft-breaking parameter, $\mu$ the Higgs
superfield mixing parameter, $\tb \equiv v_2/v_1$ denotes the ratio
of the two vacuum expectation values in the Higgs sector (see
\refse{sec:higgs}), $\MZ$ and $\MW$ are the $Z$~and $W$~boson
mass, respectively, and $\cw \equiv \cos \theta_\mathrm{w} = \MW/\MZ$ 
with $\theta_\mathrm{w}$ being the weak mixing angle, and 
$\sw = \sqrt{1 - \cw^2}$.
The mass matrix can be diagonalized with the help of a unitary
 transformation ${\matr{U}}_{\sq}$,
\begin{align}\label{transformationkompl}
\matr{D}_{\sq} &= 
\matr{U}_{\sq}\, \matr{M}_{\sq} \, 
{\matr{U}}_{\sq}^\dagger = 
\begin{pmatrix} \msqe^2 & 0 \\ 0 & \msqz^2 \end{pmatrix}~, \qquad
{\matr{U}}_{\sq}= 
\begin{pmatrix} U_{\sq_{11}}  & U_{\sq_{12}} \\  
                U_{\sq_{21}} & U_{\sq_{22}}  \end{pmatrix}~,
\end{align}
where the scalar quark masses, $\msqe$, $\msqz$, will always be mass
ordered%
\footnote{
  Because of the mass ordering, $\Sbotz \approx \SbotL$ is 
  possible, which should be remembered when choosing a set of
  independent parameters - in our numerical examples, however, $\Sbotz$
  is rather $\SbotR$ like.
}
i.e.\ $\msqe \le \msqz$, and are given by
\begin{align}
m_{\sq_{1,2}}^2 &= \edz \KL M_{\sq_L}^2 + M_{\sq_R}^2 \KR
       + m_q^2 + \edz I_q^3 \MZ^2 c_{2\be} \non \\
&\quad \mp \frac{1}{2} \sqrt{\KKL M_{\sq_L}^2 - M_{{\sq_R}}^2
       + \MZ^2 c_{2\be} (I_q^3 - 2 Q_q \sw^2) \KKR^2 + 4 m_q^2 |\Xq|^2}~.
\label{MSbot}
\end{align}

\smallskip
For the parameter and the field renormalization of the quark/squark
sector we follow the procedure described in \citere{SbotRen}. The quark
mass and the 
squark mass matrix are replaced by the renormalized mass and mass matrix,
respectively, and their counterterms, 
\begin{align}
m_q &\to m_q + \de m_q~,\\
\matr{M}_{\sq} &\to \matr{M}_{\sq} + \de\matr{M}_{\sq}~. 
\end{align}
The mass matrix counterterm $\de\matr{M}_{\sq}$ is obtained by
applying the renormalization procedure  -- replacement of the parameters
by the renormalized ones and the corresponding counterterms -- for each
parameter and expanding with respect to the introduced counterterms,
\begin{align}
\label{proc1a}
\de\matr{M}_{\sq_{11}} &= \de M_{\sq_L}^2 + 2 m_q \de m_q 
- M_Z^2 c_{2 \beta}\, Q_q \, \de \sw^2 + (I_q^3 - Q_q \sw^2) 
  ( c_{2 \beta}\, \de M_Z^2 + M_Z^2\, \de c_{2\beta})~, \\\label{proc1b}
\de\matr{M}_{\sq_{12}} &= (\Aq^*  - \mu \kappa)\, \de m_q 
+ m_q (\de \Aq^* - \mu\, \de \kappa - \kappa \, \de \mu)~, \\\label{proc1c}
\de\matr{M}_{\sq_{21}} &=\de\matr{M}_{\sq_{12}}^*~, \\\label{proc1d}
\de\matr{M}_{\sq_{22}} &= \de M_{\sq_R}^2 
+ 2 m_q \de m_q +  M_Z^2 c_{2 \beta}\, Q_q \, \de \sw^2
+ Q_q \sw^2 ( c_{2 \beta}\, \de M_Z^2+ M_Z^2\, \de c_{2 \beta})
\end{align}
with $\kappa$ given in \refeq{kappa}.

Instead of starting out with the squark mass matrix
in \refeq{Sfermionmassenmatrix} the mass matrix in terms of the squark
masses as given in \refeq{transformationkompl} can be used:
\begin{align} \label{proc2}
\matr{U}_{\sq}\, \matr{M}_{\sq} \, 
{\matr{U}}_{\sq}^\dagger &\to\matr{U}_{\sq}\, \matr{M}_{\sq} \, 
{\matr{U}}_{\sq}^\dagger + \matr{U}_{\sq}\, \de \matr{M}_{\sq} \, 
{\matr{U}}_{\sq}^\dagger =
\begin{pmatrix} \msqe^2 & Y_q \\ Y_q^* & \msqz^2 \end{pmatrix} +
\begin{pmatrix}
\de \msqe^2 & \de Y_q \\ \de Y_q^* & \de \msqz^2
\end{pmatrix}
\end{align}
where $\de \msqe^2$ and  $\de \msqz^2$ are the counterterms 
of the squark masses squared. In the mass matrix in
\refeq{transformationkompl}  as well as in the first term of the right
hand side of \refeq{proc2} the squark mixing parameter $Y_q$ vanishes as
it should at tree-level because the unitary matrix
$\matr{U}_{\sq}$ is chosen in that way to diagonalize the mass matrix
$\matr{M}_{\sq}$. However, already at one-loop level, the squark mixing
parameter $Y_q$ receives a nonvanishing counterterm $\de Y_q$ (which can
be related to the counterterms of a mixing angle and a phase; see
\citere{mhcMSSM2L}). Using \refeq{proc2} 
one can express $\de\matr{M}_{\sq}$ by the counterterms $\de \msqe^2$,
$\de \msqz^2$, and $\de Y_q$. Especially for $\de\matr{M}_{\sq_{11}}$ 
and $\de\matr{M}_{\sq_{12}}$ this yields
\begin{align}\label{proc2a}
\de\matr{M}_{{\sq}_{11}} &=
|U^*_{\sq_{11}}|^2 \de \msqe^2 + |U_{\sq_{12}}|^2\de \msqz^2 -
 U_{\sq_{22}} U^*_{\sq_{12}} \de Y_q - U_{\sq_{12}}
U^*_{\sq_{22}} \de Y_q^* \\\label{dMsq12physpar}
\de\matr{M}_{{\sq}_{12}} &=
U^*_{\sq_{11}} U_{\sq_{12}}
(\de \msqe^2 - \de \msqz^2) +
U^*_{\sq_{11}} U_{\sq_{22}} \de Y_q + U_{\sq_{12}}
U^*_{\sq_{21}} \de Y_q^*~.
\end{align}
Equation~(\ref{dMsq12physpar}) can be used together with \refeq{proc1b}
to express the counterterm for the trilinear top coupling $\de \At$ and 
the counterterm for the bottom squark mixing parameter $\de Y_b$ by the 
other counterterms (see \refeqs{deltaAt} and \eqref{dYb_mbDRbar_AbDRbar}). 

For the field renormalization of the quark and the squark fields the
following procedure is applied, 
\begin{align}
\OM q &\to (1 + \edz\dZ{q}^{L})\, \OM q~,\\
\OP q &\to (1 + \edz\dZ{q}^{R})\, \OP q~,\\
\sqi  &\to \KKL \id + \edz \dZm{\sq} \KKR_{ij} \sqj~.
\end{align}
$\dZZm{\sq}_{ij}$ with $i,j =1,2$ are the squark field
renormalization constants and
$\dZ{q}^{L}$ and $\dZ{q}^{R}$ the field renormalization constants for the
left- and right-handed quark fields, respectively.

Following this renormalization procedure yields for the renormalized
squark self-energies  
\begin{align}
\hSi_{\sq_{11}}(p^2) &= \Si_{\sq_{11}}(p^2) 
  + \edz (p^2 - \msqe^2) \KKL \dZm{\sq} + \dZm{\sq}^*\KKR_{11} 
  - \de\msqe^2~, \\
\hSi_{\sq_{12}}(p^2) &= \Si_{\sq_{12}}(p^2)
  + \edz (p^2 - \msqe^2) \bigl[ \dZm{\sq} \bigr]_{12}
  + \edz (p^2 - \msqz^2) \bigl[ \dZm{\sq}^* \bigr]_{21}
  - \de Y_q~, \\
\hSi_{\sq_{21}}(p^2) &= \Si_{\sq_{21}}(p^2)
  + \edz (p^2 - \msqe^2) \bigl[ \dZm{\sq}^* \bigr]_{12}
  + \edz (p^2 - \msqz^2) \bigl[ \dZm{\sq} \bigr]_{21}
  - \de Y_q^*~, \\
\hSi_{\sq_{22}}(p^2) &= \Si_{\sq_{22}}(p^2) 
  + \edz (p^2 - \msqz^2) \KKL \dZm{\sq} + \dZm{\sq}^*\KKR_{22}
  - \de\msqz^2~.
\end{align}
The renormalized quark self-energy, $\hSi_{q}$, can be decomposed
into left/right-handed and scalar left/right-handed parts, 
$\Si_q^{L/R}$ and $\Si_q^{SL/SR}$, respectively,
\begin{align}
\label{decomposition}
\hSi_{q} (p) &= \pslash\, \OM \hSi_q^L (p^2)
                   + \pslash\, \OP \hSi_q^R (p^2)
                   + \OM \hSi_q^{SL} (p^2) 
                   + \OP \hSi_q^{SR} (p^2)~,
\end{align}
where the components are given by
\begin{align}
\hSi_q^{L/R} (p^2) &= \Si_q^{L/R} (p^2) 
   + \frac{1}{2} (\dZ{q}^{L/R} + {\dZ{q}^{L/R}}^*)~, \\
\hSi_q^{SL} (p^2) &=  \Si_q^{SL} (p^2) 
   - \frac{m_q}{2} (\dZ{q}^L + {\dZ{q}^R}^*) - \de m_q~,  \\
\hSi_q^{SR} (p^2) &=  \Si_q^{SR} (p^2) 
   - \frac{m_q}{2} (\dZ{q}^R + {\dZ{q}^L}^*) - \de m_q~. 
\end{align}
It should be noted that 
$\wtre\hSi_q^{SR} (p^2) = (\wtre\hSi_q^{SL} (p^2))^*$ 
holds due to ${\cal CPT}$ invariance.

We now review our choice of renormalization conditions where we follow
the renormalization scheme of Ref.~\cite{SbotRen} with our favored
``$m_b, A_b$ \drbar'' scheme for the bottom quark/squark part.
However, we expand this scheme to include also external bottom quarks  
(which was not investigated in \citere{SbotRen}). 
In this case we deviate from the ``$\mb$, $\Ab$ \DRbar'' scheme and
renormalize the bottom quark mass on-shell; see below. 
The problems found in \citere{SbotRen} with this scheme
do not arise in the processes with external bottom quarks considered in
this paper as no external bottom squarks occur, and the trilinear
coupling $\Ab$ is only needed at leading order in these processes.

The original parameters that we count as parameters of the top and
bottom quark/squark sector are the soft SUSY-breaking mass parameters 
$M_{\sq_L}$, $M_{\Stop_R}$, and $M_{\Sbot_R}$, the complex trilinear 
couplings $\At \equiv |\At| e^{i \phiat}$, and 
$\Ab \equiv |\Ab| e^{i \phiab}$ and the Yukawa couplings $y_t$ and $y_b$ 
that can be chosen to be real (the Cabbibo-Kobayashi-Maskawa (CKM)
matrix is set to unity in our calculation and generation mixing effects 
are neglected). 
Consequently, there are nine parameters to be defined in the top and 
bottom quark/squark sector. Instead of using the original parameters we 
choose the top squark masses $\mste$, $\mstz$ and one bottom squark mass, 
$\msbz$, as well as the quark masses $\mt$ and $\mb$ as independent input 
parameters.%
\footnote{
  It should be noted that in the case $\Sbotz \approx \SbotL$, $\msbz$ 
  cannot be chosen as an independent parameter and the renormalization 
  scheme has to be switched to one with $\msbe$ as an input parameter.
}
Also, in the scalar top quark sector a renormalization condition is
chosen that fixes the counterterm $\de Y_t$ instead of $\de \At$. 
For the parameters of the top quark/squark sector we impose on-shell (OS)
conditions while in the bottom quark/squark sector a mixed \DRbar/OS
scheme is employed:


\begin{itemize}
\item[(i-iii)]
The two top squark masses and the one bottom squark mass are determined
via on-shell conditions,
\begin{align}
\label{dmstrencond} 
\wtre \hat{\Si}_{\Stop_{ii}}(\msti^2) &= 0 \qquad\ (i = 1,2)~, \\
\wtre \hat{\Si}_{\Sbot_{22}}(\msbz^2) &= 0~,
\end{align}
 yielding  
\begin{align}
\label{dmst} 
\de\msti^2 &= 
\wtre \Si_{\Stop_{ii}}(\msti^2) \qquad\ (i = 1,2)~, \\
\de\msbz^2 &= 
\wtre \Si_{\Sbot_{22}}(\msbz^2)~.
\end{align}
$\wtre $ denotes the real part with respect to
contributions from the loop integrals, but leaves the complex
couplings unaffected.%
\footnote{
  It should be noted that we impose later an extra renormalization 
  condition concerning the $\Sbote$-mass to solve infrared problems, 
  see \refeq{dmsbot1OS} below.
}

\item[(iv)] 
The top-quark mass is also defined on-shell, 
\begin{align}\label{mtrencond}
\wtre \hSi_t(p)\, t(p) \big|_{p^2 = m_t^2} &= 0~,
\end{align}
yielding the
one-loop  counterterm $\de \mt$,
\begin{align}
\label{dmt}
\de \mt &= \edz \wtre \KKKL 
    \mt \KKL\Si_t^L (\mt^2) + \Si_t^R (\mt^2) \KKR  
  + \KKL \Si_t^{SL} (\mt^2) + \Si_t^{SR} (\mt^2) \KKR \KKKR~,
\end{align}
referring to the Lorentz decomposition of the self-energy 
$\hSi_t(p)$, see \refeq{decomposition}.

\item[(v)] 
For the bottom quark mass we use two different definitions depending on 
whether in the considered decay channel a bottom quark appears as an external 
particle or not:

\begin{itemize}

\item[(a)] 
In the case that no external bottom quarks are involved 
the bottom quark mass is defined as \DRbar-mass with the corresponding
counterterm, 
\begin{align}
\label{dmb}
\de\mb^{\DRbar} = \edz \wtre \KKKL
  \mb \KKL \Si_b^L (\mb^2) + \Si_b^R (\mb^2) \KKR_{\rm div}
+ \KKL \Si_b^{SL} (\mb^2) + \Si_b^{SR} (\mb^2) \KKR_{\rm div} \KKKR~.
\end{align}

\item[(b)] 
If bottom quarks appear as external particles then we define the
 bottom quark mass on-shell,
\begin{align}\label{mbrencond}
\wtre \hSi_b(p)\, b(p) \big|_{p^2 = m_b^2} &= 0~,
\end{align}
 to ensure the on-shell properties of the external
 particles which yields the following counterterm:
\begin{align}
\label{dmbOS}
\de\mb^{\OS} &= \edz \wtre \KKKL
  \mb \KKL \Si_b^L (\mb^2) + \Si_b^R (\mb^2) \KKR
    + \KKL \Si_b^{SL} (\mb^2) + \Si_b^{SR} (\mb^2) \KKR \KKKR~.
\end{align}
To have consistent input in all the decay channels we calculate the on-shell
bottom quark mass $\mb^{\OS}$ starting from the \DRbar-mass,
\begin{align}
\label{eq:mbcorr}
\mb^{\OS} &= \mb^{\DRbar} + \de\mb^{\DRbar} - \de\mb^{\OS}~.
\end{align}
The value of $\mb^{\DRbar}$ is obtained as described in \refeq{eq:mbDR}. 
\end{itemize}

It should be noted that the problems found in \citere{SbotRen} with an
on-shell renormalization condition for $\mb$ (leading, e.g., to 
unphysically large contributions to $\de\Ab$) do not occur as long as no
external scalar bottom quarks appear at the same time
and the parameter $\Ab$ is only needed at leading order. 
For example, the proposed scheme would presumably
fail in the process $b \bar b \to \Sboti \aSbotj$, which, 
however, is beyond the scope of our paper.

\item[(vi,vii)] 
The complex counterterm of the non-diagonal entry of \refeq{proc2}, 
which corresponds to two separate conditions, 
is fixed as~\cite{mhcMSSM2L,dissHR,SbotRen}
\begin{align}
\de Y_t =  \edz \wtre 
    \KKKL \Si_{\Stop_{12}}(\mste^2) + \Si_{\Stop_{12}}(\mstz^2) \KKKR~. 
\end{align}

\item[(viii,ix)] 
In the scalar bottom quark sector the trilinear coupling
is defined as a \DRbar\ parameter with the counterterm,
\begin{align}
\de\Ab &= \frac{1}{\mb} \Bigl[ U_{\Sbot_{11}} U_{\Sbot_{12}}^*
         \KL  \wtre\Si_{\Sbot_{11}}(\msbe^2)\ddiv 
             -\wtre\Si_{\Sbot_{22}}(\msbz^2)\ddiv \KR  \non \\ 
&\quad  + \edz\, U_{\Sbot_{12}}^* U_{\Sbot_{21}} 
          \KL \wtre\Si_{\Sbot_{12}}(\msbe^2)\ddiv
             +\wtre\Si_{\Sbot_{12}}(\msbz^2)\ddiv \KR  \non \\
&\quad  + \edz\, U_{\Sbot_{11}} U_{\Sbot_{22}}^* 
          \KL \wtre\Si_{\Sbot_{12}}(\msbe^2)\ddiv
             +\wtre\Si_{\Sbot_{12}}(\msbz^2)\ddiv \KR^*  \non \\
&\quad - \edz(\Ab - \mu^* \tb)\, 
              \wtre \bigl\{
          \mb \KKL \Si_b^L (\mb^2) + \Si_b^R (\mb^2) \KKR_{\rm div} \non \\
&\qquad + \KKL \Si_b^{SL} (\mb^2) + \Si_b^{SR} (\mb^2) \KKR_{\rm div}
              \bigr\} \Bigr]
        + \de\mu^*\ddiv \tb + \mu^*\, \dtanb~,
\end{align}
which also counts for two separate renormalization conditions as $\Ab$
is a complex parameter. The divergent parts of $\de\mu$ and $\dtanb$ can be
extracted from the \refeq{deltamu} and \refeq{deltatanb}, respectively.
\end{itemize}
With these renormalization conditions all independent parameters in the top 
and bottom quark/squark sector are defined. The dependent parameters can be
expressed in terms of those independent ones and the same applies for the
corresponding counterterms. With respect to this renormalization scheme the
(one-loop corrected) on-shell $\Sbote$ mass, $\msbe^{\OS}$, 
differs from the mass parameter $\msbe$. 
As an external particle $\Sbote$ should fulfill the on-shell 
properties, which in turn requires that it should have the mass

\begin{align}
\label{msbotOS}
\big(\msbe^{\OS}\big)^2 = \big(\msbe^{}\big)^2 
                          + \big(\de\msbe^{\text{dep.}}\big)^2
                          - \wtre\Sigma_{\tilde{b}_{11}}(\msbe^2)
\end{align}
where $\de\msbe^{\text{dep.}}$ is the dependent mass counterterm%
\footnote{
  It can be found, up to \order{\als} in Eq.~(8.43) of \citere{dissHR}.
}
that results from imposing only the renormalization conditions (i) -- (ix).
On the other hand, using $\msbe$ for the internal $\Sbote$~squarks and 
$\msbe^{\OS}$ for the external $\Sbote$~squarks (which would formally be 
correct with respect to the considered loop-order) leads to nonvanishing
infrared (IR) singularities (for details see \citere{SbotRen}). 

To circumvent this problem we impose a further OS renormalization 
condition
\begin{align}
\label{dmsbot1OS}
\de\msbe^2 &= \wtre \Si_{\Sbot_{11}}(\msbe^2)~.
\end{align}
As now all the squark masses within one generation are renormalized as
on-shell an explicit restoration of the $SU(2)$~relation is needed. 
This is performed in requiring that the left-handed (bare) soft
SUSY-breaking mass parameter is the same in the bottom as in the 
top squark sector at the one-loop level,   
\begin{align}
M_{\sq_L}^2(\tilde{b}) + \de M_{\sq_L}^2(\tilde{b})
= M_{\sq_L}^2(\tilde{t}) + \de M_{\sq_L}^2(\tilde{t})~.
\end{align}
More precisely, we define (see also
\citeres{squark_q_V_als,stopsbot_phi_als,dr2lA})
\begin{align}
M_{\sq_L}^2(\tilde{b}) = M_{\sq_L}^2(\tilde{t}) 
  + \de M_{\sq_L}^2(\tilde{t}) - \de M_{\sq_L}^2(\tilde{b})
\label{MSbotshift}
\end{align}
with
\begin{align}
\de M_{\sq_L}^2(\sq) &= |U_{\sq_{11}}|^2 \de\msqe^2
   + |U_{\sq_{12}}|^2 \de\msqz^2
   - U_{\sq_{22}} U_{\sq_{12}}^* \de Y_q
   - U_{\sq_{12}} U_{\sq_{22}}^* \de Y_q^* - 2 \mq \de\mq \non \\
&\quad  + \MZ^2\, c_{2\be}\, Q_q\, \de \sw^2 
        - (I_q^3 - Q_q \sw^2) (c_{2\be}\, \de \MZ^2 + \MZ^2\, \de c_{2\be})~.
\label{MSbotshift-detail}
\end{align}
where $\de M_{\sq_L}^2(\sq)$ is derived with the help of \refeqs{proc1a} 
and (\ref{proc2a}). 
Now $M_{\sq_L}^2(\Sbot)$ is used in the scalar bottom mass matrix instead of 
the parameter $M_{\sq_L}^2$ in \refeq{Sfermionmassenmatrix} when calculating 
the values of $\msbe$ and $\msbz$. 
However, with this procedure, also the mass of the $\Sbotz$ squark is 
shifted, which contradicts our choice of independent parameters. 
To keep this choice, also the right-handed soft SUSY-breaking mass parameter 
$M_{\tilde{b}_R}$ receives a shift%
\footnote{
  If the mass of the $\Sbote$ squark is chosen as independent mass as 
  $\Sbotz \approx \tilde{b}_L$ then the shift of $M_{\tilde{b}_R}$ has to 
  be performed with respect to $\msbe$.
}:
\begin{align}
M_{\tilde{b}_R}^2 = \frac{\mb^2\, |\Ab^* - \mu \tb|^2}
  {M_{\sq_L}^2(\tilde{b}) + \mb^2 
   + \MZ^2\, c_{2\be} (T_b^3 - Q_b \sw^2) - \msbz^2} 
  - \mb^2 - \MZ^2\, c_{2\be}\, Q_b\, \sw^2+ \msbz^2
\label{backshift}
\end{align}
Taking into account this shift in $M_{\tilde{b}_R}$, up to one-loop order%
\footnote{In the case of a pure OS scheme 
  (see e.g.~\cite{hr,mhiggsFDalbals} for the rMSSM) 
  the shifts \refeqs{MSbotshift} and (\ref{backshift}) result in a mass 
  parameter $\msbe$, which is exactly the same as in \refeq{msbotOS}.
  This constitutes an important consistency check of these two 
  different methods.
}%
, the resulting mass parameter $\msbe$ is the same as 
the on-shell mass \refeq{msbotOS}.

In the top and bottom quark/squark sector the counterterm for the trilinear
top coupling $\de \At$ and the counterterm $\de Y_b$ are given as a
combination of the independent parameters that can be derived from the
relation of \refeqs{proc1b} and \eqref{dMsq12physpar},
\begin{align}\label{deltaAt}
\de \At &= \frac{1}{\mt}\bigl[U_{\Stop_{11}} U_{\Stop_{12}}^*
           (\de \mste^2 - \de \mstz^2)
        +  U_{\Stop_{11}} U_{\Stop_{22}}^{*} \de Y_t^*
        + U_{\Stop_{12}}^{*} U_{\Stop_{21}} \de Y_t  
        - (\At - \mu^* \cot\beta)\, \de\mt  \bigr]  \non \\
&\quad  + (\de\mu^* \cot\be - \mu^* \cot^2\be\, \dtanb)~
\end{align}
and 
\begin{align}
\de Y_b &= \frac{1}{|U_{\Sbot_{11}}|^2 - |U_{\Sbot_{12}}|^2} \Big[ 
           U_{\Sbot_{11}} U_{\Sbot_{21}}^* 
          \KL \de\msbe^2 - \de\msbz^2 \KR \non \\
&\quad  + \mb \Big( U_{\Sbot_{11}} U_{\Sbot_{22}}^* 
          \KL \de\Ab^* - \mu\, \dtanb - \tb\, \de\mu \KR 
                  - U_{\Sbot_{12}} U_{\Sbot_{21}}^* 
          \KL \de\Ab - \mu^* \dtanb - \tb\, \de\mu^* \KR \Big) \non \\
&\quad  + \KL U_{\Sbot_{11}} U_{\Sbot_{22}}^* (\Ab^* - \mu \tb)
            - U_{\Sbot_{12}} U_{\Sbot_{21}}^* (\Ab - \mu^* \tb) \KR\, 
          \de\mb^{\DRbar}
\Big]~,
\label{dYb_mbDRbar_AbDRbar}
\end{align} 
where $\dtanb$ and $\de\mu$ will be defined within the Higgs/gauge
sector in \refse{sec:higgs}, \refeq{deltatanb} and the
chargino/neutralino sector in \refse{sec:chaneu}, \refeq{deltamu},
respectively.  

\bigskip
Now, the parameter renormalization for the top and bottom quark/squark sector
is accomplished but the field renormalization still has to be done.
We determine the $Z$~factors of the quark and squark fields in the OS
scheme. In the quark sector we have
\begin{align}
\label{Zquark1}
\wtre \hSi_q(p)\, q(p) \big|_{p^2 = m_q^2} &= 0~, \\
\label{Zquark2}
\lim_{p^2 \to m_q^2} \frac{(\pslash + m_q) \wtre \hSi_q (p)}
                        {p^2 - m_q^2} q(p) &= 0~,
\end{align}
where these two equations determine not only  the quark mass
counterterms (see Eqs.~\eqref{mtrencond} and  \eqref{mbrencond}) and the real
part of the $Z$~factors  but also the difference of the imaginary
parts of the quark $Z$~factors, $\im \dZ{q}^{L} - \im \dZ{q}^{R}$
(analogously to the chargino/neutralino case in \refse{sec:chaneu}). This
leaves us the freedom to impose additionally
\begin{align}
\im \dZ{q}^{L} &= - \im \dZ{q}^{R}~.
\end{align}
With these equations we find
\begin{align}
\label{RedZq}
\re \dZ{q}^{L/R} &= - \wtre \Big\{ \Si_q^{L/R} (m_q^2)  \non \\
&\qquad + m_q^2 \KKL {\Si_q^{L}}'(m_q^2) + {\Si_q^{R}}'(m_q^2) \KKR
             + m_q \KKL {\Si_q^{SL}}'(m_q^2) + {\Si_q^{SR}}'(m_q^2) \KKR
                                 \Big\}~,  \\
\label{ImdZq}
\im \dZ{q}^{L/R} &= \pm \frac{i}{2\, \mq} 
        \wtre \KKKL \Si_q^{SR}(\mq^2) - \Si_q^{SL}(\mq^2) \KKKR
     = \pm \frac{1}{\mq} \im \KKKL \wtre \Si_q^{SL}(\mq^2) \KKKR~.
\end{align}
For the scalar quarks we demand
\begin{align}
\wtre \hSi'_{\sq_{ii}}(p^2)
           \big|_{p^2 = \msqi^2} = 0 \qquad (i = 1,2)~,
\end{align}
\begin{align}\label{residuumSqOS}
\wtre{\hat{\Si}}_{\sq_{12}}(\msqe^2) =
\wtre{\hat{\Si}}_{\sq_{21}}(\msqe^2) = 0~, \qquad
\wtre{\hat{\Si}}_{\sq_{12}}(\msqz^2) =
\wtre{\hat{\Si}}_{\sq_{21}}(\msqz^2) = 0~,
\end{align}
yielding
\begin{align}
\re \dZZm{\sq}_{ii} &= 
- \wtre \Si'_{\sq_{ii}}(p^2)\big|_{p^2 = \msqi^2} \qquad (i = 1,2)~, \\
\dZZm{\sq}_{12} &= 
+ 2 \frac{\wtre\Si_{\sq_{12}}(\msqz^2) - \de Y_q}
                       {(\msqe^2 - \msqz^2)}~, \non \\
\dZZm{\sq}_{21} &= 
- 2 \frac{\wtre\Si_{\sq_{21}}(\msqe^2) - \de Y_q^*} 
                       {(\msqe^2 - \msqz^2)}~.
\label{dZstopoffdiagOS}
\end{align}
with $\Si'(p^2) \equiv \frac{\partial \Si(p^2)}{\partial p^2}$,
$q = \{t, b\}$, and $\sq = \{\tilde{t}, \tilde{b}\}$. 
It should be noted that the on-shell
conditions leave the imaginary part of $\dZZm{\sq}_{ii}$ undefined; 
it can be (implicitly) set to zero as it does not contain any 
divergences,
\begin{align}
\im \dZZm{\sq}_{ii} &= 0 \qquad (i = 1,2)~.
\end{align}

\bigskip

The input parameters in the $b/\Sbot$ sector have to correspond to the
chosen renormalization. We start by defining the bottom mass, where the
experimental input is the SM \MSbar\ mass \cite{pdg},
\begin{align}
\label{def:mbMB}
\mb^{\MSbar}(\mb) & = 4.2 \gev~.
\end{align}
The value of $\mb^{\MSbar}(\mu_R)$ (at the renormalization scale 
$\mu_R = \mstz$) is calculated from $\mb^{\MSbar}(\mb)$ at the three-loop
level following the prescription given in~\citere{RunDec}.

The ``on-shell'' mass is connected to the \MSbar\ mass via
\begin{align}
\mb^{\os} &= \mb^{\MSbar}(\mu_R) \; 
   \KKL 1 + \frac{\als^{\MSbar}(\mu_R)}{\pi} 
   \KL \frac{4}{3} + 2\, \ln \frac{\mu_R}{\mb^{\MSbar}(\mu_R)} \KR 
   \KKR~.
\end{align}
The $\DRbar$ bottom quark mass at the scale $\mu_R$ is 
calculated iteratively from \cite{deltab2,dissHR,mhiggsFDalbals}
\begin{align}
\label{eq:mbDR}
\mb^{\DRbar} &= \frac{\mb^{\os} |1 + \db| + \de\mb^{\OS} - \de\mb^{\DRbar}}
            {|1 + \db|}
\end{align}
with an accuracy of 
$|1 - (\mb^{\DRbar})^{(n)}/(\mb^{\DRbar})^{(n-1)}| < 10^{-5}$
reached in the $n$th step of the iteration
where $\de\mb^{\DRbar}$ and $\de\mb^{\OS}$ are given in
\refeqs{dmb} and (\ref{dmbOS}), respectively.

The quantity $\db$~\cite{deltab2,deltab1,deltabc} resums the 
\order{(\als\tb)^n} and \order{(\alt\tb)^n} terms and is given by 
\begin{align}
\db &= \frac{2\als(\mt)}{3\pi} \, \tb \, M_3^* \, \mu^* \,
                          I(\msbe^2, \msbz^2, \mgl^2) \;
     + \frac{\alt(\mt)}{4\pi} \, \tb \, \At^* \, \mu^* \, 
                          I(\mste^2, \mstz^2, |\mu|^2)
\end{align}
with
\begin{align}
I(a, b, c) &= - \frac{a b \ln(b/a) + a c \ln(a/c) + b c \ln(c/b)}
                 {(a - c) (c - b) (b - a)}~.
\end{align}
Here $\alt$ is defined in terms of the top Yukawa coupling 
$y_t(\mt) = \sqrt{2} \mt(\mt)/v$ as
$\alt(\mt) = y_t^2(\mt)/(4\pi)$ with 
$v = 1/\sqrt{\sqrt{2}\, G_F} = 246.218 \gev$
and 
$\mt(\mt)\approx \mt/(1-\frac{1}{2\,\pi} \alt(\mt) 
+\frac{4}{3\,\pi}\als(\mt))$.
Setting in the evaluation of $\db$ the scale to $\mt$ was shown to
yield in general a more stable result~\cite{db2l} as long as 
two-loop corrections to $\db$ are not included%
\footnote{
  It should be noted that in Ref.~\cite{db2l} a different scale has been 
  advocated due to the emphasis on the two-loop contributions presented 
  in this paper. The plots, however, show that $\mt$ is a good scale 
  choice if only one-loop corrections are included.}.
$M_3$ is the soft SUSY-breaking parameter for the gluinos; see below. 
We have neglected any CKM mixing of the quarks.


\subsubsection{The gluino sector}
\label{sec:gluino}

The gluinos appear as external particles, for instance, in the decay 
$\Stopz \to t \gl$. Therefore, a renormalization procedure for the
gluino field and the corresponding parameters is necessary.

The Fourier transformed Lagrangian bilinear in the gluino fields is given by 
\begin{align}
\cL^{\text{bil.}}_{\gl_{\text{org}}}
 = \overline{\gl}^a_{\text{org}} \pslash\, \OM \, \gl^a_{\text{org}} +
 \overline{\gl}^a_{\text{org}} \pslash\, \OP \,\gl_{\text{org}}^a - 
\overline{\gl}_{\text{org}}^a \, M_3 \, \OM \, \gl_{\text{org}}^a
- \overline{\gl}^a_{\text{org}} \, M_3^* \, \OP \, \gl_{\text{org}}^a
\end{align}
with $M_3$ being the soft-breaking gluino mass parameter, which is in 
general complex, 
\BE
M_3 = |M_3| e^{i \phigl}~.
\end{equation}
The gluino field $\gl_{\text{org}}^a$ can be redefined using a phase
  transformation 
\begin{align}
\om_{\pm} \gl^a = e^{\mp i \frac{\phigl}{2}} \om_{\pm} \gl_{\text{org}}^a
\end{align}
such that the gluino phase $\phigl$ appears only in the gluino 
couplings, but not in the mass term with the gluino mass $\mgl = |M_3|$.

The renormalization is performed as follows~\cite{dissTF}:
\begin{align}
M_3 &\to M_3 + \de M_3\; = \; M_3 + \de |M_3|e^{i \phigl} 
                                 + i M_3 \de \phigl~, \non \\
\OM \gl^a &\to \KL 1 + \tfrac{1}{2} \dZ{\gl} \KR \OM \gl^a~, \non \\
\OP \gl^a &\to \KL 1 + \tfrac{1}{2} \dZ{\gl}^* \KR \OP \gl^a~.
\end{align}
Note that, analogous to the mixing matrix, only the gluino phase
appearing in $M_3$ is renormalized but not the one appearing due to the
redefinition of the gluino field.

The renormalized gluino self-energies read
\begin{align}
\hSi_{\gl}^{L/R}(p^2) &= \Si_{\gl}^{L/R}(p^2) 
                       + \tfrac{1}{2} (\dZ{\gl} + \dZ{\gl}^*)~, \\
\hSi_{\gl}^{SL}(p^2) &= \Si_{\gl}^{SL}(p^2) 
                       - \mgl\, \dZ{\gl} - \de M_3 e^{-i \phigl}~, \\
\hSi_{\gl}^{SR}(p^2) &= \Si_{\gl}^{SR}(p^2) 
                       - \mgl\, \dZ{\gl}^* - \de M_3^* e^{i \phigl}~.
\end{align}
We choose OS renormalization conditions for the gluino,
\begin{align}
\wtre \hSi_{\gl} (p) \gl^a(p)\big|_{p^2 = \mgl^2} &= 0 \\
\lim_{p^2 \to \mgl^2} \frac{(\pslash + \mgl)\hSi_{\gl}(p)}
                         {p^2 - \mgl^2} \gl^a(p) &= 0 
\end{align}
with $\hSi(p)$ defined according to \refeq{decomposition}. 
Because of the Majorana nature of the gluino this leads to three 
independent conditions, yielding -- with 
$\Si'(m^2) \equiv \frac{\partial \Si(p^2)}{\partial p^2}\big|_{p^2 = m^2}$ 
--
\begin{align}
\de |M_3| &= \frac{1}{2} \wtre \KKKL
    \mgl \KKL \Si_{\gl}^L (\mgl^2) + \Si_{\gl}^R (\mgl^2) \KKR
    + \KKL \Si_{\gl}^{SL} (\mgl^2) + \Si_{\gl}^{SR} (\mgl^2) \KKR \KKKR~, \\
\re \dZ{\gl} &= - \wtre \Big\{ \Si_{\gl}^L (\mgl^2) 
    + \mgl^2 \KKL {\Si_{\gl}^{L}}'(\mgl^2) + {\Si_{\gl}^{R}}'(\mgl^2) \KKR
    + \mgl \KKL {\Si_{\gl}^{SL}}'(\mgl^2) + {\Si_{\gl}^{SR}}'(\mgl^2) \KKR
                             \Big\}~,  \\
 \im \dZ{\gl} &= \frac{i}{2 \mgl}  
    \wtre \KKKL \Si_{\gl}^{SR}(\mgl^2) - \Si_{\gl}^{SL}(\mgl^2) \KKKR
    - \de \phigl~. 
\end{align}
We have then chosen $\de \phigl = 0$, which is similar to the quark case:
There, we use a real Yukawa coupling due to the possibility of redefining 
the quark fields and have a complex $Z$ factor at one-loop order that 
keeps the Yukawa coupling real also at one-loop order; 
in contrast the gluino phase still appears in the Lagrangian after the 
redefinition of the fields but this phase factor can be considered as a
``transformation matrix'' and does not obtain a counterterm. Note that in the
chargino/neutralino sector we keep the diagonal $Z$~factors real and have
complex parameters.


\subsubsection{The strong coupling constant}
\label{sec:alphas}

The strong coupling constant is renormalized as
\begin{align}
\als &\to Z_{\als} \als = (1 + \dZ{\als}) \als~.
\end{align}
For $Z_{\als}$ we are using a \DRbar\ renormalization condition yielding
\begin{align}
\dZ{\als} &= - \edz \Si^{\trans\prime}_{GG}(p^2)
               \big|_{p^2 = 0, {\rm div}}~,
\end{align}
where $\Si^{\trans\prime}_{GG}$ denotes the derivative of the transverse 
part of the gluon self-energy.  

The decoupling of the heavy particles and the running is taken into 
account in the definition of $\als$: starting point is~\cite{pdg}
\begin{align}
\als^{\MSbar,(5)}(\MZ) &= 0.1176~,
\end{align}
where the running of $\als^{\MSbar,(n_f)}(\mu_R)$ can be found in \citere{pdg}.
From the \MSbar\ value the \DRbar\ value is obtained at the two-loop 
level via~\cite{alsDRbar}
\begin{align}
\als^{\DRbar,(n_f)}(\mu_R) &= \als^{\MSbar,(n_f)}(\mu_R)\, \KKL 1 + 
      \frac{\als^{\MSbar,(n_f)}(\mu_R)}{4\,\pi} +
      \KL \frac{\als^{\MSbar,(n_f)}(\mu_R)}{\pi} \KR^2 
      \KL \frac{11}{8} - \frac{n_f}{12} \KR  \KKR~.
\end{align}
For $\mu_R > \mt$ we have $n_f = 6$.
Within the MSSM, at one-loop level, $\als$ reads 
\begin{align}
\als^{\rm MSSM}(\mu_R) &= \als^{\DRbar,(6)}(\mu_R) \, \KKL 1 + 
               \frac{\als^{\DRbar,(6)}(\mu_R)}{\pi}
               \KL \ln \frac{\mu_R}{\mgl} + 
                   \ln \frac{\mu_R}{M_{\sq}} \KR \KKR~,
\end{align}
with $M_{\sq}$ being defined as the geometric average of all 
squark masses, $M_{\sq} = \Pi_{\sq}(\msqe\msqz)^{\frac{1}{12}}$.
The log~terms originates from the decoupling of the SQCD particles 
from the running of $\als$ at lower scales 
$\mu_R \leq \mu_{\rm dec.} = \mstz$.
For simplification we have chosen $\mstz$ 
(representing the energy scale of the considered decays and as a
typical SUSY scale) also as decoupling scale.


\subsection{The Higgs and gauge boson sector of the cMSSM}
\label{sec:higgs}

The MSSM Higgs potential $\VHiggs$,
\begin{align}
\label{eq:higgspotential}
  \VHiggs &= m_1^2 H_{1i}^* H_{1i} + m_2^2 H_{2i}^* H_{2i} 
- \epsilon^{ij} 
(m_{12}^2 H_{1i} H_{2j} + {m_{12}^2}^* H_{1i}^*
  H_{2j}^*) \notag \\[.4em]
  &\quad +\tfrac{1}{8}(g_1^2+g_2^2)(H_{1i}^* H_{1i} - H_{2i}^* H_{2i})^2 
    + \tfrac{1}{2} g_2^2 | H_{1i}^* H_{2i} |^2~ 
\end{align}
with $\{i,j\}=\{1,2\}$ and $\epsilon^{12}= 1$, contains both the $U(1)$
and $SU(2)_L$ gauge coupling constants $g_1$ and $g_2$, respectively, which 
are considered to be part of the gauge boson sector as well as the soft 
SUSY-breaking parameters $m_{12}$, $\tilde m_1^2$, and $\tilde m_2^2$ (with 
$m_1^2 \equiv \tilde m_1^2 +|\mu|^2$, $m_2^2 \equiv \tilde m_2^2 + |\mu|^2$), 
which are part of the Higgs sector. For this reason we do not separate those 
two sectors but treat them within one section. The $H_{ij}$ with 
$\{i,j\}=\{1,2\}$ are the components of the two Higgs doublets that
can be decomposed in the following way:
\begin{align}
\label{eq:higgsdoublets}
\cHe = \begin{pmatrix} H_{11} \\ H_{12} \end{pmatrix} &=
\begin{pmatrix} v_1 + \tfrac{1}{\sqrt{2}} (\phi_1-i \chi_1) \\
  -\phi^-_1 \end{pmatrix}, \notag \\ 
\cHz = \begin{pmatrix} H_{21} \\ H_{22} \end{pmatrix} &= e^{i \xi}
\begin{pmatrix} \phi^+_2 \\ v_2 + \tfrac{1}{\sqrt{2}} (\phi_2+i
  \chi_2) \end{pmatrix}. 
\end{align}
Besides the vacuum expectation values $v_1$ and $v_2$, in 
\refeq{eq:higgsdoublets} a possible new phase $\xi$ between the two
Higgs doublets is introduced.

In total, the Higgs and gauge boson sector contains 7 real parameters: 
$g_1$, $g_2$, $\tilde m_1^2$, $\tilde m_2 ^2$, $v_1$, $v_2$, and $\xi$ 
and one complex one $m_{12}$. 
$\mu$ is defined within the chargino/neutralino sector; see \refse{sec:chaneu}. 
With the help of a Peccei-Quinn transformation~\cite{Peccei} $\mu$ and 
$m_{12}^2$ can be redefined~\cite{MSSMcomplphasen} such that the complex 
phase of $m_{12}^2$ vanishes.

The part of the Fourier transformed Lagrangian that is linear or bilinear in
the massive gauge boson and Higgs boson fields is the following%
\footnote{
  Corresponding to the convention used in \fa/\fc, we exchanged in 
  the charged part the positive Higgs fields with the negative ones, 
  which is in contrast to \cite{mhcMSSMlong}. 
  As we keep the definition of the matrix $\matr{M}_{\phi^\pm\phi^\pm}$ 
  used in \cite{mhcMSSMlong} the transposed matrix will appear in the 
  expression for $\matr{M}_{H^\pm G^\pm}^{\rm diag}$; see below.
}: 
\begin{align} 
\cL_{\text{Higgs}}^{\text{gauge}} &= T_{\phi_1}\,
\phi_1 +T_{\phi_2}\, \phi_2 + T_{\chi_1}\, \chi_1 + T_{\chi_2}\, \chi_2 \non \\ 
&\quad + \frac{1}{2} \begin{pmatrix} \phi_1,\phi_2,\chi_1,\chi_2
        \end{pmatrix} (p^2 \id - \matr{M}_{\phi\phi\chi\chi})
\begin{pmatrix} \phi_1 \\ \phi_2 \\ \chi_1 \\ \chi_2 \end{pmatrix} +
\begin{pmatrix} \phi^{+}_1,\phi^{+}_2  \end{pmatrix} 
(p^2 \id - \matr{M}_{\phi^\pm\phi^\pm}^{\top})
\begin{pmatrix} \phi^{-}_1 \\ \phi^{-}_2  \end{pmatrix} \non \\ 
&\quad - \frac{1}{2} Z_{\mu} \bigl[(p^2 - M_Z^2) g^{\mu \nu} 
  - (1 - \frac{1}{\xi_Z})
  p^{\mu}p^{\nu}\bigr] Z_{\nu} \non \\ 
&\quad - \frac{g_2}{\sqrt{2} \cw}i p^{\mu}(v_1 \chi_1 + v_2
      \chi_2) Z_{\mu} + i p^{\mu} M_Z Z_{\mu} G
      - \frac{1}{2} \xi_Z M_Z^2 G^2 \non \\ 
&\quad - W^+_{\mu} \bigl[(p^2 - M_W^2) g^{\mu \nu} 
  - (1 - \frac{1}{\xi_W}) p^{\mu}p^{\nu}\bigr] W^-_{\nu} \non \\ 
&\quad + \frac{g_2}{\sqrt{2}} p^{\mu}  
          \bigl[W^+_{\mu}(v_1 \phi_1^- + v_2 \phi_2^-)   
          + (v_1 \phi_1^+ + v_2\phi_2^+) W^-_{\mu}\bigr] \non \\ 
&\quad -  p^{\mu} M_W (W^+_{\mu} G^- + G^+   W^-_{\mu}) 
       - \xi_W M_W^2 G^+G^-~,
\label{L-Higgs}
\end{align}
where we have used the decomposition of \refeq{eq:higgsdoublets}.   
The coefficients of the linear terms are called tadpoles with 
$T_{\phi_i}$, $T_{\chi_i}$, $i =\{1, 2\}$ being the tadpole parameters.
$\matr{M}_{\phi\phi\chi\chi}$ and $\matr{M}_{\phi^\pm\phi^\pm}$ are the Higgs
mass matrices.
The terms containing the neutral and the charged Goldstone boson fields, 
$G$ and $G^\pm$, and the gauge parameters, $\xi_Z$ and $\xi_W$, respectively, 
are coming from the gauge-fixing part of the Lagrangian.
The interaction fields can be transformed into mass eigenstates 
(see \citeres{mhcMSSMlong,dissHR,mhiggsCPXFD1} for details, especially 
on notation) with the help of the real and orthogonal transformation 
matrices $\matr{U}_{\mathrm{n}(0)}$ and $\matr{U}_{\mathrm{c}(0)}$,
\begin{align}
\begin{pmatrix} h \\ H \\ A \\ G  \end{pmatrix} = \matr{U}_{\mathrm{n}(0)} 
\begin{pmatrix} \phi_1 \\ \phi_2 \\ \chi_1 \\ \chi_2 \end{pmatrix} \quad
&\text{with}  \quad
\matr{M}_{hHAG}^{\rm diag} = \matr{U}_{\mathrm{n}(0)}
\matr{M}_{\phi\phi\chi\chi} \matr{U}_{\mathrm{n}(0)}^\dagger~,\\
&\text{and} \quad (T_h, T_H, T_A, T_G) = 
(T_{\phi_1}, T_{\phi_2}, T_{\chi_1},T_{\chi_2}) \matr{U}_{\mathrm{n}(0)}^\dagger
\\[0.2cm]
\begin{pmatrix} H^{-} \\ G^{-}  \end{pmatrix} =
\matr{U}_{\mathrm{c}(0)} \begin{pmatrix} \phi^{-}_1 \\ 
                                       \phi^{-}_2  \end{pmatrix}\quad
&\text{with} \quad \matr{M}_{H^\pm G^\pm}^{\rm diag} =
\matr{U}_{\mathrm{c}(0)}\matr{M}_{\phi^\pm\phi^\pm}^{\top}
\matr{U}_{\mathrm{c}(0)}^\dagger~.
\end{align}

\noindent
where the diagonal elements of $\matr{M}_{hHAG}^{\rm diag}$ 
and $\matr{M}_{H^\pm G^\pm}^{\rm diag}$ are the tree-level masses denoted as
$\mh$, $\mH$, $\mA$, $\mG$ and $\MHp$, $\mGp$,
respectively. It should be noted that the tadpole parameter $T_G$ can be
expressed by $\tb$, the entries of $\matr{U}_{\mathrm{n}(0)}$ and $T_A$.

Throughout our calculation we use the 't~Hooft--Feynman gauge, 
$\xi_Z = \xi_W = 1$.
Concerning the renormalization procedure, we follow the usual approach
where the gauge-fixing terms do not receive a net contribution from the 
renormalization transformations. Accordingly, no counterterms as given
below arise from the gauge-fixing terms. 


We replace the 8 original parameters $v_1$,
$v_2$, $g_1$, $g_2$, $m_{12}^2$, $\tilde{m}_1^2$, $\tilde{m}_2^2$, and
$\xi$ by the $Z$~ and $W$~boson mass $\MZ$, $\MW$, the electric charge
$e$, $\tb$, the mass of the charged Higgs boson $\MHpm$, and the
tadpole parameters $T_h$, $T_H$, and $T_A$ (where we have chosen
$m_{12}^2$ to be real, which is always possible with the help of a
Peccei-Quinn transformation).
Details about this replacement can be found in \citere{mhcMSSMlong}.

The minimization of the Higgs potential in lowest order leads to the
requirement that the tadpole coefficients $T_{\{h,H,A\}}$ in \refeq{L-Higgs}
must vanish (the tadpole coefficient $T_G$ vanishes automatically if 
$T_A = 0$ holds, as $T_G$ can be written in terms of $T_A$). 
In particular, the condition $T_A = 0$ implies that the complex phase $\xi$ 
has to vanish, see e.g.\ \citeres{mhcMSSMlong}, so that the Higgs sector 
in lowest order is $\cp$ conserving. 

In order to derive the counterterms entering the one-loop corrections to the 
Higgs boson masses and effective couplings, we renormalize the parameters 
appearing in the linear and bilinear terms of the Higgs potential%
\footnote{
  It should be noted that in \citere{mhcMSSMlong} a slightly different
  renormalization prescription for $\tb$ had been introduced, 
  $\tb \to \tb (1 + \dtanb^{\!\!\mbox{\tiny \cite{mhcMSSMlong}}})$, 
  such that $\dtanb = \tb\; \dtanb^{\!\!\mbox{\tiny \cite{mhcMSSMlong}}}$.
}%
, %
\begin{align}
  e &\to (1 + \dZ{e}) e~, & \MHp^2 &\to \MHp^2 +
      \de m^2_{H^\pm} \notag \\
  \MZ^2 &\to \MZ^2 + \dMZsq~,  & \tadh &\to \tadh + \dtadh~, 
  \notag \\
  \MW^2 &\to \MW^2 + \dMWsq~,  & \tadH &\to \tadH +
\dtadH~, \notag \\ 
   \tanb &\to \tanb + \dtanb &
  \tadA &\to \tadA + \dtadA~.
\label{eq:PhysParamRenorm}
\end{align}
It is important that according to our renormalization procedure the 
renormalization prescription has to be applied {\em before} the 
transformation into the mass eigenstates, also for $\tb$, i.e.\ a~$\be$ 
appearing from the transformation to the fields $A$, $G$, $H^\pm$, and
$G^\pm$ does not obtain a counterterm.
For the counterterms arising from the mass matrices we use the
definitions~\cite{mhcMSSMlong}
\begin{align}
  \matr{M}_{\phi\phi\chi\chi} &\to \matr{M}_{\phi\phi\chi\chi} +
  \delta \matr{M}_{\phi\phi\chi\chi}~, \\
  \matr{M}_{\phi^\pm\phi^\pm} &\to \matr{M}_{\phi^\pm\phi^\pm} +
  \delta \matr{M}_{\phi^\pm\phi^\pm}~, 
\end{align}
with
\begin{align}
\de \matr{M}_{hHAG} &= \matr{U}_{\mathrm{n}(0)}
\, \de\matr{M}_{\phi\phi\chi\chi} \matr{U}_{\mathrm{n}(0)}^{\dagger} =
  \begin{pmatrix}
    \dmhsq  & \dmhHsq & \dmhAsq & \dmhGsq \\[.4em]
    \dmhHsq & \dmHsq  & \dmHAsq & \dmHGsq \\[.4em]
    \dmhAsq & \dmHAsq & \dmAsq  & \dmAGsq \\[.4em]
    \dmhGsq & \dmHGsq & \dmAGsq & \dmGsq
  \end{pmatrix}~, \\[.4em]
\de\matr{M}_{H^\pm G^\pm} &= \matr{U}_{\mathrm{c}(0)}\, 
\de\matr{M}_{\phi^\pm\phi^\pm}^{\top} \matr{U}_{\mathrm{c}(0)}^{\dagger} =
  \begin{pmatrix}
    \dmHpmsq & \dmGmHpsq \\[.4em]
    \dmHmGpsq & \dmGpmsq 
  \end{pmatrix}~,
\end{align}
where $\de\matr{M}_{\phi\phi\chi\chi}$ and $\de\matr{M}_{\phi^\pm\phi^\pm}$ 
denote the counterterm mass matrices that are obtained when 
replacing the parameters in $\matr{M}_{\phi\phi\chi\chi}$ and
$\matr{M}_{\phi^\pm\phi^\pm}$ by the renormalized ones and their counterterms 
using \refeq{eq:PhysParamRenorm}, expanding and taking the first-order 
expressions and applying the zeroth order relation $T_{\{h,H,A\}} = 0$; 
see \citere{mhcMSSMlong}.

As mentioned above, in contrast to what is often done in the MSSM with
real parameters, we use $\MHpm$ as an independent input parameter. 
The counterterm $\dmAsq$ in the formulas above is therefore a dependent
quantity, which has to be expressed in terms of $\dmHpmsq$ using
\begin{align}
  \dmAsq = \dmHpmsq - \dMWsq~.
\label{massct}
\end{align}

\medskip
For the field renormalization we choose to give each Higgs doublet one
single renormalization constant,
\begin{align}
\label{eq:HiggsDublettFeldren}
  \cHe \to (1 + \tfrac{1}{2} \dZ{\cHe}) \cHe~, \qquad
  \cHz \to (1 + \tfrac{1}{2} \dZ{\cHz}) \cHz~.
\end{align}
In the mass eigenstate basis, the field renormalization matrices read 
\begin{subequations}
\label{eq:higgsfeldren}
\begin{align}
  \begin{pmatrix} h \\[.4em] H \\[.4em] A \\[.4em] G \end{pmatrix} \to
\begin{pmatrix} h \\[.4em] H
    \\[.4em] A \\[.4em] G \end{pmatrix}
  + \frac{1}{2} \begin{pmatrix}
    \dZ{hh} & \dZ{hH} & 
    \dZ{hA} & \dZ{hG} \\[.4em]
     \dZ{hH} &  \dZ{HH} & 
    \dZ{HA} &  \dZ{HG} \\[.4em] 
     \dZ{hA} & \dZ{HA} & 
    \dZ{AA} &  \dZ{AG} \\[.4em] 
    \dZ{hG} &  \dZ{HG} & \dZ{AG}
    & \dZ{GG} 
  \end{pmatrix} \cdot \begin{pmatrix} h \\[.4em] H
    \\[.4em] A \\[.4em] G \end{pmatrix}
\end{align}
and
\begin{align}
\label{eq:dZHmGm}
  \begin{pmatrix} H^- \\[.4em] G^- \end{pmatrix} &\to 
\begin{pmatrix} H^- \\[.4em] G^- \end{pmatrix}
+ \frac{1}{2}
  \begin{pmatrix}
     \dZ{H^-H^+}   &   \dZ{G^- H^+} \\[.4em]
     \dZ{H^- G^+} &    \dZ{G^-G^+}
  \end{pmatrix} \cdot
  \begin{pmatrix} H^- \\[.4em] G^- \end{pmatrix}~,
\end{align}
\end{subequations}
where, according to \refeq{eq:HiggsDublettFeldren}, 
$\dZ{hh}$, \dots, $\dZ{G^-G^+}$ are not independent but can be derived via 
$\matr{U}_{\mathrm{n}(0)}\, 
\text{\bf diag}(\dZ{\cHe}, \dZ{\cHz}, \dZ{\cHe}, \dZ{\cHz})\, 
\matr{U}_{\mathrm{n}(0)}^\dagger$  
and 
$\matr{U}_{\mathrm{n}(0)}\, 
\text{\bf diag}(\dZ{\cHe},\dZ{\cHz})\, \matr{U}_{\mathrm{c}(0)}^\dagger$~,
yielding the following expressions for the field
renormalization constants in \refeq{eq:higgsfeldren}%
\footnote{
  It should be noted that \refeq{dZHpHm} is sufficient to yield
  an UV finite result for decays involving a charged Higgs boson. To
  obtain also an IR finite result and to ensure the on-shell properties 
  of the outgoing charged Higgs boson, another $Z$~factor has to be 
  taken into account; see \refeq{dhZHpHm} below.
}:
\begin{subequations}
\label{eq:FeldrenI_H1H2}
\begin{align}
\dZ{hh} &= \sinasq \dZ{\cHe} + \cosasq \dZ{\cHz}~, \\[.2em]
\dZ{AA} &= \sinbsq \dZ{\cHe} + \cosbsq \dZ{\cHz}~, \\[.2em]
\dZ{hH} &= \sina \cosa (\dZ{\cHz} - \dZ{\cHe})~, \\[.2em]
\dZ{AG} &= \sinb \cosb (\dZ{\cHz} - \dZ{\cHe})~, \\[.2em]
\dZ{HH} &= \cosasq \dZ{\cHe} + \sinasq \dZ{\cHz}~, \\[.2em]
\dZ{GG} &= \cosbsq \dZ{\cHe} + \sinbsq \dZ{\cHz}~, \\[.2em]
\label{dZHpHm}
\dZ{H^- H^+} &= \dZ{AA}
                ~, \\[.2em]
\dZ{H^- G^+} &= \dZ{G^- H^+} = \dZ{AG}
                ~, \\[.2em]
\dZ{G^- G^+} &= \dZ{GG}
                ~.
\end{align}
\end{subequations}
For the field renormalization constants of the $\cp$-violating
self-energies it follows,
\begin{align}
  \dZ{hA} = \dZ{hG} = \dZ{HA} = \dZ{HG} = 0~,
\label{eq:dZzero}
\end{align}
which is related to the fact that the Higgs potential is $\cp$ conserving 
in lowest order.

In the case of a decay to a neutral or charged Higgs boson some
self-energy transitions on the external Higgs leg have to be taken into
account. In order to define the various counterterm contributions, we
list here the respective renormalized self-energies (taking already
\refeq{eq:dZzero} into account). For a scalar-vector self-energy we
use $\Si^\mu_{\rm SV}(p^\mu) = p^\mu \Si_{\rm SV}(p^2)$, where $p^\mu$
is the momentum of the incoming scalar or vector particle. Then we have
the following renormalized self-energies:
\begin{align}
\hSi_{hG}(p^2) &= \Si_{hG}(p^2) 
                 - \de m_{hG}^2~, \\
\hSi_{HG}(p^2) &= \Si_{HG}(p^2) 
                 - \de m_{HG}^2~, \\
\hSi_{AG}(p^2) &= \Si_{AG}(p^2) + \dZ{AG} (p^2 - \edz \mA^2) 
                 - \de m_{AG}^2~, \\
\hSi_{H^-G^+}(p^2) &= \Si_{H^-G^+}(p^2) + \dZ{H^-G^+} (p^2 - \edz \MHp^2) 
                     - \de m_{H^-G^+}^2~, \\
\hSi_{hZ}(p^2) &= \Si_{hZ}(p^2)
~, \label{hZmix} \\
\hSi_{HZ}(p^2) &= \Si_{HZ}(p^2)
~, \label{HZmix}\\
\hSi_{AZ}(p^2) &= \Si_{AZ}(p^2) - \de m_{AZ}^2~, \\
\hSi_{H^-W^+}(p^2) &= \Si_{H^-W^+}(p^2) - \de m_{H^-W^+}^2~.
\end{align}
with the mass counterterms expressed as~\cite{mhcMSSMlong} 
\begin{align}
\de m_{hG}^2 &= \frac{e\, \Cba}{2 \MZ \sw \cw}\, \dtadA~, \\ 
\de m_{HG}^2 &= - \frac{e\, \Sba}{2 \MZ \sw \cw}\, \dtadA~,\\
\de m_{AG}^2 &= \frac{e\, \KL \Sba\, \dtadH - \Cba\, \dtadh \KR}
                   {2 \MZ \sw \cw}
                - \CQb\, (\MHp^2 - \MW^2)\, \dtanb~, \\
\de m_{H^-G^+}^2 &= \frac{e\, \KL \Sba\, \dtadH - \Cba\, \dtadh 
                   + i\, \dtadA \KR}{2 \MZ \sw \cw} 
                   - \CQb\, \MHp^2\, \dtanb~, \\
\de m_{AZ}^2 &= + i \MZ \cos\beta\, [\cos\beta\, \dtanb
      +\edz \sin\beta (\dZ{\cHz} - \dZ{\cHe})]~, \\[.4em]
\de m_{H^-W^+}^2 &= - \MW \cos\beta\, [\cos\beta\, \dtanb
      +\edz \sin\beta (\dZ{\cHz} - \dZ{\cHe})]~.
\end{align}
$\al$ denotes the angle that diagonalizes the $\cp$-even Higgs boson mass
matrix at the tree-level. 
It should be  noted that according to \refeq{L-Higgs}, no mixing of the
$\cp$-even Higgs fields and the $Z$~boson fields occurs at tree-level. 
Consequently, there are no counterterm contributions to this mixing at 
one-loop level; see \refeqs{hZmix} and (\ref{HZmix}).%
\footnote{
  The other renormalized Higgs boson self-energies, 
  $\hSi_{hh}$, $\hSi_{hH}$, $\hSi_{HH}$, $\hSi_{AA}$, $\hSi_{GG}$,
  $\hSi_{hA}$, $\hSi_{HA}$, and $\hSi_{H^-H^+}$ and the corresponding mass
  matrix counter\-term contributions, $\de \mh^2$, $\de \mH^2$, $\de m_{hH}^2$,
  $\de m_G^2$, $\de m_{hA}^2$, $\de m_{HA}^2$,
  are not explicitly needed in our calculation (employing
  {\tt FeynHiggs}, these self-energy contributions are automatically
  taken care of). They are given by Eqs.~(64) and (53) of 
  \citere{mhcMSSMlong}, respectively, considering 
  $\dtanb^{\!\!\mbox{\tiny \cite{mhcMSSMlong}}} = \dtanb/\tb$.
}


\medskip
In the following we list our renormalization conditions and the resulting
counterterms:
\begin{itemize}
\item[(i-iii)] We impose on-shell renormalization conditions
for the masses of the SM gauge bosons and the charged Higgs boson,
\BE
\wtre\hSi^{\trans}_{ZZ}(\MZ^2) = 0~, \qquad 
\wtre\hSi^{\trans}_{WW}(\MW^2) = 0~, \qquad
\wtre\hSi_{H^+H^-}(\MHp^2) = 0~.
\label{eq:rencond1}
\end{equation}
The gauge boson self-energies $\Si^{\trans}$ are the transverse parts of
the full self-energies. \refeq{eq:rencond1} yields for the mass counterterms,
\begin{align}
\label{eq:mass_osdefinition}
\dMZsq = \wtre \se{ZZ}^{\trans}(\MZ^2)~, \quad 
\dMWsq = \wtre \se{WW}^{\trans}(\MW^2)~, \quad 
\dmHpmsq = \wtre \se{H^+ H^-}(\MHp^2)~.
\end{align}
\item[(iv)]
The electric charge is defined via the standard on-shell conditions 
requiring that no corrections occur to the electron-positron-photon
vertex with on-shell external particles at zero photon momentum. 
This yields $\dZ{e}$ expressed through the photon and photon--$Z$
self-energies, 
\begin{align}
\dZ{e} &= \frac{1}{2} \Sip_{\ga\ga}(0)
          +\frac{\sw}{\cw} \frac{\Si_{\ga Z}(0)}{\MZ^2}~.
\end{align}

\item[(v-vii)]
The tadpole parameters are renormalized such that the complete one-loop
tadpole contributions vanish, 
\BE
T_{\{h,H,A\}}^{(1)} + \de T_{\{h,H,A\}} = 0~,
\end{equation}
where $T_{\{h,H,A\}}^{(1)}$ denote the contributions coming from
the genuine one-loop tadpole graphs. These conditions lead to
\begin{align}
  \dtadh = -\tadh^{(1)}~, \qquad \dtadH = -\tadH^{(1)}~, \qquad 
  \dtadA = -\tadA^{(1)}~. 
\end{align}

\item[(viii)]
The last parameter that has to be defined is $\tb$. We do that
together with the Higgs boson field renormalization constants $\dZ{\cHe}$
and $\dZ{\cHz}$.

A convenient choice that avoids large (and unphysically) higher-order
corrections in the (c)MSSM Higgs sector is a \drbar\ renormalization of 
$\dZ{\cHe}$, $\dZ{\cHz}$, and $\dtanb$~\cite{tbren1}, 
\begin{subequations}
\label{eq:deltaZHiggs}
\begin{align}
\label{dZH1}
\dZ{\cHe} &\equiv \dZ{\cHe}^{\DRbar}
          = - \re \Sip_{HH}(0)\big|_{\al = 0, \rm{div}}~, \\
\label{dZH2}
\dZ{\cHz} &\equiv \dZ{\cHz}^{\DRbar} 
          = - \re \Sip_{hh}(0)\big|_{\al = 0, \rm{div}}~, \\
\label{deltatanb}
\dtanb &\equiv \dtanb^{\DRbar}
       = \tfrac{1}{2} \tb \KL \dZ{\cHz} - \dZ{\cHe} \KR~,
\end{align}
\end{subequations}
i.e.\ the counterterms in \refeqs{eq:deltaZHiggs} contribute only via
divergent parts%
\footnote{
  The divergences in \refeqs{dZH1}, (\ref{dZH2}) are momentum 
  independent.
}%
, and the finite result depends on the renormalization scale $\mu_R$.
For the setting of $\mu_R$ see \refse{sec:paraset}.

The \DRbar\ renormalization of the parameter $\tb$, which is manifestly
process indepen\-dent, is convenient since there is no obvious relation of
this parameter to a specific physical observable that would favor a
particular on-shell definition. Furthermore, the
$\DRbar$~renormalization of $\tb$ has been shown to yield stable
numerical results~\cite{mhiggsf1lA,tbren1,tbren2}. This
scheme is also gauge independent at the one-loop level within
the class of $R_\xi$ gauges~\cite{tbren2}.
\end{itemize}

Finally, the field renormalization constants of the gauge bosons have to
be determined. Applying an on-shell condition for the gauge boson
fields, the field renormalization constants can be derived as
\begin{align}
\label{eq:mass_osdef}
\de Z_{ZZ} = -\wtre \Sigma_{ZZ}^{\trans \prime}(\MZ^2)~, \quad 
\de Z_{WW} = -\wtre \Sigma_{WW}^{\trans \prime}(\MW^2)~.
\end{align}

In other sectors, one may need the following counterterms expressed by
independent ones:
\begin{alignat}{2}
\de\sw &= \frac{1}{2} \frac{\cw^2}{\sw} \KL \frac{\de\MZ^2}{\MZ^2} 
                     -\frac{\de\MW^2}{\MW^2} \KR~, \qquad 
&\de\cw &= -\frac{\cw}{2} \KL \frac{\de\MZ^2}{\MZ^2} 
                             -\frac{\de\MW^2}{\MW^2} \KR~,\\
\de\Sbe &= \CDb\, \dtanb~,  &\de\Cb &= -\Sbe\CQb\, \dtanb~.
\end{alignat}

We have checked (at the one-loop level) that the following
Slavnov-Taylor identities~\cite{kwdiss,baro,higgsdecays} hold:
\begin{align}
\hSi_{hG}(p^2) - \frac{i\,p^2}{\MZ} \hSi_{hZ}(p^2) &= 0~, \\
\hSi_{HG}(p^2) - \frac{i\,p^2}{\MZ} \hSi_{HZ}(p^2) &= 0~, \\
\hSi_{AG}(p^2) 
- \frac{i\,p^2}{\MZ} \hSi_{AZ}(p^2) + (p^2 - \mA^2) f_0(p^2) &= 0~, \\
\hSi_{GG}(p^2) - \frac{2\,i\,p^2}{\MZ} \hSi_{GZ}(p^2)
- \frac{p^2}{\MZ^2} \hSi_{ZZ}^L(p^2) &= 0~, \\
\hSi_{H^-G^+}(p^2) 
- \frac{p^2}{\MW} \hSi_{H^-W^+}(p^2) + (p^2 - \MHp^2) f_\pm(p^2) &= 0~, \\
\hSi_{G^-G^+}(p^2) - \frac{2\,p^2}{\MW} \hSi_{G^-W^+}(p^2) 
- \frac{p^2}{\MW^2} \hSi_{WW}^L(p^2) &= 0~,
\end{align}
where $\Si^L$ denotes the longitudinal part of the self-energy and
\begin{align}
f_0(p^2) &= -\frac{\al}{16 \,\pi \,\sw^2 \,\cw^2} \Sba\Cba
  \KKL B_0(p^2, \mh^2, \MZ^2) - B_0(p^2, \mH^2, \MZ^2) \KKR~, \\
f_\pm(p^2) &= -\frac{\al}{16 \,\pi \,\sw^2} \Sba\Cba
  \KKL B_0(p^2, \mh^2, \MW^2) - B_0(p^2, \mH^2, \MW^2) \KKR~.
\end{align}
The definition for the $B_0$ function can be found in \citere{denner}.
The Slavnov-Taylor identities also hold for the unrenormalized self-energies
(where the tadpole contributions must not be neglected).

\medskip
The Higgs boson field renormalization constants are necessary to
render the one-loop calculations of partial decay widths with external
Higgs bosons UV finite. The \DRbar~scheme for the field 
renormalization constants is used in the calculation of the Higgs masses
within {\tt FeynHiggs} in order to avoid the possible occurrence of unphysical
threshold effects. As the results of {\tt FeynHiggs} are used within the
numerical evaluation in Sec.~\ref{sec:numeval}, it is appropriate
and consistent to follow the same renormalization procedure. As always,
Higgs bosons appearing as external particles in a physical process
have to obey proper on-shell conditions. 
A vertex with an external on-shell Higgs boson $h_n$ ($n = 1,2,3$),
$\Ga_{h_n}$, is obtained from the tree-level vertices $\Ga_h$, $\Ga_H$,
and $\Ga_A$ via the complex matrix $\matr{Z}$~\cite{mhcMSSMlong},
\begin{align}
\Ga_{h_n} &= [\matr{Z}]_{n1} \Ga_h +
             [\matr{Z}]_{n2} \Ga_H +
             [\matr{Z}]_{n3} \Ga_A + \ldots ~,
\label{eq:zfactors123}
\end{align}
where the ellipsis represents contributions from the mixing with the
Goldstone boson and the $Z$~boson; see \refse{sec:calc}.
It should be noted that the transformation with $\matr{Z}$ is not a
unitary transformation; see \citere{mhcMSSMlong} for details.

Also the charged Higgs boson appearing as an external particle in a
$\Stop$~decay has to obey the proper on-shell conditions. The
corrections to the charged Higgs boson propagator lead to an extra
$Z$~factor, 
\begin{align}\label{ZHpHm_hut}
\hat Z_{H^-H^+} = 
   \KKL 1 + \re \hSip_{H^-H^+}(p^2)\big|_{p^2 = \MHp^2} \KKR^{-1}~.
\end{align}
Expanding both sides of \refeq{ZHpHm_hut} up to one-loop order and
using $\hat Z_{H^-H^+} = 1 + \de\hat  Z_{H^-H^+}$, leads to
\begin{align}
\label{dhZHpHm}
\de\hat Z_{H^-H^+} = - \re\hSip_{H^-H^+}(p^2)\big|_{p^2 = \MHp^2} = 
- \re\Sip_{H^-H^+}(\MHp^2) - \dZ{H^-H^+}~.
\end{align}
Analogous to the procedure for the neutral Higgs bosons; see
\citere{mhcMSSMlong},  
\begin{align}
\sqrt{\hat Z_{H^-H^+}} = 1 + \frac{1}{2} \de\hat Z_{H^-H^+} 
\end{align}
has to be applied to a process with an external charged Higgs boson. Within
the presented calculations, for the charged Higgs 
bosons, we include contributions from $\sqrt{\hat Z_{H^-H^+}}$
strictly at the one-loop level i.e.\ the correction coming from 
$\edz \de\hat Z_{H^-H^+}$ of \refeq{dhZHpHm} multiplied by the corresponding
tree-level vertex contribution.%
\footnote{
  In our calculational set up we add \refeq{dhZHpHm} to \refeq{dZHpHm}.
}
~As for the neutral Higgs bosons, there are contributions from the 
mixing with the Goldstone boson and the
$W$~boson,
which we deal with separately calculating the mixing explicitly and
strictly at one-loop order. 
The $Z$~factor $\hat Z_{H^-H^+}$ is UV finite by definition. However, it
contains IR divergences that cancel with (IR divergent) soft
photon contributions from the one-loop diagrams; see \refse{sec:calc}.


\subsection{The chargino/neutralino sector of the cMSSM}
\label{sec:chaneu}

The chargino/neutralino sector contains two soft SUSY-breaking gaugino mass
parameters $M_1$ and $M_2$ corresponding to the bino and the wino fields,
respectively, as well as the Higgs superfield mixing parameter $\mu$,
which, in general, can be complex.%
\footnote{
  Often, $M_2$ is chosen to be real, which is possible without loss of 
  generality as not all the possible phases of the MSSM Lagrangian are 
  physical and there is a certain freedom of choice;
  see the discussion in \refse{sec:full1Lphiat}.
}
The gauge boson masses and $\tb$ that also appear in this sector have
already been defined within the context of the Higgs and gauge boson
sector; see \refse{sec:higgs}. For our calculation we also need 
to renormalize the chargino and neutralino fields.

The starting point for the renormalization procedure of the
chargino/neutralino sector is the part of the Fourier transformed MSSM
Lagrangian that is bilinear in the chargino and neutralino fields,
\begin{align}
\cL^{\text{bil.}}_{\cham{},\tilde{\chi}^0} &= 
  \overline{\cham{i}}\, \pslash\, \OM \cham{i} 
+ \overline{\cham{i}}\, \pslash\, \OP \cham{i} 
- \overline{\cham{i}}\, [\matr{V}^* \matr{X}^\top \matr{U}^\dagger]_{ij} \,
  \OM \cham{j} 
- \overline{\cham{i}}\, [\matr{U} \matr{X}^* \matr{V}^{\top}]_{ij} \,
  \OP \cham{j} \non \\
&+ \frac{1}{2} \KL
  \overline{\neu{k}}\, \pslash\, \OM \neu{k} 
+ \overline{\neu{k}}\, \pslash\, \OP \neu{k} 
- \overline{\neu{k}}\, [\matr{N}^*\matr{Y} \matr{N}^\dagger]_{kl} \,
  \OM \neu{l} 
- \overline{\neu{k}}\, [\matr{N} \matr{Y}^* \matr{N}^{\top}]_{kl} \,
  \OP \neu{l} \KR~, 
\end{align}
already expressed in terms of the chargino and neutralino mass eigenstates
$\cham{i}$ and $\neu{k}$, respectively, 
and $i,j = 1,2$ and $k,l = 1,2,3,4$.
The mass eigenstates can be determined via unitary 
transformations where the corresponding matrices diagonalize the chargino and
neutralino mass matrix, $\matr{X}$ and $\matr{Y}$, respectively. 

In the chargino case, two $2 \times 2$ matrices $\matr{U}$ and
$\matr{V}$ are necessary for the diagonalization of the chargino mass
matrix%
\footnote{
  Corresponding to the convention used in \fa/\fc, we express the chargino
  part in terms of negative chargino fields, which is in contrast to
  \cite{dissTF}. As we keep the commonly used definition of the matrix
  $\matr{X}$  the transposed matrix appears in the expression for
  $\matr{M}_{\cham{}}$. 
}%
~$\matr{X}$, 
\begin{align}
\matr{M}_{\cham{}} = \matr{V}^* \, \matr{X}^\top \, \matr{U}^{\dagger} =
  \begin{pmatrix} m_{\tilde{\chi}^\pm_1} & 0 \\ 
                  0 & m_{\tilde{\chi}^\pm_2} \end{pmatrix}  \quad
\text{with} \quad
  \matr{X} =
  \begin{pmatrix}
    \MTwo & \sqrt{2} \sinb \MW \\
    \sqrt{2} \cosb \MW & \mu
  \end{pmatrix}~,
\end{align}
where $\matr{M}_{\cham{}}$ is the diagonal mass matrix with the chargino
masses $\mcha{1}, \mcha{2}$ as entries, which are determined as the
(real and positive) singular values of $\matr{X}$. 
The singular value decomposition of $\matr{X}$ also yields results for 
$\matr{U}$ and~$\matr{V}$. 
Using the transformation matrices $\matr{U}$ and $\matr{V}$, the interaction
Higgsino and Wino spinors $\tilde{H}^-_1$, $\tilde{H}^+_2$ and
$\tilde{W}^\pm$, which are two component Weyl spinors, can be transformed into
the mass eigenstates 
\begin{align}
\cham{i} = 
\begin{pmatrix} \psi^L_i
   \\ \overline{\psi^R_i} \end{pmatrix}
\quad \text{with} \quad \psi^L_i = U_{ij} \begin{pmatrix} \tilde{W}^-
  \\ \tilde{H}^-_1 \end{pmatrix}_j \quad \text{and} \quad
 \psi^R_i = V_{ij} \begin{pmatrix} \tilde{W}^+
  \\ \tilde{H}^+_2 \end{pmatrix}_j
\end{align}
where the $i$th mass eigenstate can be expressed in terms of either 
the Weyl spinors $\psi^L_i$ and $\psi^R_i$  or the Dirac spinor $\cham{i}$.

In the neutralino case, as the neutralino mass matrix $\matr{Y}$ is
symmetric, one $4 \times 4$~matrix is sufficient for the diagonalization
\begin{align}
\matr{M}_{\neu{}} = \matr{N}^* \, \matr{Y} \, \matr{N}^{\dagger} =
\text{\bf diag}(m_{\neu{1}}, m_{\neu{2}}, m_{\neu{3}}, m_{\neu{4}})
\end{align}
with
\begin{align}
\matr{Y} &=
  \begin{pmatrix}
    \MOne                  & 0                & -\MZ \, \sw \cosb
    & \MZ \, \sw \sinb \\ 
    0                      & \MTwo            & \quad \MZ \, \cw \cosb
    & -\MZ \, \cw \sinb \\ 
    -\MZ \, \sw \cosb      & \MZ \, \cw \cosb & 0
    & -\mu             \\ 
    \quad \MZ \, \sw \sinb & -\MZ \, \cw \sinb & -\mu              & 0
  \end{pmatrix}~.
\end{align}
The unitary 4$\times$4 matrix $\matr{N}$ and the physical neutralino
masses $\mneu{k}$ ($k = 1,2,3,4$) result from a numerical Takagi 
factorization \cite{Takagi} of $\matr{Y}$. 
Starting from the original bino/wino/Higgsino basis, the mass
eigenstates can be determined with the help of the transformation matrix
$\matr{N}$, 
\begin{align}
\neu{k} = \begin{pmatrix} \psi^0_k \\[.2em] \overline{\psi^0_k} 
\end{pmatrix} 
\qquad \text{with} \qquad 
\psi^0_k = N_{kl}
     \begin{pmatrix} \tilde{B}^0 \\ \tilde{W}^0 \\ 
                     \tilde{H}^0_1 \\ \tilde{H}^0_2 \end{pmatrix}_l~
\end{align}
where $\psi^0_k$ denotes the two-component Weyl spinor and $\neu{k}$
the four-component Majorana spinor of the $k$th neutralino field.

\newcommand{\chapm}{\cham}
Concerning the renormalization we follow the prescription of
\citere{dissTF}. The following replacements of the parameters and the
fields are performed according to the multiplicative renormalization
procedure:
\begin{align}
M_1 \; &\to \; M_1 + \de M_1 ~, \\
M_2 \; &\to \; M_2 + \de M_2 ~, \\
\mu \; &\to \; \mu + \de \mu ~, \\
\OM \chapm{i} \; &\to \; \KKL \id + \edz \dZm{\chapm{}}^L \KKR_{ij}
                         \OM \chapm{j} \qquad (i,j = 1,2)~, \\
\OP \chapm{i} \; &\to \; \KKL \id + \edz \dZm{\chapm{}}^R \KKR_{ij}
                         \OP \chapm{j} \qquad (i,j = 1,2)~, \\
\OM \neu{k} \; &\to \; \KKL \id + \edz \dZm{\neu{}}^{} \KKR_{kl}
                       \OM \neu{l} \qquad (k,l = 1,2,3,4)~, \\
\OP \neu{k} \; &\to \; \KKL \id + \edz \dZm{\neu{}}^* \KKR_{kl}
                       \OP \neu{l} \qquad (k,l = 1,2,3,4)~.
\label{dZNeuR}
\end{align}
It should be noted that the parameter counterterms are complex
counterterms that each need two renormalization conditions to be fixed.
The transformation matrices are not renormalized, so that, using the notation 
  of replacing a matrix by its renormalized matrix and a counterterm matrix 
\begin{align}
\matr{X} &\to \matr{X} + \de\matr{X} ~, \\
\matr{Y} &\to \matr{Y} + \de\matr{Y} ~
\end{align}
with
\begin{align}\label{deltaX}
\de\matr{X} &= 
  \begin{pmatrix} \de M_2 & \sqrt{2}\, \de(\MW \Sbe) \\
                  \sqrt{2}\, \de(\MW \Cb) & \de \mu
  \end{pmatrix}~, \\[.4em]
\de\matr{Y} &= 
  \begin{pmatrix} 
      \de M_1 & 0 & -\de(\MZ\sw\Cb) & \de(\MZ\sw\Sbe) \\
      0 & \de M_2 & \de(\MZ\cw\Cb) & -\de(\MZ\cw\Sbe) \\
      -\de(\MZ\sw\Cb) & \de(\MZ\cw\Cb) & 0 & -\de\mu  \\
      \de(\MZ\sw\Sbe) & -\de(\MZ\cw\Sbe) & -\de\mu & 0
  \end{pmatrix}~,
\end{align}
the replacements of the matrices $\matr{M}_{\cham{}}$ and $\matr{M}_{\neu{}}$
can be expressed as
\begin{align}
\matr{M}_{\cham{}} &\to \matr{M}_{\cham{}} + \de\matr{M}_{\cham{}}
   = \matr{M}_{\cham{}} + \matr{V}^* \de\matr{X}^\top \matr{U}^\dagger \\
\label{Mneu}
\matr{M}_{\neu{}} &\to \matr{M}_{\neu{}} + \de\matr{M}_{\neu{}}
   = \matr{M}_{\neu{}} + \matr{N}^* \de\matr{Y} \matr{N}^\dagger~. 
\end{align}

\noindent
Now the renormalized self-energies are given by
\begin{align}\label{renSEcha_L}
\KKL \hSi_{\chapm{}}^L(p^2) \KKR_{ij} &=
  \KKL \Si_{\chapm{}}^L(p^2) \KKR_{ij} 
+ \edz \KKL \dZm{\chapm{}}^L + \dZm{\chapm{}}^{L\dagger} \KKR_{ij}~, \\
\KKL \hSi_{\chapm{}}^R(p^2) \KKR_{ij} &=
  \KKL \Si_{\chapm{}}^R(p^2) \KKR_{ij} 
+ \edz \KKL \dZm{\chapm{}}^R + \dZm{\chapm{}}^{R\dagger} \KKR_{ij}~, \\
\KKL \hSi_{\chapm{}}^{SL}(p^2) \KKR_{ij} &=
  \KKL \Si_{\chapm{}}^{SL}(p^2) \KKR_{ij}
  - \KKL \edz \dZm{\chapm{}}^{R\dagger} \matr{M}_{\cham{}}
        + \edz \matr{M}_{\cham{}} \dZm{\chapm{}}^L
        + \de\matr{M}_{\cham{}} \KKR_{ij}~, \\
\KKL \hSi_{\chapm{}}^{SR}(p^2) \KKR_{ij} &=
  \KKL \Si_{\chapm{}}^{SR}(p^2) \KKR_{ij}
  - \KKL \edz \dZm{\chapm{}}^{L\dagger} \matr{M}_{\cham{}}^\dagger
       + \edz \matr{M}_{\cham{}}^\dagger \dZm{\chapm{}}^R
       + \de\matr{M}_{\cham{}}^\dagger \KKR_{ij}~, \\[.4em]
\KKL \hSi_{\neu{}}^L(p^2) \KKR_{kl} &=
  \KKL \Si_{\neu{}}^L(p^2) \KKR_{kl} 
+ \edz \KKL \dZm{\neu{}} + \dZm{\neu{}}^{\dagger} \KKR_{kl}~, \\
\KKL \hSi_{\neu{}}^R(p^2) \KKR_{kl} &=
  \KKL \Si_{\neu{}}^R(p^2) \KKR_{kl} 
+ \edz \KKL \dZm{\neu{}}^* + \dZm{\neu{}}^{\top} \KKR_{kl}~, \\
\KKL \hSi_{\neu{}}^{SL}(p^2) \KKR_{kl} &=
  \KKL \Si_{\neu{}}^{SL}(p^2) \KKR_{kl}
  - \KKL \edz \dZm{\neu{}}^{\top} \matr{M}_{\neu{}}
       + \edz \matr{M}_{\neu{}} \dZm{\neu{}}
        + \de\matr{M}_{\neu{}} \KKR_{kl}~, \\
\label{renSEneu_SR}
\KKL \hSi_{\neu{}}^{SR}(p^2) \KKR_{kl} &=
  \KKL \Si_{\neu{}}^{SR}(p^2) \KKR_{kl}
  - \KKL \edz \dZm{\neu{}}^{\dagger} \matr{M}_{\neu{}}^{\dagger} 
       + \edz \matr{M}_{\neu{}}^{\dagger} \dZm{\neu{}}^*
       + \de\matr{M}_{\neu{}}^\dagger \KKR_{kl}~.
\end{align}

Instead of choosing the three complex parameters $M_1$, $M_2$, and $\mu$ 
as independent parameters, we impose on-shell conditions for the two 
chargino masses and the mass of the lightest neutralino and extract the 
expressions for the counterterms of  $M_1$, $M_2$, and $\mu$, accordingly.
In a recent analysis~\cite{onshellCNmasses} it was emphasized that in the 
case of the renormalization of two chargino and one neutralino mass, always 
the most binolike neutralino has to be renormalized in order to find a 
numerically stable result.
Also, in \citere{Baro} the problem of large unphysical contributions due 
to a non-binolike lightest neutralino is discussed. 
In our numerical set up, see \refse{sec:numeval}, the lightest neutralino 
is always rather binolike.  
On the other hand, it would be trivial to change our prescription from 
the lightest neutralino to any other neutralino. 
In \citere{onshellCNmasses} it was also suggested that the numerically 
most stable result is obtained via the renormalization of one chargino 
and two neutralinos. 
However, in our approach, this choice leads to IR divergences, since the 
chargino mass changes (from the tree-level mass to the one-loop pole mass) 
by a finite shift due to the renormalization procedure. 
Using the shifted mass for the external particle but the tree-level mass 
for internal particles results in IR divergences.
On the other hand, in general, inserting the shifted chargino mass 
everywhere yields UV divergences.
Consequently, we stick to our choice of imposing on-shell conditions 
for the two charginos and one neutralino. The conditions read
\begin{align}
\label{mcha-OS_org}
\Bigl(\KKL \wtre \hSi_{\cham{}} (p)\KKR_{ii} 
     \cham{i}(p)\Bigr)\Big|_{p^2 = \mcha{i}^2} &= 0 \qquad  (i = 1,2)~, \\
\label{mneu-OS_org}
\quad \Bigl(\KKL\wtre \hSi_{\neu{}} (p)\KKR_{11} 
      \neu{1}(p)\Bigr)\Big|_{p^2 = \mneu{1}^2} &= 0~.
\end{align}
These conditions can be rewritten in terms of six equations defining six
real parameters and field renormalization constants or three complex ones,
\begin{align}
\label{mcha-OS}
\wtre \KKL \mcha{i} \KL \hSi_{\cham{}}^{L}(\mcha{i}^2)
                       +\hSi_{\cham{}}^{R}(\mcha{i}^2) \KR 
                       +\hSi_{\cham{}}^{SL}(\mcha{i}^2)
                       +\hSi_{\cham{}}^{SR}(\mcha{i}^2) 
      \KKR_{ii} &= 0~, \\
\label{mcha-OS-2}
\wtre \KKL \mcha{i} \KL \hSi_{\cham{}}^{L}(\mcha{i}^2)
                       -\hSi_{\cham{}}^{R}(\mcha{i}^2) \KR 
                       -\hSi_{\cham{}}^{SL}(\mcha{i}^2)
                       +\hSi_{\cham{}}^{SR}(\mcha{i}^2) 
      \KKR_{ii} &= 0~, \\
\label{mneu-OS}
\wtre \KKL \mneu{1} \KL \hSi_{\neu{}}^{L}(\mneu{1}^2)
                       +\hSi_{\neu{}}^{R}(\mneu{1}^2) \KR
                       +\hSi_{\neu{}}^{SL}(\mneu{1}^2)
                       +\hSi_{\neu{}}^{SR}(\mneu{1}^2) 
      \KKR_{11} &= 0~, \\
\label{mneu-OS-2}
\wtre \KKL \mneu{1} \KL \hSi_{\neu{}}^{L}(\mneu{1}^2)
                       -\hSi_{\neu{}}^{R}(\mneu{1}^2) \KR
                       -\hSi_{\neu{}}^{SL}(\mneu{1}^2)
                       +\hSi_{\neu{}}^{SR}(\mneu{1}^2) 
      \KKR_{11} &= 0~.
\end{align}
For the further determination of the field renormalization constants we
also impose 
\begin{align}
\lim_{p^2 \to \mcha{i}^2} 
\Bigl(\frac{(\pslash\, + \mcha{i}) \bigl[ \wtre \hSi_{\cham{}}(p)\bigr]_{ii}}
           {p^2 - \mcha{i}^2} \cham{i}(p)\Bigr) &= 0 \qquad (i = 1,2)~, \\ 
 \lim_{p^2 \to \mneu{1}^2} 
\Bigl(\frac{(\pslash\, + \mneu{1})\bigl[ \wtre \hSi_{\neu{}}(p)\bigr]_{11}}
           {p^2 - \mneu{1}^2}\neu{1}(p)\Bigr) &= 0~.
\end{align}
This leads to the following set of equations:
\begin{align}
\non
\wtre\Bigl[\edz \KL \hSi_{\cham{}}^{L}(\mcha{i}^2)
                  + \hSi_{\cham{}}^{R}(\mcha{i}^2) \KR
                  + \mcha{i}^2 \KL \hSi_{\cham{}}^{L'}(\mcha{i}^2)
                  + \hSi_{\cham{}}^{R'}(\mcha{i}^2) \KR \quad\ &\\
                  + \mcha{i} \KL \hSi_{\cham{}}^{SL'}(\mcha{i}^2)
                  + \hSi_{\cham{}}^{SR'}(\mcha{i}^2) \KR 
      \Bigr]_{ii} &= 0~, \label{Zchadiag-OS}\\
\wtre \KKL \hSi_{\cham{}}^{L}(\mcha{i}^2)
                       -\hSi_{\cham{}}^{R}(\mcha{i}^2) 
      \KKR_{ii} &= 0~, \label{Zchadiag-OS-2}\\
\non
\wtre \Bigl[\edz \KL \hSi_{\neu{}}^{L}(\mneu{1}^2)
                   + \hSi_{\neu{}}^{R}(\mneu{1}^2) \KR 
                   + \mneu{1}^2 \KL \hSi_{\neu{}}^{L'}(\mneu{1}^2) 
                   + \hSi_{\neu{}}^{R'}(\mneu{1}^2) \KR \quad\ &\\
                   + \mneu{1} \KL \hSi_{\neu{}}^{SL'}(\mneu{1}^2)
                   + \hSi_{\neu{}}^{SR'}(\mneu{1}^2) \KR
      \Bigr]_{11} &= 0~, \label{Zneudiag-OS}\\
\wtre \KKL \hSi_{\neu{}}^{L}(\mneu{1}^2)
                       -\hSi_{\neu{}}^{R}(\mneu{1}^2)
      \KKR_{11} &= 0~, \label{Zneudiag-OS-2}
\end{align}
where we have used again the shorthand
$\Si'(m^2) \equiv \frac{\partial \Si}{\partial p^2}\big|_{p^2 = m^2}$. 
It should be noted that \refeq{Zneudiag-OS-2} is already fulfilled due to the 
Majorana nature of the neutralinos. 
Inserting \refeqs{renSEcha_L}~--~(\ref{renSEneu_SR}) for the
renormalized self-energies in \refeqs{mcha-OS}~--~(\ref{mneu-OS-2})
and solving for 
$\KKL \de\matr{M}_{\cham{}}\KKR_{ii}$ and $\KKL \de\matr{M}_{\neu{}} \KKR_{11}$
results in 
\begin{align}
\re\KKL \de\matr{M}_{\cham{}} \KKR_{ii} &= \frac{1}{2}
 \wtre \KKL \mcha{i} \KL \Si_{\chapm{}}^L(\mcha{i}^2) 
                        + \Si_{\chapm{}}^R(\mcha{i}^2) \KR
                        + \Si_{\chapm{}}^{SL}(\mcha{i}^2)
                        + \Si_{\chapm{}}^{SR}(\mcha{i}^2) \KKR_{ii}~, \\
\im\KKL\de\matr{M}_{\cham{}}\KKR_{ii} &= \frac{i}{2}
   \wtre \KKL \Si_{\chapm{}}^{SR}(\mcha{i}^2) 
            - \Si_{\chapm{}}^{SL}(\mcha{i}^2) \KKR_{ii}
 - \frac{1}{2} \mcha{i} \im \KKL \dZm{\cham{}}^L
           - \dZm{\cham{}}^R \KKR_{ii}~, \label{eq:imdZcha}\\[.4em]
\re\KKL\de\matr{M}_{\neu{}}\KKR_{11} &= \frac{1}{2}
   \wtre \KKL \mneu{1} \KL \Si_{\neu{}}^L(\mneu{1}^2)
                         + \Si_{\neu{}}^R(\mneu{1}^2) \KR
                         + \Si_{\neu{}}^{SL}(\mneu{1}^2)
                         + \Si_{\neu{}}^{SR}(\mneu{1}^2) \KKR_{11}~, \\
\im\KKL\de\matr{M}_{\neu{}}\KKR_{11} &= \frac{i}{2}
   \wtre \KKL \Si_{\neu{}}^{SR}(\mneu{1}^2)
             -\Si_{\neu{}}^{SL}(\mneu{1}^2) \KKR_{11}
   -\mneu{1} \im \dZZm{\neu{}}_{11}~.\label{eq:imdZneu}
\end{align}
where we have used already \refeqs{Zchadiag-OS-2}~and~(\ref{Zneudiag-OS-2}).
Using \refeqs{deltaX}~--~(\ref{Mneu}), these conditions lead
to~\cite{dissTF,diplTF} 
\begin{align}
\de M_1 &= \frac{1}{(N_{11}^*)^2} \Big(
  2 N_{11}^* \KKL N_{13}^* \de(\MZ\sw\Cb) - N_{14}^* \de(\MZ\sw\Sbe) \KKR
  \non \\
&\qquad - N_{12}^* \KKL 2 N_{13}^* \de(\MZ\cw\Cb)
                      -2 N_{14}^* \de(\MZ\cw\Sbe) + N_{12}^* \de M_2 \KKR
  \non \\
&\qquad + 2 N_{13}^* N_{14}^* \de\mu
        + \KKL \mneu{1} \KL \wtre \Si_{\neu{}}^L(\mneu{1}^2) 
                    - i \im \dZm{\neu{}} \KR
        + \wtre \Si_{\neu{}}^{SL}(\mneu{1}^2) \KKR_{11} \Big)~, \\
\de M_2 &= \frac{1}{2 \KL U_{11}^* U_{22}^* V_{11}^* V_{22}^*
                         -U_{12}^* U_{21}^* V_{12}^* V_{21}^* \KR } \times
  \non \\
&\hspace*{-0.7cm} \Big( 
       U_{22}^* V_{22}^*\KKL \mcha{1} 
       \KL \wtre \Si_{\chapm{}}^L(\mcha{1}^2) 
          + \wtre \Si_{\chapm{}}^R(\mcha{1}^2) 
          - i\, \im \{ \dZm{\chapm{}}^L - \dZm{\cham{}}^R \} \KR
       + 2 \wtre\Si_{\chapm{}}^{SL}(\mcha{1}^2) \KKR_{11}
       \non \\
& \hspace*{-0.7cm} - U_{12}^* V_{12}^* \KKL \mcha{2} 
       \KL \wtre \Si_{\chapm{}}^L(\mcha{2}^2) 
          + \wtre \Si_{\chapm{}}^R(\mcha{2}^2) 
          - i\, \im \{ \dZm{\chapm{}}^L - \dZm{\cham{}}^R \} \KR
       + 2 \wtre\Si_{\chapm{}}^{SL}(\mcha{2}^2) \KKR_{22} \non \\
&\qquad 
+ 2 (U_{12}^* U_{21}^* - U_{11}^* U_{22}^*) V_{12}^* V_{22}^*
          \,\de(\sqrt{2}\MW\Sbe) \non \\
&\qquad 
+ 2 U_{12}^* U_{22}^* (V_{12}^* V_{21}^* - V_{11}^* V_{22}^*)
          \,\de(\sqrt{2}\MW\Cb) \Big)~, \\ 
\de \mu &= \frac{1}{2 \KL U_{11}^* U_{22}^* V_{11}^* V_{22}^*
                         -U_{12}^* U_{21}^* V_{12}^* V_{21}^* \KR } \times
  \non \\
&\hspace*{-0.7cm} \Big( 
       U_{11}^* V_{11}^* \KKL \mcha{2} 
       \KL \wtre \Si_{\chapm{}}^L(\mcha{2}^2) 
          + \wtre \Si_{\chapm{}}^R(\mcha{2}^2) 
          - i\, \im \{ \dZm{\chapm{}}^L - \dZm{\cham{}}^R \} \KR
       + 2 \wtre\Si_{\chapm{}}^{SL}(\mcha{2}^2) \KKR_{22} \non \\
&\hspace*{-0.7cm} - U_{21}^* V_{21}^* \KKL \mcha{1} 
       \KL \wtre \Si_{\chapm{}}^L(\mcha{1}^2) 
          + \wtre \Si_{\chapm{}}^R(\mcha{1}^2) 
          - i\, \im \{ \dZm{\chapm{}}^L - \dZm{\cham{}}^R \} \KR
       + 2 \wtre\Si_{\chapm{}}^{SL}(\mcha{1}^2) \KKR_{11} \non \\
&\qquad + 2 (U_{12}^* U_{21}^* - U_{11}^* U_{22}^*) V_{11}^* V_{21}^*
          \,\de(\sqrt{2}\MW\Cb) \non \\
&\qquad + 2 U_{11}^* U_{21}^* (V_{12}^* V_{21}^* - V_{11}^* V_{22}^*)
          \,\de(\sqrt{2}\MW\Sbe) \Big)~.
\label{deltamu}
\end{align}
Equations (\ref{Zchadiag-OS})~--~(\ref{Zneudiag-OS})
define the real part of the diagonal field renormalization constants of 
the chargino fields and of the lightest neutralino field. 
We generalize the latter result for the diagonal field renormalization 
constants of the other neutralino fields imposing \refeq{Zneudiag-OS}
also for the components $k = 2,3,4$ -- though we do not define them 
fully on-shell; see below.
The imaginary parts of the diagonal field renormalization 
constants are still undefined. 
For the definition of the imaginary parts, we use \refeqs{eq:imdZcha}
and (\ref{eq:imdZneu}), 
where the latter one is generalized for the components $k = 2,3,4$ 
and is imposed to hold also for those neutralinos. Now, for the
charginos and the lightest neutralino \refeqs{eq:imdZcha} and
(\ref{eq:imdZneu}) already define
the imaginary parts of $\KKL\de\matr{M}_{\cha{}}\KKR_{ii}$ 
($i = 1,2$), and $\KKL\de\matr{M}_{\neu{}}\KKR_{11}$, 
which
means that a further condition for the imaginary part of the field
renormalization constants of the charginos and the lightest neutralino is
required, and we just set them to zero (see below \refeqs{eq:dZcha} 
and \eqref{eq:dZneu})
which is possible as all divergences are absorbed by other counterterms.
The off-diagonal field renormalization constants are fixed by the
condition that 
\begin{align}
\Bigl(\KKL \wtre \hSi_{\cham{}} (p)\KKR_{ij} 
    \cham{j}(p)\Bigr)\Big|_{p^2 = \mcha{j}^2} &= 0 \qquad (i,j = 1,2)~,
\\
\Bigl(\KKL\wtre \hSi_{\neu{}} (p)\KKR_{kl} 
    \neu{l}(p)\Bigr)\Big|_{p^2 = \mneu{l}^2} &= 0 \qquad (k,l = 1,2,3,4)~.
\end{align}
Finally, this yields for the field renormalization constants~\cite{dissTF},
\begin{align}
\re \KKL \dZm{\chapm{}}^{L/R} \KKR_{ii} &=
        - \wtre \Big[ \Si_{\chapm{}}^{L/R}(\mcha{i}^2) \\
&\qquad + \mcha{i}^2 \KL \Si_{\chapm{}}^{L'}(\mcha{i}^2)
                       + \Si_{\chapm{}}^{R'}(\mcha{i}^2) \KR
        + \mcha{i} \KL \Si_{\chapm{}}^{SL'}(\mcha{i}^2)
                    +  \Si_{\chapm{}}^{SR'}(\mcha{i}^2) \KR
      \Big]_{ii}~, \non \\
\im \KKL \dZm{\chapm{}}^{L/R} \KKR_{ii} &= 
       \pm \frac{1}{\mcha{i}} \KKL \frac{i}{2} 
    \wtre \KKKL \Si_{\chapm{}}^{SR}(\mcha{i}^2) 
              - \Si_{\chapm{}}^{SL}(\mcha{i}^2) \KKKR
    - \im \de\matr{M}_{\cham{}} \KKR_{ii} := 0~,\label{eq:dZcha} 
\\
\KKL \dZm{\chapm{}}^{L/R} \KKR_{ij} &= \frac{2}{\mcha{i}^2 - \mcha{j}^2} 
  \wtre \Big[ \mcha{j}^2 \Si_{\chapm{}}^{L/R}(\mcha{j}^2) 
             +\mcha{i} \mcha{j} \Si_{\chapm{}}^{R/L}(\mcha{j}^2)
\label{dZcha_ij} \\
&\qquad + \mcha{i} \Si_{\chapm{}}^{SL/SR}(\mcha{j}^2)
        + \mcha{j} \Si_{\chapm{}}^{SR/SL}(\mcha{j}^2)
        - \mcha{i/j} \de\matr{M}_{\cham{}}
        - \mcha{j/i} \de\matr{M}_{\cham{}}^{\dagger} \Big]_{ij}~, \non \\
\re \KKL \dZm{\neu{}}^{} \KKR_{kk} &= 
  -\wtre \Big[  \Si_{\neu{}}^L(\mneu{k}^2) \\
&\qquad  + \mneu{k}^2 \KL \Si_{\neu{}}^{L'}(\mneu{k}^2)
                         +\Si_{\neu{}}^{R'}(\mneu{k}^2) \KR
         + \mneu{k}   \KL \Si_{\neu{}}^{SL'}(\mneu{k}^2)
                         +\Si_{\neu{}}^{SR'}(\mneu{k}^2) \KR
       \Big]_{kk}~, \non \\
\im \KKL \dZm{\neu{}}^{} \KKR_{kk} &= 
         \frac{1}{\mneu{k}} \KKL \frac{i}{2} 
    \wtre \KKKL \Si_{\neu{}}^{SR}(\mneu{k}^2) \label{eq:dZneu}
              - \Si_{\neu{}}^{SL}(\mneu{k}^2) \KKKR
    - \im \de\matr{M}_{\neu{}} \KKR_{kk} \stackrel{k=1}{:=} 0~, \\
\KKL \dZm{\neu{}}^{} \KKR_{kl} &= \frac{2}{\mneu{k}^2 - \mneu{l}^2}
 \wtre \Big[ \mneu{l}^2 \Si_{\neu{}}^L(\mneu{l}^2) 
            +\mneu{k}\mneu{l} \Si_{\neu{}}^R(\mneu{l}^2) \non \\
&\qquad + \mneu{k} \Si_{\neu{}}^{SL}(\mneu{l}^2)
        + \mneu{l} \Si_{\neu{}}^{SR}(\mneu{l}^2)
        - \mneu{k} \de\matr{M}_{\neu{}} 
        - \mneu{l} \de\matr{M}_{\neu{}}^\dagger \Big]_{kl}~.
\end{align}
The \refeqs{mcha-OS_org}, (\ref{mneu-OS_org}) result in three on-shell
masses in the neutralino/chargino sector. Therefore the 
three neutralino masses $\mneu{2,3,4}$, on 
the other hand, require a finite shift for their on-shell value. We
have checked that this shift (for the scenarios under investigation in
\refse{sec:numeval}) is numerically small and does not change our
results. Consequently, these shifts, though formally necessary, are
not further taken into account to simplify the numerical evaluation.

\bigskip
The field renormalization constants for squark, quark, gluino, gauge 
boson as well as chargino and neutralino fields that have been 
derived in Sec.~\ref{sec:color}~--~\ref{sec:chaneu} are constructed via 
the multiplicative renormalization procedure in a symmetry conserving way, 
absorb the divergences accordingly and are defined via on-shell 
renormalization conditions. In the presence of complex phases and 
nonvanishing absorptive parts of the self-energy type corrections, 
further wave function corrections may arise that are not part of the 
field renormalization constant but can be taken into account by additional 
$Z$~factors; see the Appendix.


\section{Calculation of loop diagrams}
\label{sec:calc}

\begin{figure}[t!]
\vspace{2em}
\begin{center}
\includegraphics[width=0.9\textwidth]{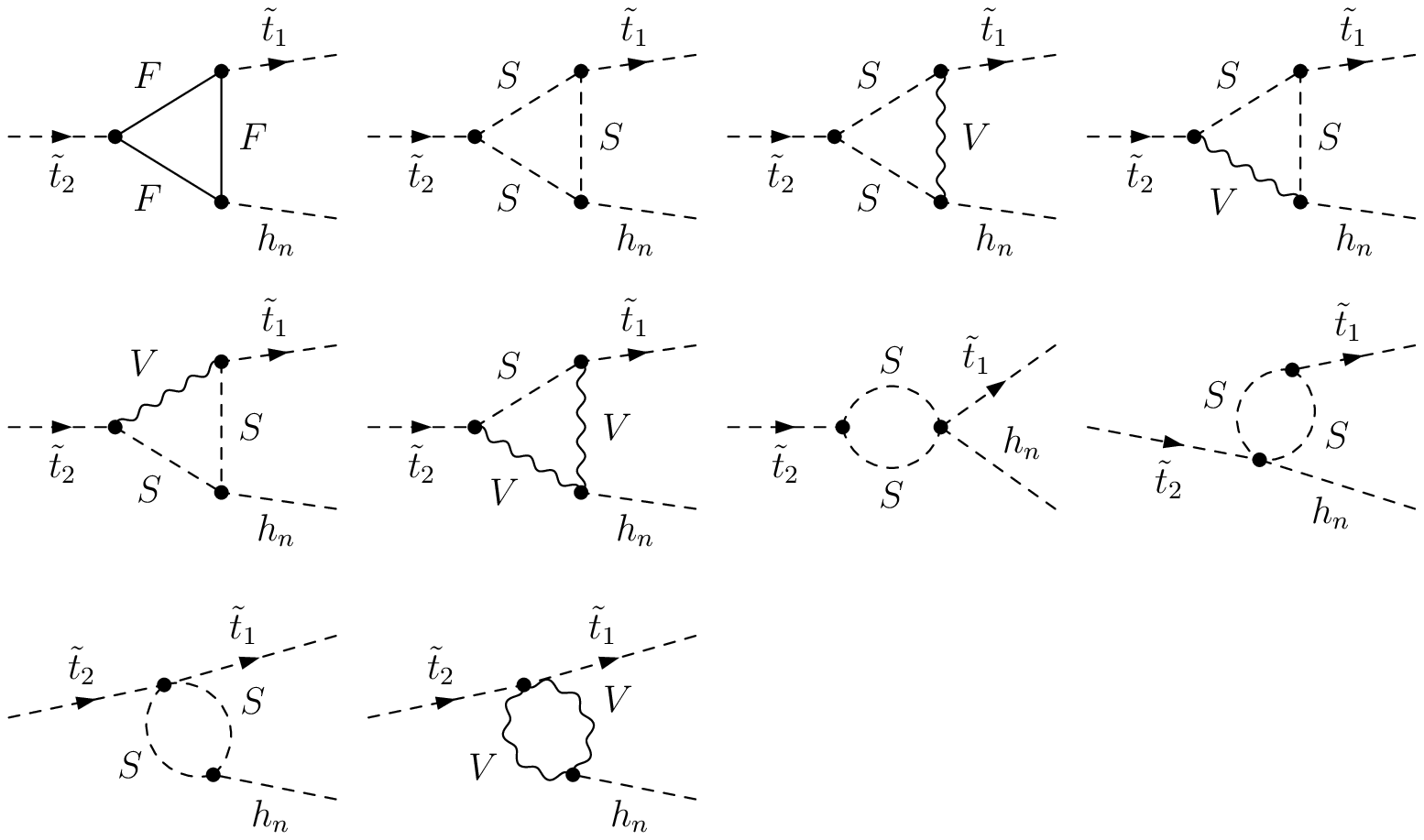}
\caption{
  Generic Feynman diagrams for the decay $\decayhn$ ($n = 1, 2, 3$).
  $F$ can be a SM fermion, chargino, neutralino, or gluino; 
  $S$ can be a sfermion or a Higgs boson; 
  $V$ can be a $\ga$, $Z$, $W^\pm$, or $g$. 
  Not shown are the diagrams with a $Z$--$h_n$ or $G$--$h_n$ transition 
  contribution on the external Higgs boson leg.
}
\label{fig:fdstophiggs}
\end{center}
\end{figure}

\begin{figure}[htb!]
\begin{center}
\includegraphics[width=0.9\textwidth]{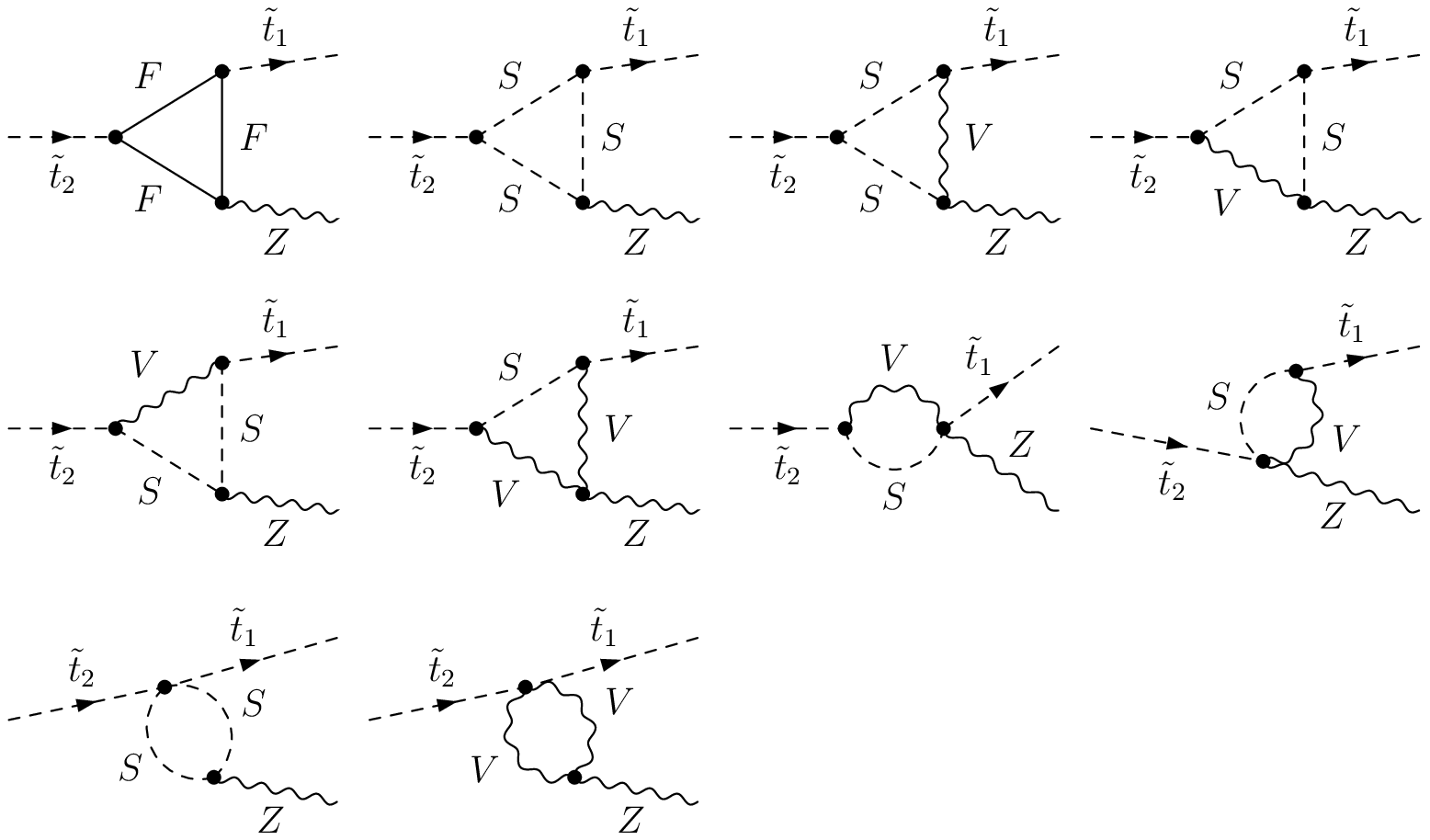}
\caption{
  Generic Feynman diagrams for the decay $\decayZ$.
  $F$ can be a SM fermion, chargino, neutralino, or gluino; 
  $S$ can be a sfermion or a Higgs boson;
  $V$ can be a $\ga$, $Z$, $W^\pm,$ or $g$. 
}
\label{fig:fdstopZ}
\end{center}
\vspace{2em}
\end{figure}

\begin{figure}[htb!]
\begin{center}
\includegraphics[width=0.9\textwidth]{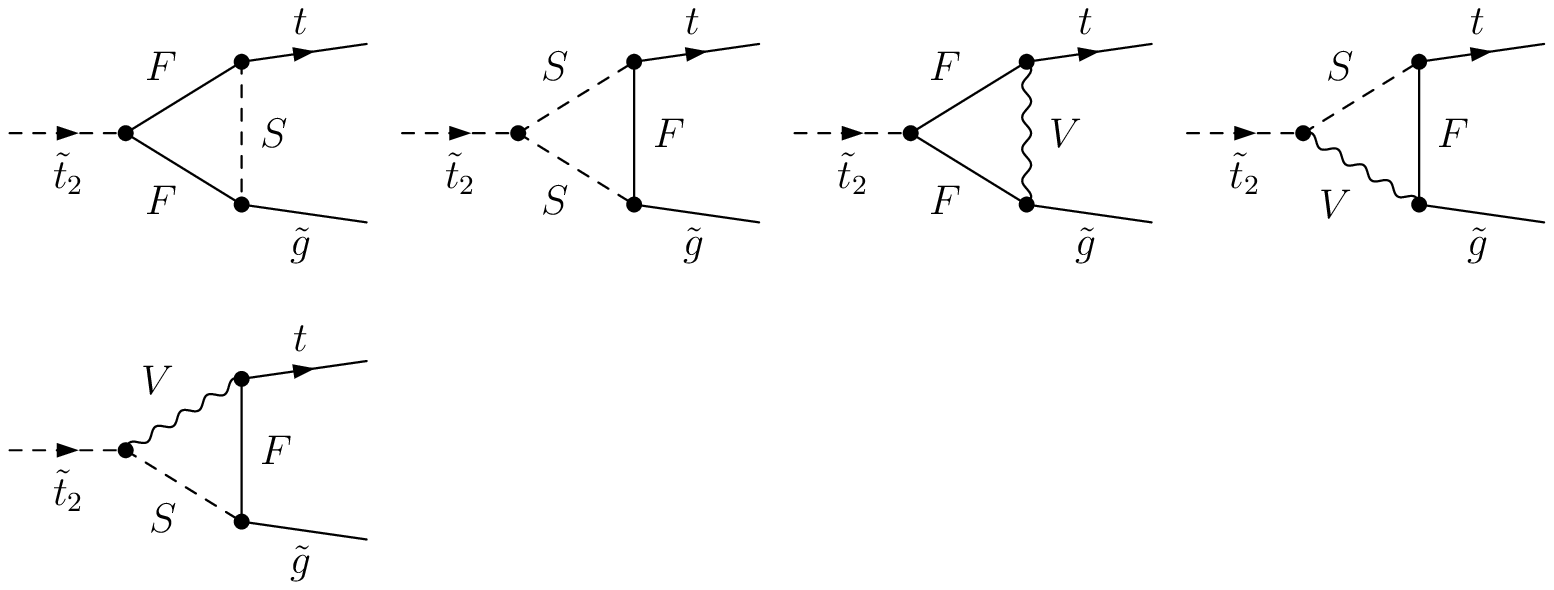}
\caption{
  Generic Feynman diagrams for the decay $\decaygl$.
  $F$ can be a SM fermion, chargino, neutralino, or gluino;
  $S$ can be a sfermion or a Higgs boson; 
  $V$ can be a $\ga$, $Z$, $W^\pm$, or $g$. 
}
\label{fig:fdtopgl}
\end{center}
\vspace{2em}
\end{figure}

\begin{figure}[htb!]
\begin{center}
\includegraphics[width=0.9\textwidth]{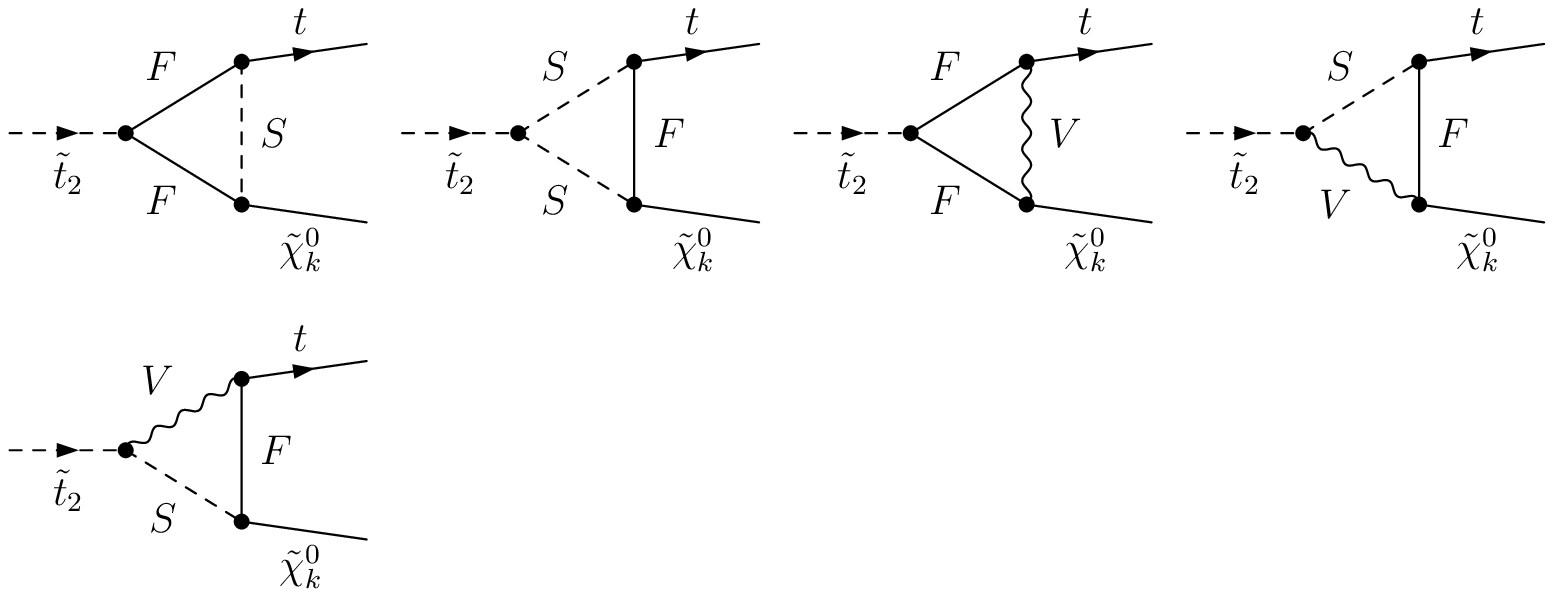}
\caption{
  Generic Feynman diagrams for the decay $\decayNk$ ($k = 1,2,3,4$).
  $F$ can be a SM fermion, chargino, neutralino, or gluino;
  $S$ can be a sfermion or a Higgs boson;
  $V$ can be a $\ga$, $Z$, $W^\pm$, or $g$. 
}
\label{fig:fdtopneu}
\end{center}
\end{figure}

\begin{figure}[htb!]
\begin{center}
\includegraphics[width=0.9\textwidth]{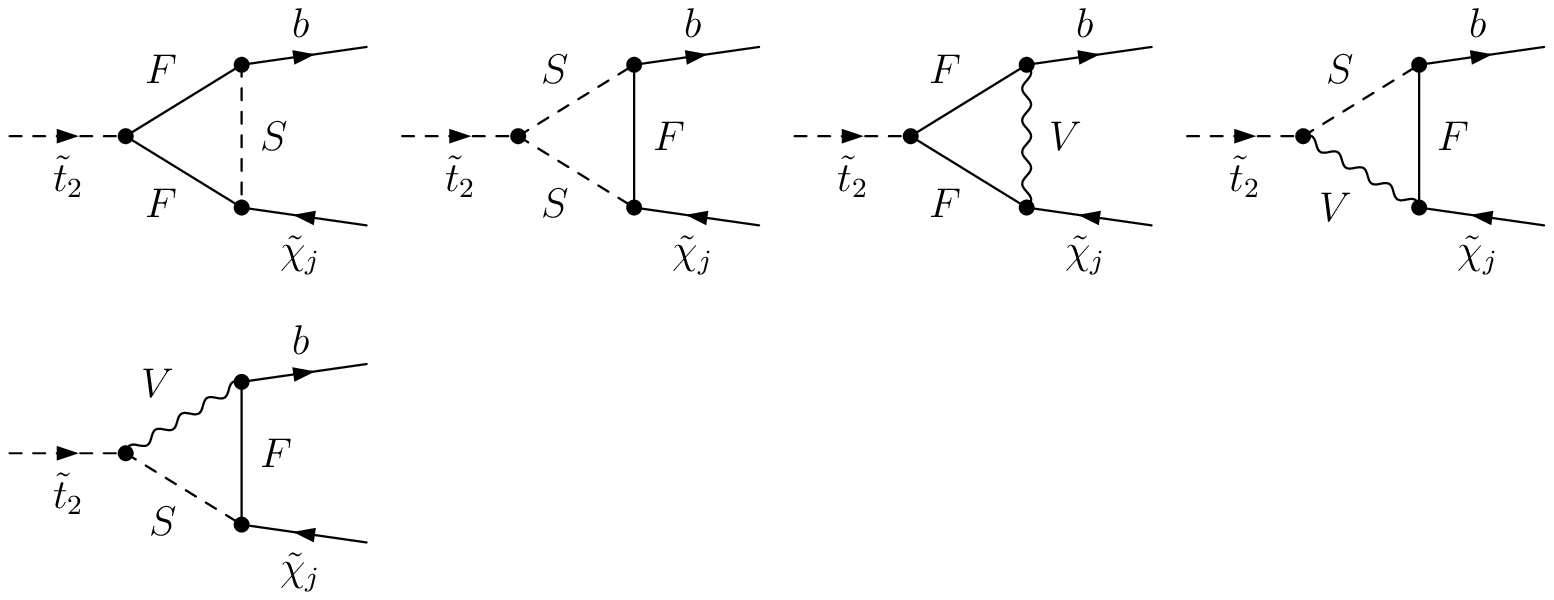}
\caption{
  Generic Feynman diagrams for the decay $\decayCj$ ($j = 1,2$).
  $F$ can be a SM fermion, chargino, neutralino, or gluino; 
  $S$ can be a sfermion or a Higgs boson;
  $V$ can be a $\ga$, $Z$, $W^\pm$, or $g$. 
}
\label{fig:fdbotcha}
\end{center}
\end{figure}

\begin{figure}[htb!]
\begin{center}
\includegraphics[width=0.9\textwidth]{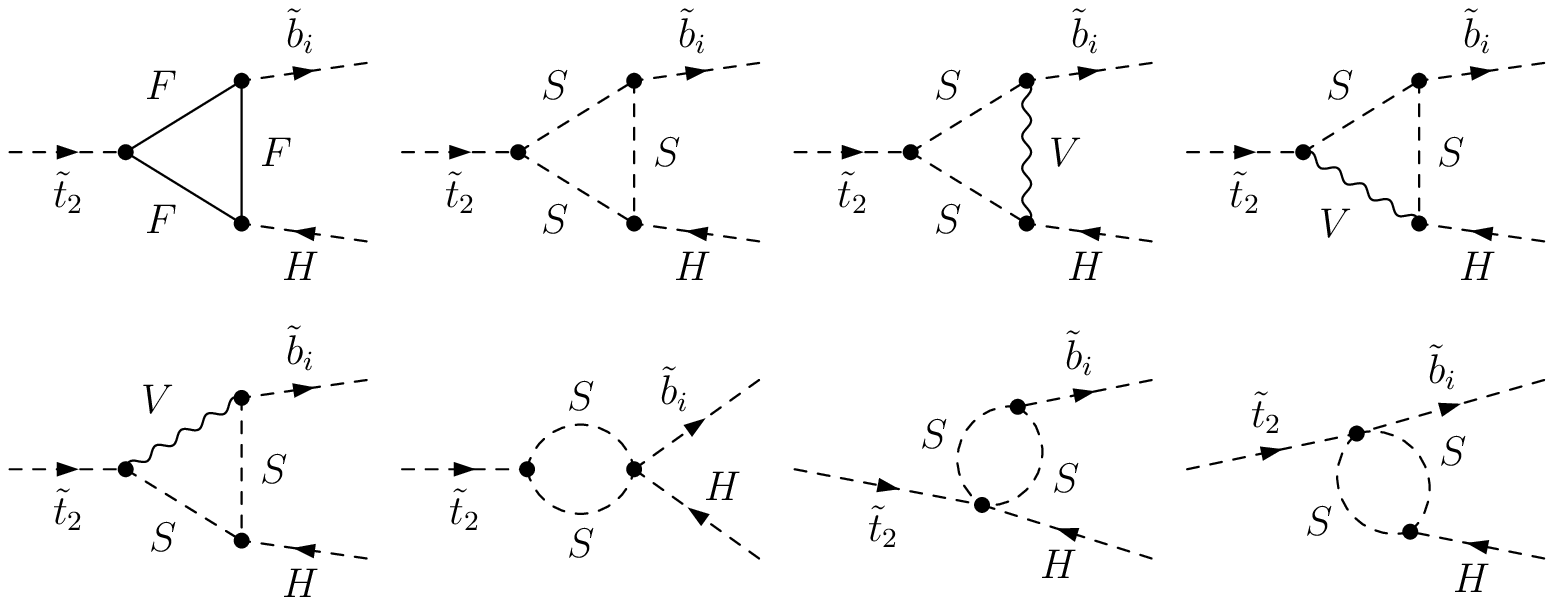}
\caption{
  Generic Feynman diagrams for the decay $\decaySbiH$ ($i = 1,2$).
  $F$ can be a SM fermion, chargino, neutralino, or gluino;
  $S$ can be a sfermion or a Higgs boson;
  $V$ can be a $\ga$, $Z$, $W^\pm$, or $g$. 
  Not shown are the diagrams with a $W^+$--$H^+$ or $G^+$--$H^+$ transition
  contribution on the external Higgs boson leg. 
}
\label{fig:fdsbotHpm}
\end{center}
\vspace{2em}
\end{figure}

\begin{figure}[htb!]
\begin{center}
\includegraphics[width=0.9\textwidth]{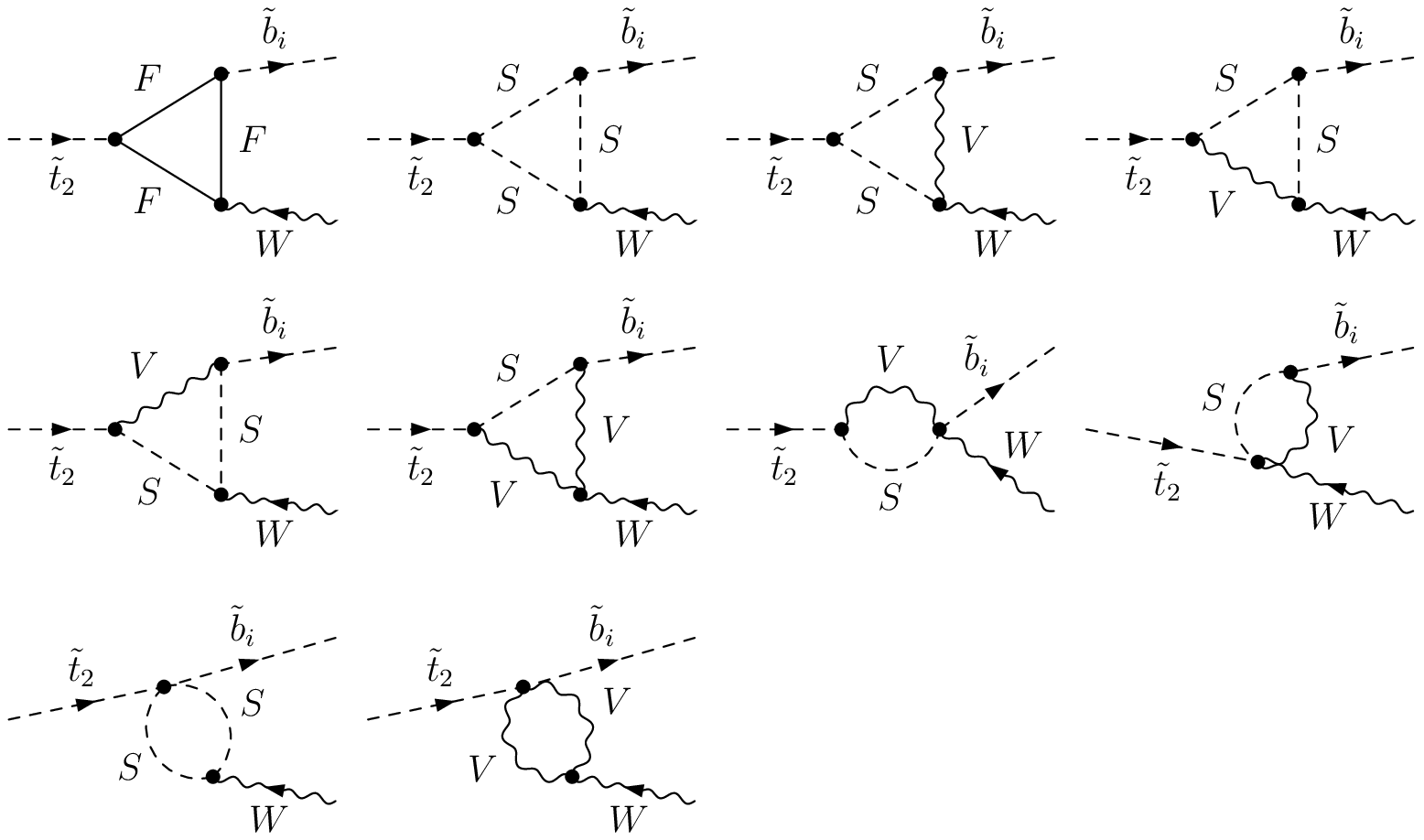}
\caption{
  Generic Feynman diagrams for the decay $\decaySbiW$ ($i = 1,2$).
  $F$ can be a SM fermion, chargino, neutralino, or gluino;
  $S$ can be a sfermion or a Higgs boson;
  $V$ can be a $\ga$, $Z$, $W^\pm$, or $g$. 
}
\label{fig:fdsbotW}
\end{center}
\end{figure}

In this section we give some details about the calculation of the
higher-order corrections to the partial decay widths of scalar quarks. 
Sample diagrams are shown in \reffis{fig:fdstophiggs} -- \ref{fig:fdsbotW}. 
Not shown are the diagrams for real (hard and soft) photon and gluon
radiation. They are obtained from the corresponding tree-level diagrams
by attaching a photon (gluon) to the electrically (color) charged
particles. The internal generically depicted particles in
\reffis{fig:fdstophiggs} -- \ref{fig:fdsbotW} are labeled as follows:
$F$ can be a SM fermion $f$, chargino $\cha{j}$, neutralino 
$\neu{k}$, or gluino $\gl$; 
$S$ can be a sfermion $\sfi$ or a Higgs boson $h_n$; 
$V$ can be a photon $\ga$, gluon $g$, or a massive SM gauge boson, 
$Z$ or $W^\pm$. 
For internally appearing Higgs bosons no higher-order
corrections to their masses or couplings are taken into account; 
these corrections would correspond to effects beyond one-loop order.%
\footnote{
  We found that using loop corrected Higgs boson masses 
  in the loops leads to a UV divergent result.
}%
~For external Higgs bosons, as described in
\refse{sec:higgs}, the appropriate $\matr{Z}$~factors are applied and
on-shell masses (including higher-order corrections) are used.

Also not shown are the diagrams with a gauge/Goldstone--Higgs boson
self-energy contribution on the external Higgs boson leg. 
They appear in the decay $\Stopz \to \Stope h_n$, \reffi{fig:fdstophiggs}, 
with a $Z/G$--$h_n$ transition and in the decay $\Stopz \to \Sboti H^+$, 
\reffi{fig:fdsbotHpm}, with a $W^+$/$G^+$--$H^+$ transition.%
\footnote{
  From a technical point of view, the $W^+$/$G^+$--$H^+$ transitions 
  have been absorbed into the respective counterterms, while the 
  $Z/G$--$h_n$ transitions have been calculated explicitly.
}%
~The corresponding self-energy diagram belonging to the process 
$\Stopz \to \Stope Z$ or $\Stopz \to \Sboti W^+$, respectively, yields a
vanishing contribution for external on-shell gauge bosons due  
to $\eps \cdot p = 0$ for $p^2 = \MZ^2$ ($p^2 = \MW^2$), 
where $p$ denotes the external momentum and $\eps$ the polarization
vector of the gauge boson.

Furthermore, in general, in \reffis{fig:fdstophiggs} --
\ref{fig:fdsbotW} we have  omitted diagrams with self-energy type
corrections of external (on-shell) particles. 
While the contributions from the real parts of the loop functions are
taken into account via the renormalization constants defined by on-shell
renormalization conditions, the contributions coming from the imaginary
part of the loop functions can result in an additional (real) correction
if multiplied by complex parameters (such as $\At$).
In the analytical and numerical evaluation, these diagrams have been taken 
into account via the prescription described in the Appendix. 
The impact of these contributions will be discussed in 
\refse{sec:numeval}.

Within our one-loop calculation we neglect finite width effects that 
can help to cure threshold singularities. 
Consequently, in the close vicinity of those thresholds our calculation 
does not give a reasonable result. 
Switching to a complex mass scheme \cite{complexmassscheme} would be 
another possibility to cure this problem, but its application is beyond 
the scope of our paper.

Finally it should be noted that the purely loop induced decay 
channels $\Stopz \to \Stope \ga/g$ yield exactly zero due to the fact that 
the decay width is proportional to $\eps \cdot p$ and the photon/gluon 
is on-shell, i.e. $\eps \cdot p = 0$.

The diagrams and corresponding amplitudes have been obtained with 
\fa~\cite{feynarts}. 
The model file, including the MSSM counterterms, 
is largely based on \citere{dissTF}, however adjusted to
match exactly the renormalization prescription described in
\refse{sec:renorm}. 
The further evaluation has been performed with \fc~\cite{formcalc}.

\subsubsection*{Ultraviolet divergences}

As regularization scheme for the UV divergences we
have used constrained differential renormalization~\cite{cdr}, 
which has been shown to be equivalent to 
dimensional reduction~\cite{dred} at the \onel\ level~\cite{formcalc}. 
Thus the employed regularization scheme preserves SUSY~\cite{dredDS,dredDS2}
and guarantees that the SUSY relations are kept intact, e.g. that the 
gauge couplings of the SM vertices and the Yukawa couplings of the 
corresponding SUSY vertices also coincide to \onel\ order in the SUSY limit. 
Therefore no additional shifts, which might occur when using a different 
regularization scheme, arise.
All UV divergences cancel in the final result.

\subsubsection*{Infrared divergences}

The IR divergences from diagrams with an internal photon or gluon have
to cancel with the ones from the corresponding real soft radiation. 
In the case of QED we have included the soft photon contribution
following the description given in \citere{denner}. 
In the case of QCD we have modified this prescription by replacing the
product of electric charges by the appropriate combination of color
charges (linear combination of $C_A$ and $C_F$ times $\als$).
The IR divergences arising from the diagrams involving a $\ga$ 
(or a $g$) are regularized by introducing a photon (or gluon) 
mass parameter, $\la$. 
While for the QED part this procedure always works, in the QCD part 
due to its non-Abelian character this method can fail. 
However, since no triple or quartic gluon vertices appear, 
$\la$ can indeed be used as a regulator
(the appearance of the non-Abelian gluino-gluino-gluon vertex does not
pose a problem here~\cite{stop_top_gl_als}).
All IR divergences, i.e.\ all divergences in the limit
$\la \to 0$, cancel once virtual and real diagrams for one decay
channel are added.

Special care has to be taken in the decay modes involving scalar bottom
quarks. Using tree-level sbottom masses yields a cancellation of 
IR divergences to all orders for all $\Stopz$ decay modes. However,
inserting the one-loop corrected sbottom masses (see \refse{sec:topbottom}),
as required for consistency, we found cancellation to all orders of the 
related IR divergences, except for the decay modes
$\decaySbiW$. Within these decays the tree-level relation
required by the $SU(2)$ symmetry $M_{\sq_L}(\Stop) = M_{\sq_L}(\Sbot)$,
corresponding to
\begin{align}
\label{stopsbotrel}
|U_{\Sbot_{11}}|^2 \msbe^2 + |U_{\Sbot_{12}}|^2 \msbz^2 &=
|U_{\Stop_{11}}|^2 \mste^2 + |U_{\Stop_{12}}|^2 \mstz^2 
- \mt^2 + \mb^2 - \MW^2 \CZb~,
\end{align}
has to be fulfilled to yield a cancellation of all IR divergences.%
\footnote{
  Equation (\ref{stopsbotrel}) has been deduced via 
  $M_{\sq_L}^2 = |U_{\sq_{11}}|^2 \msqe^2 + |U_{\sq_{12}}|^2 \msqz^2
  - M_Z^2 c_{2\be} (I_q^3 - Q_q \sw^2) - m_q^2$.
}
On the other hand, the requirement of on-shell sbottom masses as well as
an intact $SU(2)$ relation at the one-loop level leads to the necessity
of a shift in the scalar bottom masses; see \refeq{MSbotshift}.
Therefore \refeq{stopsbotrel} is ``violated'' at the one-loop level, 
introducing a two-loop IR divergence in $\Ga(\decaySbiW)$. 
In order to eliminate this two-loop IR divergence we introduced a 
counterterm in the $\Stopz\Sboti W$ vertex,
\begin{align}
\label{ir}
\dZ{\rm ir} &=
\KL 
  \KKL |U_{\Sbot_{11}}|^2 \msbe^2 + |U_{\Sbot_{12}}|^2 \msbz^2 \KKR
- \KKL |U_{\Sbot_{11}}|^2 \msbe^2 + |U_{\Sbot_{12}}|^2 \msbz^2 \KKR_{\rm shift}
\KR \times {\rm IR~div}
\end{align}
to restore the tree-level $SU(2)$~relation.
The left term in \refeq{ir} contains only ``tree-level'' values, while
the index ``shift'' refers to inserting the one-loop masses and mixing matrices.
The IR~divergence has been taken from Eq.~(B.5) of \citere{dittmaier}
(it can also be found in \citere{beenakker}), and reads (in our case)
\begin{align}
{\rm IR~div} &= \frac{\al}{6 \pi \MW} \KKL
  \frac{2\, x_t\, \ln(x_t)}{\mstz (1 - x_t^2)} 
  \ln\KL\frac{\mstz \MW}{\la^2}\KR
- \frac{x_b\, \ln(x_b)}{\msbi (1 - x_b^2)} 
  \ln\KL\frac{\msbi \MW}{\la^2}\KR \KKR
\end{align}
with
\begin{align}
x_t &= \frac{\sqrt{1 - 4\,\mstz \MW/(\msbi^2 + i0 - (\MW - \mstz)^2)} - 1}
            {\sqrt{1 - 4\,\mstz \MW/(\msbi^2 + i0 - (\MW - \mstz)^2)} + 1}~, \\
x_b &= \frac{\sqrt{1 - 4\,\msbi \MW/(\mstz^2 + i0 - (\MW - \msbi)^2)} - 1}
            {\sqrt{1 - 4\,\msbi \MW/(\mstz^2 + i0 - (\MW - \msbi)^2)} + 1}~,
\end{align}
where $i0$ denotes an infinitesimally small imaginary part.
After including this tree-level relation restoring counterterm we find
an IR finite result to all orders as required.

We have furthermore checked that our result does not depend on $\De E$
defining the energy cut that separates the soft from the hard
radiation. Our numerical results have been obtained for 
$\De E = 10^{-5} \times \mstz$.

\subsubsection*{Tree-level formulas}

For completeness we show here also the formulas that have been
used to calculate the tree-level decay widths:
\begin{align}
\Gamma^{\rm tree}(\decayhn) &= \frac{|C(\Stopz, \Stope, h_n)|^2\,
                              \la^{1/2}(\mstz^2,\mste^2,m_{h_n}^2)}
                              {16\, \pi\, \mstz^3}\qquad (n = 1,2,3)~, \\
\Gamma^{\rm tree}(\decayZ) &= \frac{|C(\Stopz, \Stope, Z)|^2\,
                             \la^{3/2}(\mstz^2,\mste^2,\MZ^2)}
                             {16\, \pi\, \MZ^2\, \mstz^3}~, \\
\Gamma^{\rm tree}(\decaygl) &= \Big[ \KL |C(\Stopz, t, \gl)_L|^2 
       + |C(\Stopz, t, \gl)_R|^2 \KR (\mstz^2 - \mt^2 - \mgl^2) \non \\
&\qquad - 4\, \re \{C(\Stopz, t, \gl)_L^*\, C(\Stopz, t, \gl)_R \}\,
       \mt\, \mgl \Big] \times \non \\
&\qquad \frac{4}{3}\, \frac{\la^{1/2}(\mstz^2,\mt^2,\mgl^2)}
                          {16\, \pi\, \mstz^3}~, \\
\Gamma^{\rm tree}(\decayNk) &= \Big[ \KL |C(\Stopz, t, \neu{k})_{L}|^2
       + |C(\Stopz, t, \neu{k})_R|^2 \KR (\mstz^2 - \mt^2 - \mneu{k}^2) \non \\
&\qquad - 4\, \re \{C(\Stopz, t, \neu{k})_L^*\, C(\Stopz, t, \neu{k})_R \}\,
       \mt\, \mneu{k} \Big] \times \non \\
&\qquad \frac{\la^{1/2}(\mstz^2,\mt^2,\mneu{k}^2)}
            {16\, \pi\, \mstz^3}\qquad (k = 1,2,3,4)~, \\
\Gamma^{\rm tree}(\decayCpj) &= \Big[ \KL |C(\Stopz, b, \chap{j})_{L}|^2
       + |C(\Stopz, b, \chap{j})_R|^2 \KR (\mstz^2 - \mb^2 - \mcha{j}^2) \non \\
&\qquad - 4\, \re \{C(\Stopz, b, \chap{j})_L^*\, C(\Stopz, b, \chap{j})_R \}\,
       \mb\, \mcha{j} \Big] \times \non \\
&\qquad \frac{\la^{1/2}(\mstz^2,\mb^2,\mcha{j}^2)}
            {16\, \pi\, \mstz^3}\qquad (j = 1,2)~, \\
\Gamma^{\rm tree}(\decaySbiH) &= \frac{|C(\Stopz, \Sboti, H^+)|^2\,
                                \la^{1/2}(\mstz^2,\msbi^2,\MHp^2)}
                                {16\, \pi\, \mstz^3}\qquad (i = 1,2)~, \\
\Gamma^{\rm tree}(\decaySbiW) &= \frac{|C(\Stopz, \Sboti, W^+)|^2\,
                                \la^{3/2}(\mstz^2,\msbi^2,\MW^2)}
                                {16\, \pi\, \MW^2\, \mstz^3}\qquad (i = 1,2)~,
\end{align}
where $\la(x,y,z) = (x - y - z)^2 - 4yz$ and the couplings $C(a, b, c)$ 
can be found in the \fa~model files~\cite{feynarts-mf}.
$C(a, b, c)_{L,R}$ denote the part of the coupling that is 
proportional to $(\id \mp \ga_5)/2$.


\newcommand{\SE}{S1}
\newcommand{\SZ}{S2}

\section{Numerical analysis}
\label{sec:numeval}

In this section we present a numerical analysis of all 15 decay
channels. In the various figures below we show the partial decay 
widths and their relative correction at the tree-level (``tree'') 
and at the one-loop level (``full''), 
\begin{align}
\Ga^{\rm tree} &\equiv \Ga^{\rm tree}(\decayxy)~, \\
\Ga^{\rm full} &\equiv \Ga^{\rm full}(\decayxy)~, \\
\de\Ga/\Ga^{\rm tree} &\equiv \frac{\Ga^{\rm full} - \Ga^{\rm tree}}
                                  {\Ga^{\rm tree}}~,
\label{Garel}
\end{align}
where xy denotes the specific final state.
The total decay width is defined as the sum of all 15 partial
decay widths,  
\begin{align}
\Ga_{\rm tot}^{\rm tree} \equiv \sum_{{\rm xy}} \Ga^{\rm tree}(\decayxy)~, \quad
\Ga_{\rm tot}^{\rm full} \equiv \sum_{{\rm xy}} \Ga^{\rm full}(\decayxy)~, \quad
\de\Ga_{\rm tot}/\Ga_{\rm tot}^{\rm tree} \equiv 
\frac{\Ga_{\rm tot}^{\rm full} - \Ga_{\rm tot}^{\rm tree}}
     {\Ga_{\rm tot}^{\rm tree}}~.
\end{align}
We also show the absolute and relative changes of the branching ratios,
\begin{align}
\br^{\rm tree} &\equiv \frac{\Ga^{\rm tree}(\decayxy)}
                           {\Ga_{\rm tot}^{\rm tree}}~, \\
\br^{\rm full} &\equiv \frac{\Ga^{\rm full}(\decayxy)}
                           {\Ga_{\rm tot}^{\rm full}}~, \\
\de\br/\br &\equiv \frac{\br^{\rm full} - \br^{\rm tree}}{\br^{\rm full}}~. 
\label{brrel}
\end{align}
The last quantity is crucial to analyze the impact of the one-loop
corrections on the phenomenology at the LHC and the ILC.


\subsection{Parameter settings}
\label{sec:paraset}

The renormalization scale $\mu_R$ has been set to the mass of the 
decaying particle, i.e.\ $\mu_R = \mstz$.
The SM parameters are chosen as follows; see also \cite{pdg}%
\footnote{
  Using the most up-to-date values from \citere{pdg-new} would have a
  negligible impact on our numerical results.
}%
:
\begin{itemize}

\item Fermion masses (on-shell masses, if not indicated differently)
\index{leptonmasses}:
\begin{align}
m_e    &= 0.51099891\mev~, & m_{\nu_e}     &= 0\mev~, \non \\
m_\mu  &= 105.658367\mev~, & m_{\nu_{\mu}}  &= 0\mev~, \non \\
m_\tau &= 1776.84\mev~,    & m_{\nu_{\tau}} &= 0\mev~, \non \\
m_u &= 53.8\mev~,          & m_d &= 53.8\mev~, \non \\ 
m_c &= 1.27\gev~,          & m_s &= 104\mev~, \non \\
m_t &= 171.2\gev~,         & m_b(m_b) &= 4.2\gev~.
\end{align}
According to \citere{pdg}, $m_s$ is an estimate of a so-called 
"current quark mass" in the \MSbar\ scheme at the scale $\mu \approx 2\gev$. 
$m_c$ and $m_b$ are the "running" masses in the \MSbar\ scheme.
The top quark mass as well as the lepton masses are defined OS.
$m_u$ and $m_d$ are effective parameters, calculated through the hadronic
contributions to
\begin{align}
\Delta\alpha_{\text{had}}^{(5)}(M_Z) &= 
      \frac{\alpha}{\pi}\sum_{f = u,c,d,s,b}
      Q_f^2 \Bigl(\ln\frac{M_Z^2}{m_f^2} - \frac 53\Bigr)~.
\end{align}

\item The CKM matrix has been set to unity.

\item Gauge boson masses\index{gaugebosonmasses}:
\begin{align}
M_Z = 91.1876\gev~, \qquad M_W = 80.398\gev~.
\end{align}

\item Coupling constants\index{couplingconstants}:
\begin{align}
\alpha = \frac{e^2}{4 \pi} = 1/137.0359895~, \qquad
\alpha_s(M_Z) = 0.1176~,
\end{align}
where the running and decoupling of $\alpha_s$ can be found 
in \refse{sec:alphas}.
\end{itemize}

The Higgs sector quantities (masses, mixings, etc.) have been
evaluated using {\tt FeynHiggs} (version 2.6.5)
\cite{feynhiggs,mhiggslong,mhiggsAEC,mhcMSSMlong}.%
\footnote{
  As default value within {\tt FeynHiggs}, $\mu_R = \mt$ is used. 
  Furthermore we have neglected the (in our case small) corrections of
  \order{\alb\als, \alt\alb, \alb^2} (via a small modification in the code)
  and used the top pole mass for the evaluation of the MSSM Higgs boson 
  sector quantities.
}

We will show the results for some representative numerical examples. 
The parameters are chosen according to the two scenarios, \SE\ and \SZ, 
shown in \refta{tab:para}, but with one of the parameters varied.
The scenarios are defined such that {\em all} decay modes are open
simultaneously to permit an analysis of all channels, i.e.\ not picking
specific parameters for each decay.
We will start with a variation of $\mstz$, and show later the
results for varying $\phiat$.
The scenarios are in agreement with the 
MSSM Higgs boson searches at LEP~\cite{LEPHiggsSM,LEPHiggsMSSM}. 
Too small values of the lightest Higgs boson mass would be reached for 
$\tb \lsim 9.4\ (4.6)$
within \SE\ (\SZ) as given in \refta{tab:para}.%
\footnote{
  While in these scenarios we are not aiming to yield a light Higgs boson
  mass value around $\sim 125 \gev$, it should be noted that such a value is
  in principle in agreement with not too heavy scalar top
  quarks~\cite{Mh125our} that can be produced via $e^+e^- \to \aStope\Stopz$ 
  at the ILC(1000).
}
In order to avoid completely unrealistic spectra, 
the following exclusion limits \cite{pdg} hold in our two scenarios%
\footnote{
  The relatively light scalar quark masses and especially the light gluino
  mass are potentially in conflict with recent SUSY searches at the
  LHC~\cite{susy11} (although no fully model-independent results have been
  published). However, as stressed above, the parameters are chosen
  to be able to analyze as many decay modes as possible {\em simultaneously}. 
  For a {\em realistic} collider analysis these bounds~\cite{susy11} will
  have to be taken into account.
}%
:
\begin{align}
\mste &> 95 \gev, \;
\msbe > 89 \gev, \;
m_{\sq} > 379 \gev, \;
m_{\tilde{e}_1} > 73 \gev, \non \\
\mneu{1} &> 46 \gev, \;
\mcha{1} > 94 \gev, \;
\mgl > 308 \gev .
\end{align}

\begin{table}[htb!]
\renewcommand{\arraystretch}{1.5}
\BC
\begin{tabular}{|c||c|c|c|c|c|c|c|c|c|c|c|}
\hline
Scen.\ & $\tb$ & $\MHp$ & $\mstz$ & $\mste$  & $\msbz$ 
& $\mu$ & $\At$ & $\Ab$ & $M_1$ & $M_2$ & $M_3$ 
\\ \hline\hline
\SE & 20 & 150 & 650 & $0.4\, \mstz$ & $0.7\, \mstz$ & 
200 &  800 &  400 & 200 & 300 & 350 
\\ \hline
\SZ & 20 & 180 & 1200 & $0.6\, \mstz$ & $0.8\, \mstz$ &
300 & 1800 & 1600 & 150 & 200 & 400  
\\ \hline
\end{tabular}
\caption{
  MSSM parameters for the initial numerical investigation; 
  all parameters (except of $\tb$) are in GeV. 
  We always set $\mb^{\MSbar}(\mb) = 4.2 \gev$.
  In our analysis $M_{\sq_L} (= M_{\tilde{l}_L})$, 
  $M_{\Stop_R} (= M_{\tilde{u}_R} = M_{\tilde{c}_R})$, and 
  $M_{\Sbot_R} (= M_{\tilde{d}_R} = M_{\tilde{s}_R} = M_{\tilde{l}_R})$ 
  are chosen such that the values of $\mste$, $\mstz$, and 
  $\msbz$ are realized.
  For the $\Sbot$~sector the shifts in $M_{\sq_{L,R}}(\Sbot)$ as defined 
  in \refeqs{MSbotshift} and \eqref{backshift} are taken into account.
  The values for $\At$ and $\Ab\, (= A_{\tau})$ are chosen such that 
  charge- or color-breaking minima are avoided~\cite{ccb}.
}
\label{tab:para}
\EC
\renewcommand{\arraystretch}{1.0}
\end{table}

\begin{table}[htb!]
\renewcommand{\arraystretch}{1.5}
\BC
\begin{tabular}{|c|c||r|r|r|r|}
\hline
Scen. & $\tb$ & $\mste$~~ & $\mstz$~~ & $\msbe$~~ & $\msbz$~~  
\\ \hline\hline
    &  2 & 260.000 &  650.000 & 305.436 & 455.000 
\\ \cline{2-6}
\SE & 20 & 260.000 &  650.000 & 333.572 & 455.000
\\ \cline{2-6}
    & 50 & 260.000 &  650.000 & 329.755 & 455.000 
\\ \hline\hline 
    &  2 & 720.000 & 1200.000 & 769.801 & 960.000      
\\ \cline{2-6}
\SZ & 20 & 720.000 & 1200.000 & 783.300 & 960.000
\\ \cline{2-6}
    & 50 & 720.000 & 1200.000 & 783.094 & 960.000
\\ \hline
\end{tabular}
\caption{
  The stop and sbottom masses in S1 and S2 and at different $\tb$ for the 
  numerical investigation; all masses are in GeV and rounded to 1~MeV.
}
\label{tab:squark}
\EC
\renewcommand{\arraystretch}{1.0}
\end{table}

A few examples of the scalar top and bottom quark masses in \SE\ and
\SZ\ are shown in \refta{tab:squark}. The values of $\mstz$ allow copious
production of the heavier stop at the LHC.
For other choices of the
gluino mass, $\mgl > \mstz$, which would leave no visible effect for
most of the decay modes of the $\Stopz$, the heavier stop could also be
produced in gluino decays at the LHC. 
Furthermore, in \SE\ (even for the nominal value of $\mstz$ as given in
\refta{tab:para}) the production of $\Stopz$ at the ILC(1000), i.e.\ with 
$\sqrt{s} = 1000 \gev$, via $e^+e^- \to \aStope\Stopz$ will be possible,
with all the subsequent decay modes (\ref{ststphi}) -- (\ref{stbcha})
being open. The clean environment of the ILC would permit a detailed
study of the scalar top decays.
For the lowest values shown in the plots below, 
$\mstz \gsim 570 \gev$, we find (via a tree-level calculation)
$\si(e^+e^- \to \aStope\Stopz) \approx 1.5~{\rm fb}$, 
i.e.\ an integrated
luminosity of $\sim 1\, \iab$ would yield about~1500~$\Stopz$. 
This number drops to $\sim 280~\Stopz$ for the masses shown in
\refta{tab:squark}. The ILC environment would result in an accuracy of
the relative branching ratio~(\refeq{brrel}) close to the statistical
uncertainty: a BR of 30\% could be determined to $\sim 5\%$ for the
lowest $\mstz$ values and to about 11\% for the values given in
\refta{tab:squark}. Depending on the combination of allowed decay
channels a determination of the branching ratios at the few percent
level might be achievable in the high-luminosity running of the ILC(1000).

The numerical results we will show in the next subsections are of
course dependent on choice of the SUSY parameters. Nevertheless, they
give an idea of the relevance of the full one-loop corrections.
As an example, the largest decay width is $\Ga(\decaygl)$, dominating the 
total decay width, $\Ga_{\rm tot}$, and thus the various branching 
ratios. For other choices of $\mgl$ with $\mgl > \mstz$ the corrections 
to the decay widths would stay the same, but the branching ratios would 
look very different. 
Channels (and their respective one-loop corrections) that may look 
unobservable due to the smallness of their BR in the plots shown below, 
could become important if other channels are kinematically forbidden.


\subsection{Full one-loop results for varying \boldmath{$\mstz$}}
\label{sec:full1L}

The results shown in this and the following subsections consist of 
``tree'', which denotes the tree-level value and of ``full'', which is
the partial decay width including {\em all} one-loop 
corrections as described in \refse{sec:calc}.
We start the numerical analysis with partial decay widths of $\Stopz$
evaluated as a function of $\mstz$, 
starting at $\mstz = 570 \gev$ up to $\mstz = 3 \tev$, which
roughly coincides with the reach of the LHC for high-luminosity running.
The upper panels contain the results for the absolute
value of the various  partial decay widths, $\Ga(\decayxy)$ (left) and
the relative correction from the full one-loop contributions
(right). The lower panels show the same 
results for $\br(\decayxy)$.

Since in this section all parameters are chosen to be real, no
contributions from absorptive parts of self-energy type corrections 
on external legs can contribute. 
This will be different in \refse{sec:full1Lphiat}.

\begin{figure}[htb!]
\begin{center}
\begin{tabular}{c}
\includegraphics[width=0.49\textwidth,height=7.5cm]{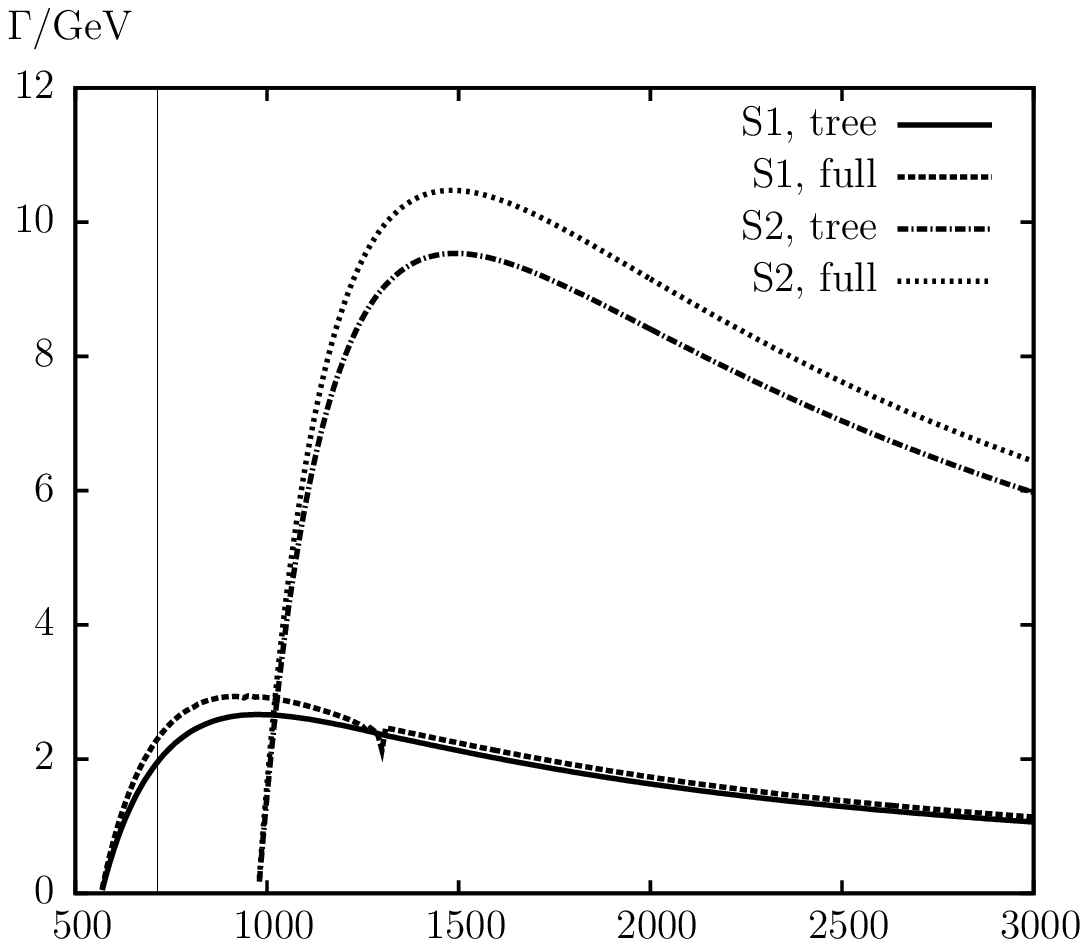}
\hspace{-4mm}
\includegraphics[width=0.49\textwidth,height=7.5cm]{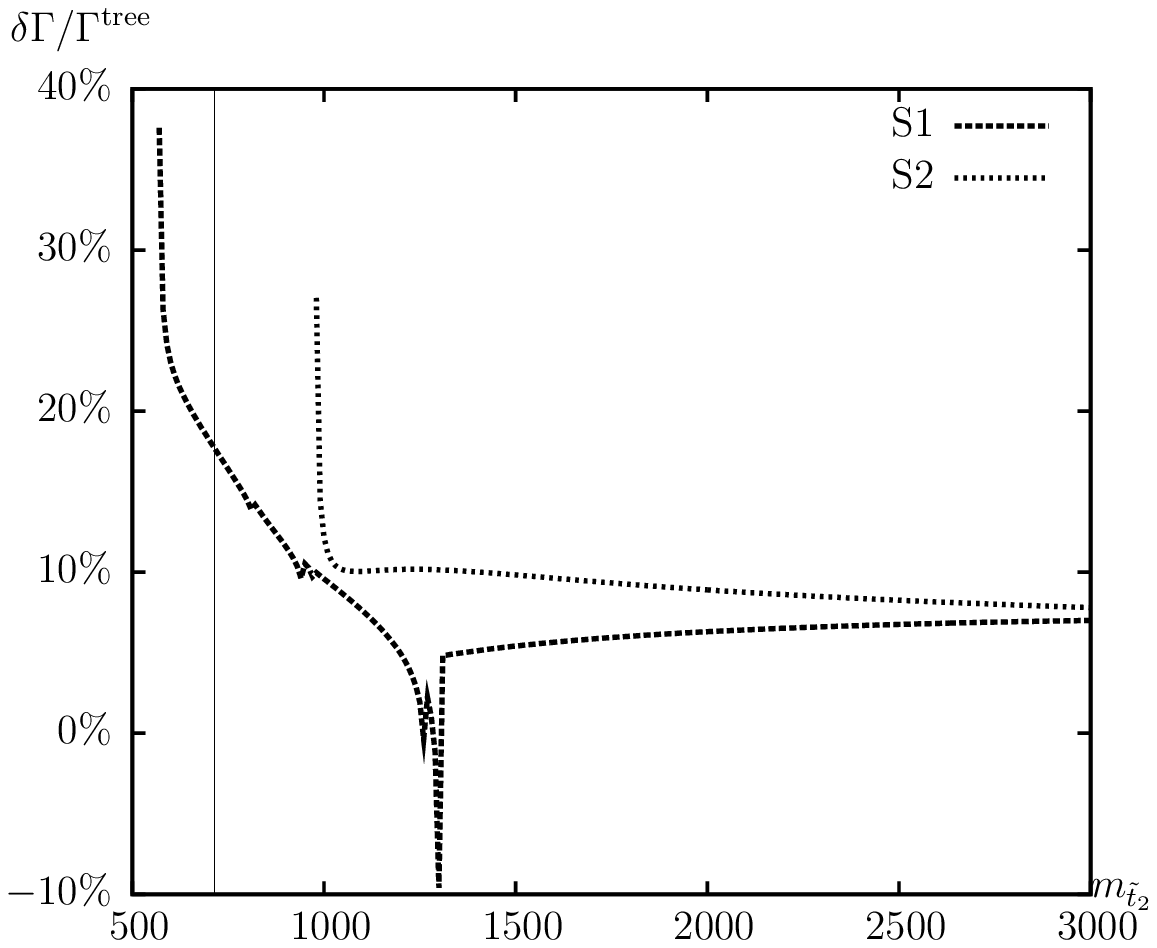} 
\\[4em]
\includegraphics[width=0.49\textwidth,height=7.5cm]{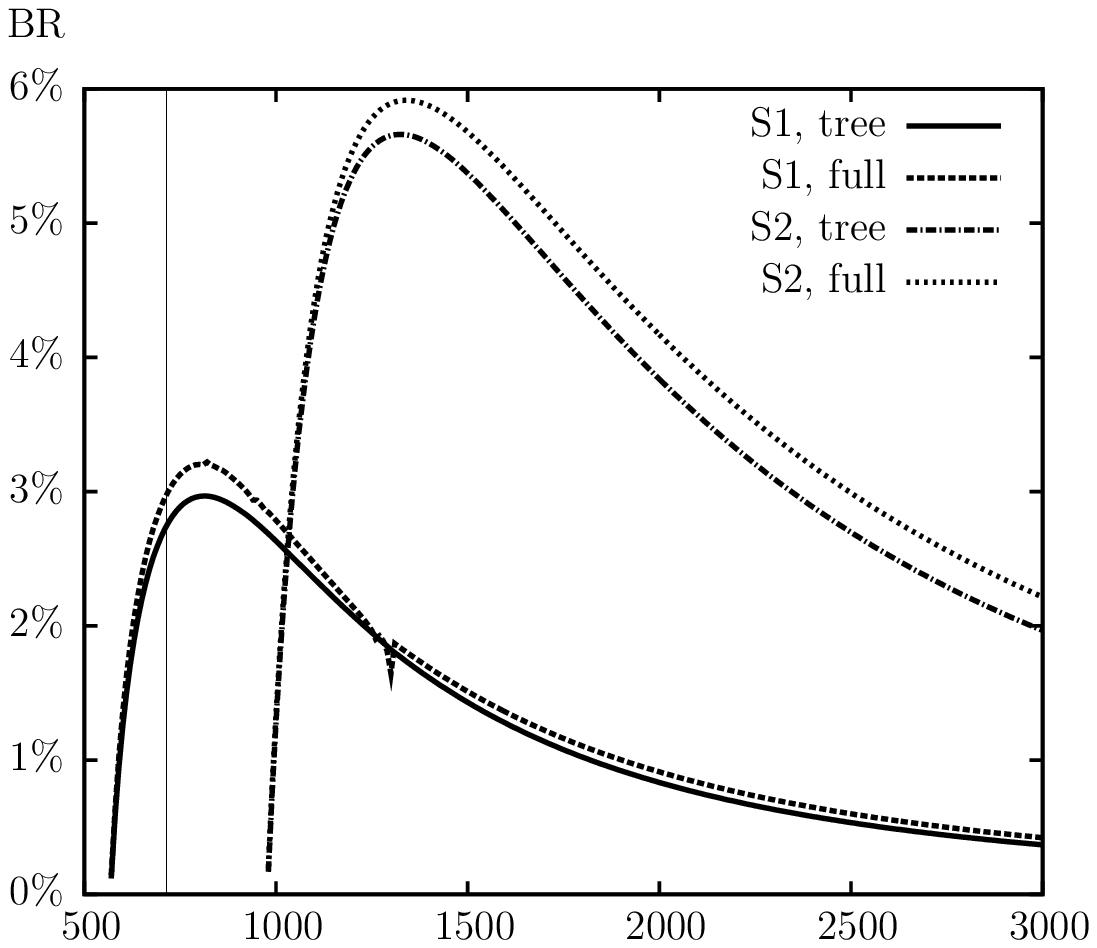}
\hspace{-4mm}
\includegraphics[width=0.49\textwidth,height=7.5cm]{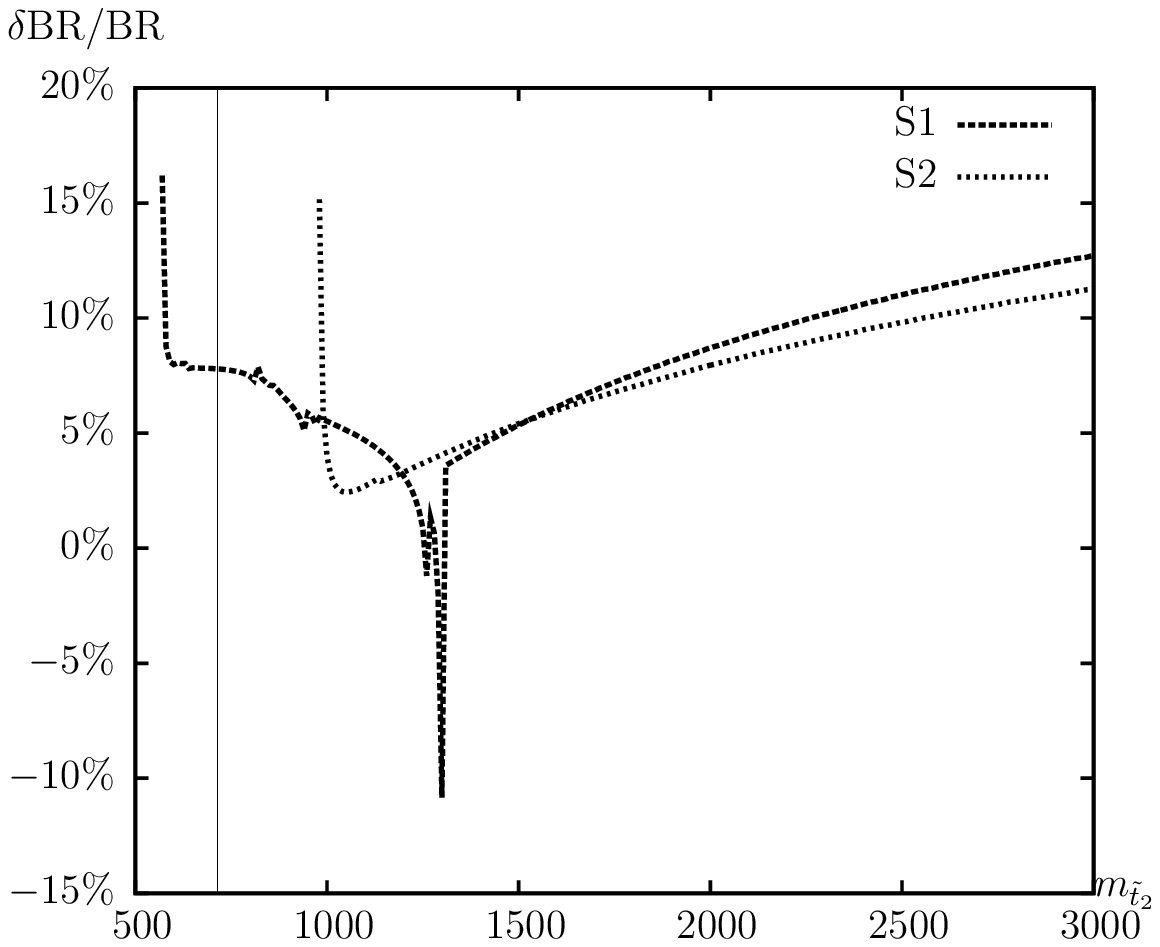}
\end{tabular}
\vspace{2em}
\caption{
  $\Ga(\decayh)$. Tree-level and full one-loop corrected partial decay widths 
  are shown with the parameters chosen according to \SE\ and \SZ\ 
  (see \refta{tab:para}), with $\mstz$ varied.
  The upper left plot shows the partial decay width; the upper right 
  plot shows the corresponding relative size of the corrections.
  The lower left plot shows the BR; the lower right plot shows 
  the relative correction of the BR.
  The vertical lines indicate where $\mstz + \mste = 1000 \gev$, 
  i.e.\ the maximum reach of the ILC(1000).
}
\label{fig:mst2.st2st1h1}
\end{center}
\end{figure}

\begin{figure}[htb!]
\begin{center}
\begin{tabular}{c}
\includegraphics[width=0.49\textwidth,height=7.5cm]{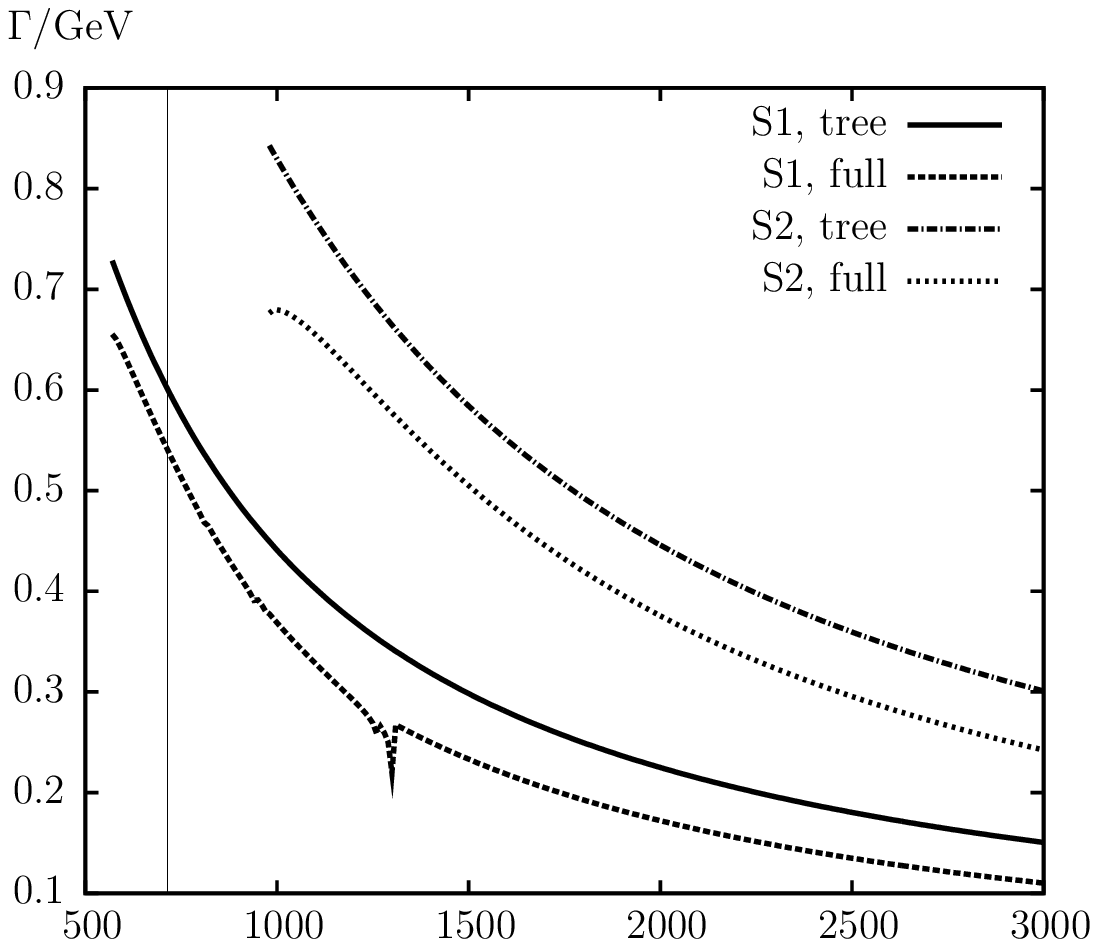}
\hspace{-4mm}
\includegraphics[width=0.49\textwidth,height=7.5cm]{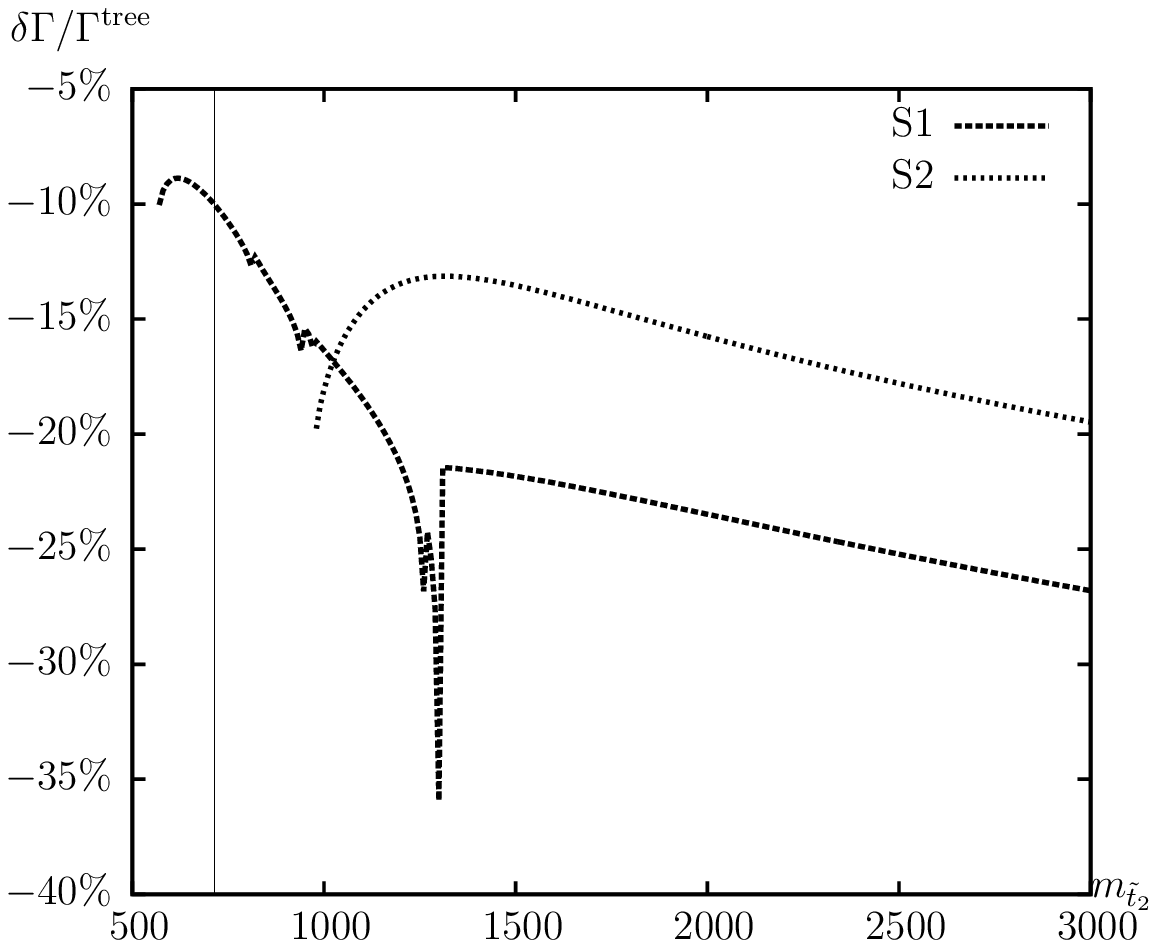} 
\\[4em]
\includegraphics[width=0.49\textwidth,height=7.5cm]{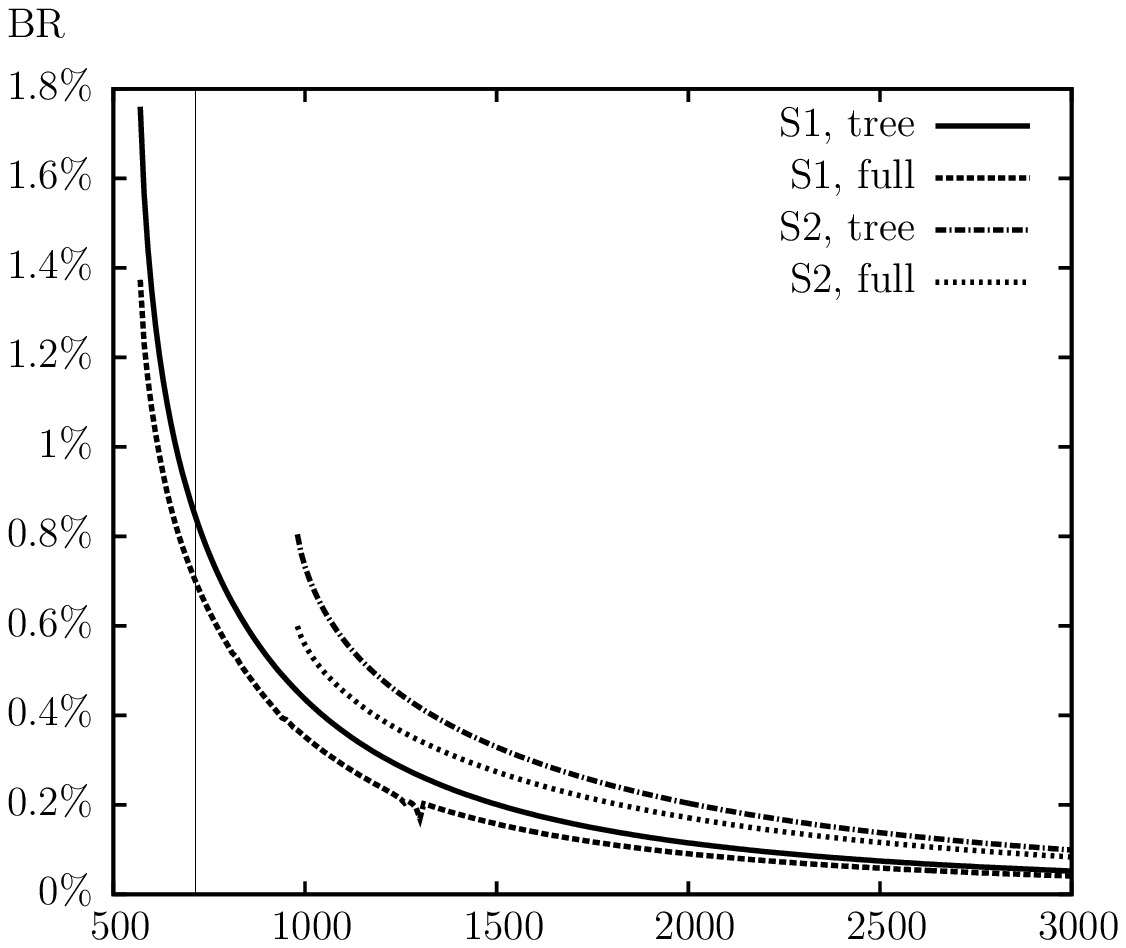}
\hspace{-4mm}
\includegraphics[width=0.49\textwidth,height=7.5cm]{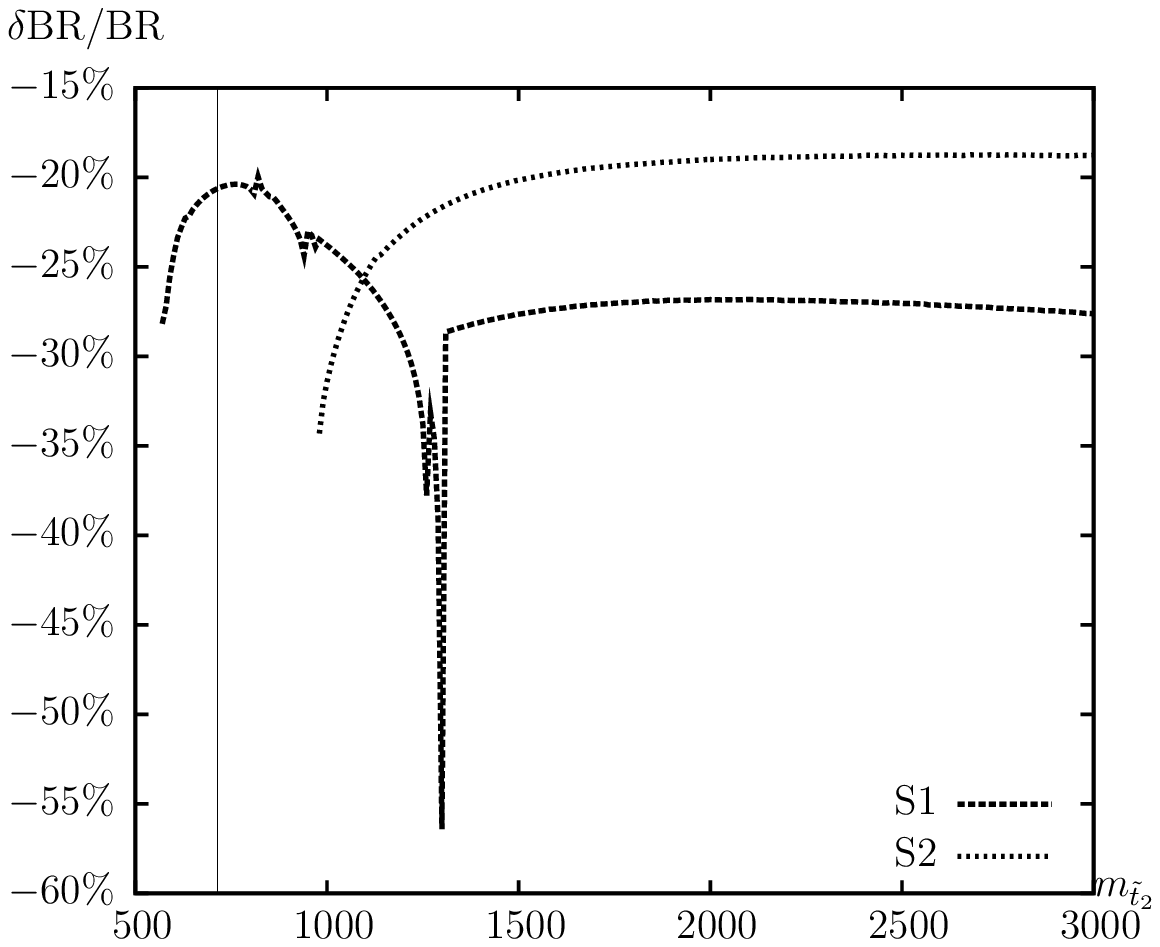}
\end{tabular}
\vspace{2em}
\caption{
  $\Ga(\decayH)$. Tree-level and full one-loop corrected partial decay widths 
  are shown with the parameters chosen according to \SE\ and \SZ\ 
  (see \refta{tab:para}), with $\mstz$ varied.
  The upper left plot shows the partial decay width; the upper right 
  plot shows the corresponding relative size of the corrections.
  The lower left plot shows the BR; the lower right plot shows 
  the relative correction of the BR.
  The vertical lines indicate where $\mstz + \mste = 1000 \gev$, 
  i.e.\ the maximum reach of the ILC(1000).
}
\label{fig:mst2.st2st1h2}
\end{center}
\end{figure}

\begin{figure}[htb!]
\begin{center}
\begin{tabular}{c}
\includegraphics[width=0.49\textwidth,height=7.5cm]{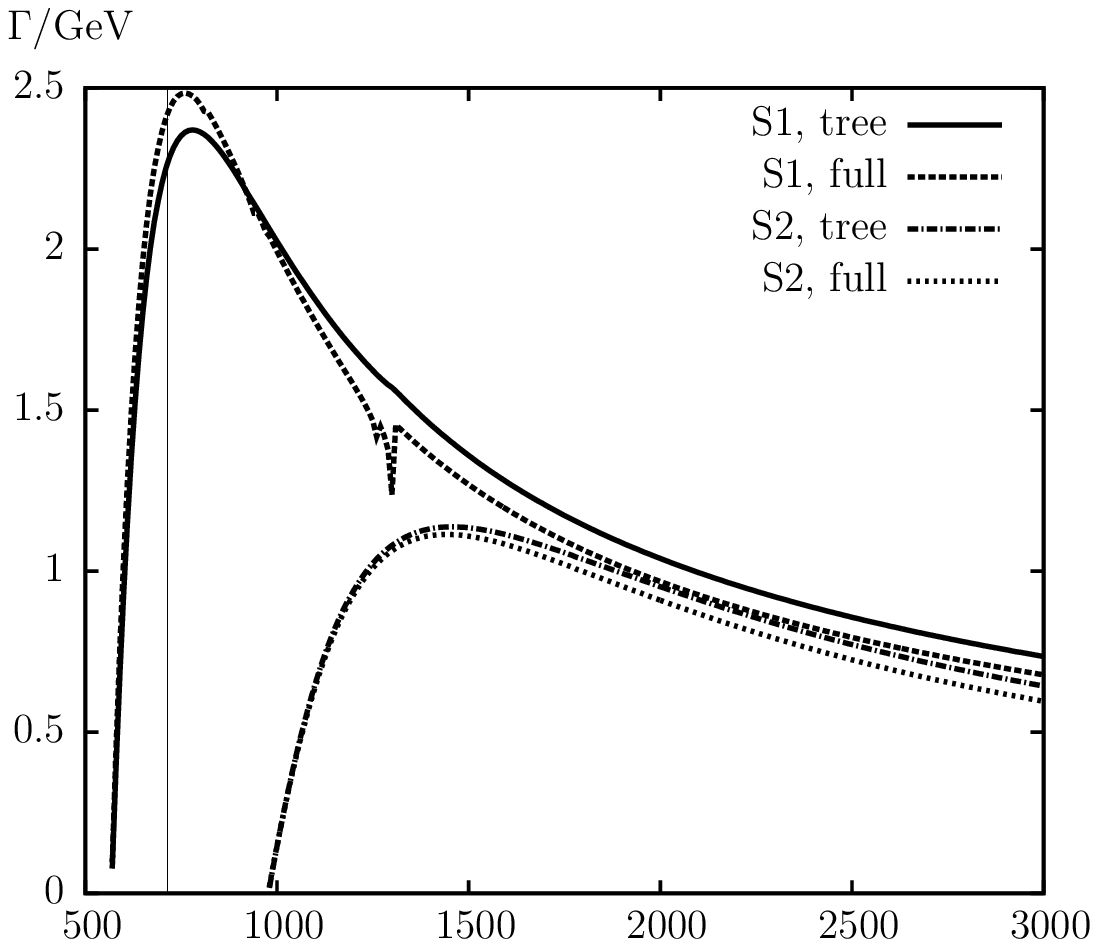}
\hspace{-4mm}
\includegraphics[width=0.49\textwidth,height=7.5cm]{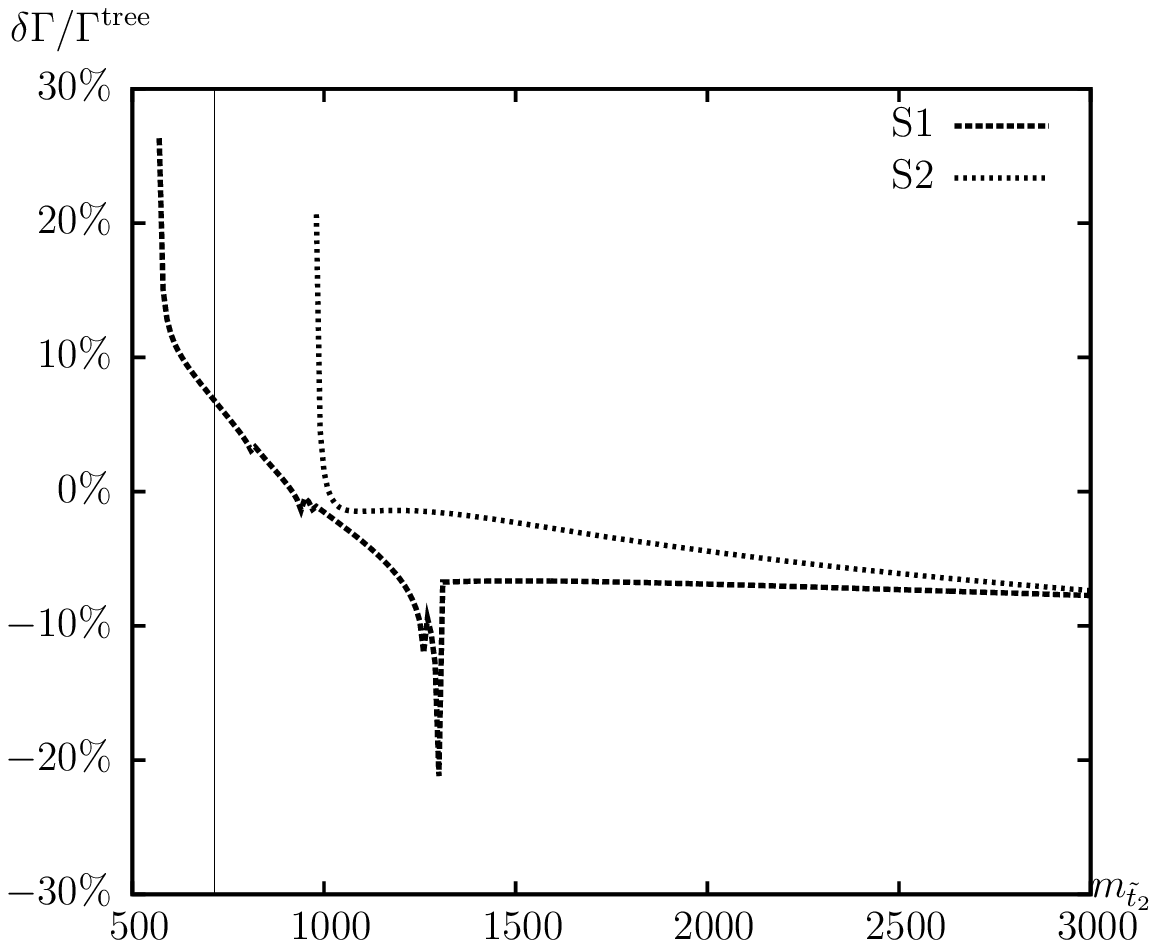} 
\\[4em]
\includegraphics[width=0.49\textwidth,height=7.5cm]{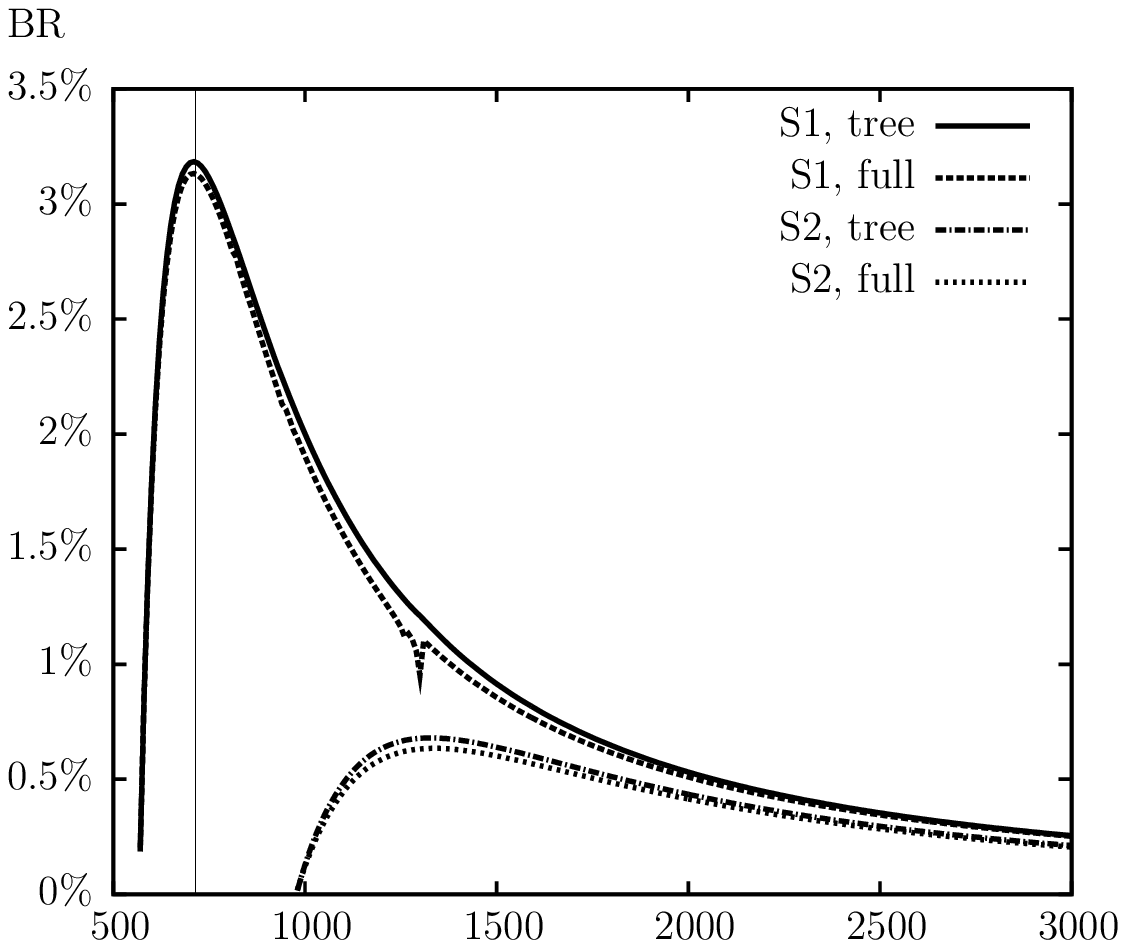}
\hspace{-4mm}
\includegraphics[width=0.49\textwidth,height=7.5cm]{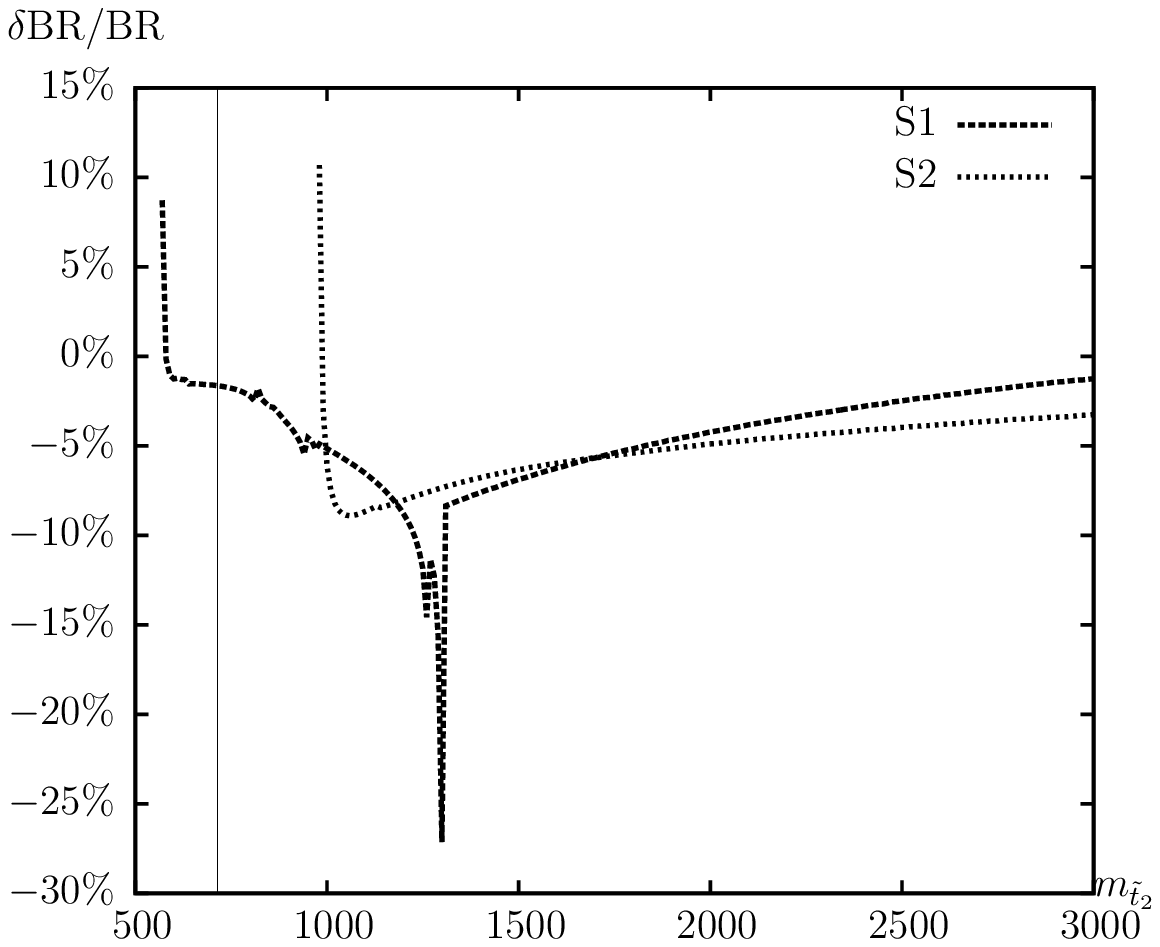}
\end{tabular}
\vspace{2em}
\caption{
  $\Ga(\decayA)$. Tree-level and full one-loop corrected partial decay widths 
  are shown with the parameters chosen according to \SE\ and \SZ\ 
  (see \refta{tab:para}), with $\mstz$ varied.
  The upper left plot shows the partial decay width; the upper right 
  plot shows the corresponding relative size of the corrections. 
  The lower left plot shows the BR; the lower right plot shows 
  the relative correction of the BR.
  The vertical lines indicate where $\mstz + \mste = 1000 \gev$, 
  i.e.\ the maximum reach of the ILC(1000).
}
\label{fig:mst2.st2st1h3}
\end{center}
\end{figure}

In \reffis{fig:mst2.st2st1h1} -- \ref{fig:mst2.st2st1h3} we show the 
results for the process $\decayhn$ ($n = 1,2,3$) as a function of $\mstz$. 
These are of particular interest for LHC 
analyses~\cite{stopstophiggs-LHC, Higgsincascades} 
(as emphasized in the Introduction).
The dips at $\mstz \approx 819, 948, 971, 1264, 1303 \gev$ 
(for all three figures) in the scenario \SE\ are effects
due to the thresholds $\mt + \mneu{1,2,3,4} = \mste$ and $\mt + \mgl = \mste$ 
(in this order) of the self-energies $\Si_{\Stop_{11,21}}(\mste^2)$ 
in the renormalization constants $\dZZm{\Stop}_{11,21}$, $\de Y_t$,
and $\de\mste^2$.
One can see that the size of the corrections of the  partial decay widths
is especially large very close to the production threshold%
\footnote{
  It should be noted that a calculation very close to the production 
  threshold requires the inclusion of additional (nonrelativistic) 
  contributions, which is beyond the scope of this paper. 
  Consequently, very close to the production threshold our calculation 
  (at the tree- or loop-level) does not provide a very accurate 
  description of the decay width.
}
from which on the considered decay mode is kinematically possible. 
Away from this threshold relative corrections of 
$\sim +10\%, -20\%, -5\%$ are found for $h_1, h_2, h_3$, respectively. 
In (all) the plots the value of $\mstz$ for which 
$\mste + \mstz = 1000 \gev$ is shown as a vertical line, 
i.e.\ the region where the heavier stop can be produced at the
ILC(1000). In these regions the size 
of the corrections amounts up to $\sim +20\%, -10\%, +10\%$
for the three neutral Higgs bosons. 
The BRs are at the few percent level for all three channels for 
the two numerical scenarios. The relative change in the BRs for the 
masses accessible at the ILC(1000) are about $+8\%$, $-21\%$, $-1\%$
for $h_1$, $h_2$, $h_3$, respectively. For lager masses, only accessible at
the LHC, the one-loop corrections are around $+10\%$, $-25\%$, 
and $-5\%$. Depending on the MSSM parameters (and the channels 
kinematically allowed) the one-loop contributions presented here can be
relevant for analyses at the ILC as well as at the LHC.

\begin{figure}[htb!]
\begin{center}
\begin{tabular}{c}
\includegraphics[width=0.49\textwidth,height=7.5cm]{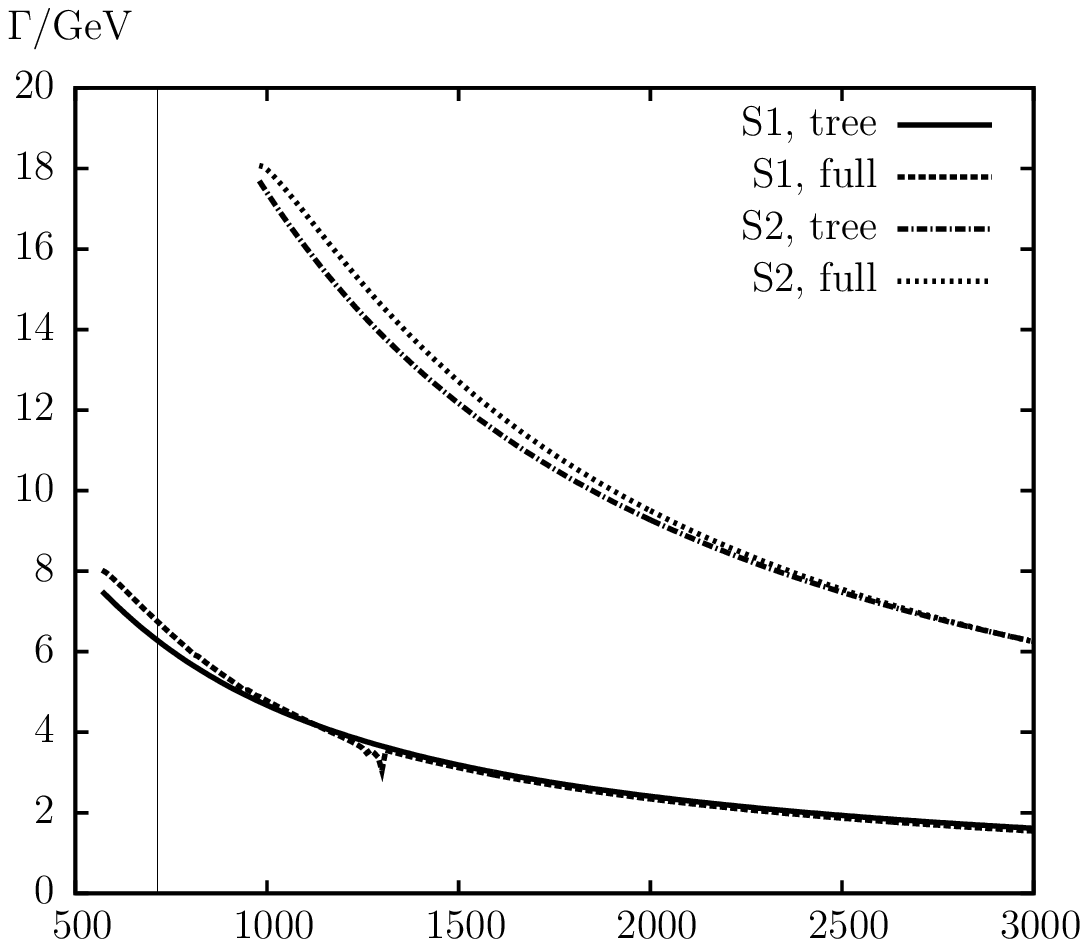}
\hspace{-4mm}
\includegraphics[width=0.49\textwidth,height=7.5cm]{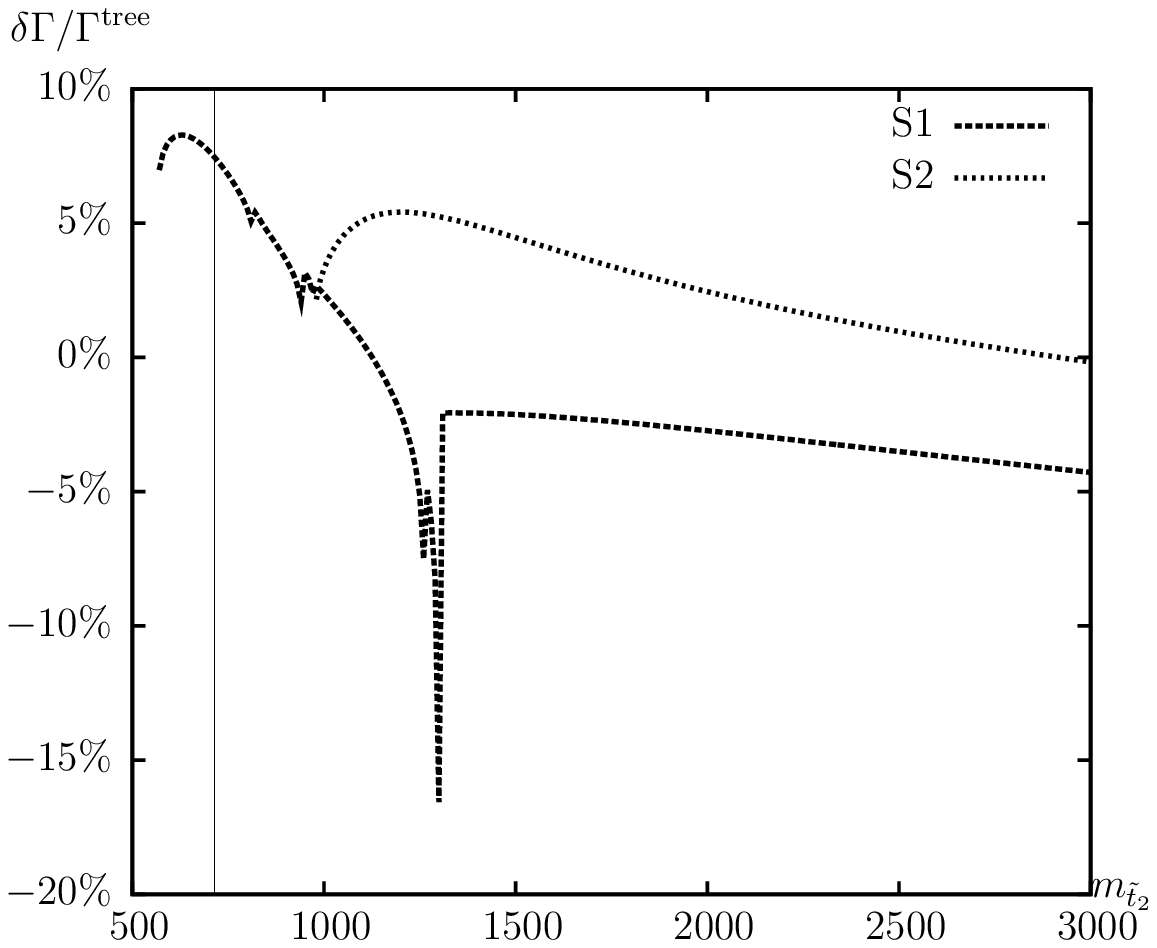} 
\\[4em]
\includegraphics[width=0.49\textwidth,height=7.5cm]{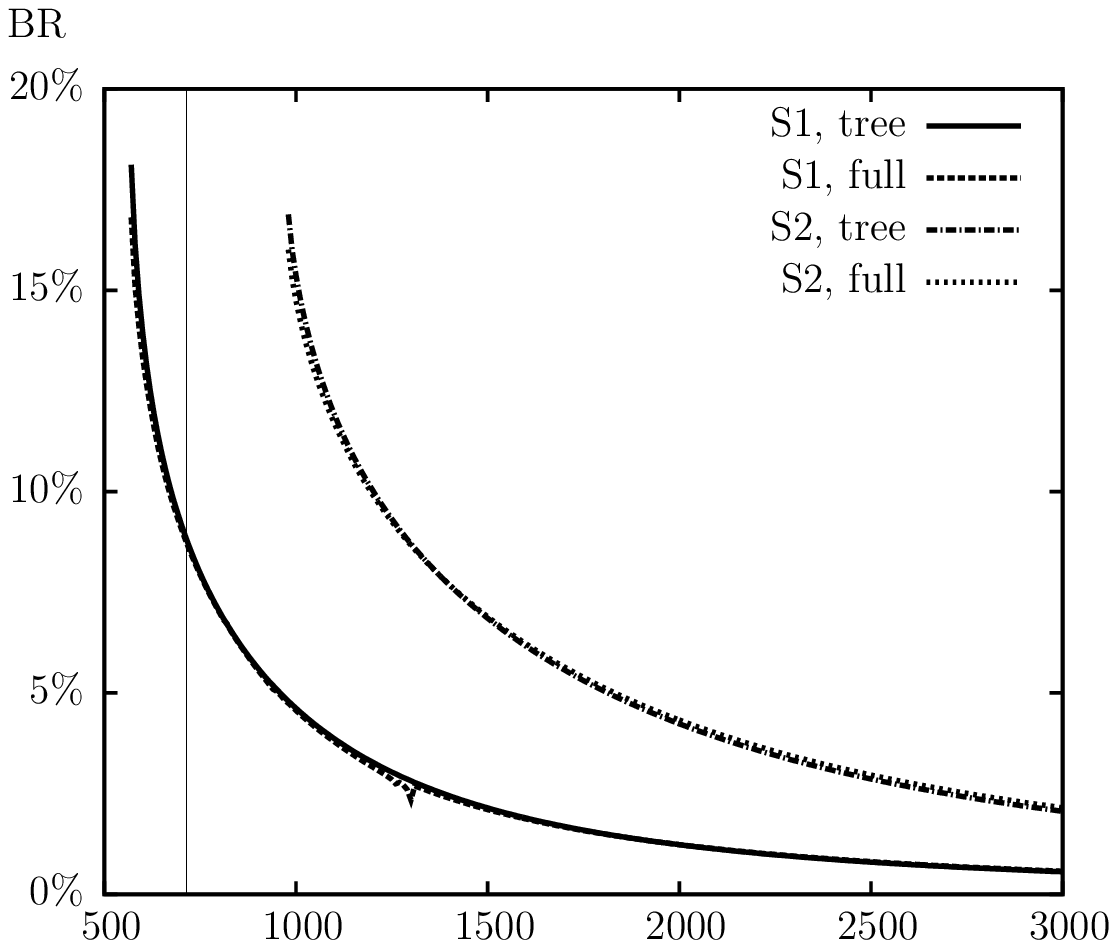}
\hspace{-4mm}
\includegraphics[width=0.49\textwidth,height=7.5cm]{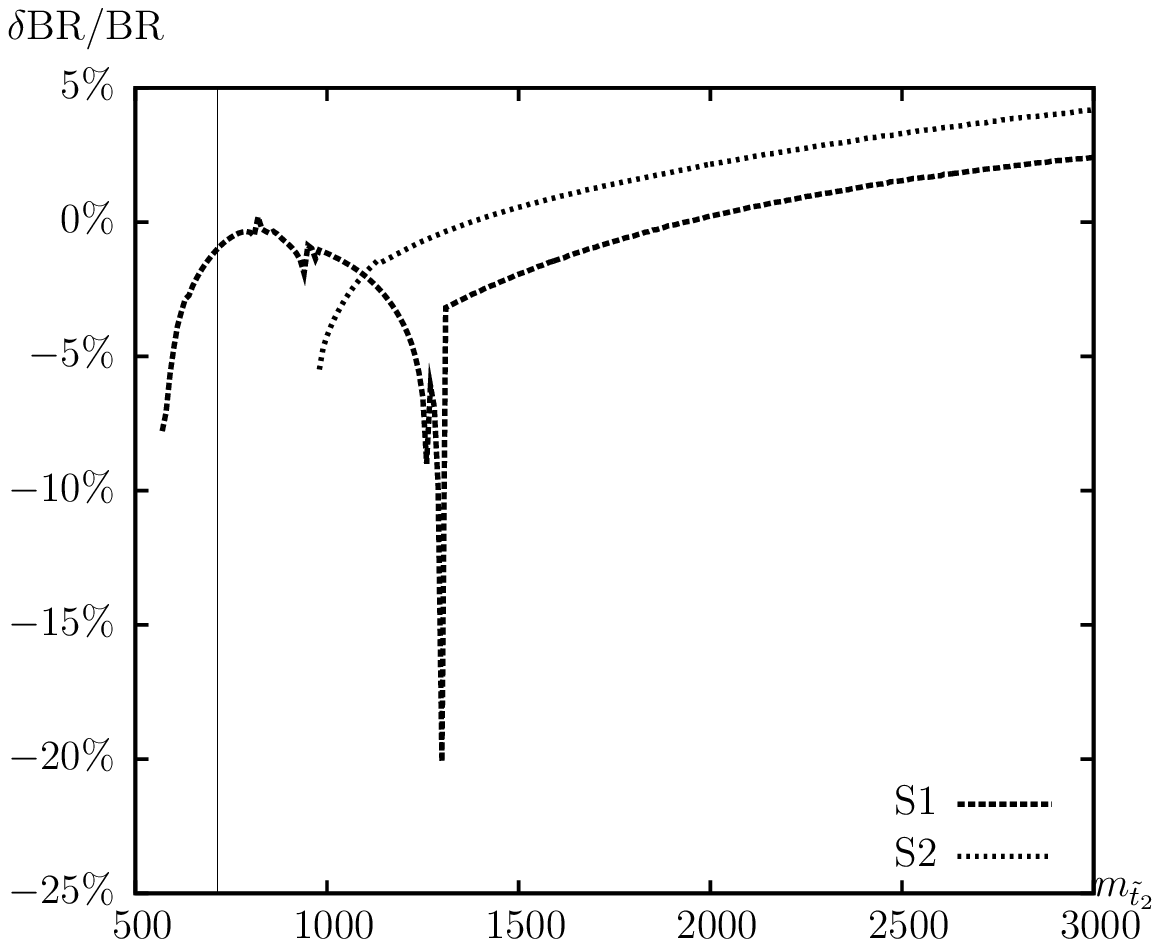}
\end{tabular}
\vspace{2em}
\caption{
  $\Ga(\decayZ)$. Tree-level and full one-loop corrected partial decay widths 
  are shown with the parameters chosen according to \SE\ and \SZ\ 
  (see \refta{tab:para}), with $\mstz$ varied.
  The upper left plot shows the partial decay width; the upper right 
  plot shows the corresponding relative size of the corrections. 
  The lower left plot shows the BR; the lower right plot shows 
  the relative correction of the BR.
  The vertical lines indicate where $\mstz + \mste = 1000 \gev$, 
  i.e.\ the maximum reach of the ILC(1000).
}
\label{fig:mst2.st2st1Z}
\end{center}
\end{figure}

Next, in \reffi{fig:mst2.st2st1Z} we show results for the decay 
$\Ga(\decayZ)$. The dips due to the thresholds in $\dZZm{\Stop}_{11,21}$, 
$\de Y_t$, and $\de\mste^2$ are the same as before. 
The relative corrections to the partial decay width in \SE\ range between 
+8\% at low $\mstz$, i.e.\ in the ``ILC(1000) regime'',  
to $-5\%$ at large $\mstz$, with the exception of the region close to
thresholds. Within \SZ\ the relative corrections stay below $\sim 5\%$.
The $\br(\decayZ)$ is larger than $15\%$ for the smallest $\mstz$
values in the two scenarios. This drops below $3\%$ for 
$\mstz \gsim 2.5 \tev$. The relative change for masses accessible at the
ILC(1000) is found at the few percent level.

\begin{figure}[htb!]
\begin{center}
\begin{tabular}{c}
\includegraphics[width=0.49\textwidth,height=7.5cm]{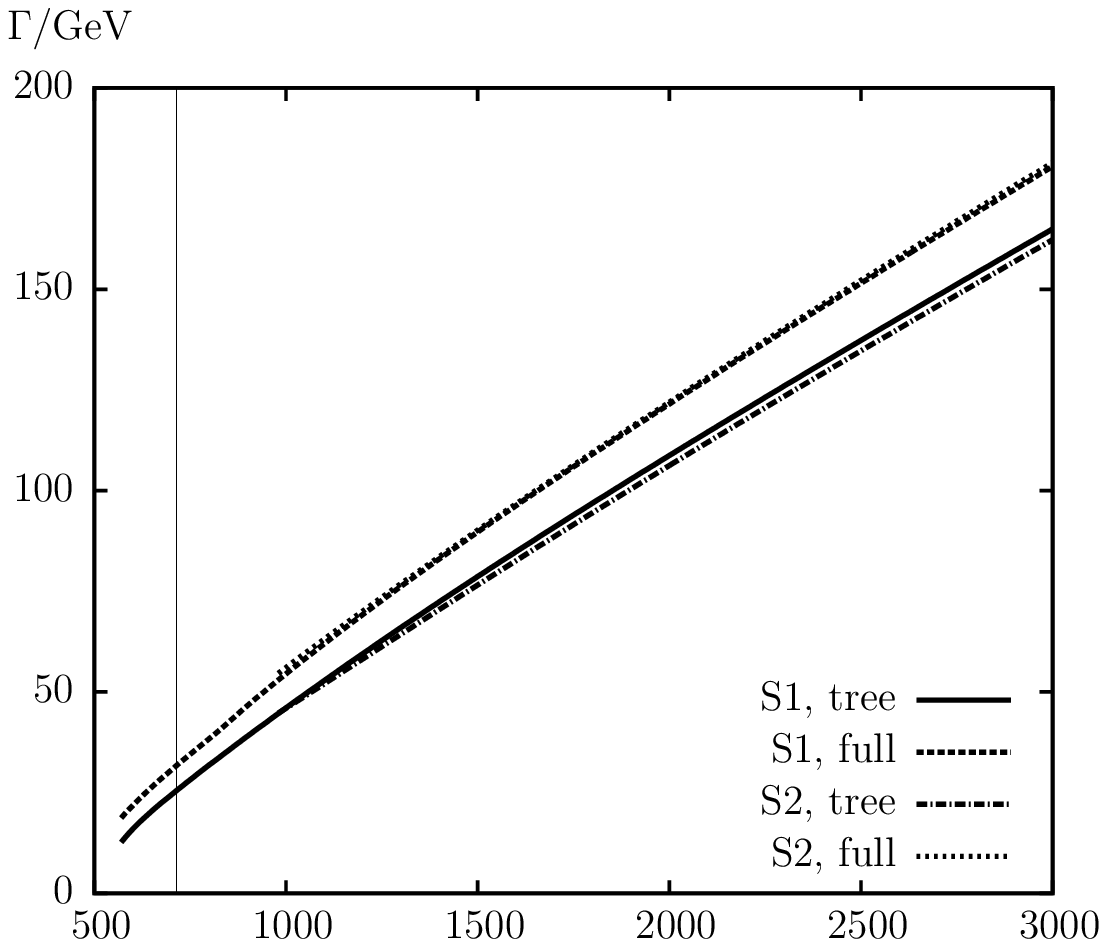}
\hspace{-4mm}
\includegraphics[width=0.49\textwidth,height=7.5cm]{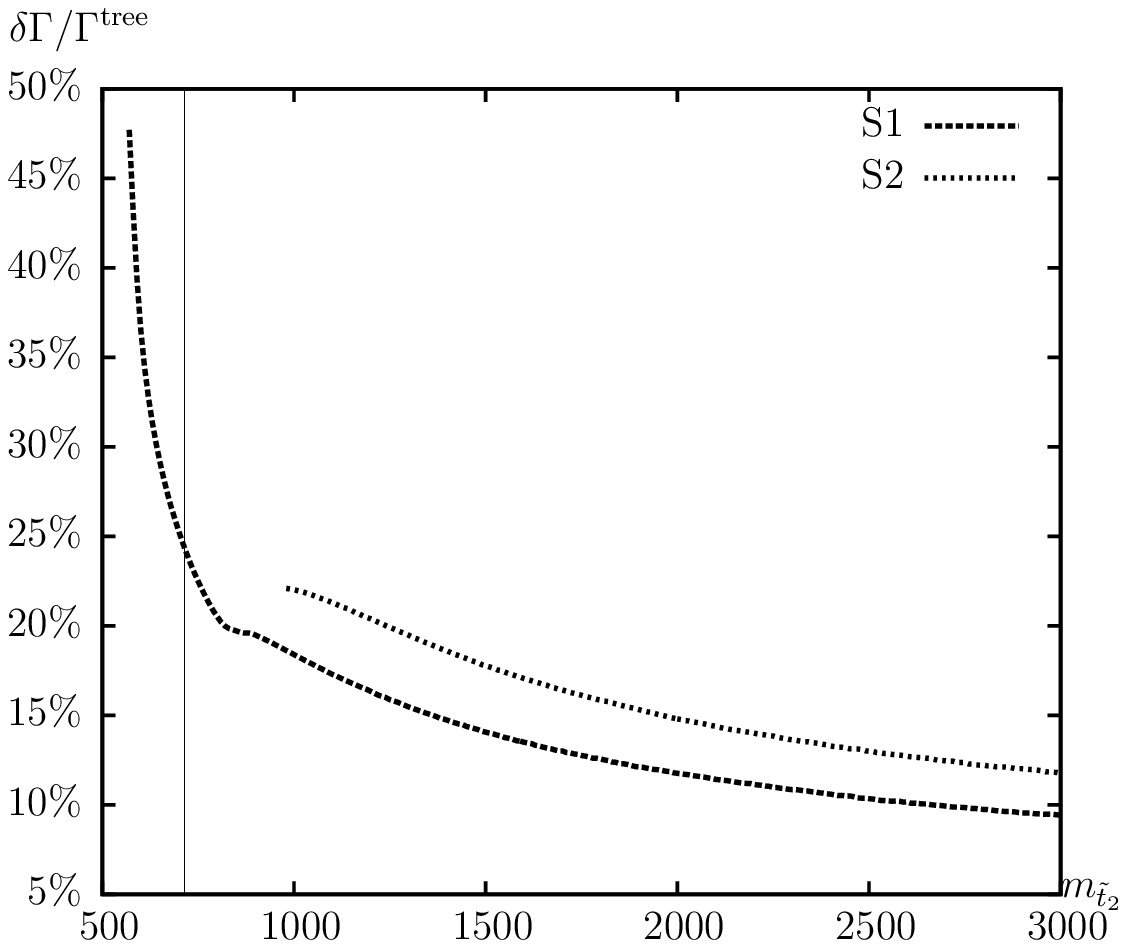}
\\[4em]
\includegraphics[width=0.49\textwidth,height=7.5cm]{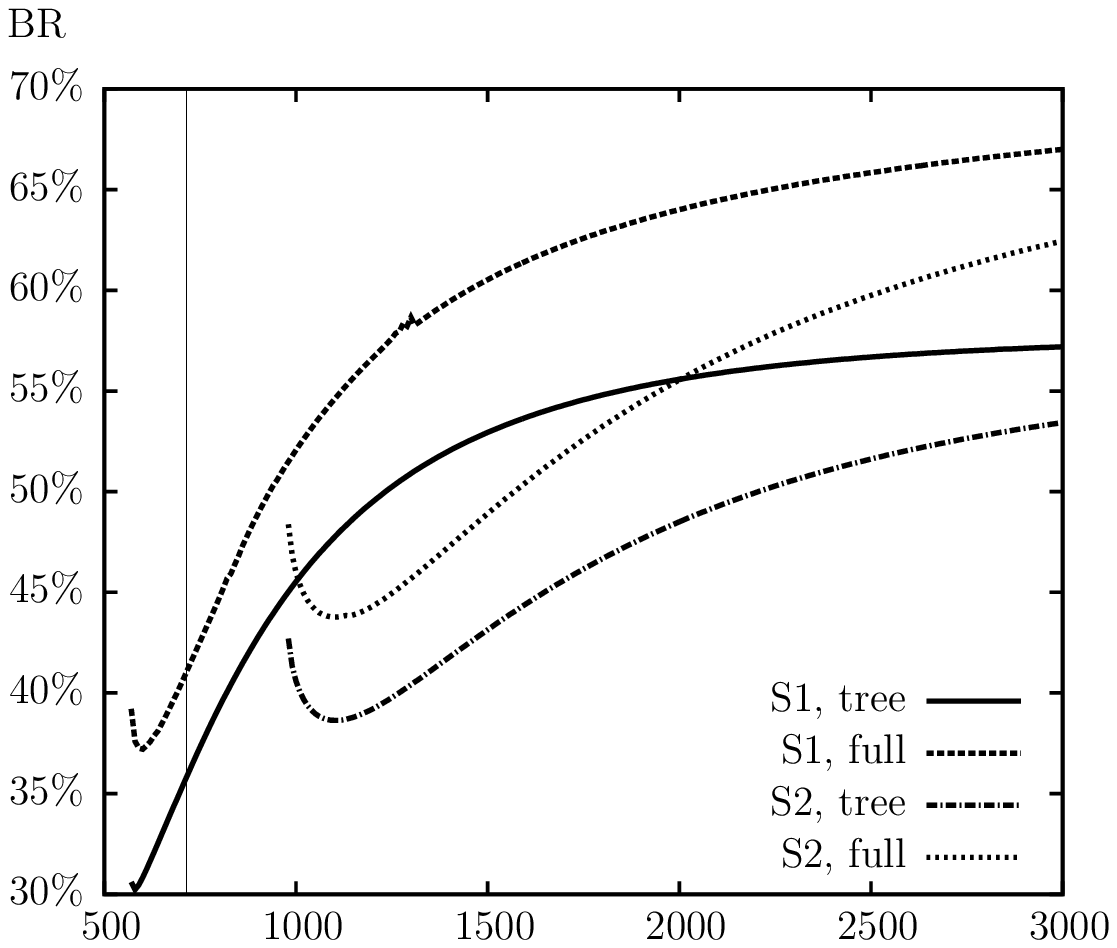}
\hspace{-4mm}
\includegraphics[width=0.49\textwidth,height=7.5cm]{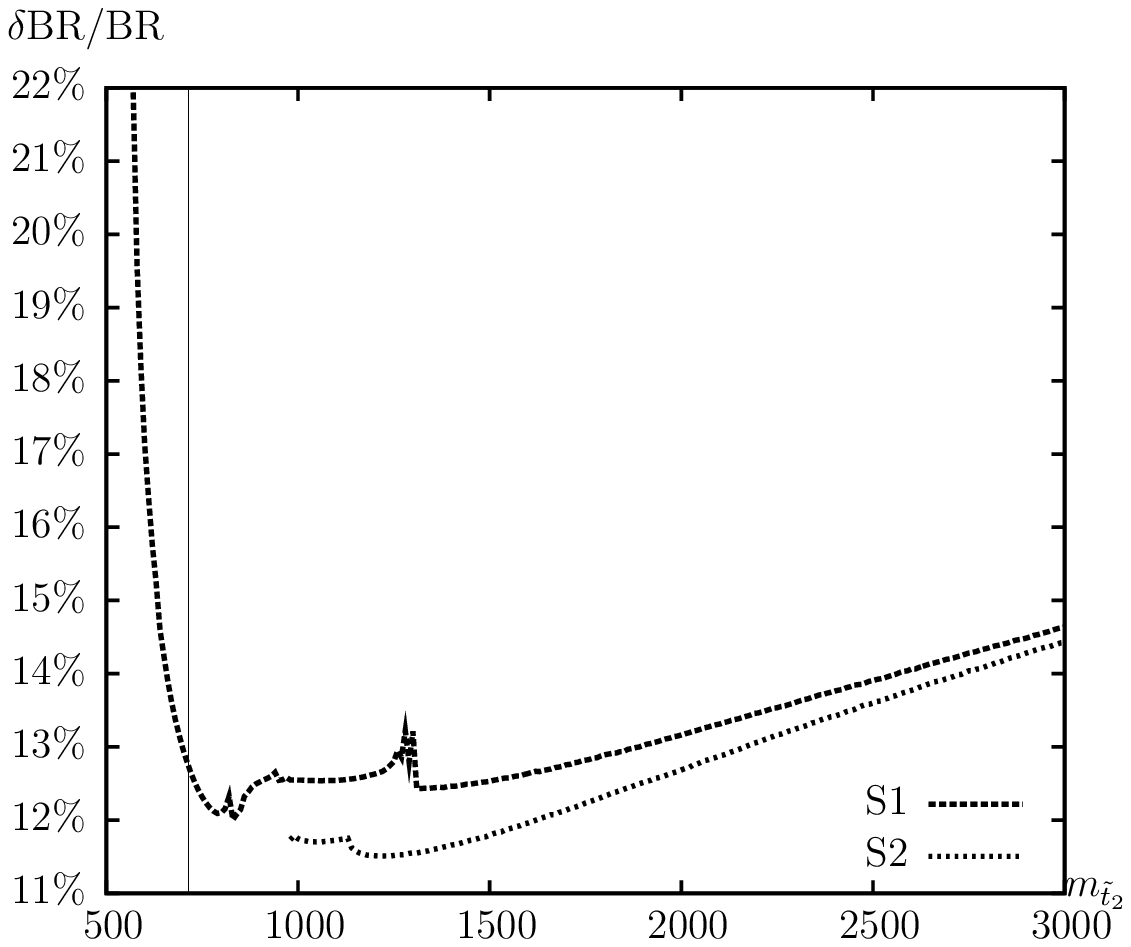}
\end{tabular}
\vspace{2em}
\caption{
  $\Ga(\decaygl)$. Tree-level and full one-loop corrected partial decay widths 
  are shown with the parameters chosen according to \SE\ and \SZ\ 
  (see \refta{tab:para}), with $\mstz$ varied.
  The upper left plot shows the partial decay width; the upper right 
  plot shows the corresponding relative size of the corrections.
  The lower left plot shows the BR; the lower right plot shows 
  the relative correction of the BR.
  The vertical lines indicate where $\mstz + \mste = 1000 \gev$, 
  i.e.\ the maximum reach of the ILC(1000).
}
\label{fig:mst2.st2tgl}
\end{center}
\end{figure}

The results for the decay $\decaygl$ are presented in
\reffi{fig:mst2.st2tgl}. We see that for the relative corrections of the
partial decay width up to 48\% (22\%) are reached 
for $\mstz = 570\ (980) \gev$ in \SE\ (\SZ), i.e.\ at the smallest
possible value and decrease for increasing $\mstz$. It should be noted
that in this case the hard and soft QCD radiation can be very large and
the two compensate each other. 
The BR turns out to be very large and growing with $\mstz$, 
where values larger than $50\%$ are found. 
Within \SE\ the relative corrections can reach up to $+20\%$ in the
production threshold region and 
are larger than $+12\%$ in the parameter space accessible at the
ILC(1000). For large $\mstz$ the corrections range between 
$+11\%$ and $+15\%$ in the two scenarios.

\begin{figure}[htb!]
\begin{center}
\begin{tabular}{c}
\includegraphics[width=0.49\textwidth,height=7.5cm]{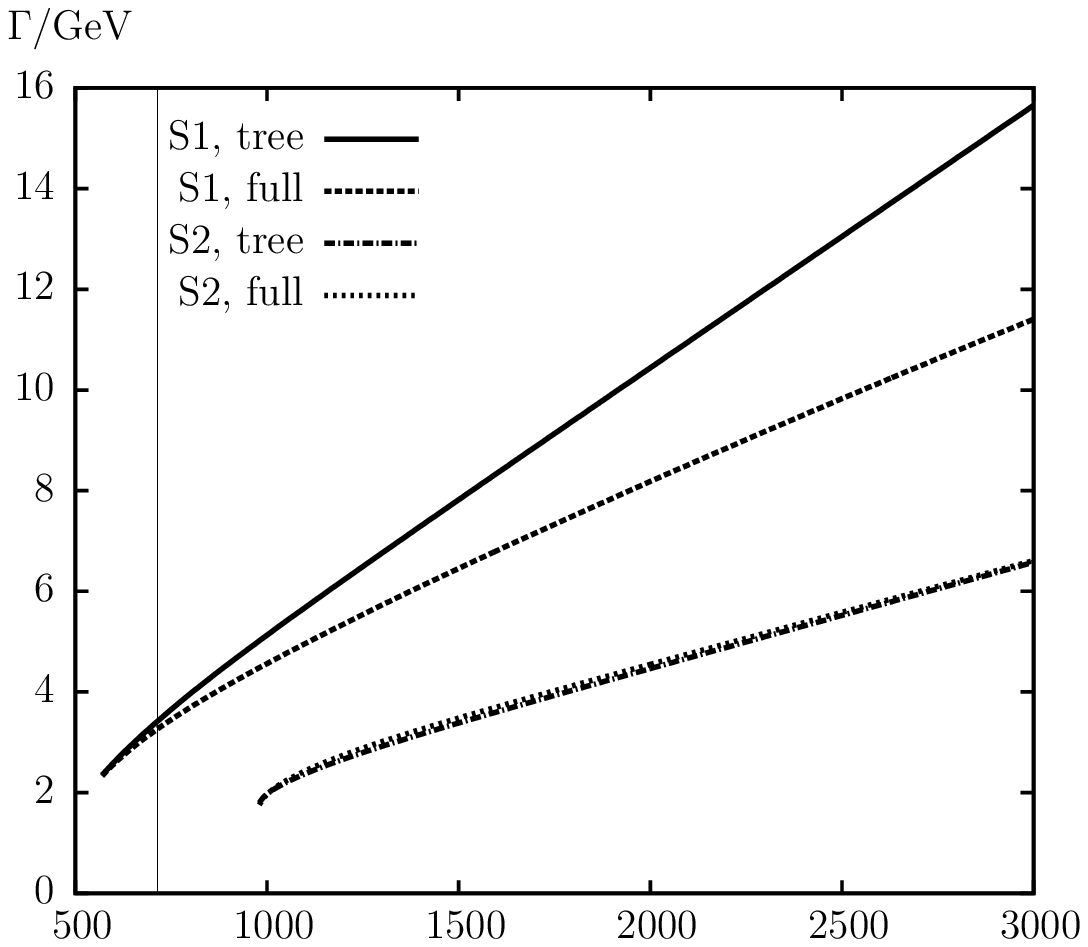}
\hspace{-4mm}
\includegraphics[width=0.49\textwidth,height=7.5cm]{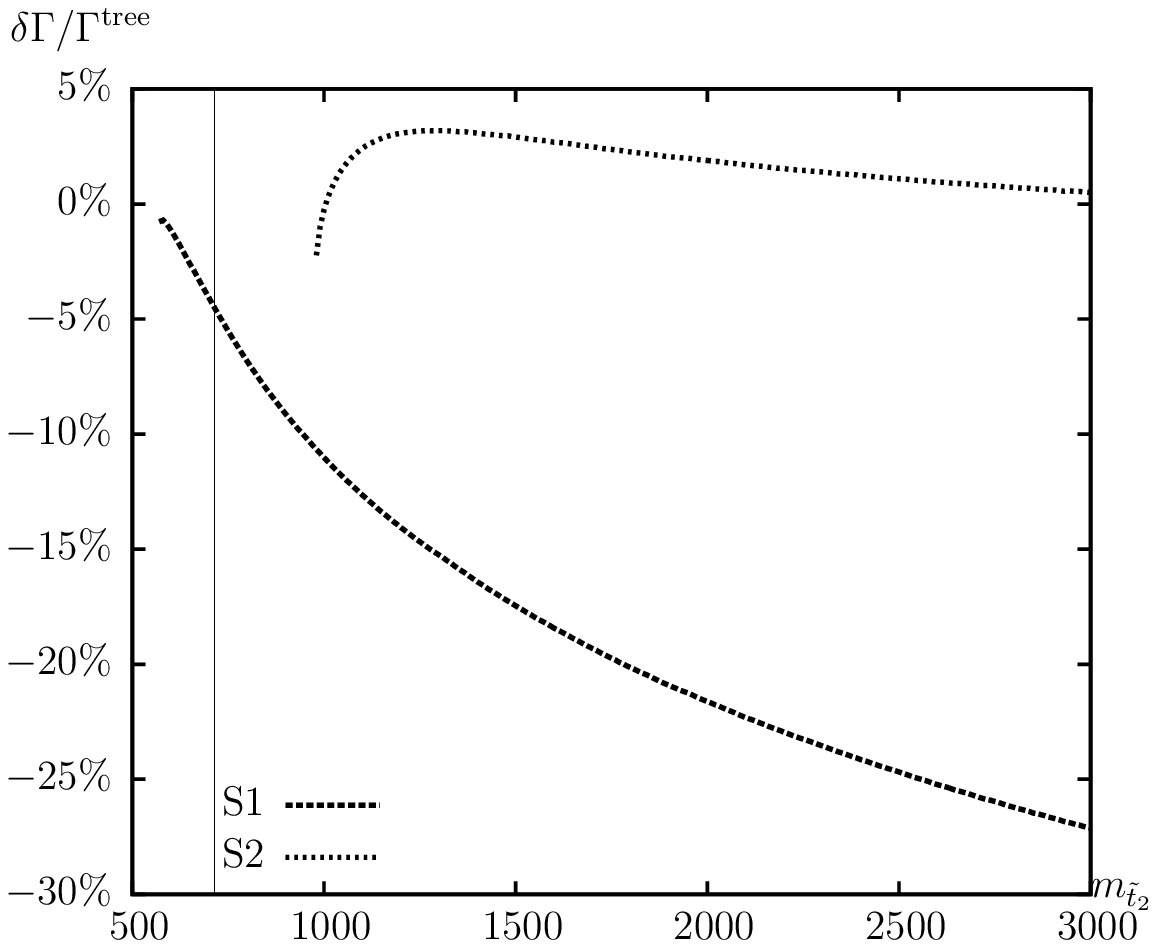}
\\[4em]
\includegraphics[width=0.49\textwidth,height=7.5cm]{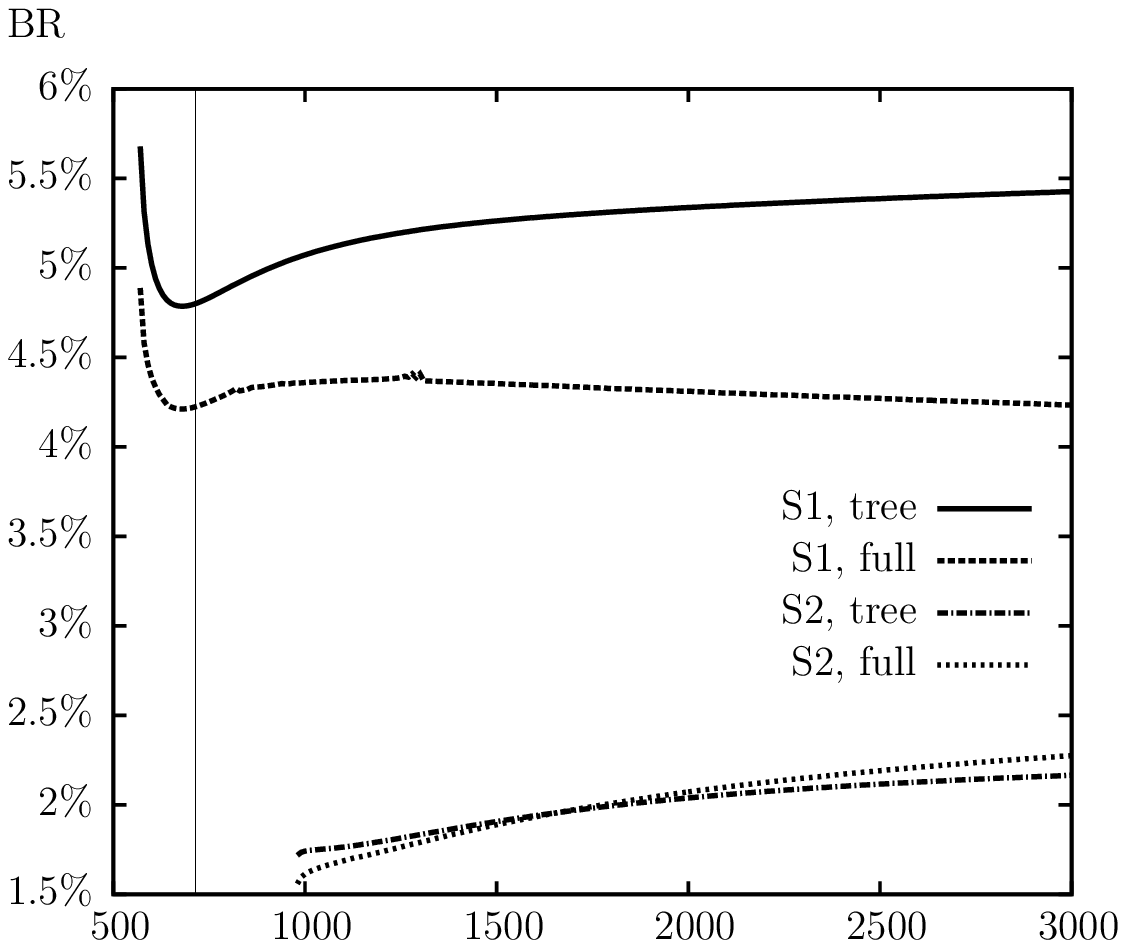}
\hspace{-4mm}
\includegraphics[width=0.49\textwidth,height=7.5cm]{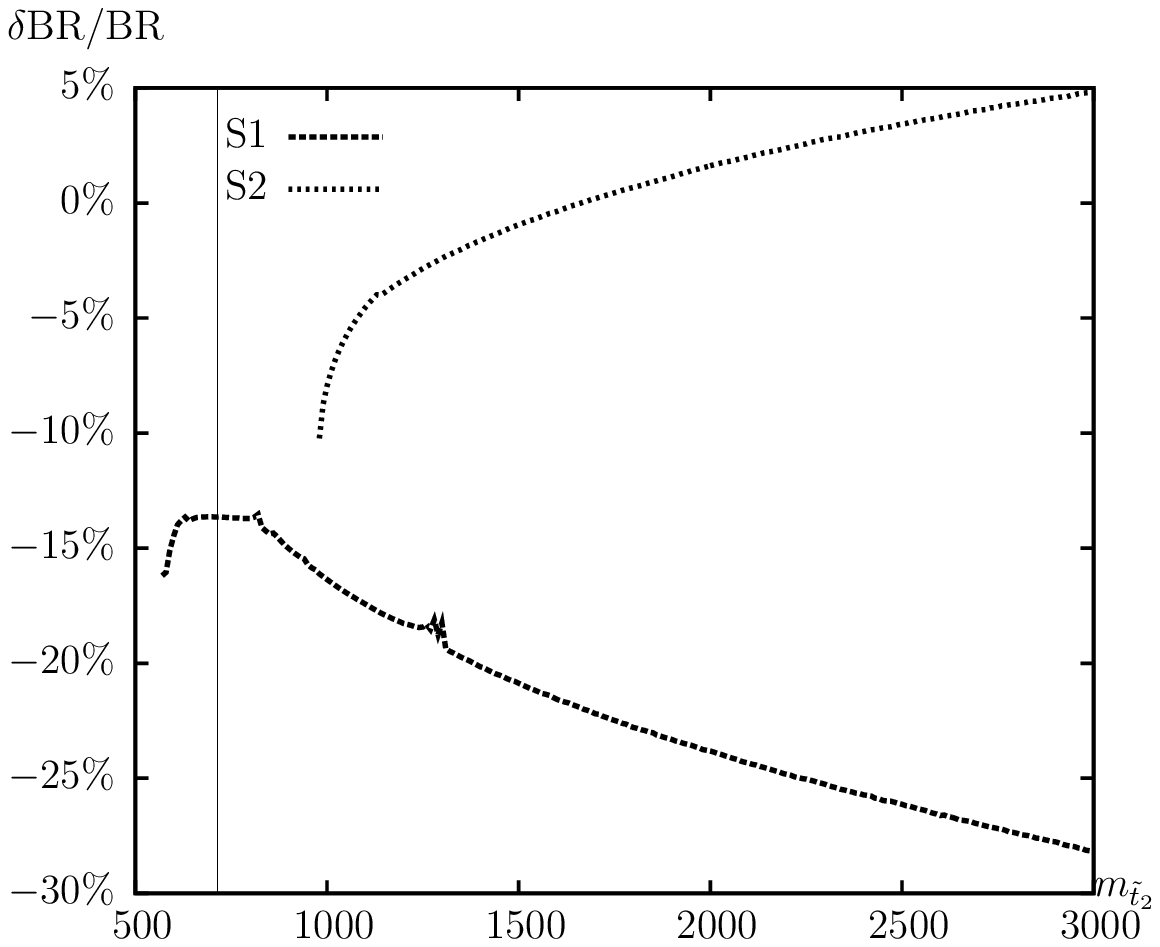}
\end{tabular}
\vspace{2em}
\caption{
  $\Ga(\decayNe)$. Tree-level and full one-loop corrected partial decay widths 
  are shown with the parameters chosen according to \SE\ and \SZ\ 
  (see \refta{tab:para}), with $\mstz$ varied.
  The upper left plot shows the partial decay width; the upper right 
  plot shows the corresponding relative size of the corrections.
  The lower left plot shows the BR; the lower right plot shows 
  the relative correction of the BR.
  The vertical lines indicate where $\mstz + \mste = 1000 \gev$, 
  i.e.\ the maximum reach of the ILC(1000).
}
\label{fig:mst2.st2tneu1}
\end{center}
\end{figure}

\begin{figure}[htb!]
\begin{center}
\begin{tabular}{c}
\includegraphics[width=0.49\textwidth,height=7.5cm]{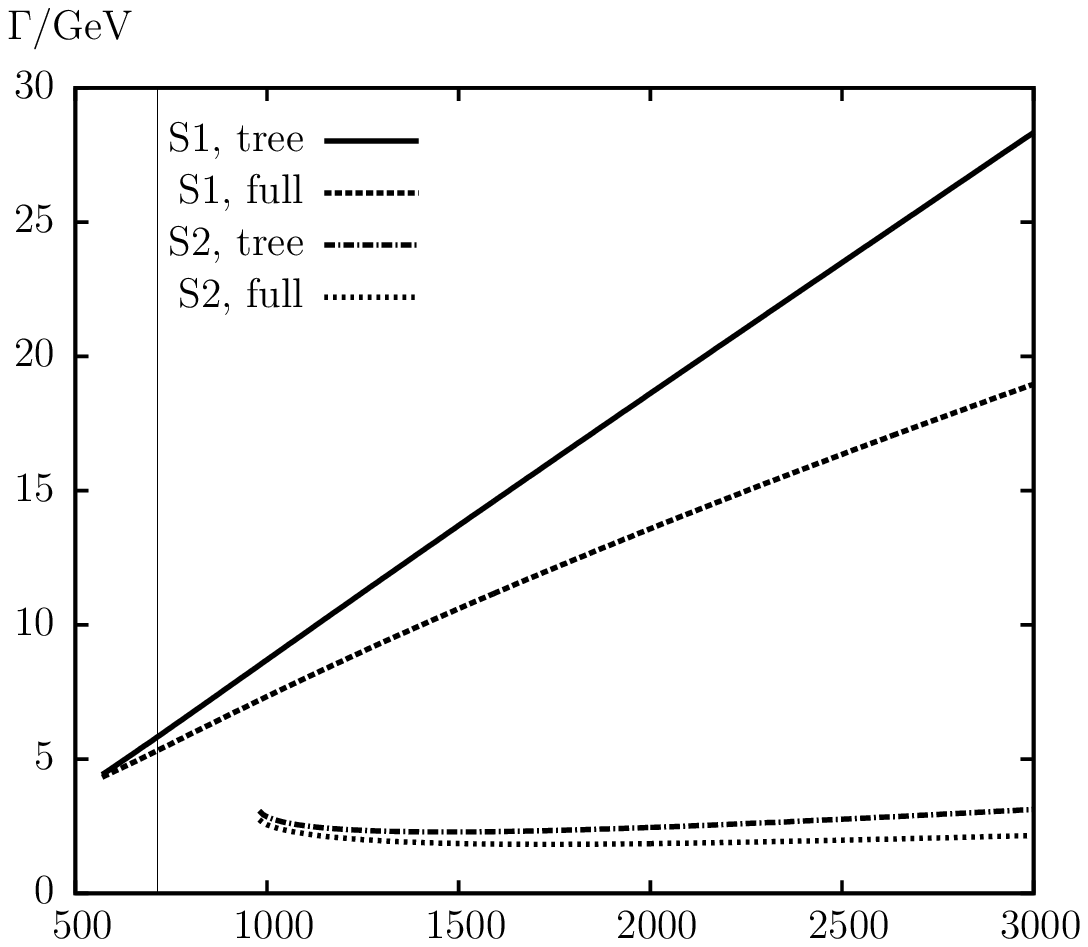}
\hspace{-4mm}
\includegraphics[width=0.49\textwidth,height=7.5cm]{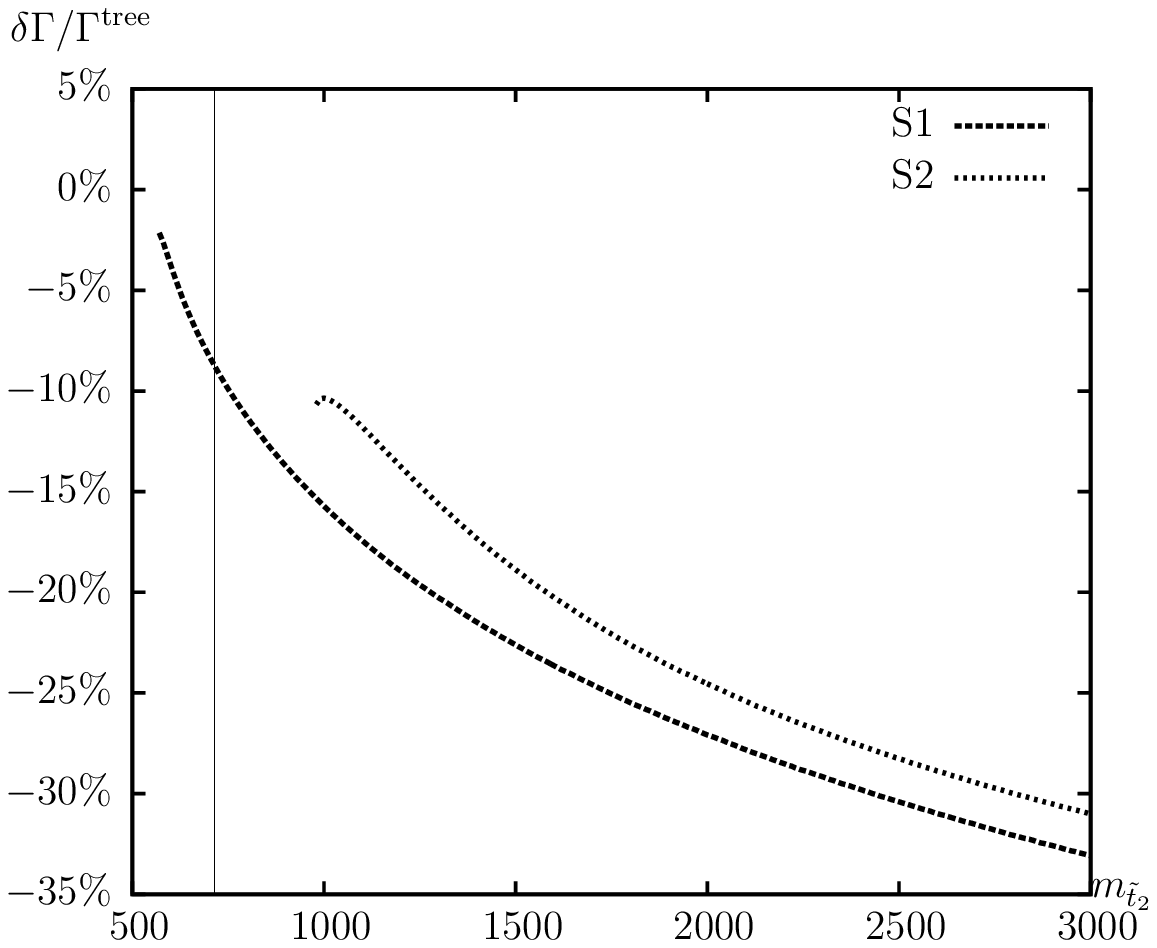}
\\[4em]
\includegraphics[width=0.49\textwidth,height=7.5cm]{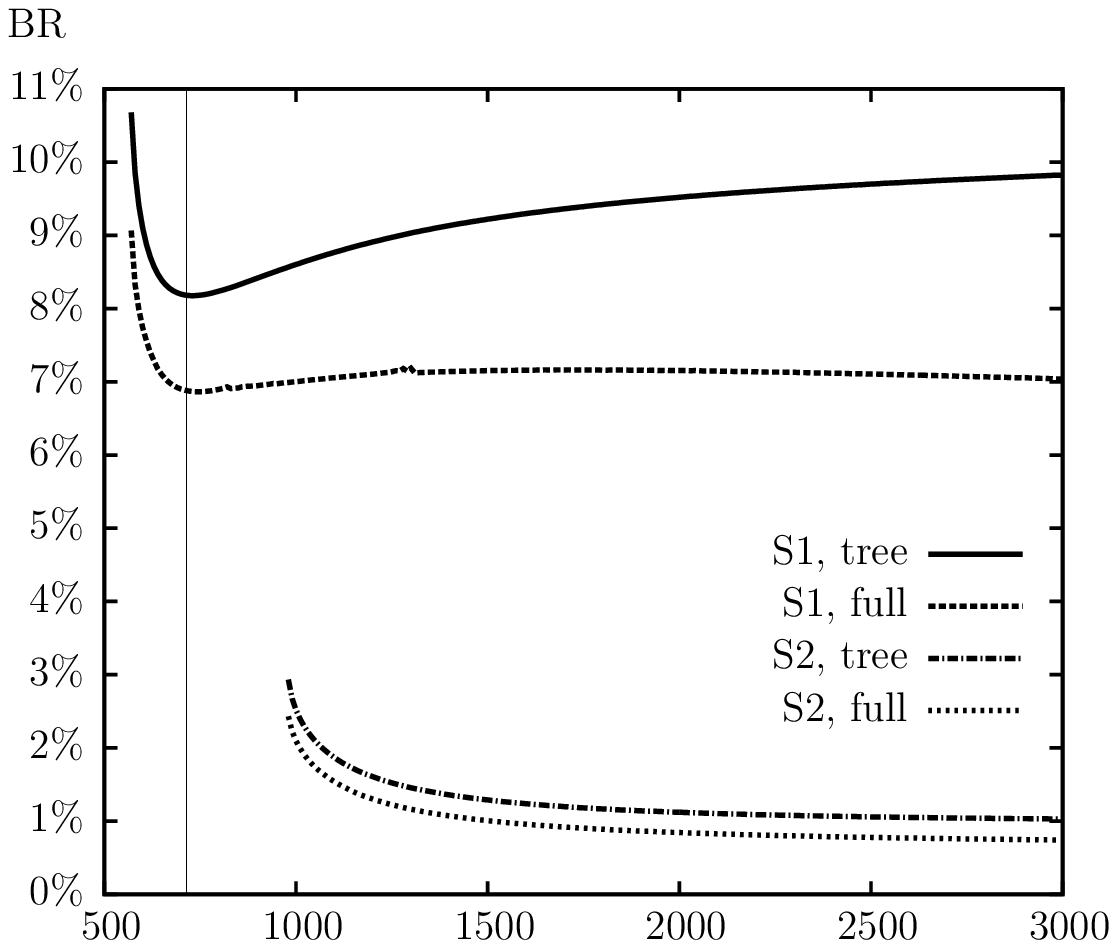}
\hspace{-4mm}
\includegraphics[width=0.49\textwidth,height=7.5cm]{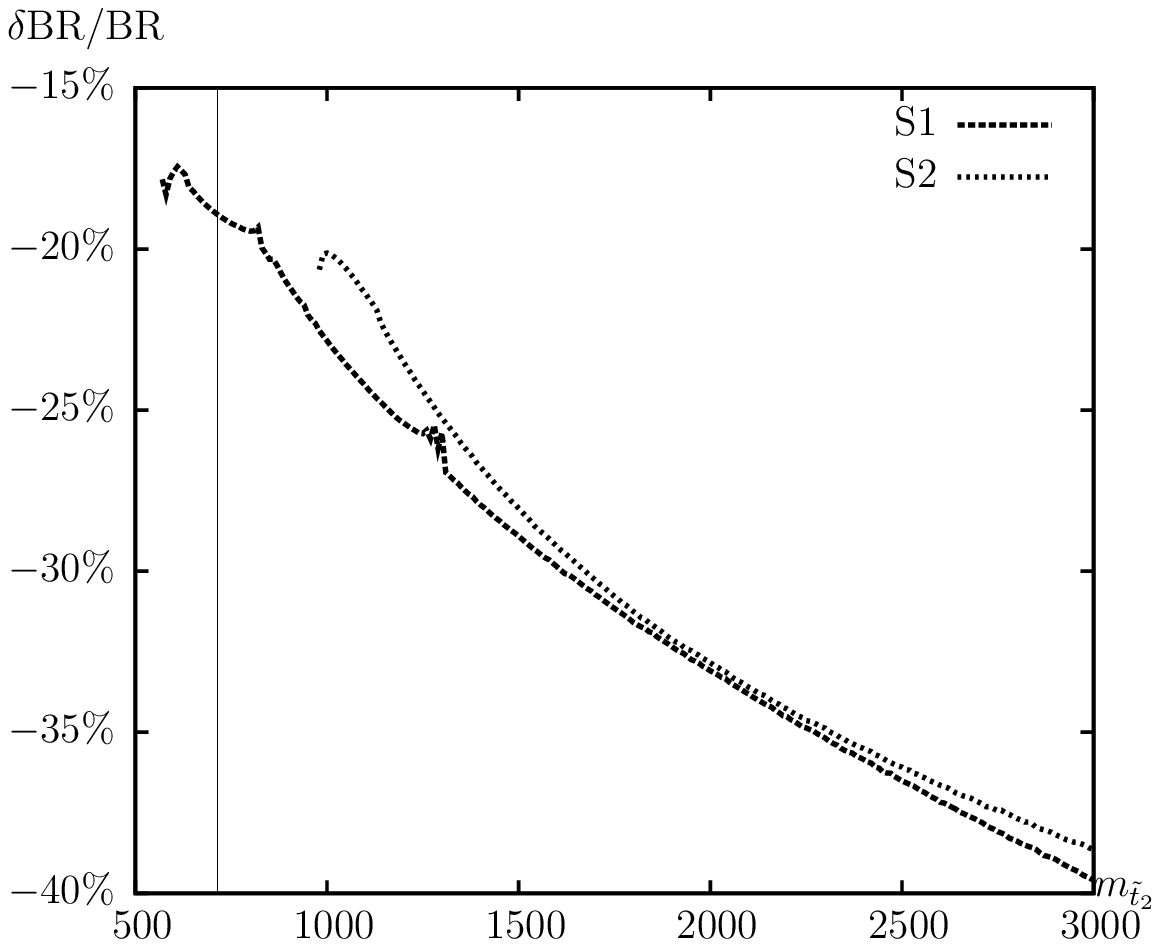}
\end{tabular}
\vspace{2em}
\caption{
  $\Ga(\decayNz)$. Tree-level and full one-loop corrected partial decay widths 
  are shown with the parameters chosen according to \SE\ and \SZ\ 
  (see \refta{tab:para}), with $\mstz$ varied.
  The upper left plot shows the partial decay width; the upper right 
  plot shows the corresponding relative size of the corrections.
  The lower left plot shows the BR; the lower right plot shows 
  the relative correction of the BR.
  The vertical lines indicate where $\mstz + \mste = 1000 \gev$, 
  i.e.\ the maximum reach of the ILC(1000).
}
\label{fig:mst2.st2tneu2}
\end{center}
\end{figure}

\begin{figure}[htb!]
\begin{center}
\begin{tabular}{c}
\includegraphics[width=0.49\textwidth,height=7.5cm]{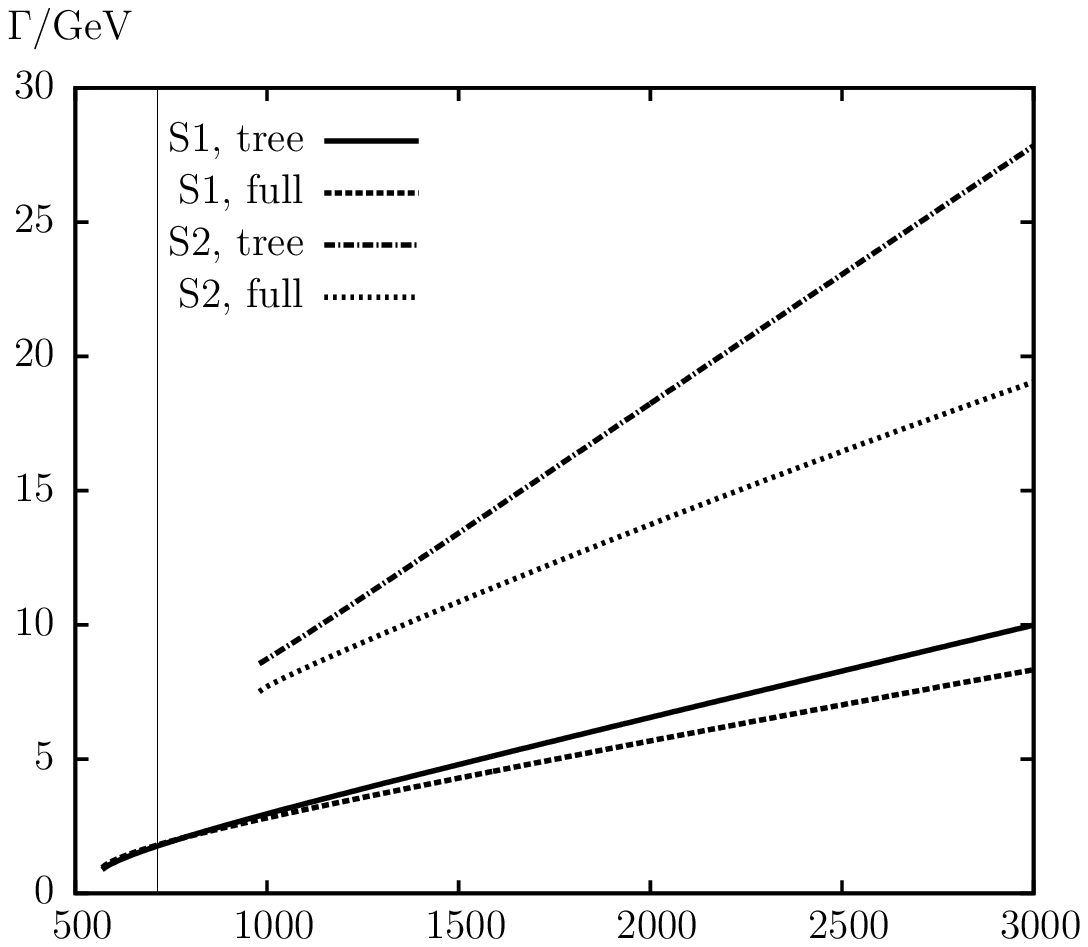}
\hspace{-4mm}
\includegraphics[width=0.49\textwidth,height=7.5cm]{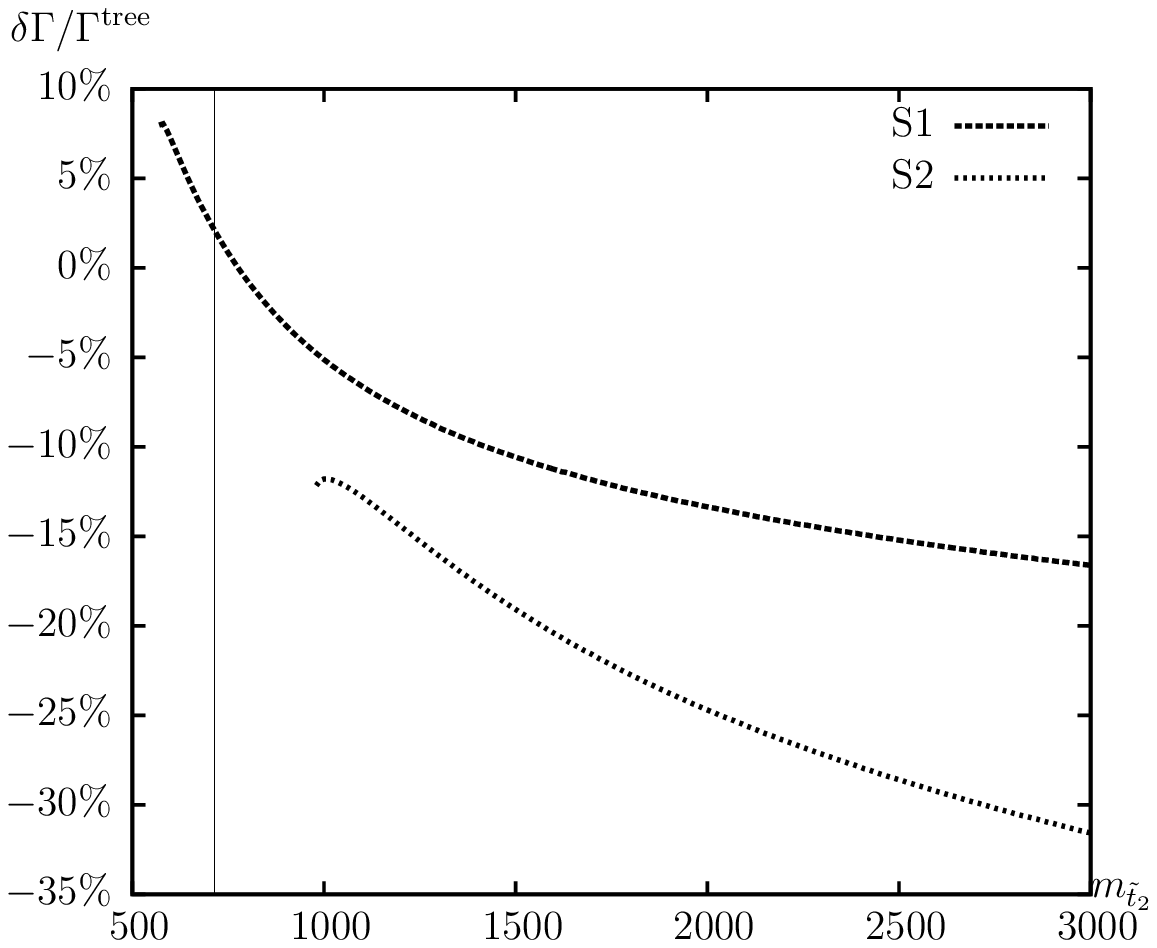}
\\[4em]
\includegraphics[width=0.49\textwidth,height=7.5cm]{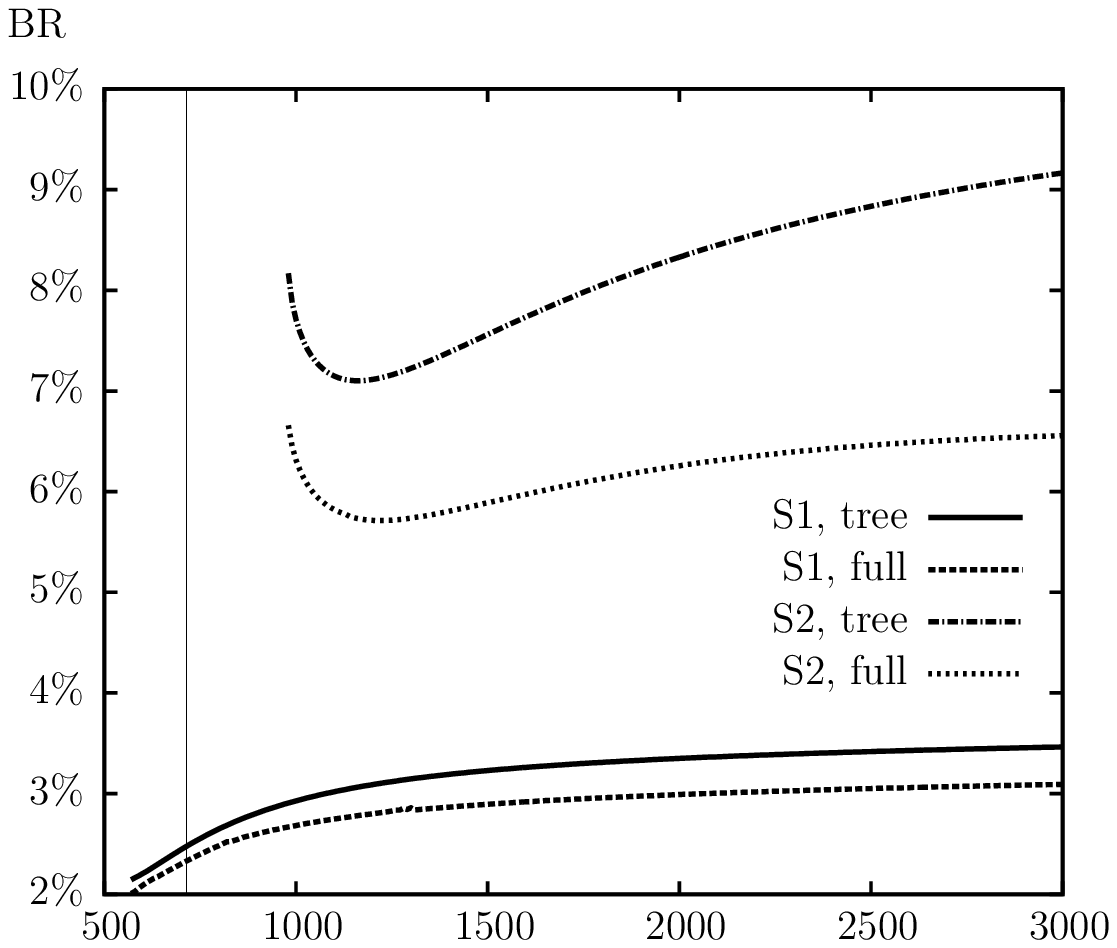}
\hspace{-4mm}
\includegraphics[width=0.49\textwidth,height=7.5cm]{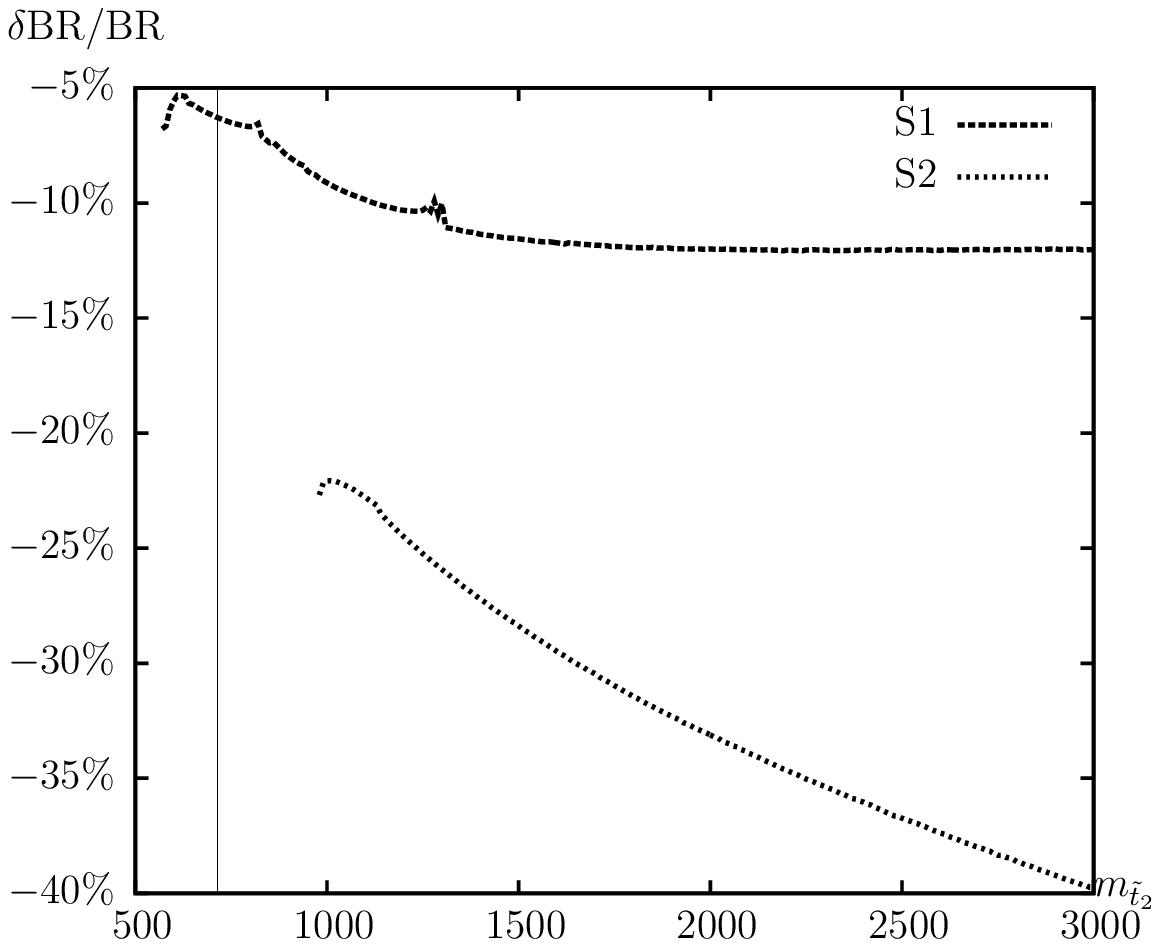}
\end{tabular}
\vspace{2em}
\caption{
  $\Ga(\decayNd)$. Tree-level and full one-loop corrected partial decay widths 
  are shown with the parameters chosen according to \SE\ and \SZ\ 
  (see \refta{tab:para}), with $\mstz$ varied.
  The upper left plot shows the partial decay width; the upper right 
  plot shows the corresponding relative size of the corrections.
  The lower left plot shows the BR; the lower right plot shows 
  the relative correction of the BR.
  The vertical lines indicate where $\mstz + \mste = 1000 \gev$, 
  i.e.\ the maximum reach of the ILC(1000).
}
\label{fig:mst2.st2tneu3}
\end{center}
\end{figure}

\begin{figure}[htb!]
\begin{center}
\begin{tabular}{c}
\includegraphics[width=0.49\textwidth,height=7.5cm]{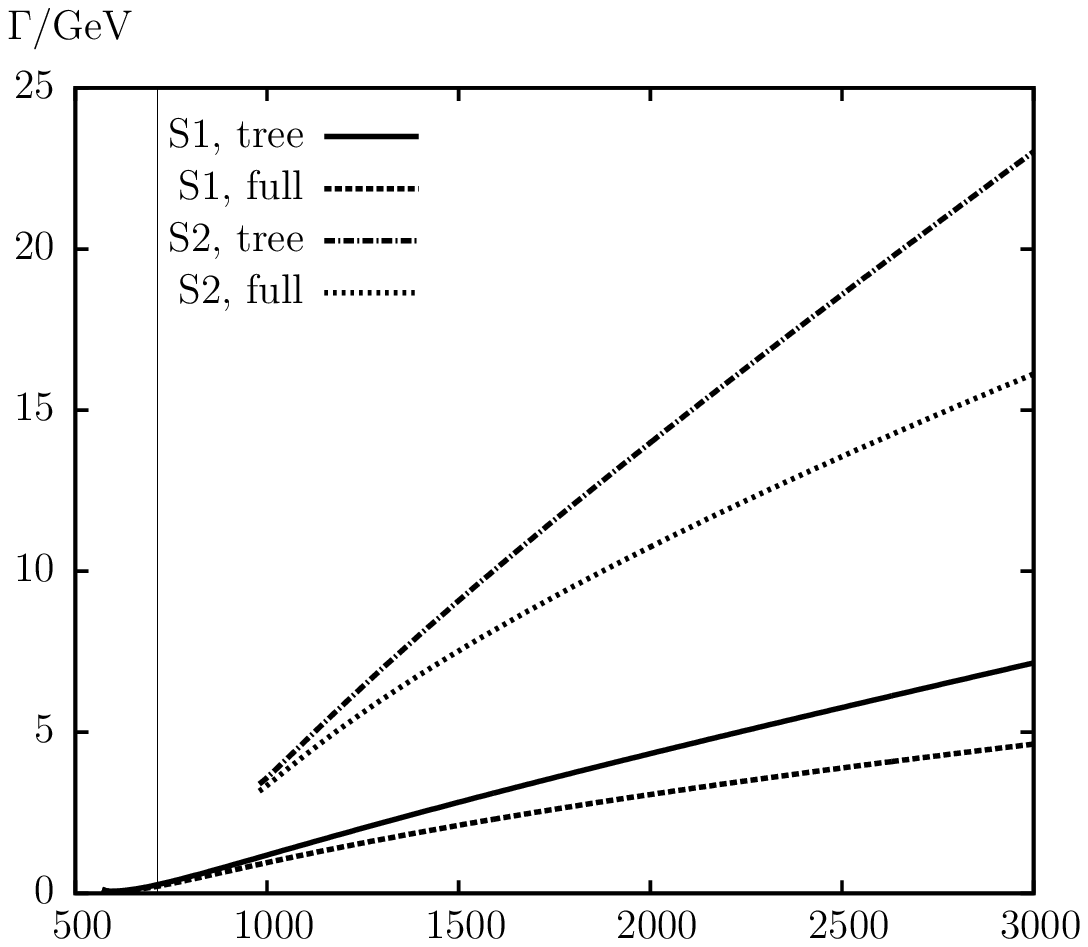}
\hspace{-4mm}
\includegraphics[width=0.49\textwidth,height=7.5cm]{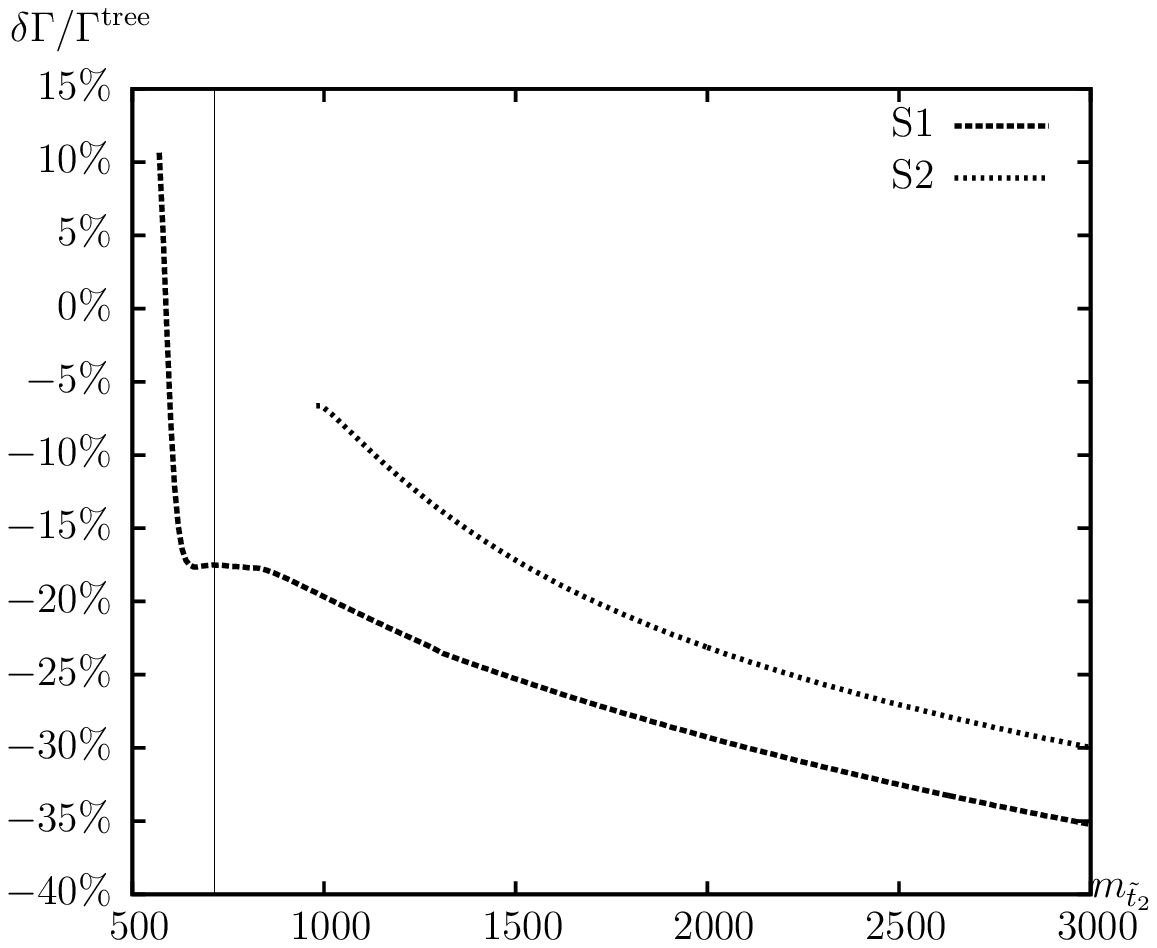}
\\[4em]
\includegraphics[width=0.49\textwidth,height=7.5cm]{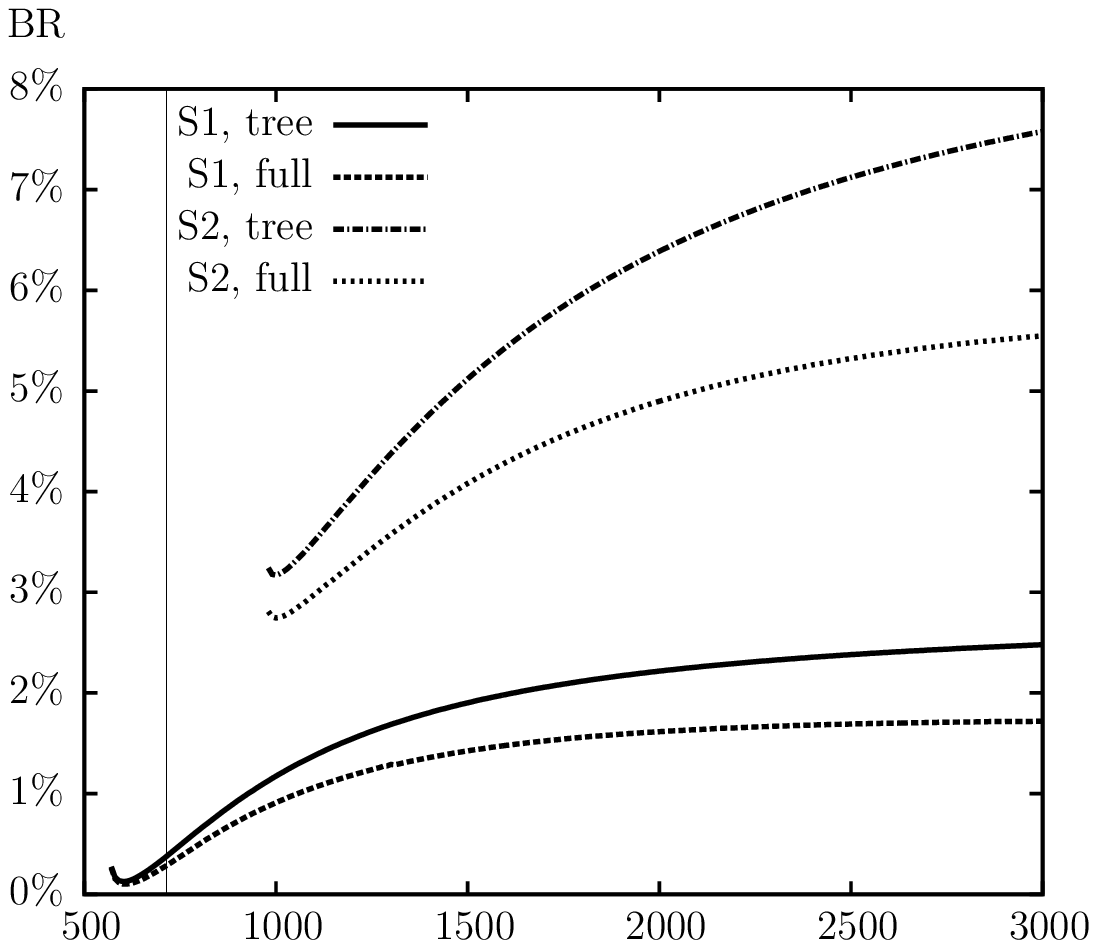}
\hspace{-4mm}
\includegraphics[width=0.49\textwidth,height=7.5cm]{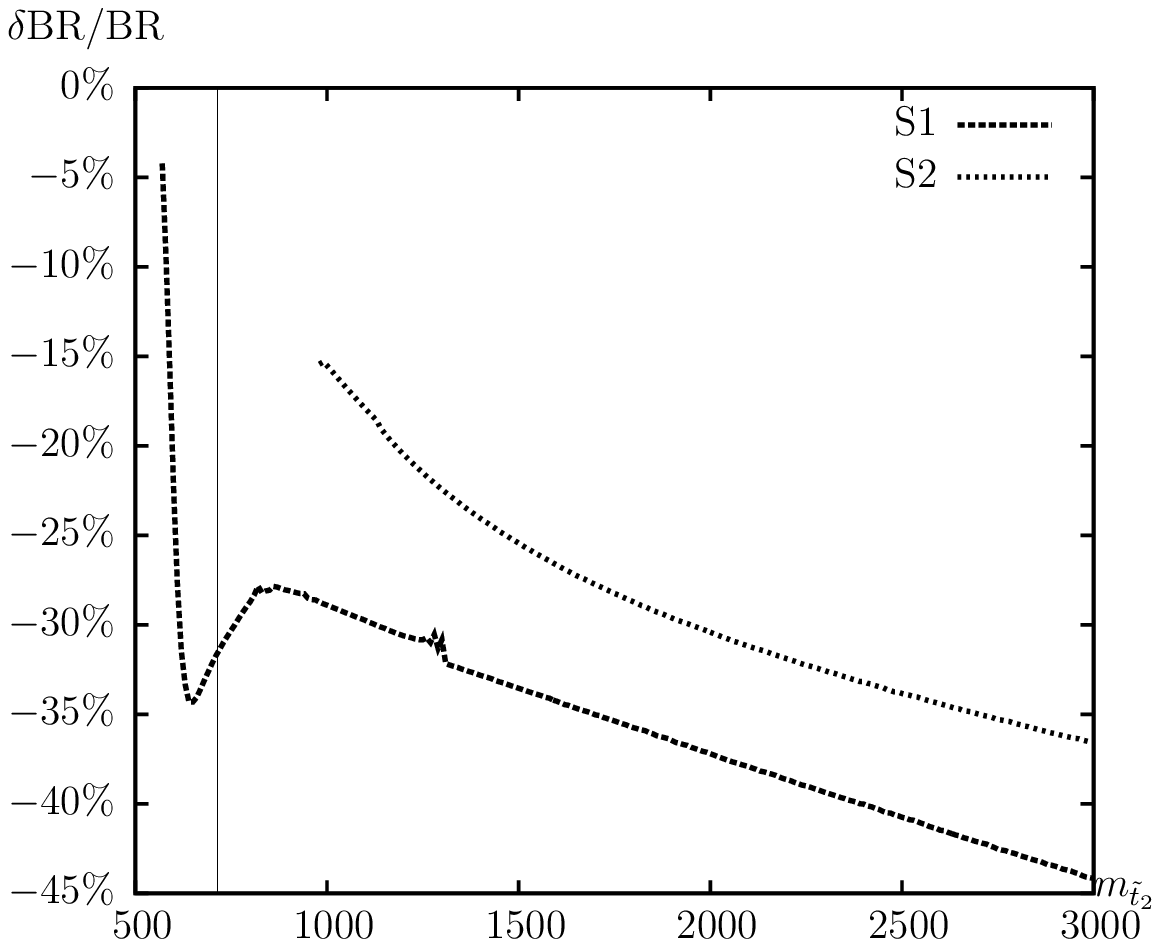}
\end{tabular}
\vspace{2em}
\caption{
  $\Ga(\decayNv)$. Tree-level and full one-loop corrected partial decay widths 
  are shown with the parameters chosen according to \SE\ and \SZ\ 
  (see \refta{tab:para}), with $\mstz$ varied.
  The upper left plot shows the partial decay width; the upper right 
  plot shows the corresponding relative size of the corrections.
  The lower left plot shows the BR; the lower right plot shows 
  the relative correction of the BR.
  The vertical lines indicate where $\mstz + \mste = 1000 \gev$, 
  i.e.\ the maximum reach of the ILC(1000).
}
\label{fig:mst2.st2tneu4}
\end{center}
\end{figure}

Now we turn to the decays $\decayNk$ ($k = 1,2,3,4$), with the results shown
in \reffis{fig:mst2.st2tneu1} -- \ref{fig:mst2.st2tneu4}. 
Since $\mu$, $M_1$, and $M_2$ are roughly of the same order, the four
states are a mixture of gauginos and Higgsinos. Consequently, the
partial decay widths are found to be roughly the same. The larger
partial decay widths for the decay modes $\decayNk$ with 
$k = 1,2$ ($k = 3,4$) are found in \SE\ (\SZ) and are $\sim 15-25 \gev$.
For \SE\ we find relative one-loop corrections ranging between
0\% and $\sim -30\%$ for $\Ga(\decayNe)$ and $\Ga(\decayNz)$, where the
smaller values are reached for small $\mstz$. $\Ga(\decayNd)$ receives
one-loop contributions between +8\% and $-16\%$, while for $\Ga(\decayNv)$
we find $-18\%$ to $-35\%$ with the exception of very small $\mstz$, 
where the partial decay width itself is negligible. 
Within \SZ\ the corrections stay at the few percent level for 
$\Ga(\decayNe)$, while for $\Ga(\decayNz)$, $\Ga(\decayNd)$ and 
$\Ga(\decayNv)$ they range between $-10\%$ and $-30\%$.
Following the size of the partial decay widths, also the branching ratios
are roughly the same for the four decay modes. The relative changes in
the BRs for $\mstz + \mste \lsim 1000 \gev$ are $\sim -15\%$, $-18\%$, 
$-6\%$, and $-30\%$ for $k = 1,2,3,4$, respectively. Especially, for
this parameter range, for the 
decays to the two lighter neutralinos the branching ratios are $\sim 5\%$
and $\sim 10\%$, i.e.\ the one-loop corrections can be crucial to match
the anticipated ILC precision.

\begin{figure}[htb!]
\begin{center}
\begin{tabular}{c}
\includegraphics[width=0.49\textwidth,height=7.5cm]{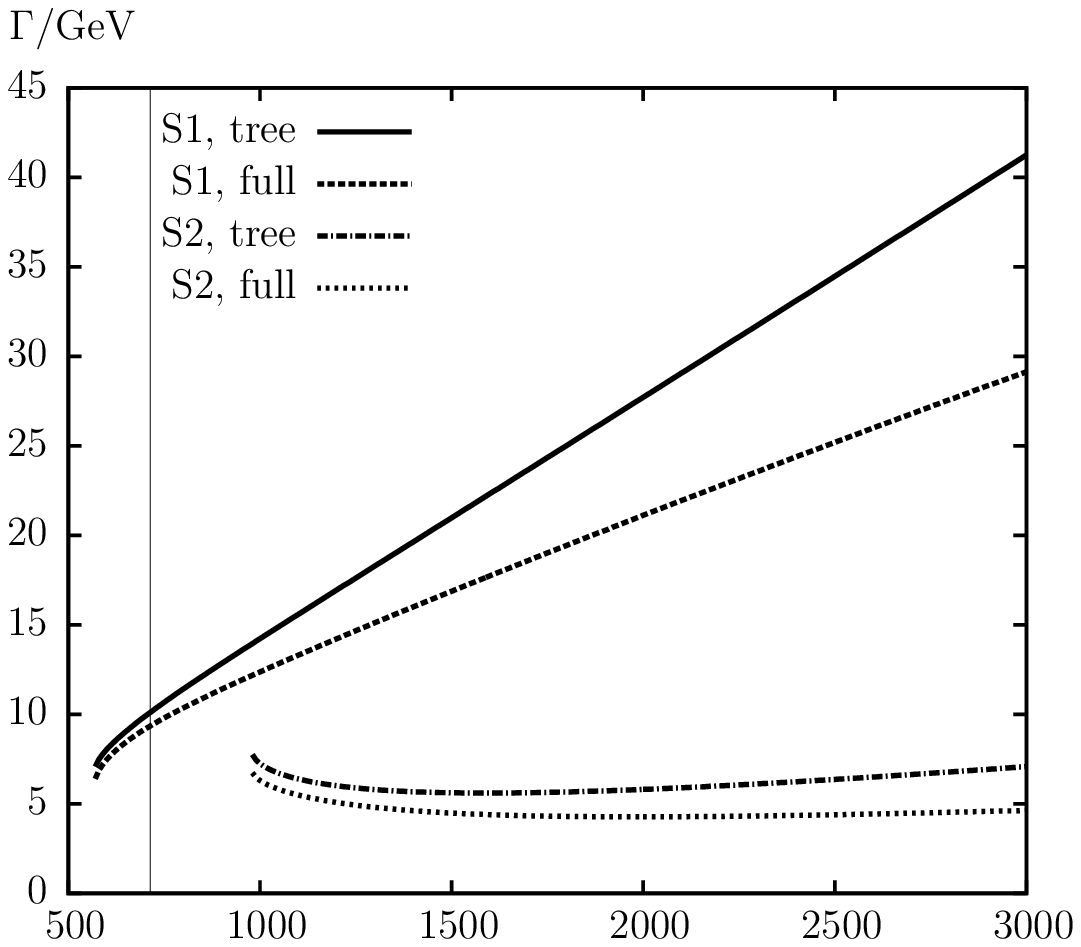}
\hspace{-4mm}
\includegraphics[width=0.49\textwidth,height=7.5cm]{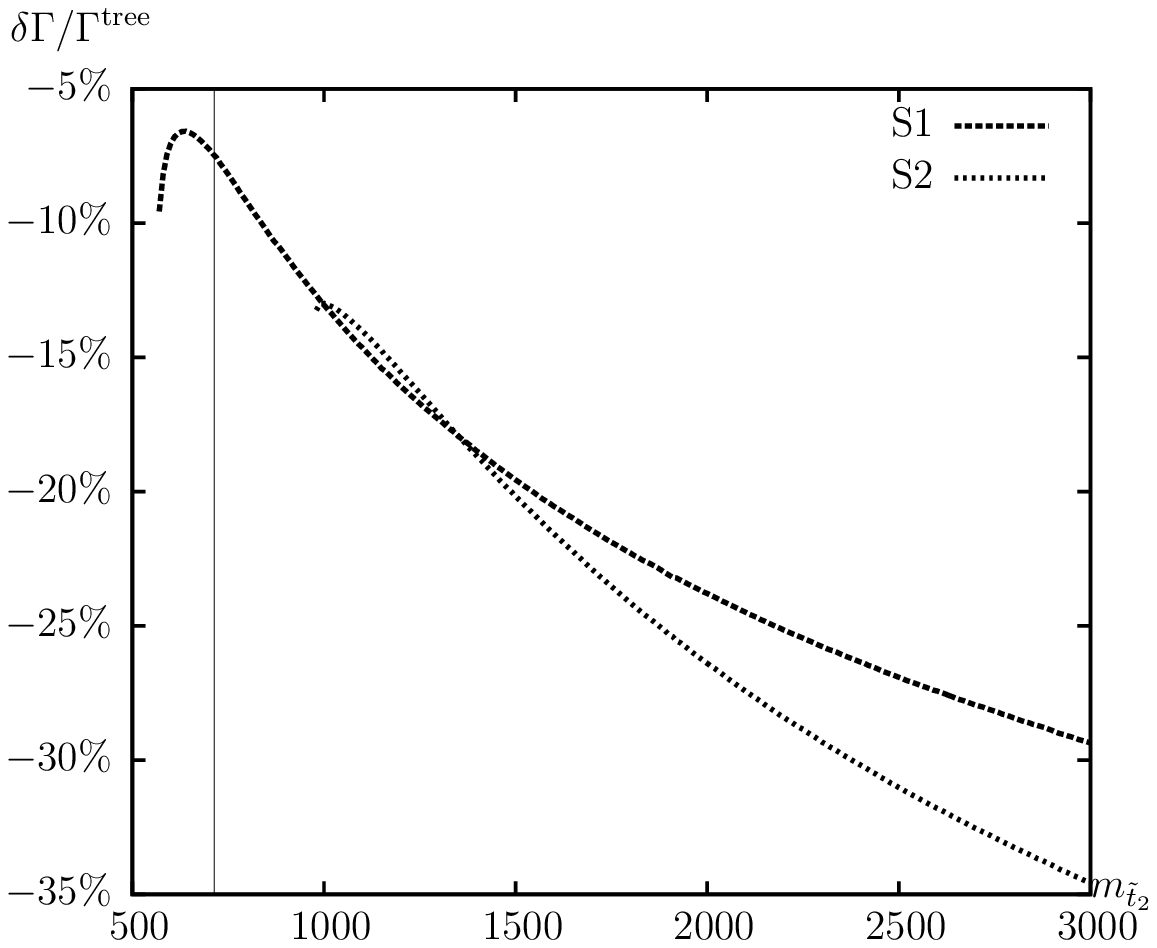}
\\[4em]
\includegraphics[width=0.49\textwidth,height=7.5cm]{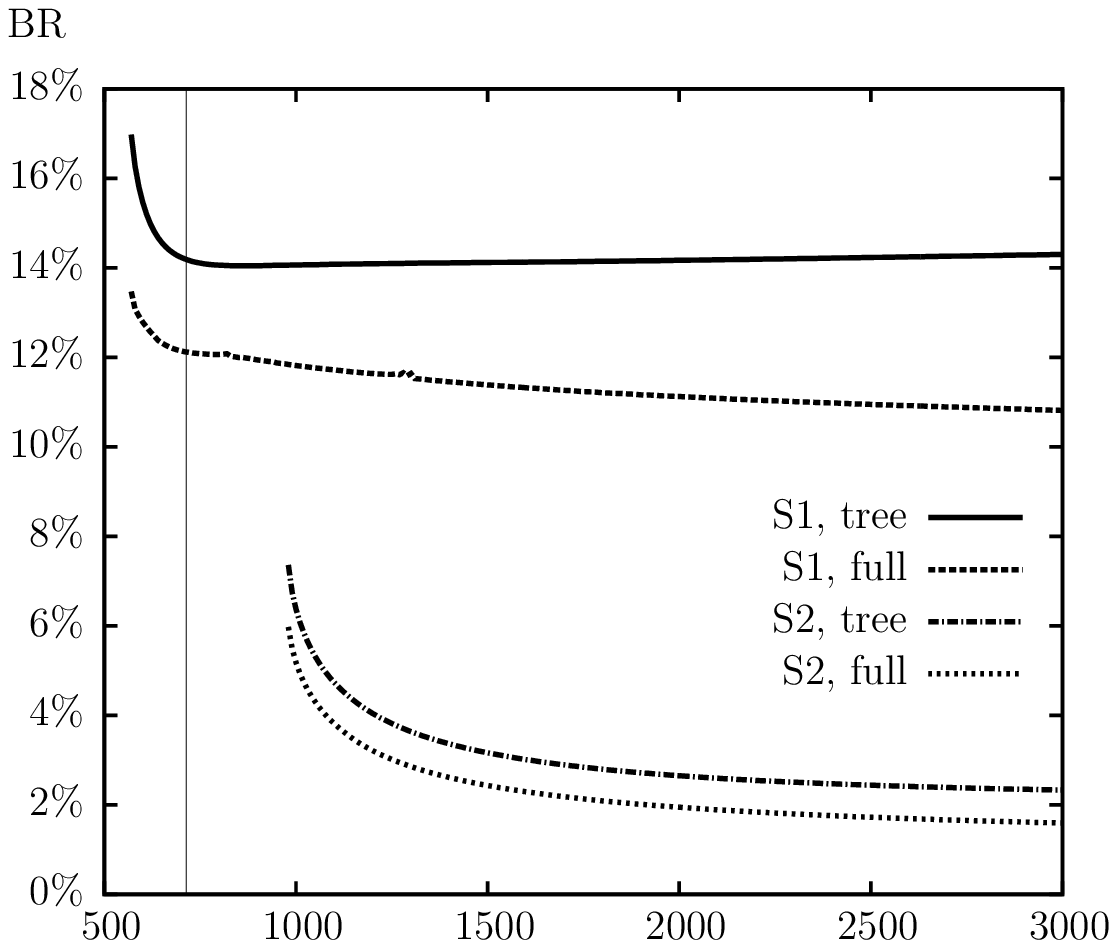}
\hspace{-4mm}
\includegraphics[width=0.49\textwidth,height=7.5cm]{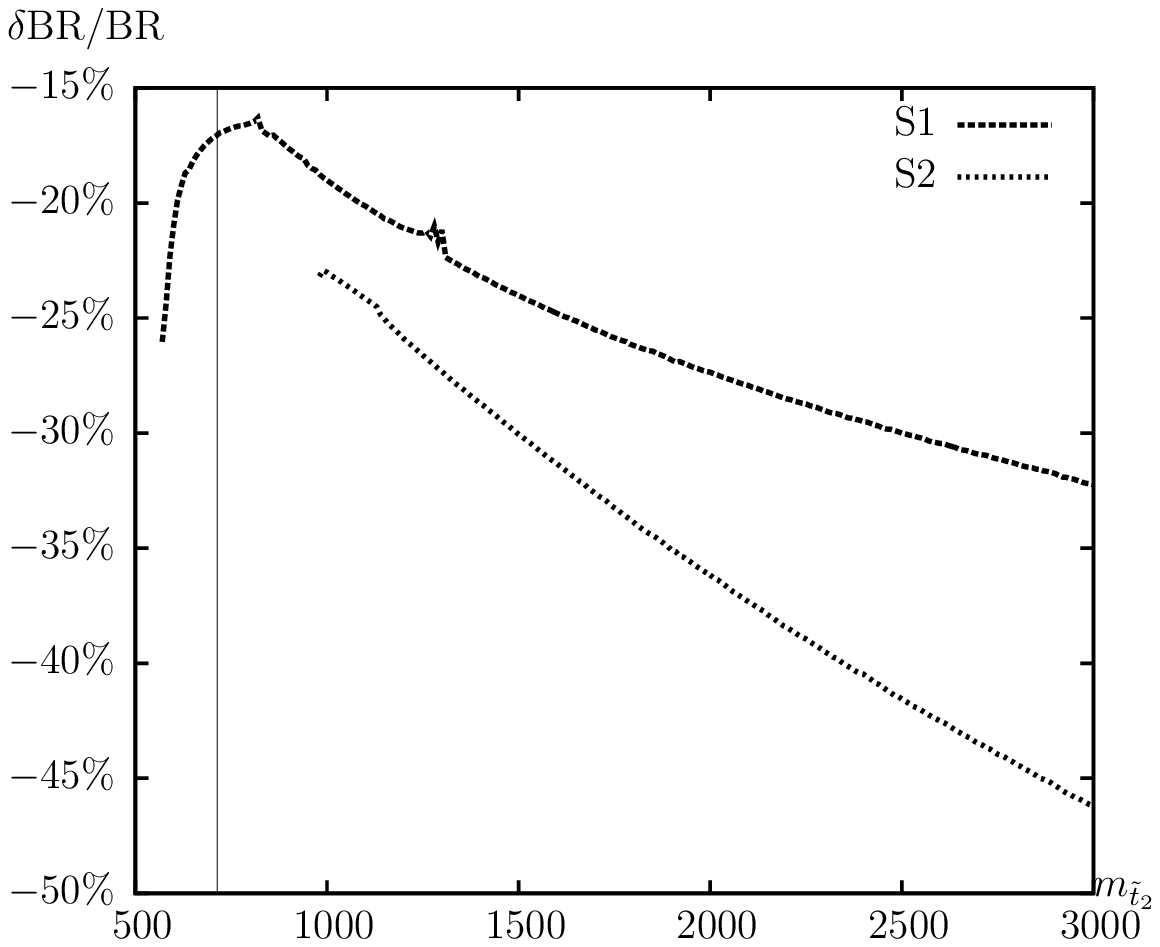}
\end{tabular}
\vspace{2em}
\caption{
  $\Ga(\decayCpe)$. Tree-level and full one-loop corrected partial decay widths 
  are shown with the parameters chosen according to \SE\ and \SZ\ 
  (see \refta{tab:para}), with $\mstz$ varied.
  The upper left plot shows the partial decay width; the upper right 
  plot shows the corresponding relative size of the corrections.
  The lower left plot shows the BR; the lower right plot shows 
  the relative correction of the BR.
  The vertical lines indicate where $\mstz + \mste = 1000 \gev$, 
  i.e.\ the maximum reach of the ILC(1000).
}
\label{fig:mst2.st2bcha1}
\end{center}
\end{figure}

\begin{figure}[htb!]
\begin{center}
\begin{tabular}{c}
\includegraphics[width=0.49\textwidth,height=7.5cm]{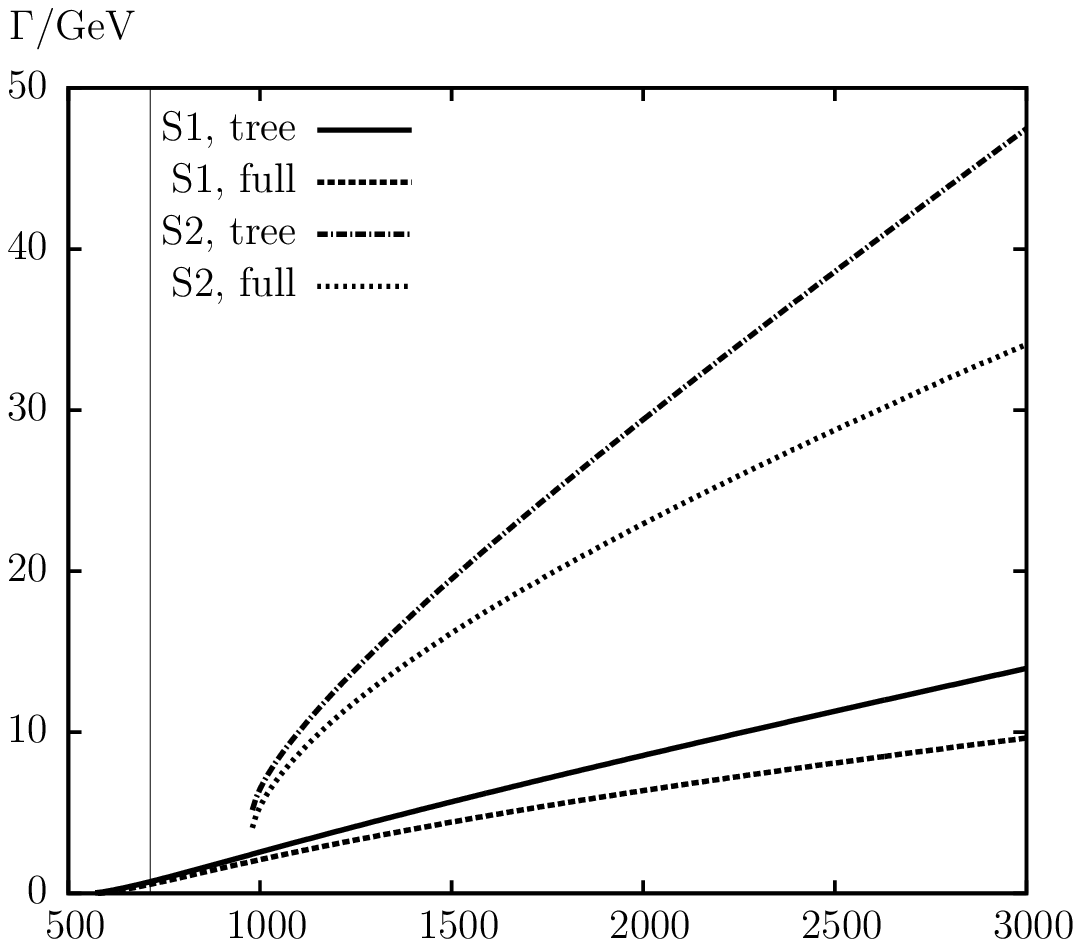}
\hspace{-4mm}
\includegraphics[width=0.49\textwidth,height=7.5cm]{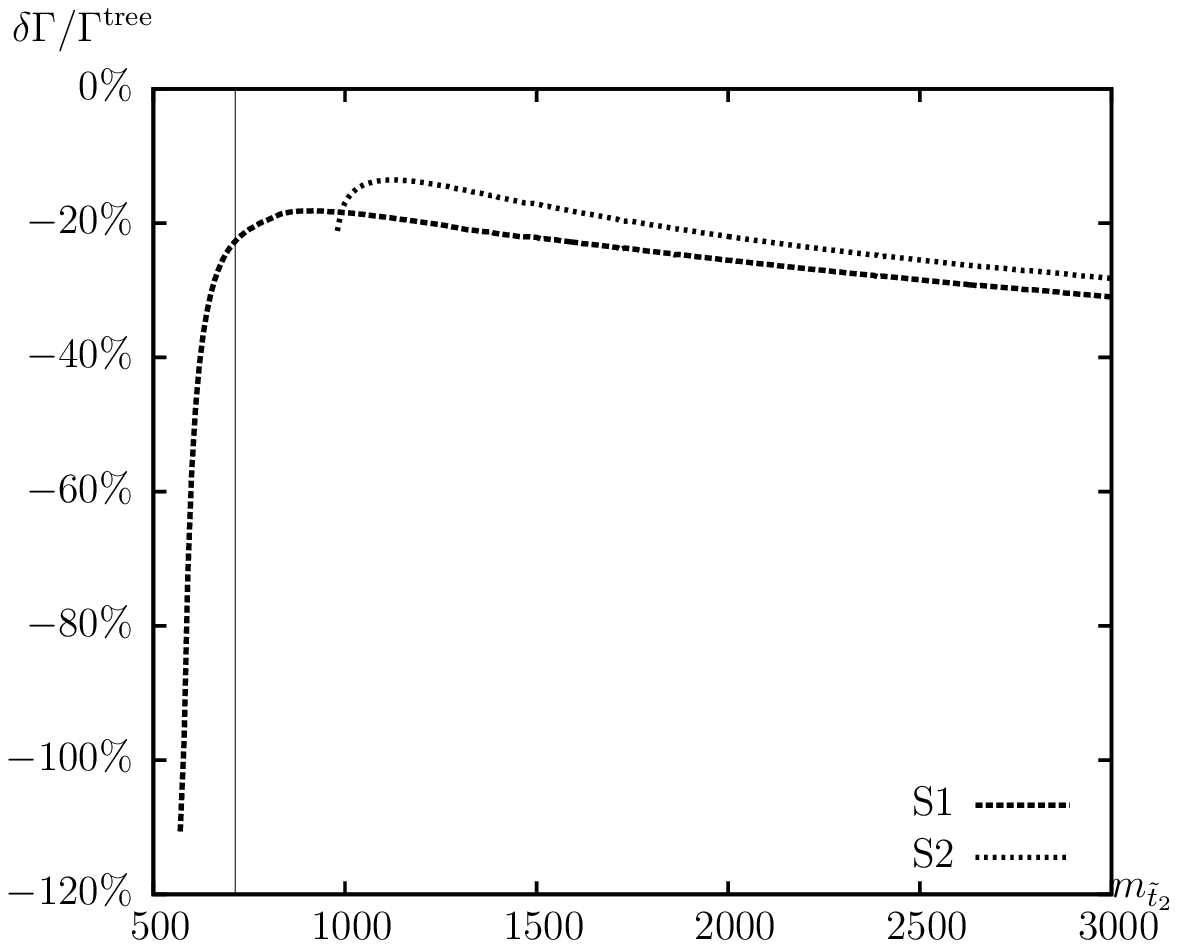}
\\[4em]
\includegraphics[width=0.49\textwidth,height=7.5cm]{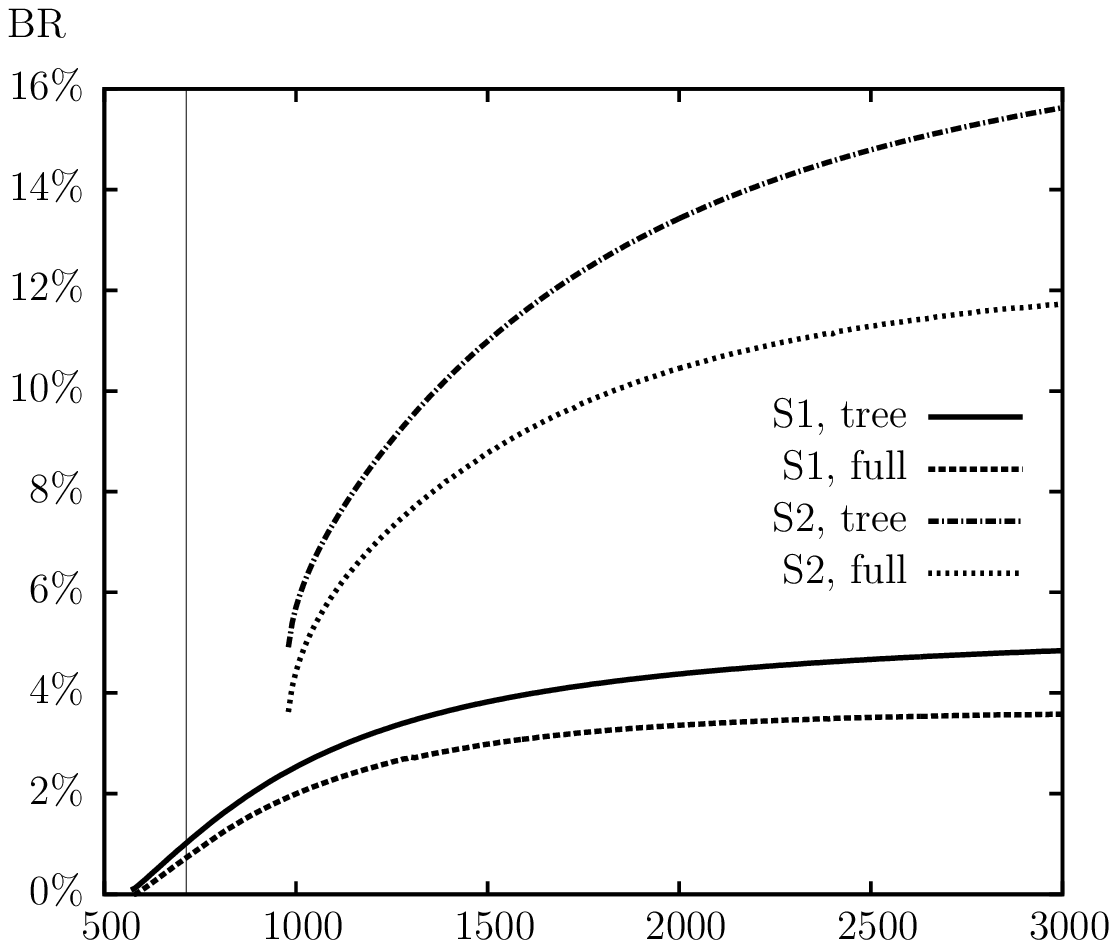}
\hspace{-4mm}
\includegraphics[width=0.49\textwidth,height=7.5cm]{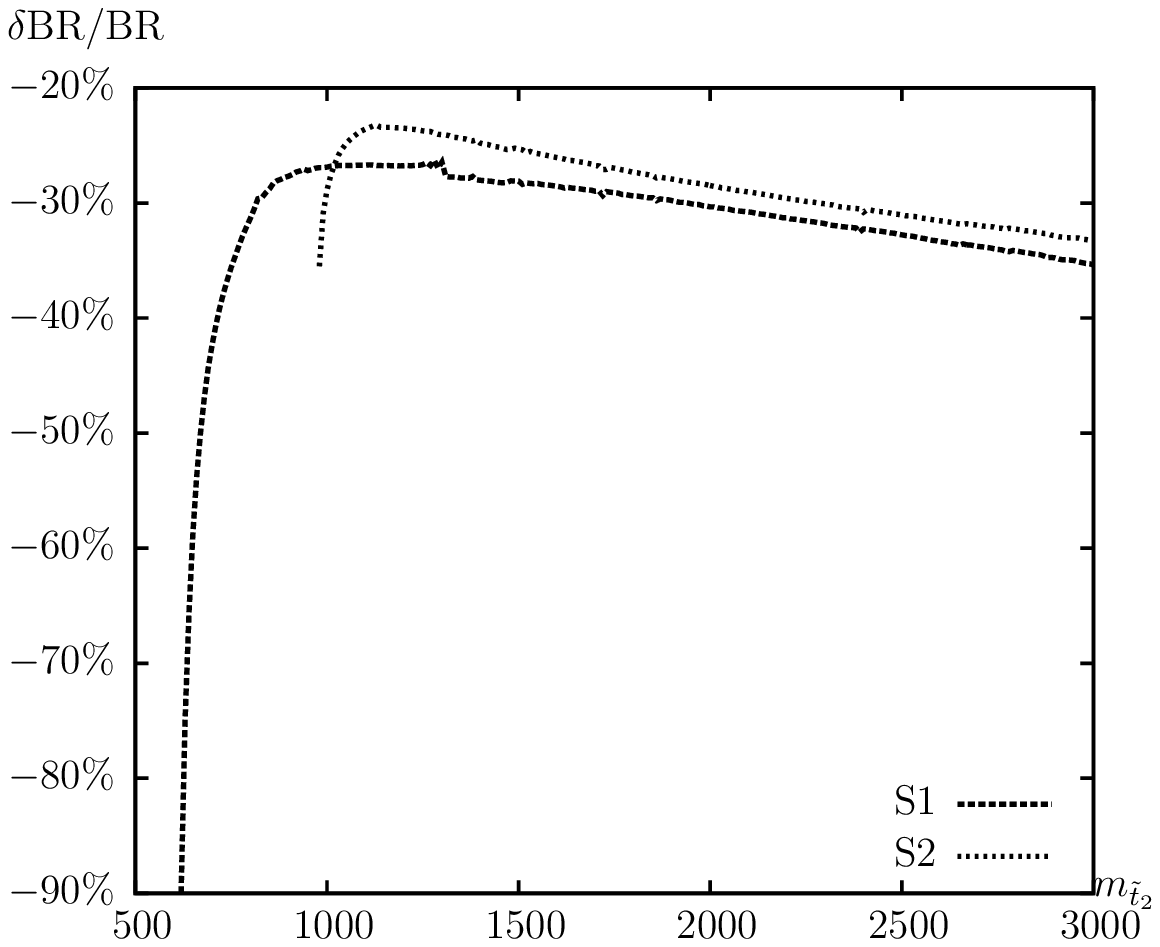}
\end{tabular}
\vspace{2em}
\caption{
  $\Ga(\decayCpz)$. Tree-level and full one-loop corrected partial decay widths 
  are shown with the parameters chosen according to \SE\ and \SZ\ 
  (see \refta{tab:para}), with $\mstz$ varied.
  The upper left plot shows the partial decay width; the upper right 
  plot shows the corresponding relative size of the corrections. 
  The lower left plot shows the BR; the lower right plot shows 
  the relative correction of the BR.
  The vertical lines indicate where $\mstz + \mste = 1000 \gev$, 
  i.e.\ the maximum reach of the ILC(1000).
}
\label{fig:mst2.st2bcha2}
\end{center}
\end{figure}

Next in \reffis{fig:mst2.st2bcha1}, \ref{fig:mst2.st2bcha2} we present
the results for $\decayCpj$ ($j = 1,2$). 
The size of the partial decay widths and branching ratios for 
$\decayCpe$ ($\decayCpz$) are roughly the same as for $\decayNk$ with 
$k = 1,2$ ($k = 3,4$).
For $\Ga(\decayCpe)$ we find relative corrections starting at $-5\%$ at low 
$\mstz$ in \SE\ down to $\sim -35\%$ at high $\mstz$ in both scenarios. 
The partial decay width $\Ga(\decayCpz)$ is very small in \SE\ for 
$\mste + \mstz < 1000 \gev$. 
Because of this smallness, and additionally pronounced due to 
the vicinity of the production threshold, the relative size of the 
corrections becomes huge and is not reliable anymore.
For higher $\mstz$ values we find relative corrections between 
$-20\%$ $(-10\%)$ to $-30\%$ in \SE\ (\SZ).
A large branching ratio of $\sim 14\%$ in the ILC(1000) accessible
regime is reached in \SE\ in the decay $\decayCpe$, where the one-loop
corrections are $\sim -20\%$. 
Again the one-loop corrections can be crucial to match the ILC precision.

\begin{figure}[htb!]
\begin{center}
\begin{tabular}{c}
\includegraphics[width=0.49\textwidth,height=7.5cm]{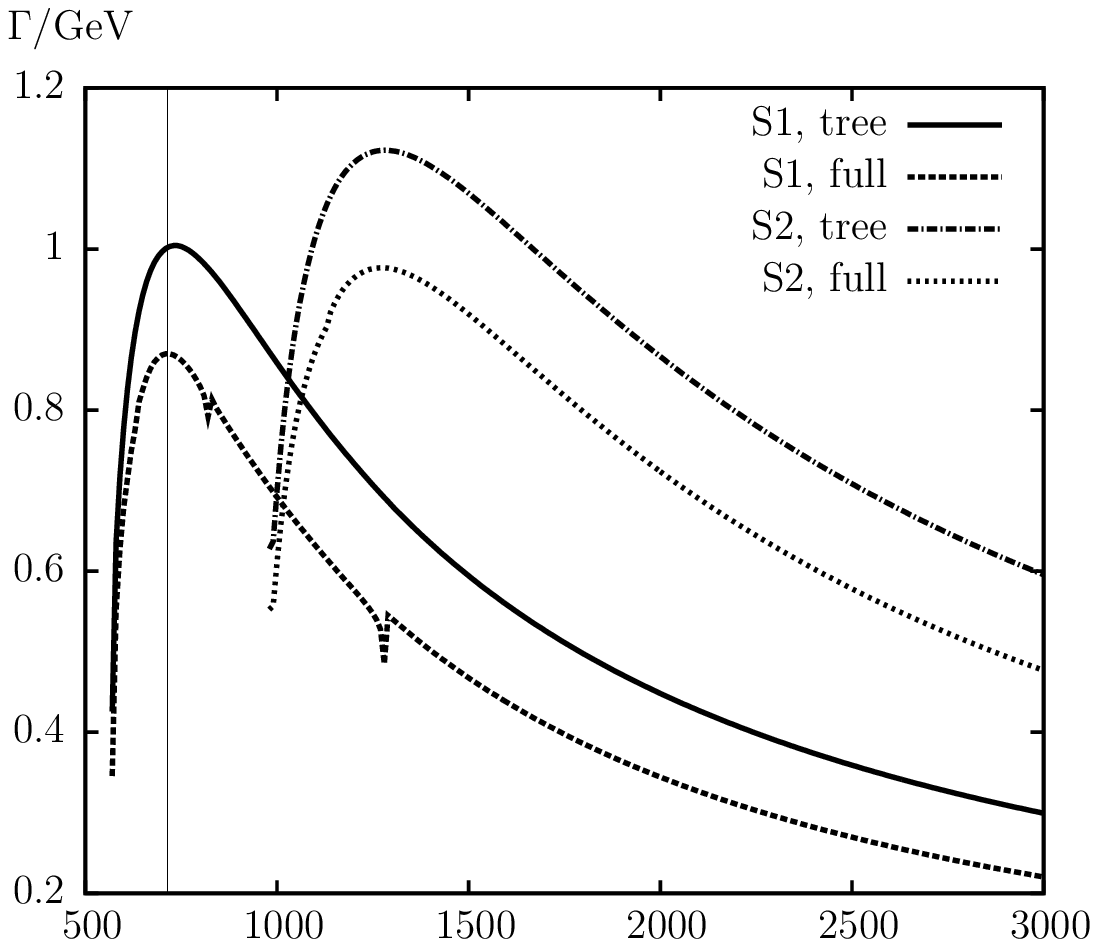}
\hspace{-4mm}
\includegraphics[width=0.49\textwidth,height=7.5cm]{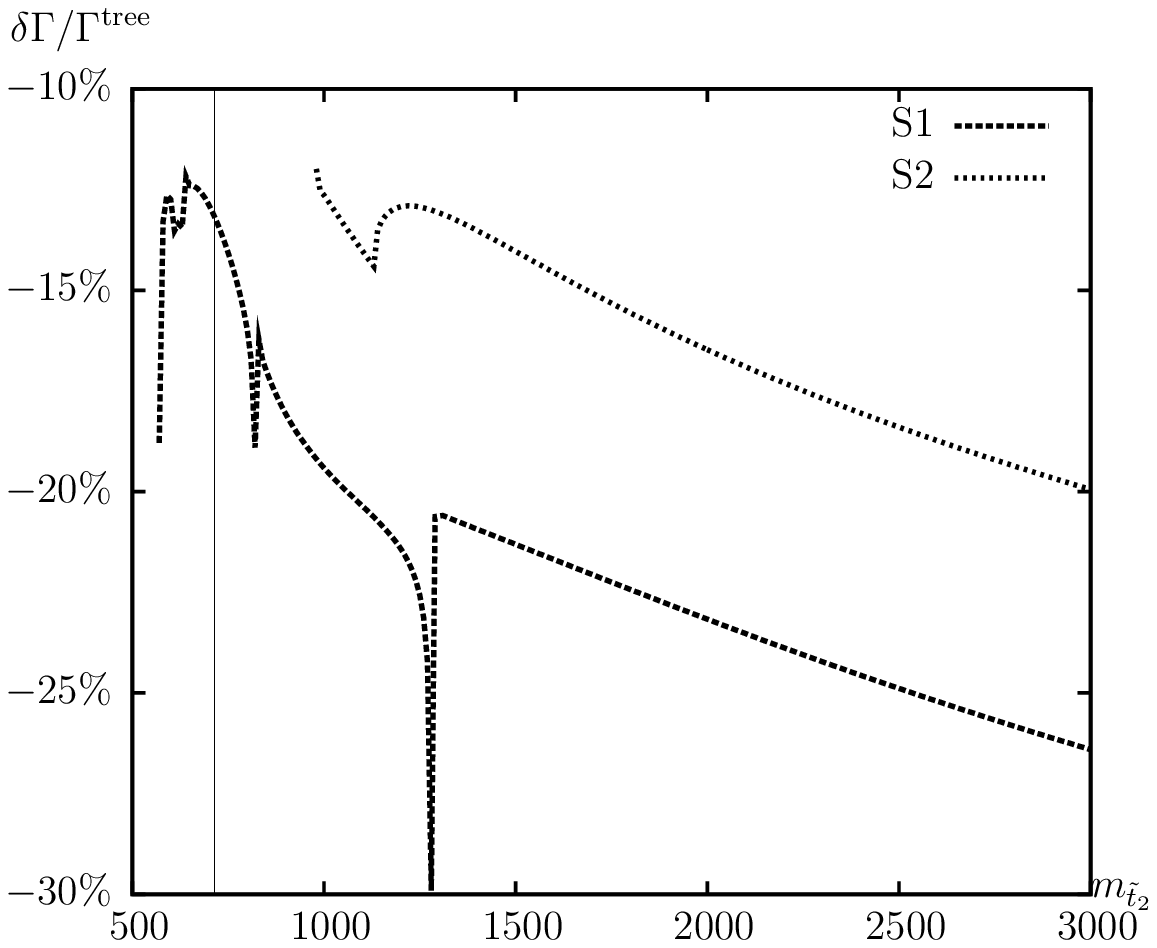}
\\[4em]
\includegraphics[width=0.49\textwidth,height=7.5cm]{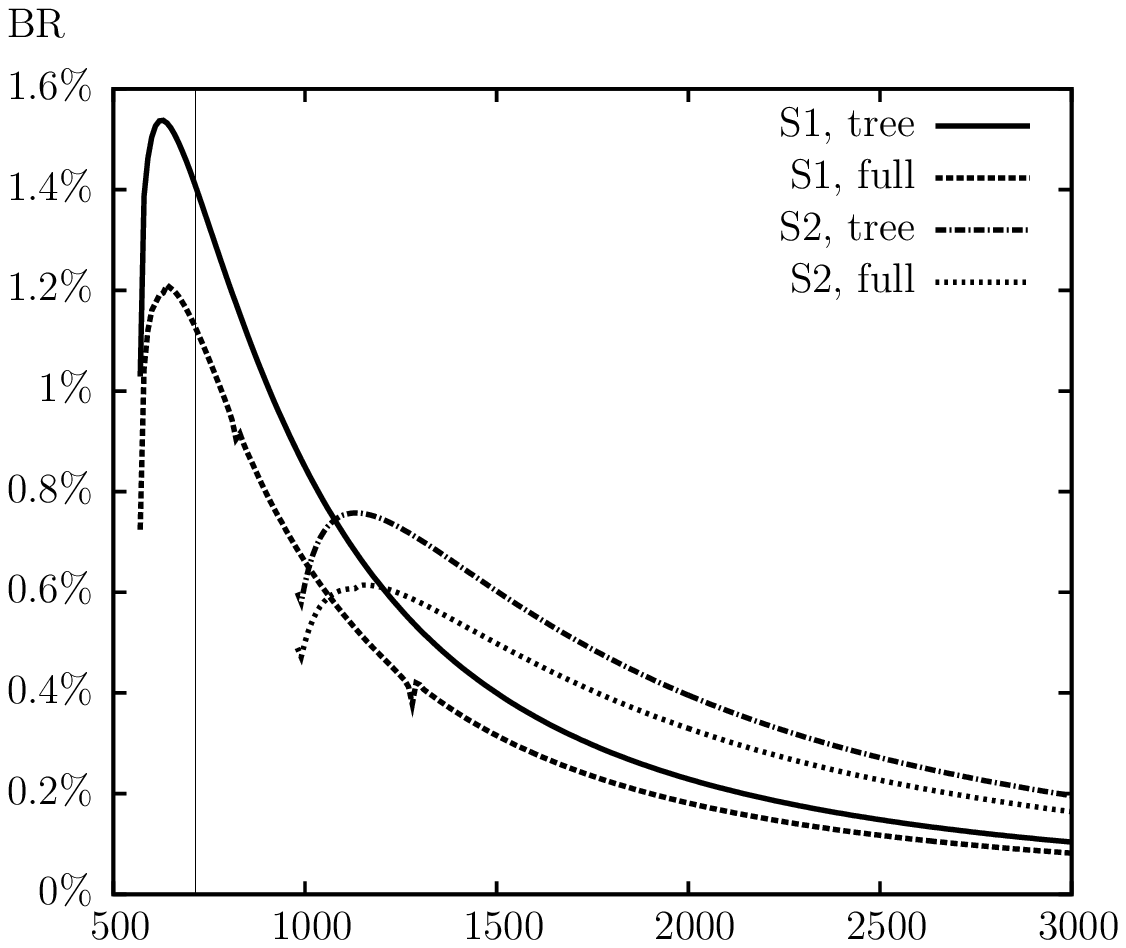}
\hspace{-4mm}
\includegraphics[width=0.49\textwidth,height=7.5cm]{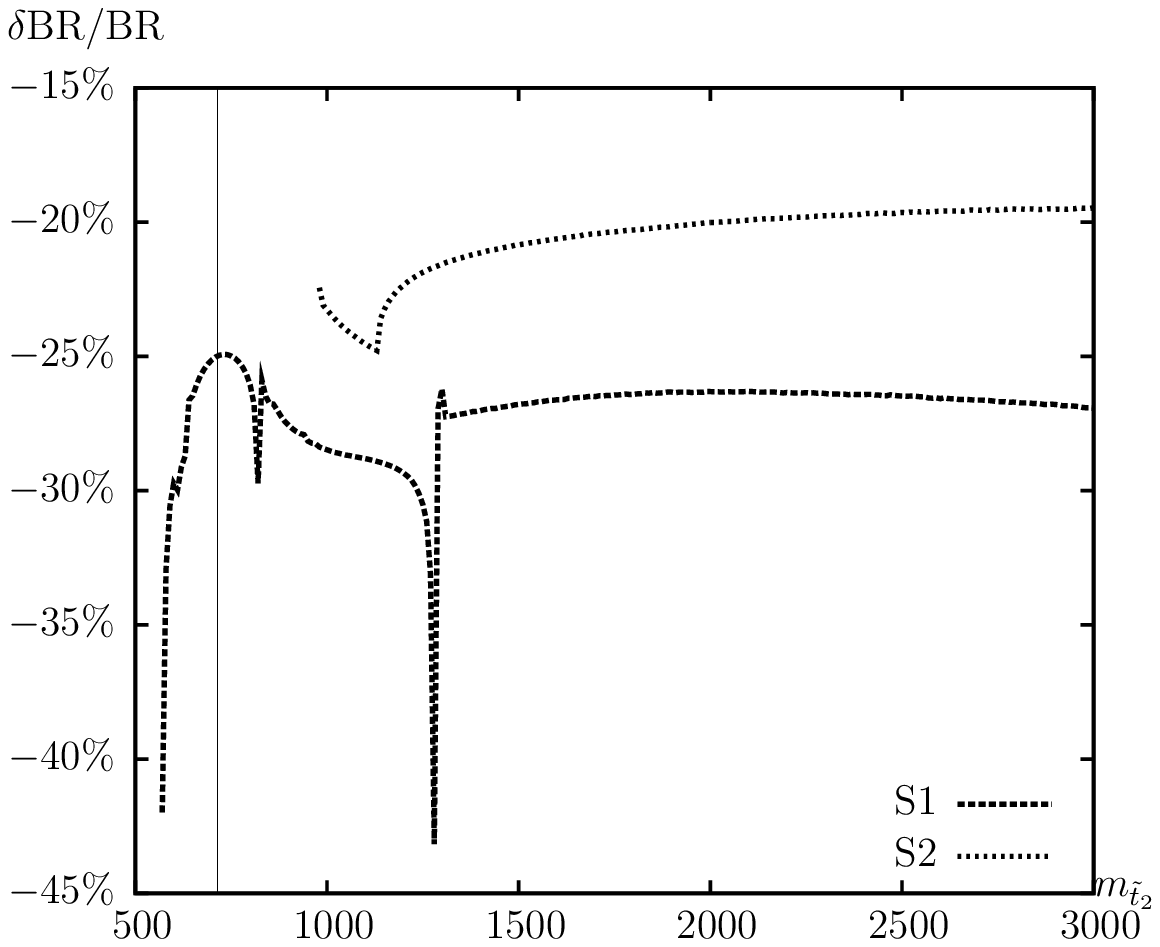}
\end{tabular}
\vspace{2em}
\caption{
  $\Ga(\decaySbeH)$. Tree-level and full one-loop corrected partial decay widths 
  are shown with the parameters chosen according to \SE\ and \SZ\ 
  (see \refta{tab:para}), with $\mstz$ varied.
  The upper left plot shows the partial decay width; the upper right 
  plot shows the corresponding relative size of the corrections.
  The lower left plot shows the BR; the lower right plot shows 
  the relative correction of the BR.
  The vertical lines indicate where $\mstz + \mste = 1000 \gev$, 
  i.e.\ the maximum reach of the ILC(1000).
}
\label{fig:mst2.st2sb1H}
\end{center}
\end{figure}

\begin{figure}[htb!]
\begin{center}
\begin{tabular}{c}
\includegraphics[width=0.49\textwidth,height=7.5cm]{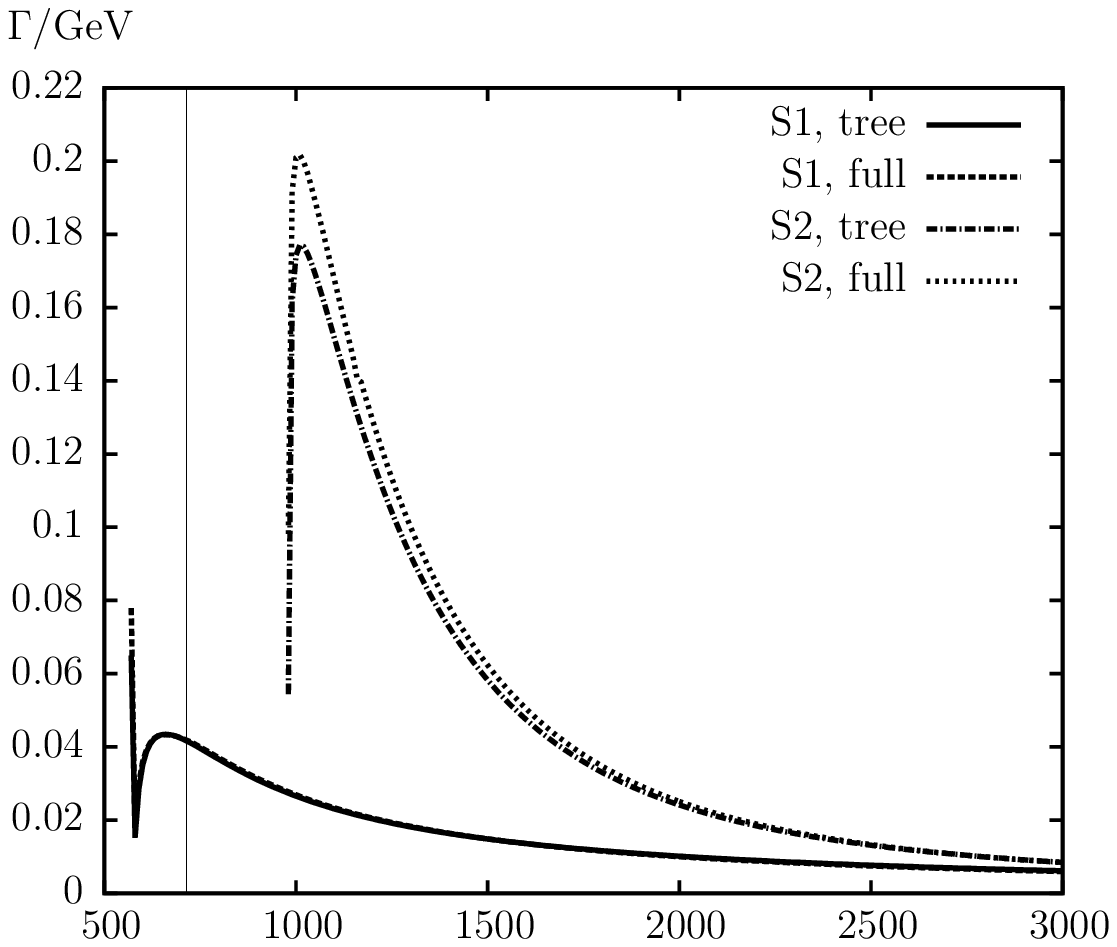}
\hspace{-4mm}
\includegraphics[width=0.49\textwidth,height=7.5cm]{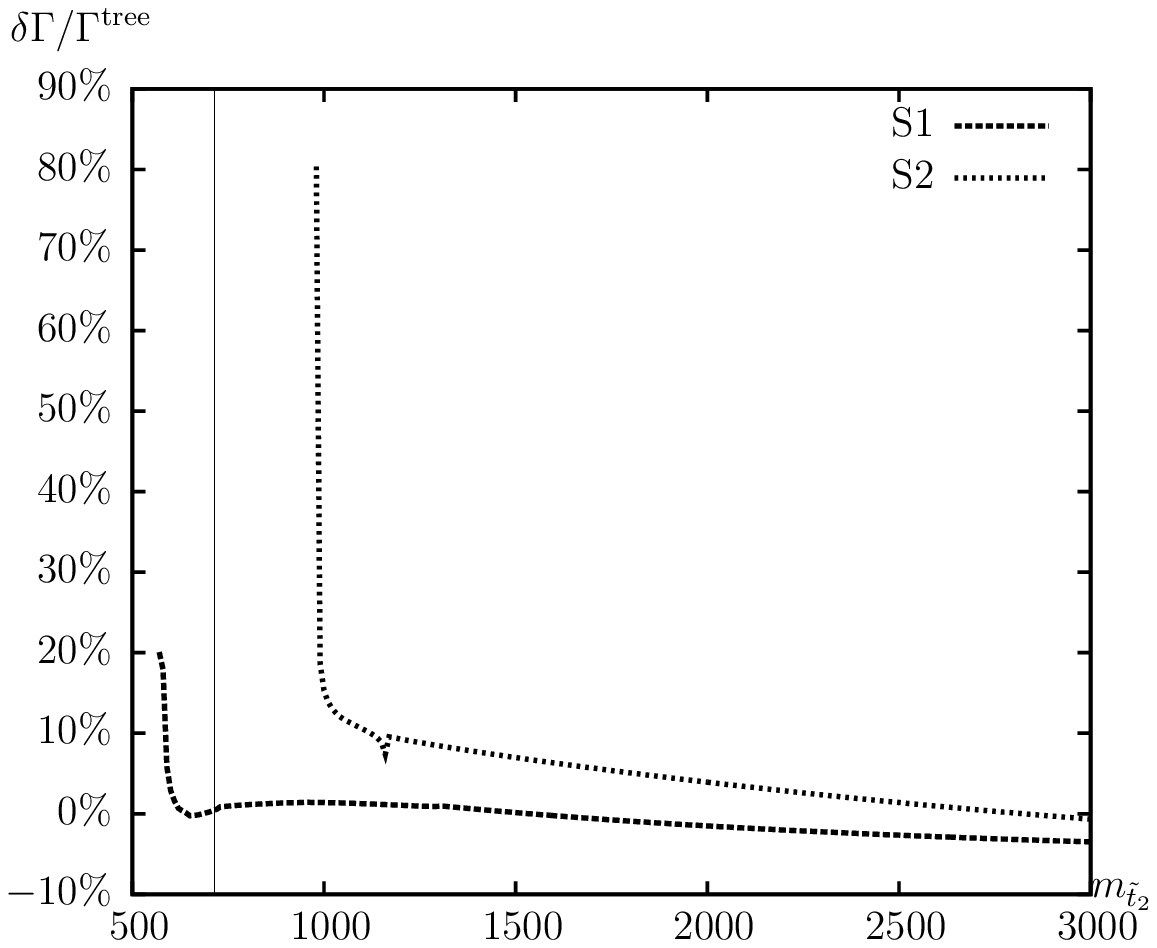}
\\[4em]
\includegraphics[width=0.49\textwidth,height=7.5cm]{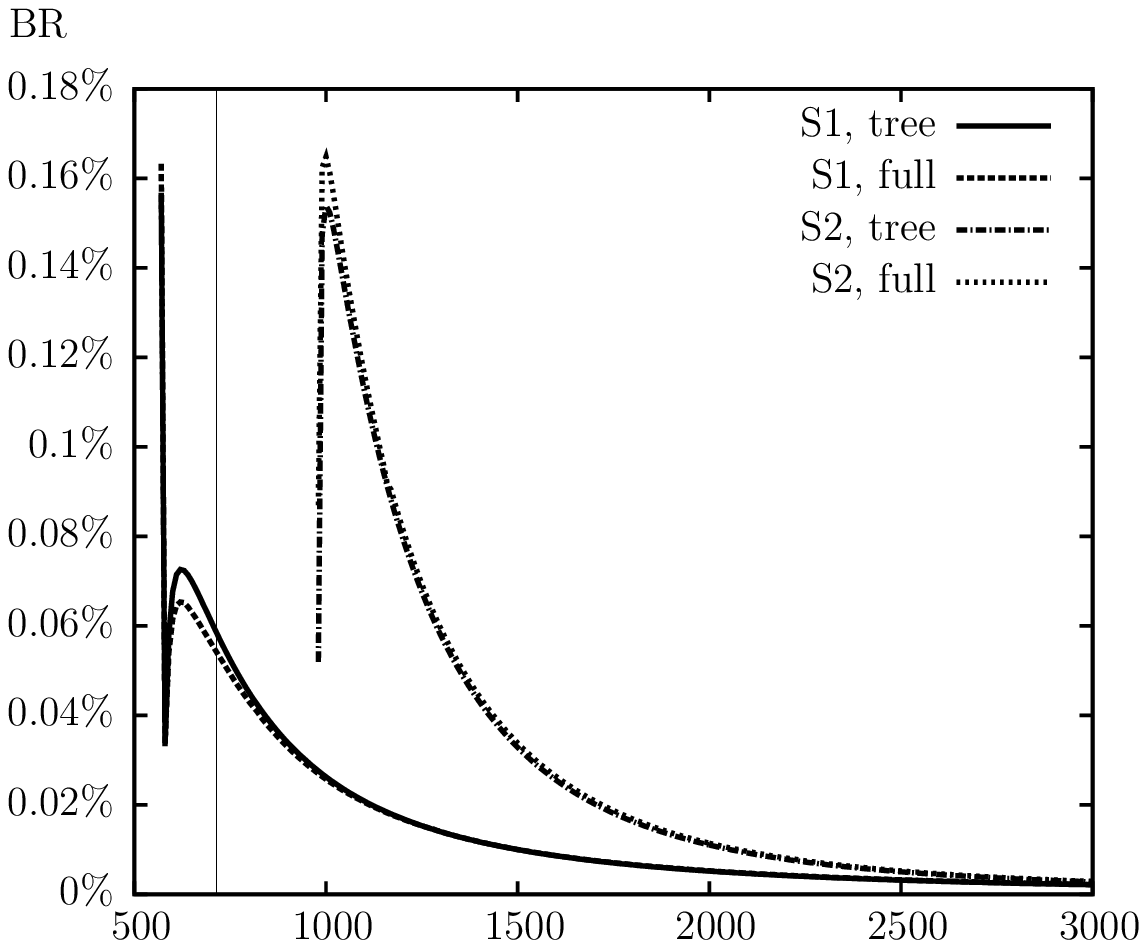}
\hspace{-4mm}
\includegraphics[width=0.49\textwidth,height=7.5cm]{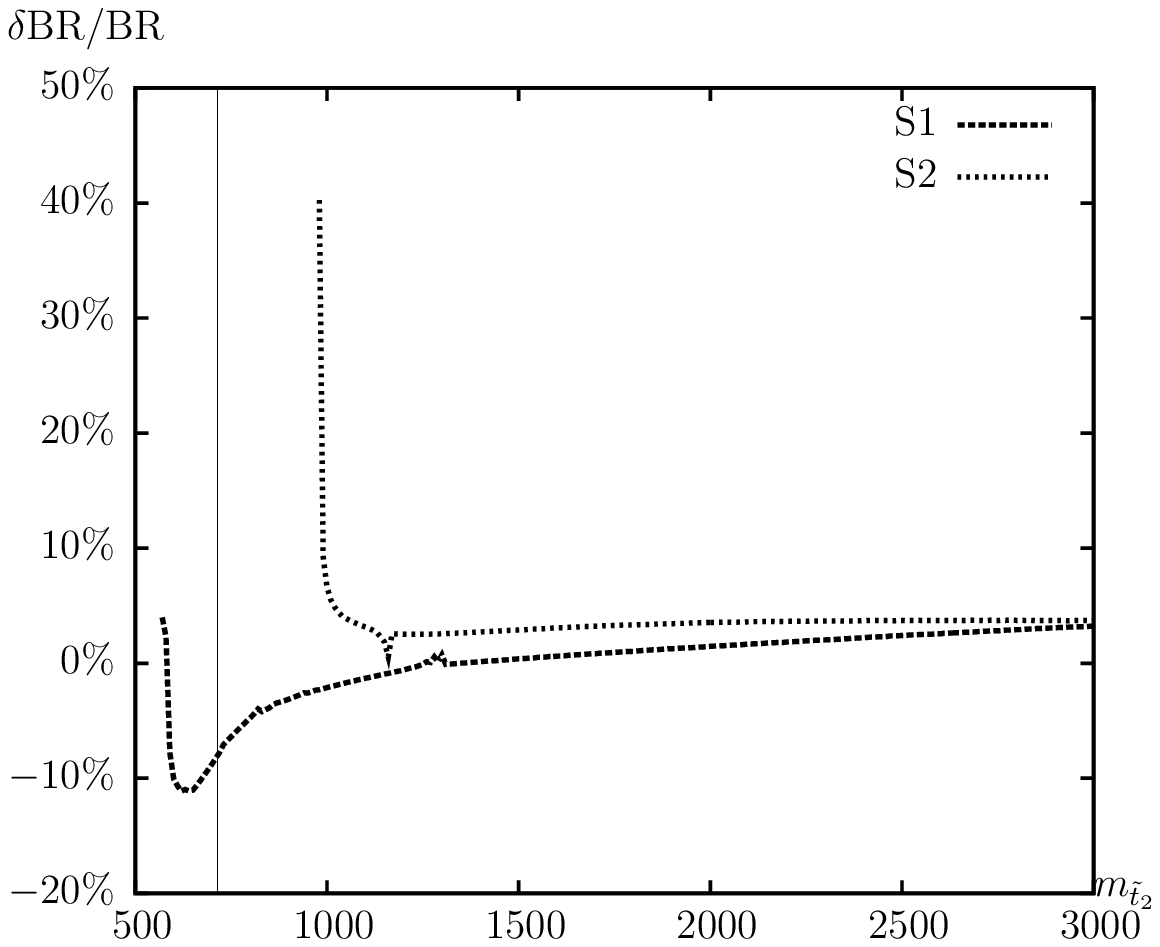}
\end{tabular}
\vspace{2em}
\caption{
  $\Ga(\decaySbzH)$. Tree-level and full one-loop corrected partial decay widths 
  are shown with the parameters chosen according to \SE\ and \SZ\ 
  (see \refta{tab:para}), with $\mstz$ varied.
  The upper left plot shows the partial decay width; the upper right 
  plot shows the corresponding relative size of the corrections.
  The lower left plot shows the BR; the lower right plot shows 
  the relative correction of the BR.
  The vertical lines indicate where $\mstz + \mste = 1000 \gev$, 
  i.e.\ the maximum reach of the ILC(1000).
}
\label{fig:mst2.st2sb2H}
\end{center}
\end{figure}

We now turn to the decay modes $\decaySbiH$ ($i = 1,2$). 
Results are shown in \reffis{fig:mst2.st2sb1H}, \ref{fig:mst2.st2sb2H}, 
(which have been used for the investigations in \citere{SbotRen}).
In \reffi{fig:mst2.st2sb1H} several peaks and dips can be observed.
Within \SE\ the first (fourth) peak at 
$\mstz \approx 571\ (638) \gev$ stems from the threshold 
$\mste + \MHp\ (\MW) = \msbe$ in the self-energies 
$\Si_{\Sbot_{11,21}}(\msbe^2)$ entering the renormalization constants 
$\dZZm{\Sbot}_{11,21}$.
The second dip at $\mstz \approx 596 \gev$ comes from the
threshold  $\mgl + \mb = \msbe$.
The third and the fifth dip come from the same threshold%
\footnote{
  It should be remembered that $\msbe$ changes its value when the value of 
  $\mstz$ is changed.
}%
~$\mt + \mcha1 = \msbe$ at $\mstz \approx 601, 823 \gev$, 
respectively.%
\footnote{
  For these two different input parameters ($\mstz \approx 601, 823 \gev$) 
  we get coincidentally $\msbe \approx 349.14 \gev$.
}%
~The sixth dip at $\mstz \approx 1282 \gev$ comes from the
threshold  $\mt + \mcha2 = \msbe$.
Within \SZ\ the peak/dip at $\mstz \approx 1130 \gev$ 
is the threshold $\mste + \MW = \msbe$.
The other peaks/dips do not appear as the values for the  stop masses
are different. 
Also in \reffi{fig:mst2.st2sb2H} some peaks/dips appear.
Within \SE\ the first peak/dip is visible at 
$\mstz \approx 571 \gev$, due to the threshold 
$\mste + \MHp = \msbe$ (in the self-energy $\Si_{\Sbot_{21}}(\msbe^2)$ 
entering the renormalization constant $\dZZm{\Sbot}_{21}$).
Within \SZ\ the ``apparently single'' dip is in reality two dips at 
$\mstz \approx 1163\ (1164) \gev$ coming from the thresholds
$\msbe + \mA\, (\mH) = \msbz$ (it should be noted that the 
internal Higgs boson masses are tree-level masses).

The absolute value of the  partial decay widths is relatively small, staying
below $\sim 1.2\ (0.2) \gev$ for $\Ga(\decaySbeH)$ ($\Ga(\decaySbzH)$). 
The relative size of the one-loop corrections to $\Ga(\decaySbeH)$ ranges
between $\sim -12\%$ and $-27\%$ ($-20\%$) in \SE\ (\SZ). 
For $\Ga(\decaySbzH)$ very large corrections are found for the smallest
$\mstz$ values, dropping to values close to zero for larger masses.
Because of the small partial decay widths also the branching ratios are 
at or below the $1\%$~level. Only if other channels were kinematically
suppressed, these decays could play a relevant role, and the one-loop
effects could be expected at the level of one-loop contributions to the
partial decay widths itself.

\begin{figure}[htb!]
\begin{center}
\begin{tabular}{c}
\includegraphics[width=0.49\textwidth,height=7.5cm]{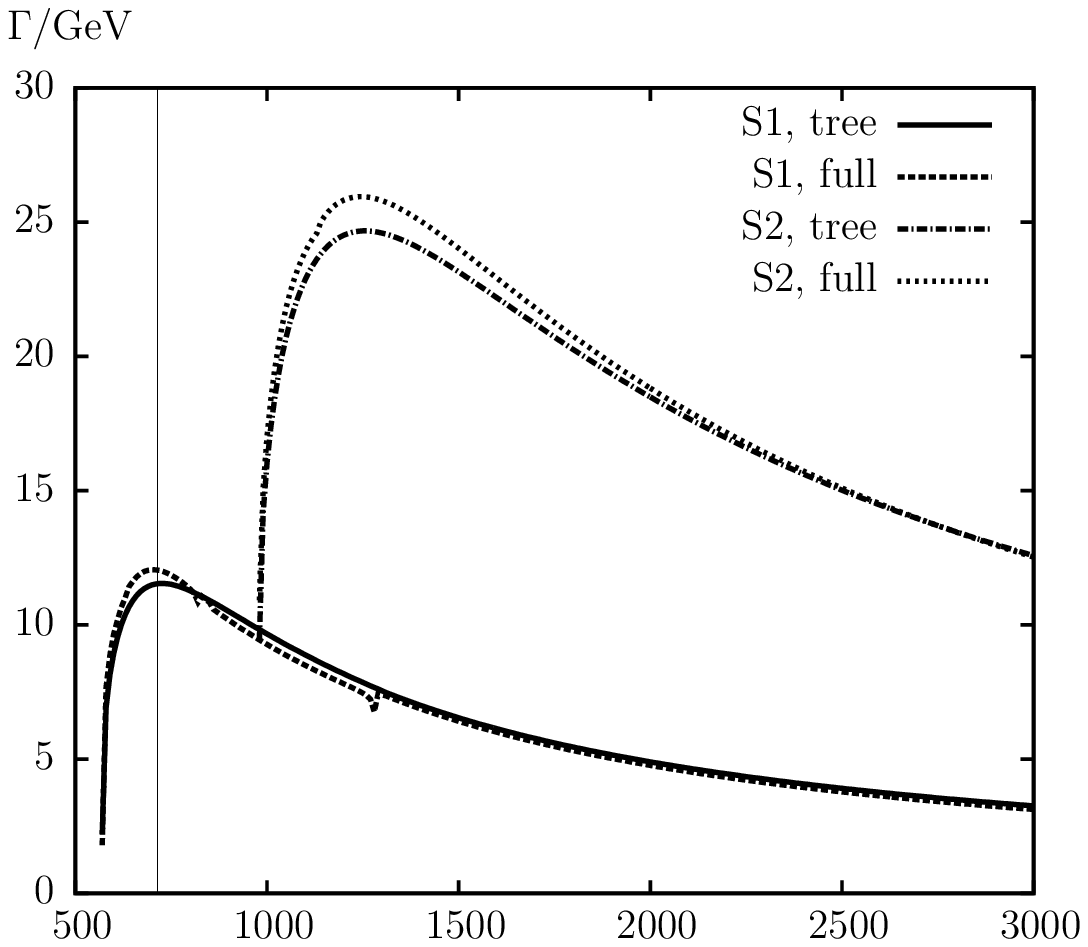}
\hspace{-4mm}
\includegraphics[width=0.49\textwidth,height=7.5cm]{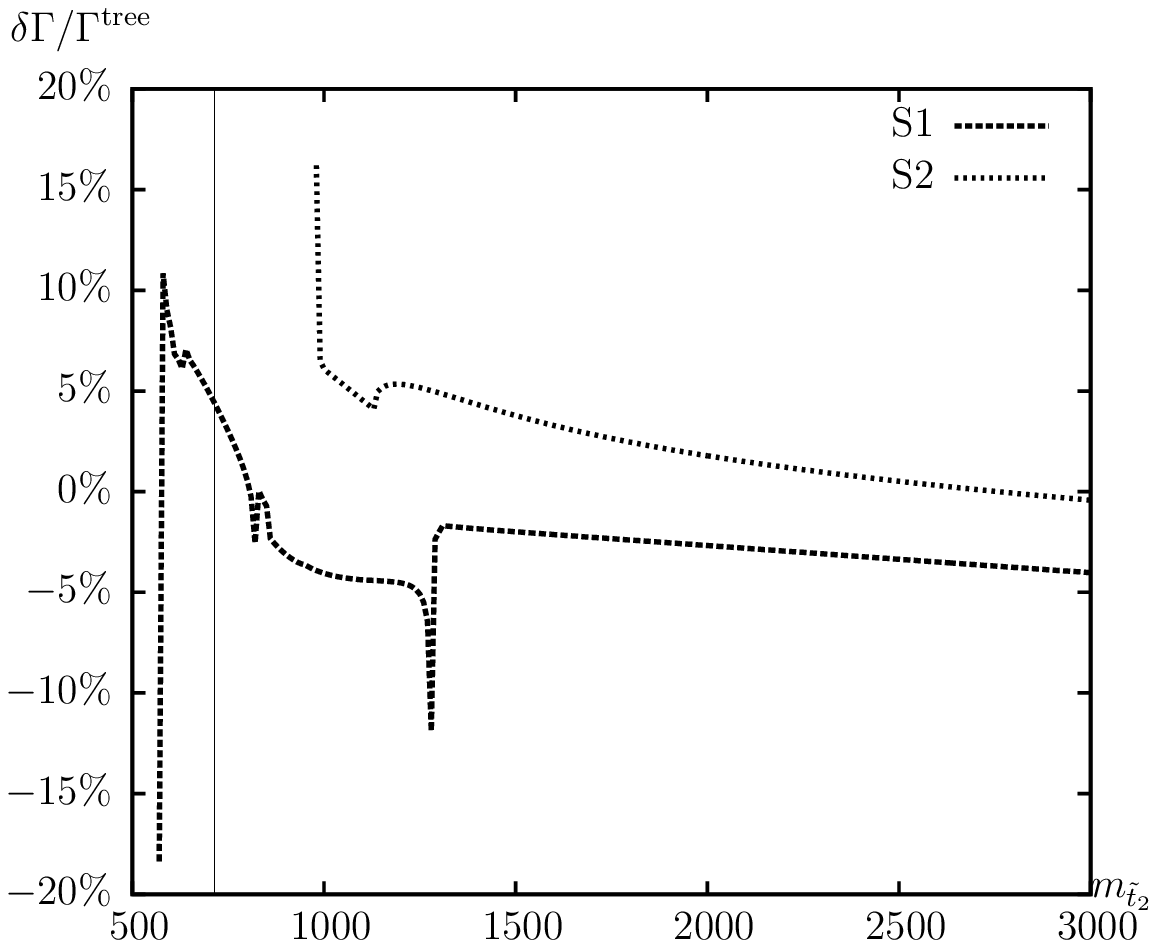}
\\[4em]
\includegraphics[width=0.49\textwidth,height=7.5cm]{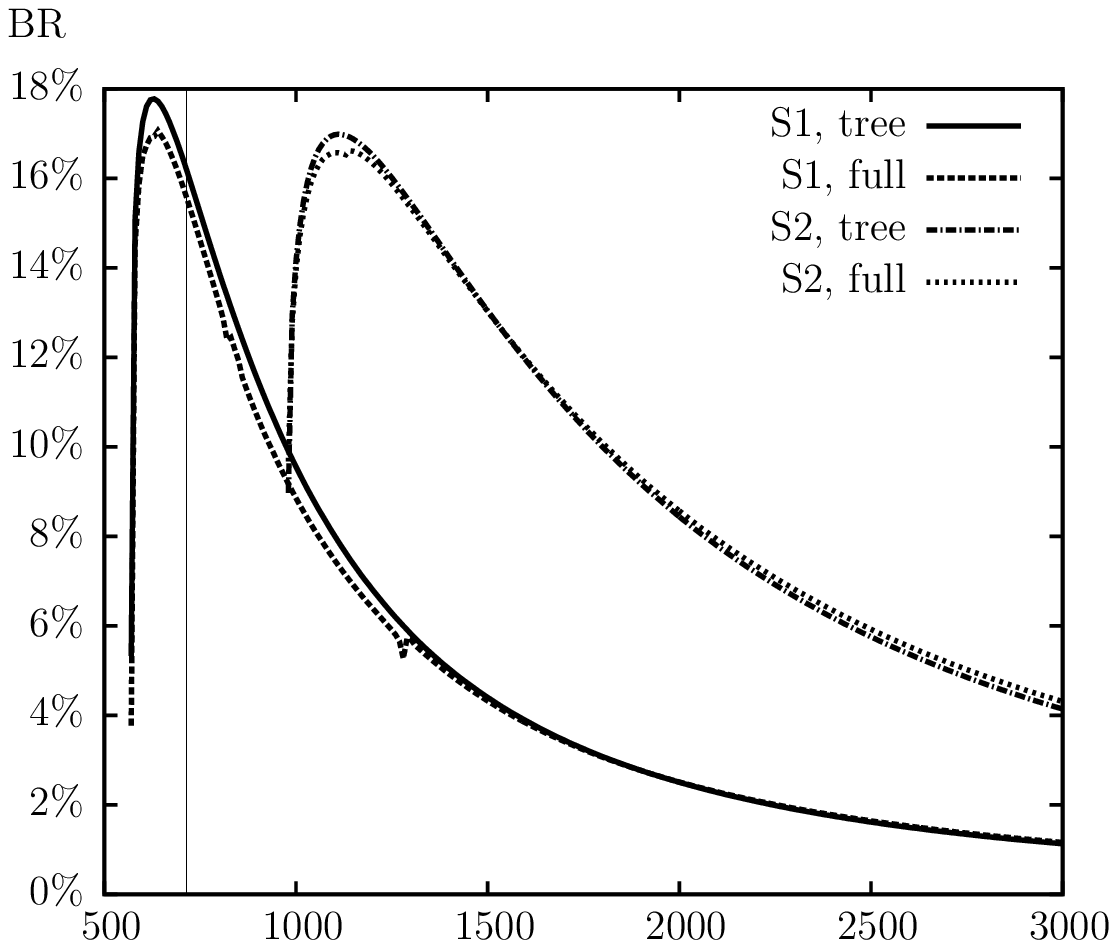}
\hspace{-4mm}
\includegraphics[width=0.49\textwidth,height=7.5cm]{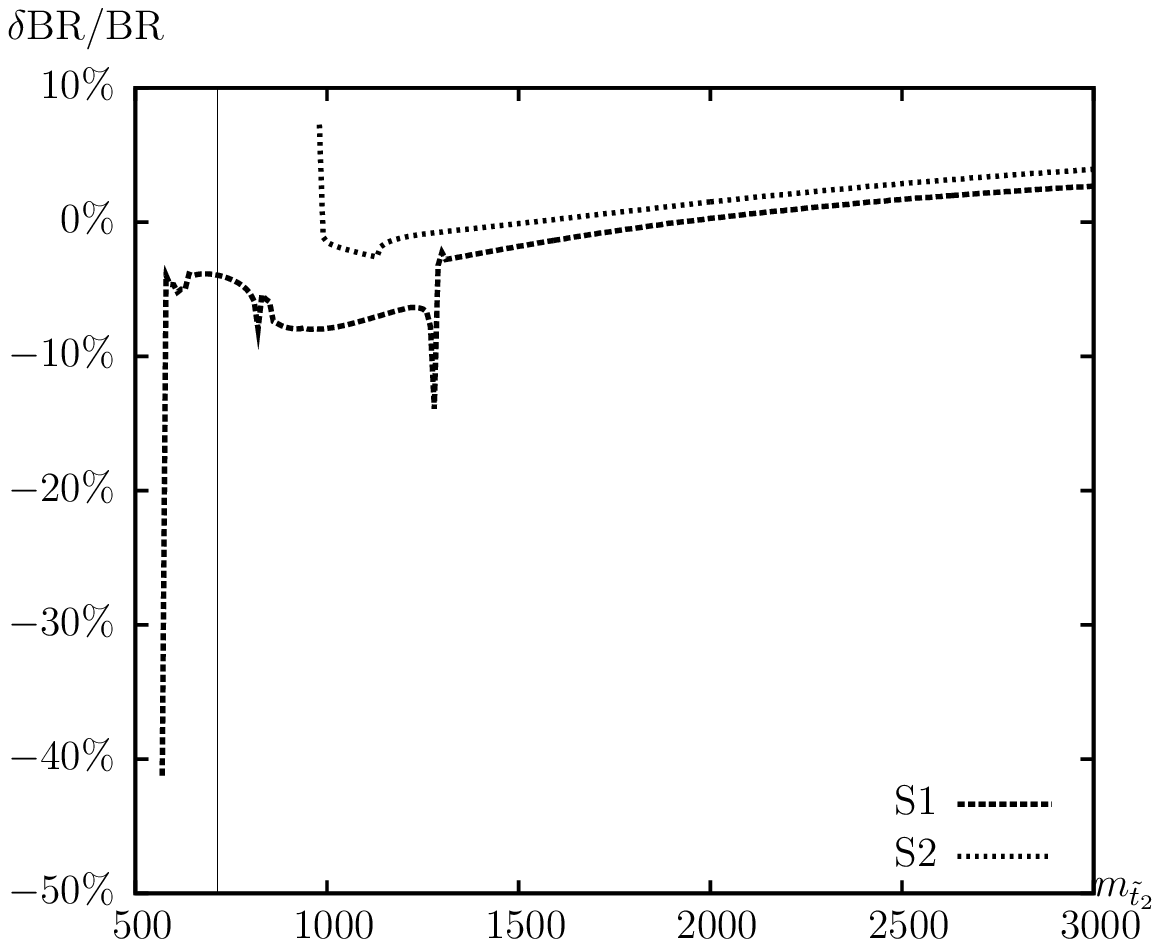}
\end{tabular}
\vspace{2em}
\caption{
  $\Ga(\decaySbeW)$. Tree-level and full one-loop corrected partial decay widths 
  are shown with the parameters chosen according to \SE\ and \SZ\ 
  (see \refta{tab:para}), with $\mstz$ varied.
  The upper left plot shows the partial decay width; the upper right 
  plot shows the corresponding relative size of the corrections.
  The lower left plot shows the BR; the lower right plot shows 
  the relative correction of the BR.
  The vertical lines indicate where $\mstz + \mste = 1000 \gev$, 
  i.e.\ the maximum reach of the ILC(1000).
}
\label{fig:mst2.st2sb1W}
\end{center}
\end{figure}

\begin{figure}[htb!]
\begin{center}
\begin{tabular}{c}
\includegraphics[width=0.49\textwidth,height=7.5cm]{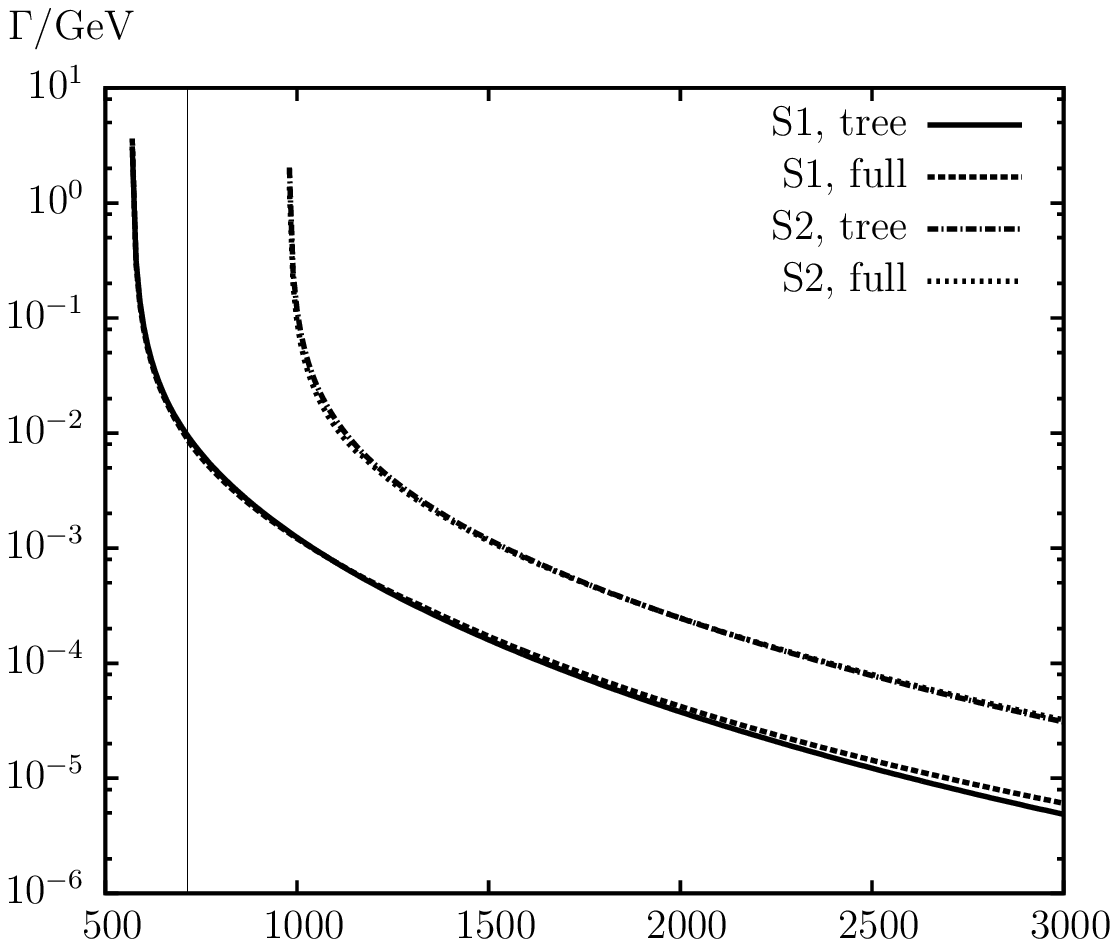}
\hspace{-4mm}
\includegraphics[width=0.49\textwidth,height=7.5cm]{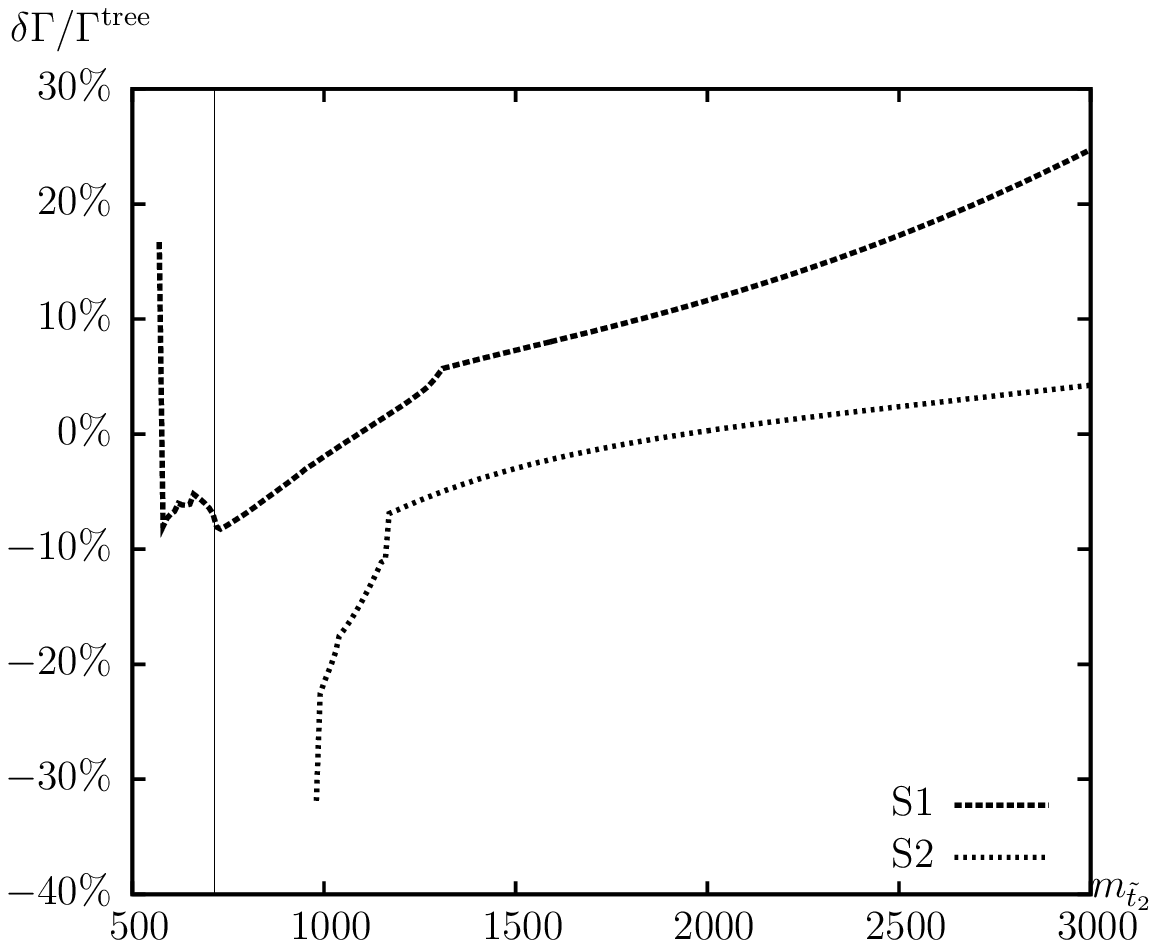}
\\[4em]
\includegraphics[width=0.49\textwidth,height=7.5cm]{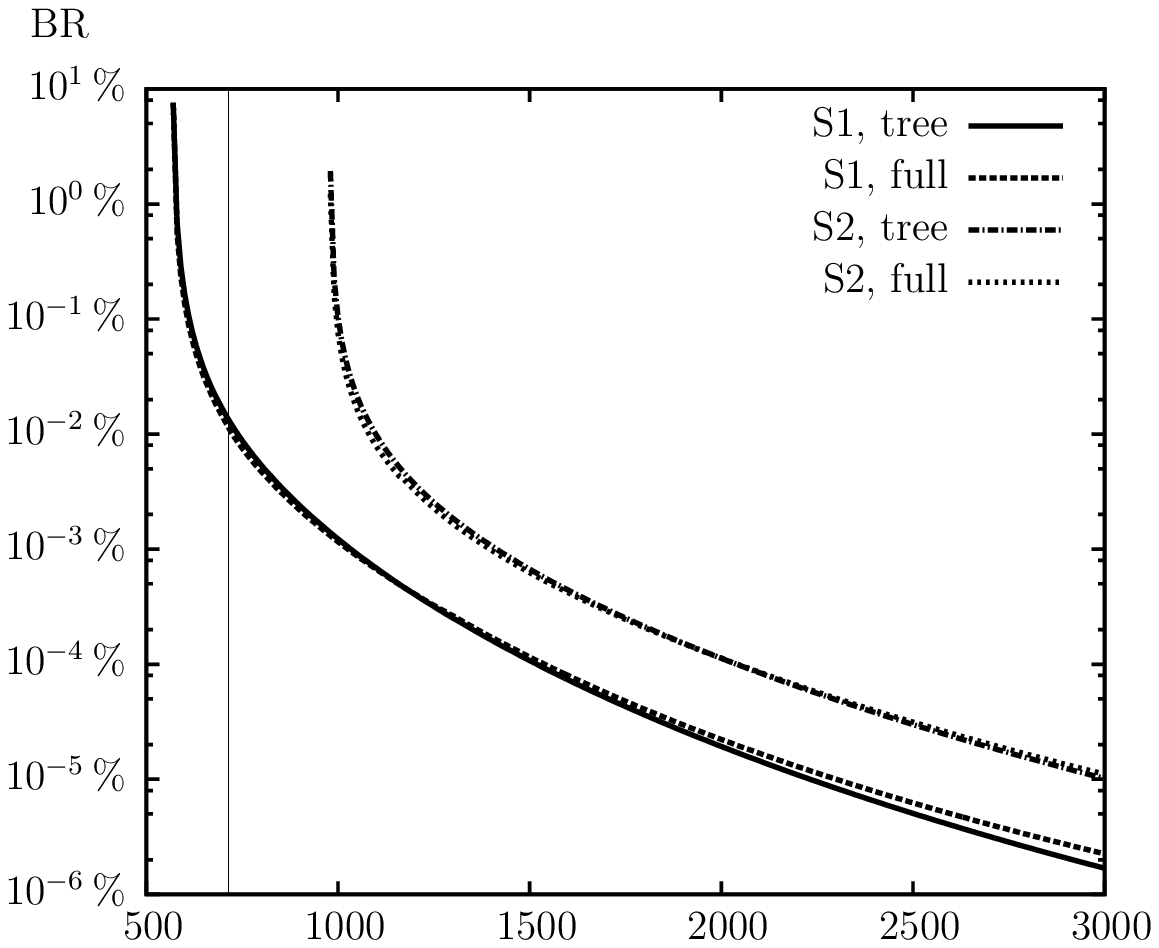}
\hspace{-4mm}
\includegraphics[width=0.49\textwidth,height=7.5cm]{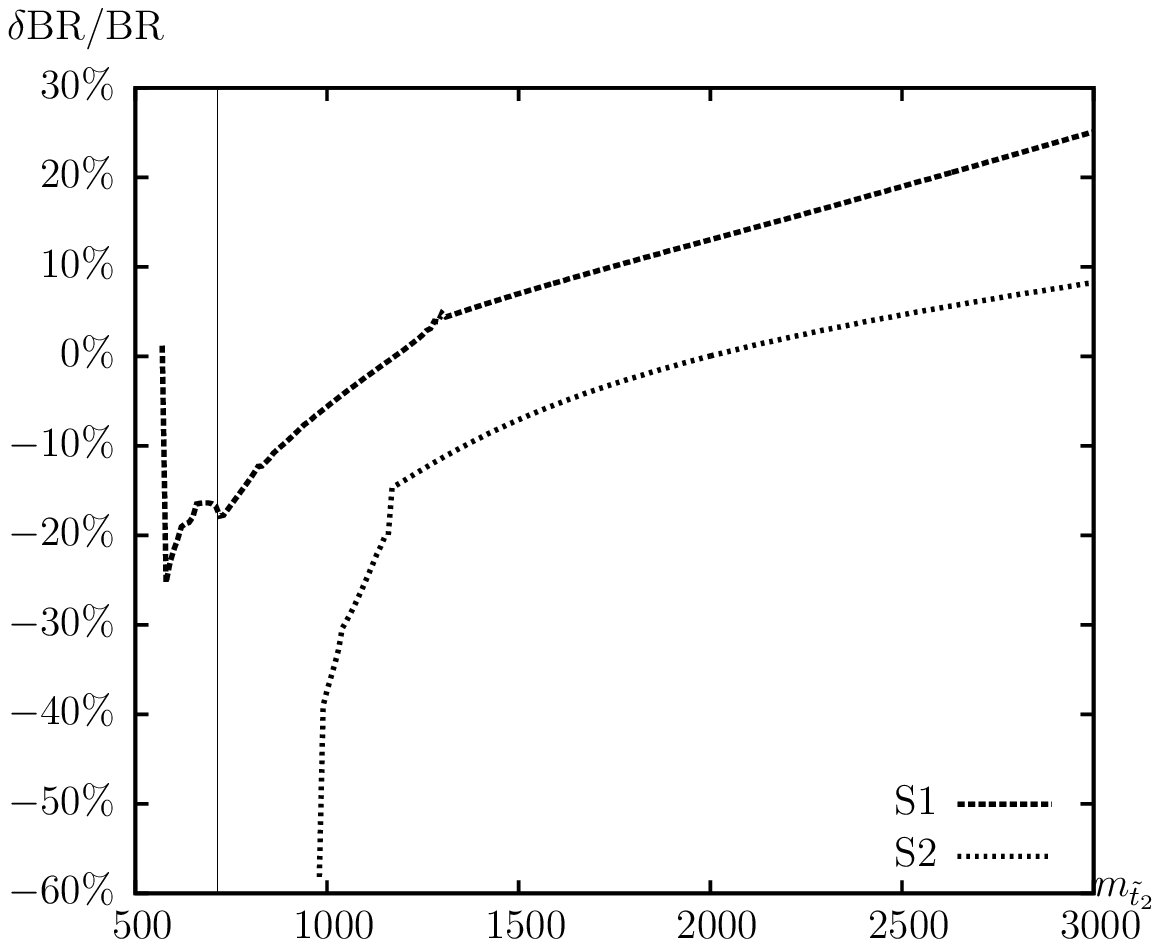}
\end{tabular}
\vspace{2em}
\caption{
  $\Ga(\decaySbzW)$. Tree-level and full one-loop corrected partial decay widths 
  are shown with the parameters chosen according to \SE\ and \SZ\ 
  (see \refta{tab:para}), with $\mstz$ varied.
  The upper left plot shows the partial decay width; the upper right 
  plot shows the corresponding relative size of the corrections.
  The lower left plot shows the BR; the lower right plot shows 
  the relative correction of the BR.
  The vertical lines indicate where $\mstz + \mste = 1000 \gev$, 
  i.e.\ the maximum reach of the ILC(1000).
}
\label{fig:mst2.st2sb2W}
\end{center}
\end{figure}

Results for the other decay modes involving scalar top and bottom 
quarks, $\decaySbiW$ ($i = 1,2$), are shown in 
\reffis{fig:mst2.st2sb1W}, \ref{fig:mst2.st2sb2W}. 
Also these decay modes have been analyzed in detail in \citere{SbotRen}. 
The peaks/dips are the same ones as for the decays $\decaySbiH$. 
On top of that due to different renormalization constants
entering the calculation one observes the following: within
\SE\ peaks appear at $\mstz \approx 618\ (656, 657) \gev$ due to 
$\msbe + \MZ\ (\mA, \mH) = \msbz$, and at $\mstz \approx 721 \gev$
due to $\mt + \mcha2 = \mstz$. The ``knee'' at $\mstz \approx 1303 \gev$
results from $\mt + \mgl = \mste$ in $\Si_{\Stop_{12}}(\mste^2)$ entering
the renormalization constant $\de Y_t$. 
Within \SZ\ one hardly visible dip can be found at
$\mstz \approx 1039 \gev$ from $\msbe + \MZ = \msbz$. The ``knee'' at
$\mstz \approx 1163\ (1164) \gev$ is the same one as in 
\reffi{fig:mst2.st2sb2H}.
The absolute size of $\Ga(\decaySbeW)$ is found to be between 
$\sim 5 \gev$ and $\sim 25 \gev$, depending on $\mstz$ and the scenario.
$\Ga(\decaySbzW)$, on the other hand, is found to be tiny for nearly 
all $\mstz$ values. 
However, the smallness of $\Ga(\decaySbzW)$ is a purely coincidental effect. 
A slightly smaller $\msbz$ would yield a width of \order{0.1 \gev}.
The relative corrections to $\Ga(\decaySbeW)$ range between 
+10\% (+15\%) and -5\% (0\%) for \SE\ (\SZ). For $\Ga(\decaySbzW)$ we
find corrections of $\sim -10\%$ (-30\%) to +18\% (+3\%) in \SE\ (\SZ),
where it has to be kept in mind that the partial decay width itself is
tiny in our scenarios \SE\ and \SZ.
The branching ratio for $\decaySbeW$ can reach up to $\sim 18\%$ in
the parameter range with $\mstz + \mste \lsim 1000 \gev$. Here the
relative one-loop effect on the BR is $\sim -5\%$ and could be important
to reach the ILC precision.

\newpage

\newpage

\newpage

\newpage

\newpage

\newpage

\newpage

\newpage

\newpage

\newpage

\newpage

\newpage

\newpage

\newpage

\newpage

\clearpage
\newpage


\subsection{Full one-loop results for varying \boldmath{$\phiat$}}
\label{sec:full1Lphiat}

In this subsection we analyze the various partial decay widths%
\footnote{
  Again we note, that we do not investigate the decays of $\aStopz$ here, 
  which would correspond to an analysis of $\cp$ asymmetries, 
  which is beyond the scope of this paper.
}
and branching ratios as a function of $\phiat$. 
The other parameters are chosen according to \refta{tab:para}. 
Thus, within \SE\ we have $\mste + \mstz = 910 \gev$, i.e.\ 
the production channel $e^+e^- \to \aStope\Stopz$ 
is open at the ILC(1000). 
Consequently, the accuracy of the prediction of the various partial decay 
widths and branching ratios should be at the same level (or better) as 
the anticipated ILC precision.
It should be noted that the tree-level prediction depends on $\phiat$ 
via the stop mixing matrix.

When performing an analysis involving complex parameters
it should be noted that the results for physical observables are
affected only 
by certain combinations of the complex phases of the 
parameters $\mu$, the trilinear couplings $\At$, $\Ab$, \ldots, and the
gaugino mass parameters $M_1$, $M_2$,
$M_3$~\cite{MSSMcomplphasen,SUSYphases}.
It is possible, for instance, to rotate the phase $\varphi_{M_2}$ away.
Experimental constraints on the (combinations of) complex phases 
arise, in particular, from their contributions to electric dipole 
moments of the electron and the neutron (see \citeres{EDMrev2,EDMPilaftsis} 
and references therein), of the deuteron~\cite{EDMRitz} and of heavy 
quarks~\cite{EDMDoink}.
While SM contributions enter only at the three-loop level, due to its
complex phases the MSSM can contribute already at one-loop order.
Large phases in the first two generations of sfermions
can only be accommodated if these generations are assumed to be very
heavy~\cite{EDMheavy} or large cancellations occur~\cite{EDMmiracle};
see, however, the discussion in \citere{EDMrev1}. 
A recent review can be found in \citere{EDMrev3}.
Accordingly (using the convention that $\phiMz = 0$, as done in this paper), 
in particular, the phase $\phimu$ is tightly constrained~\cite{plehnix}, 
while the bounds on the phases of the third generation trilinear couplings 
are much weaker.
The phases of $\mu$, $\At$, and $\Ab$ enter only in the combinations 
$(\varphi_{A_{t,b}} + \varphi_{\mu})$ (or in different combinations
together with phases of $M_1$ or $M_3$). 
Setting $\phimu = 0$ (see above) as well as $\phiMe = \phigl = 0$ 
(we do not consider these phases in this paper)
leaves us with $\phiat$ and $\phiab$ as the only complex valued
parameters. The dependence on $\phiab$ on the  partial decay widths
involving scalar bottom quarks has been analyzed in detail in
\citere{SbotRen}. Consequently, we focus on a complex~$\At$ and keep
$\Ab$ real. 

Since now a complex $\At$ can appear in the couplings, contributions 
from absorptive parts of self-energy type corrections on external legs can
arise, and their impact will be discussed%
\footnote{
  In a slight abuse of the language ``full'' still refers to  
  corrections without absorptive contributions.
}.
The corresponding formulas for an inclusion of these absorptive 
contributions via finite wave function correction factors can be found 
in the Appendix.

\begin{figure}[htb!]
\begin{center}
\begin{tabular}{c}
\includegraphics[width=0.49\textwidth,height=7.5cm]{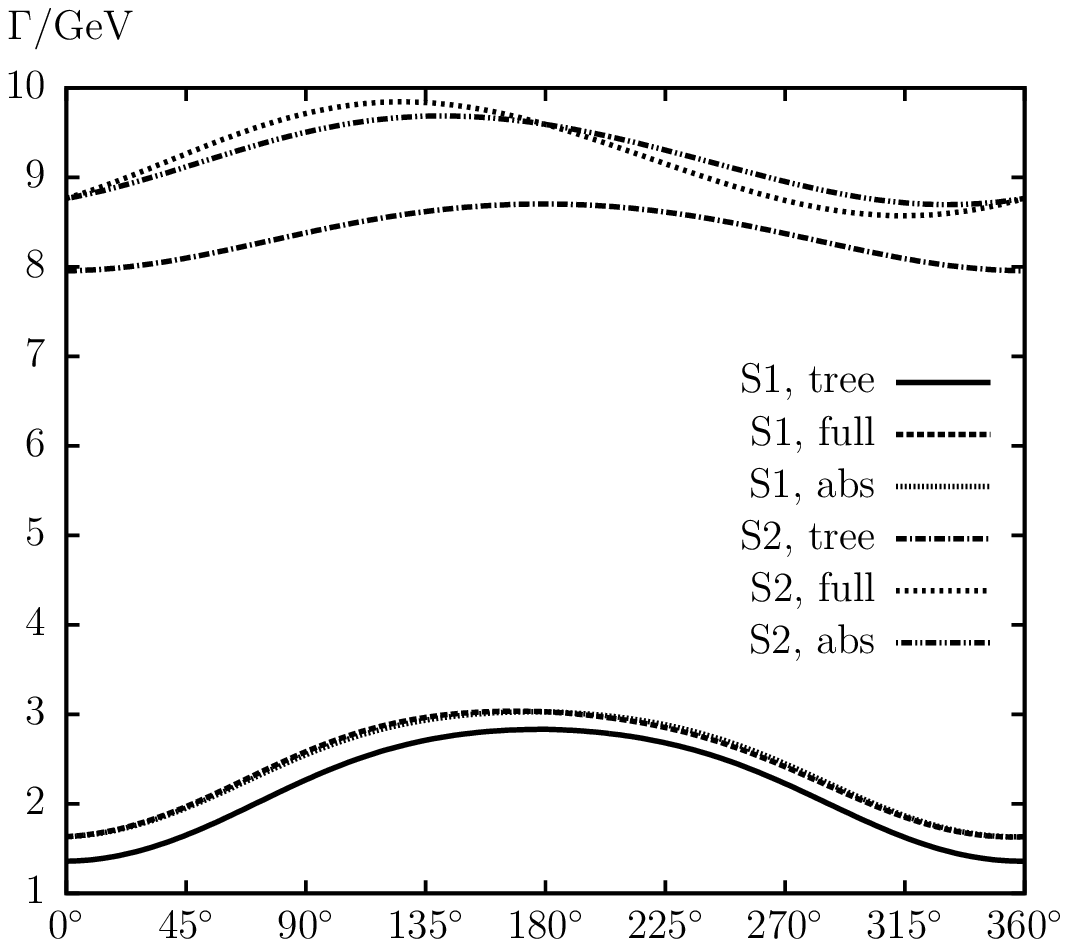}
\hspace{-4mm}
\includegraphics[width=0.49\textwidth,height=7.5cm]{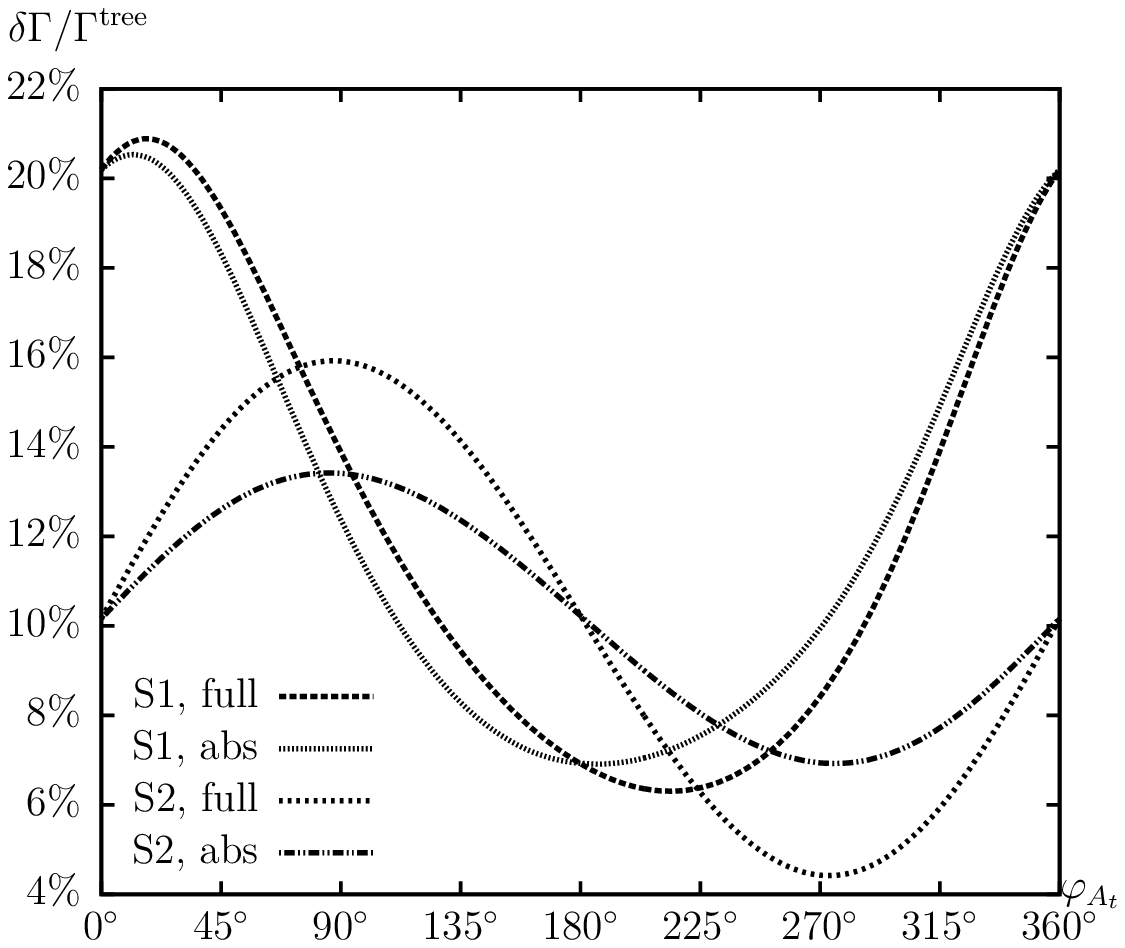}
\\[4em]
\includegraphics[width=0.49\textwidth,height=7.5cm]{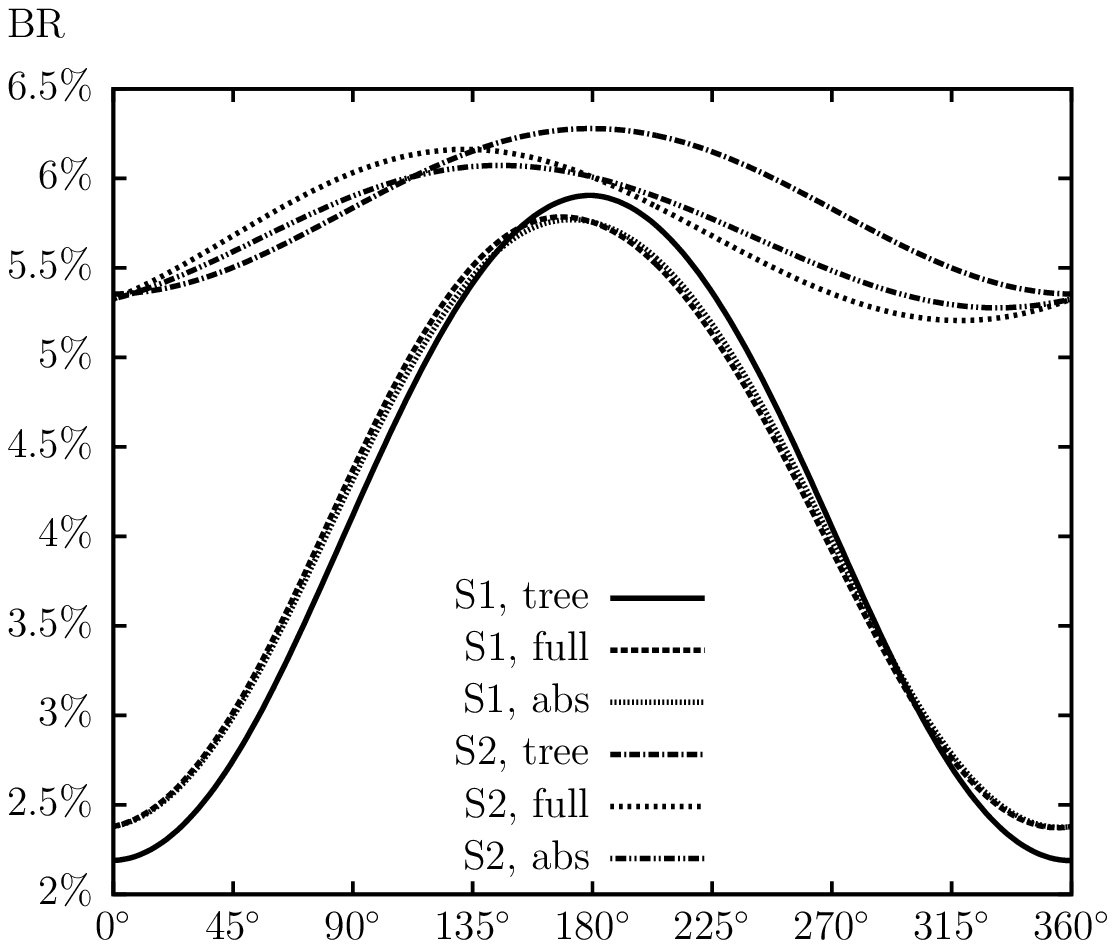}
\hspace{-4mm}
\includegraphics[width=0.49\textwidth,height=7.5cm]{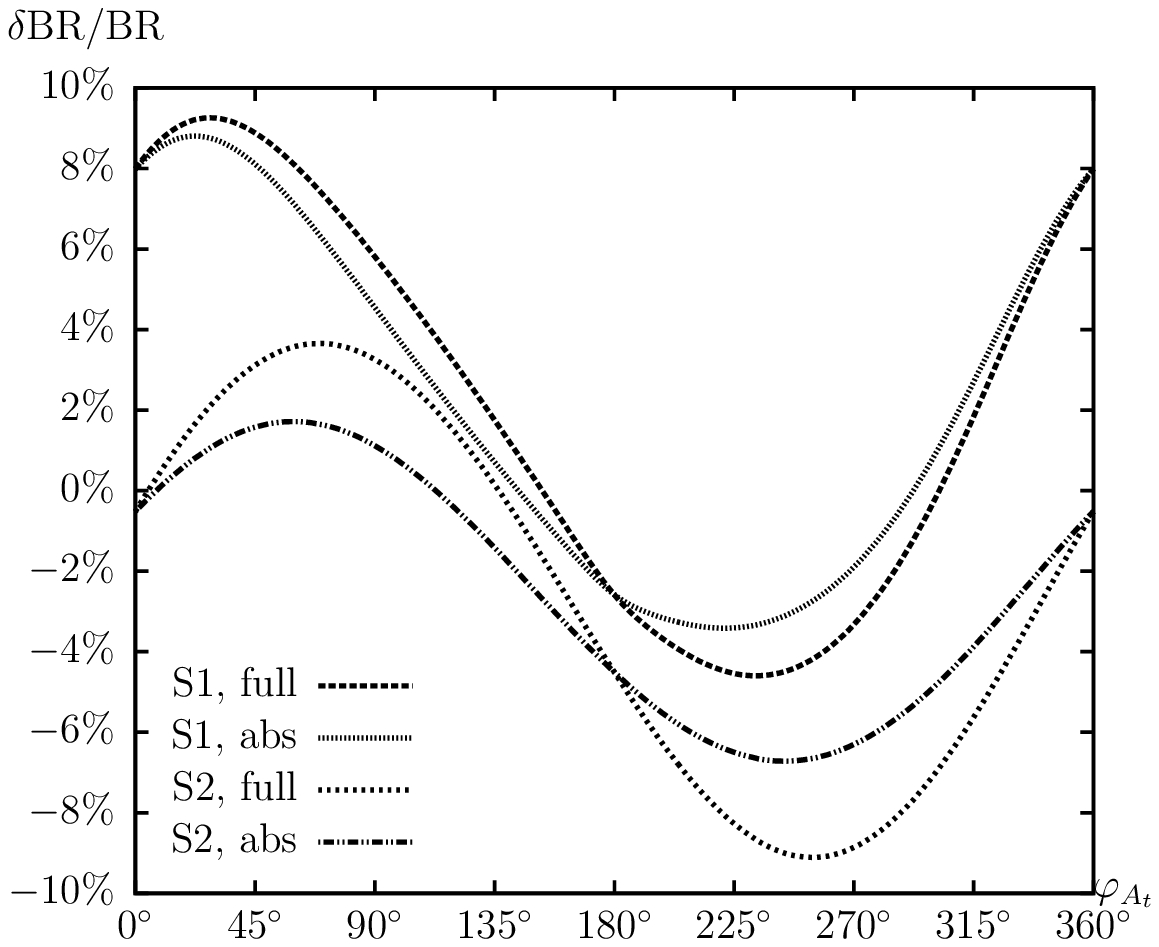}
\end{tabular}
\vspace{2em}
\caption{$\Ga(\decayh)$. 
  Tree-level (``tree'') and full one-loop (``full'') corrected 
  partial decay widths are shown. Also shown are the full one-loop
  corrected partial decay  
  widths including absorptive contributions (``abs''). 
  The parameters are chosen according to \SE\ and \SZ\ (see \refta{tab:para}), 
  with $\phiat$ varied.
  The upper left plot shows the partial decay width; 
  the upper right plot shows the corresponding relative size of the corrections. 
  The lower left plot shows the BR; 
  the lower right plot shows the relative correction of the BR.
}
\label{fig:PhiAt.st2st1h1}
\end{center}
\end{figure}

\begin{figure}[htb!]
\begin{center}
\begin{tabular}{c}
\includegraphics[width=0.49\textwidth,height=7.5cm]{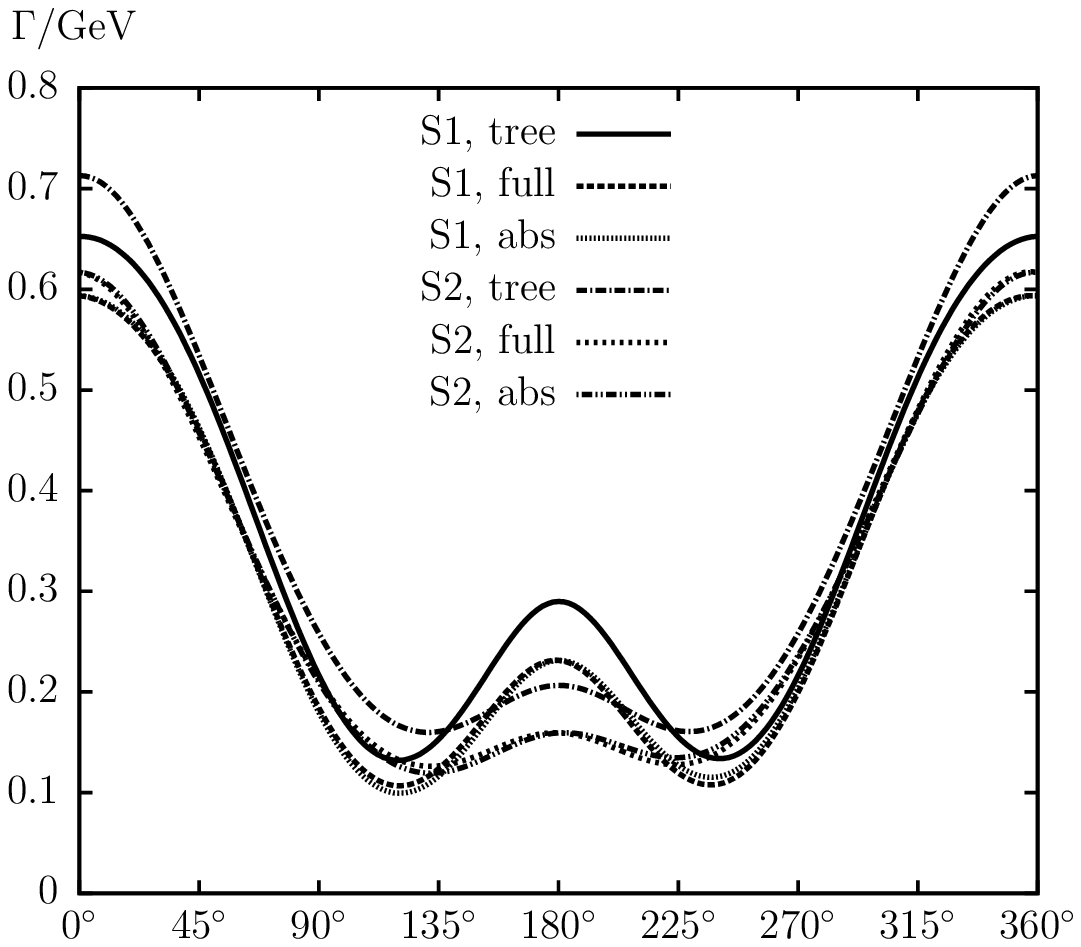}
\hspace{-4mm}
\includegraphics[width=0.49\textwidth,height=7.5cm]{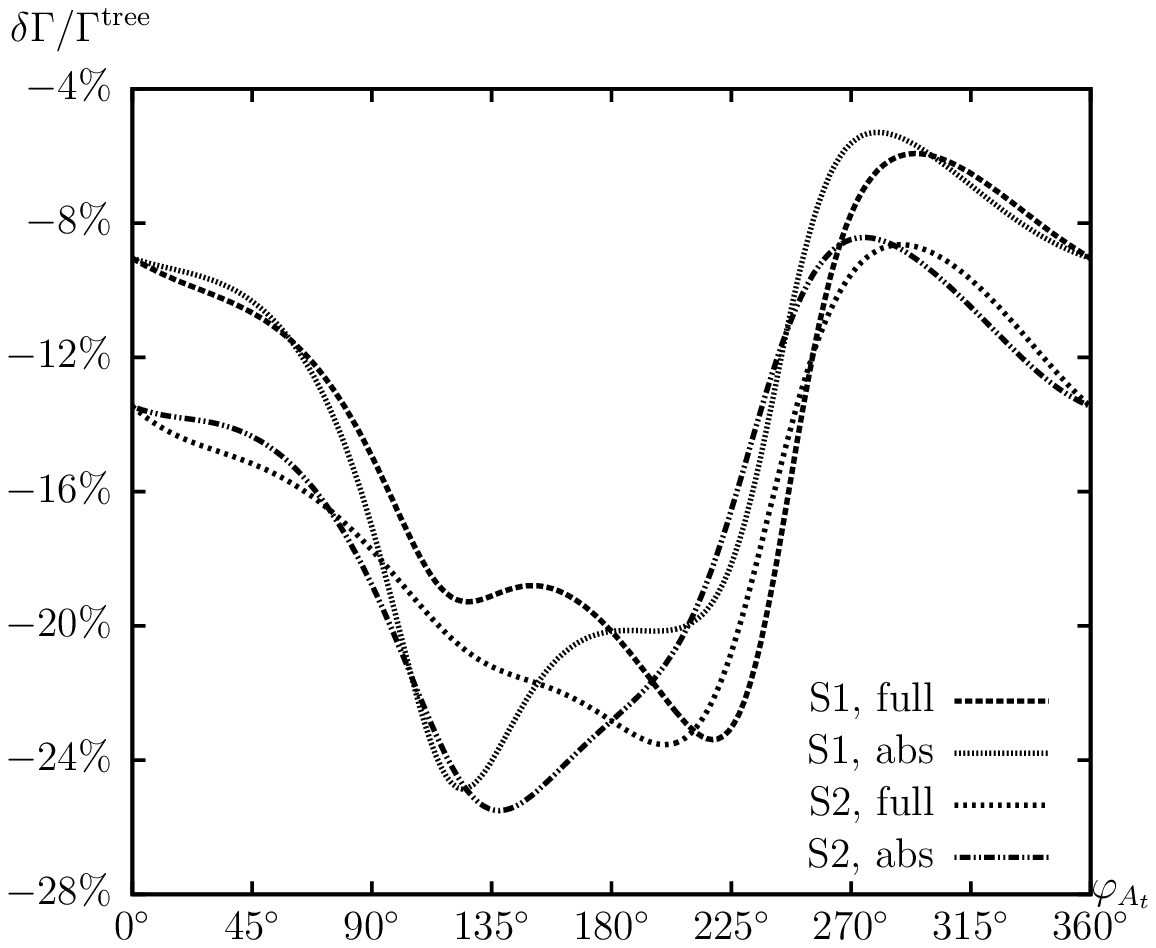}
\\[4em]
\includegraphics[width=0.49\textwidth,height=7.5cm]{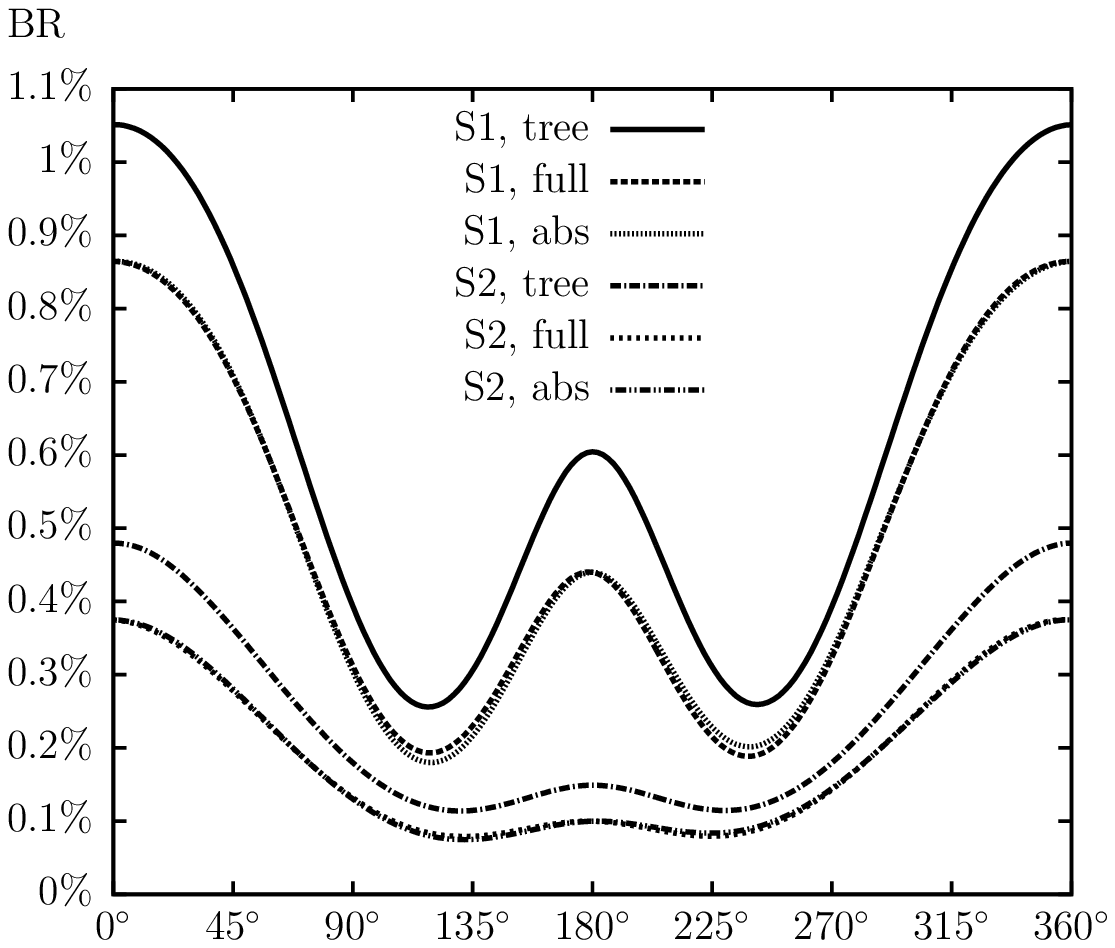}
\hspace{-4mm}
\includegraphics[width=0.49\textwidth,height=7.5cm]{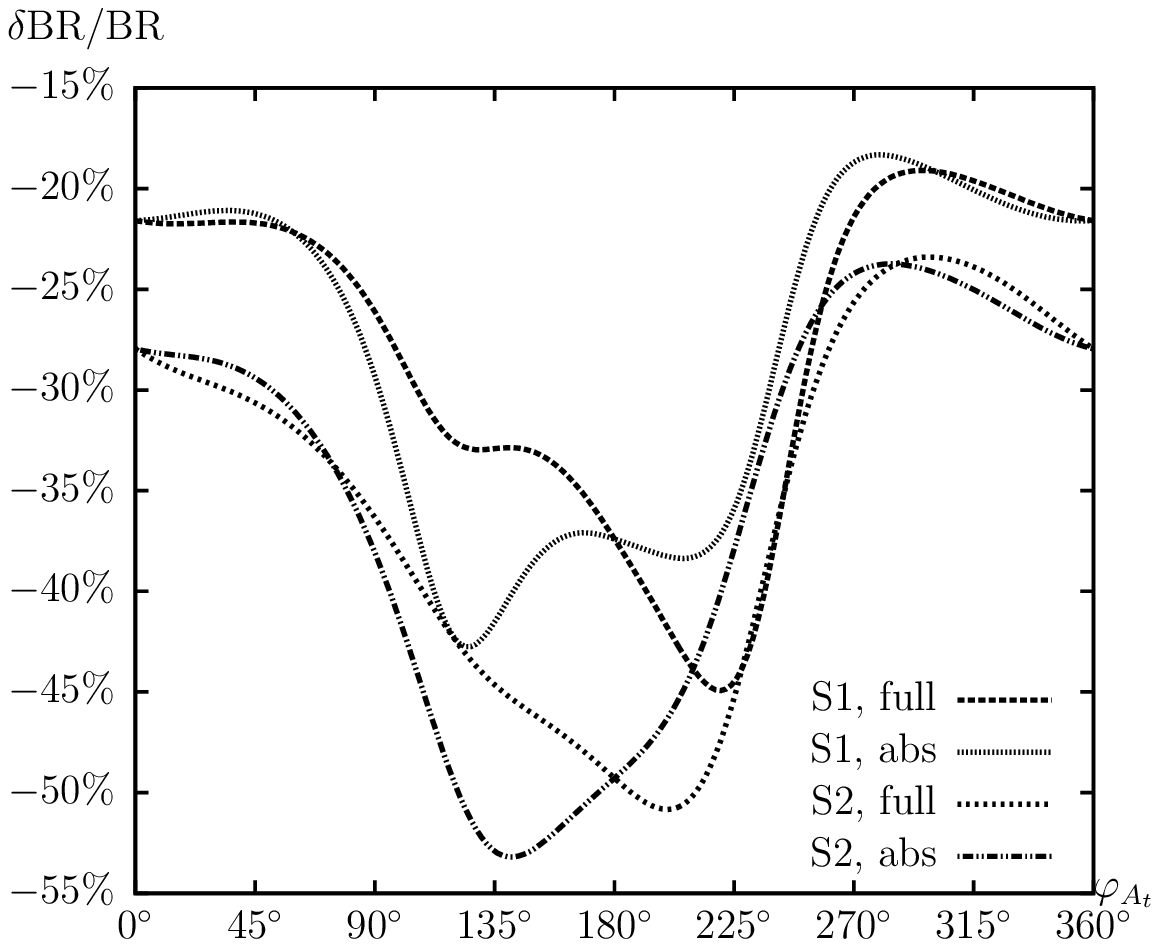}
\end{tabular}
\vspace{2em}
\caption{$\Ga(\decayH)$. 
  Tree-level (``tree'') and full one-loop (``full'') corrected 
  partial decay widths are shown. Also shown are the full one-loop
  corrected partial decay widths including absorptive contributions
  (``abs''). The parameters are chosen according to \SE\ and \SZ\ (see
  \refta{tab:para}), with $\phiat$ varied.
  The upper left plot shows the partial decay width; 
  the upper right plot shows the corresponding relative size of the corrections. 
  The lower left plot shows the BR; 
  the lower right plot shows the relative correction of the BR.
}
\label{fig:PhiAt.st2st1h2}
\end{center}
\end{figure}

\begin{figure}[htb!]
\begin{center}
\begin{tabular}{c}
\includegraphics[width=0.49\textwidth,height=7.5cm]{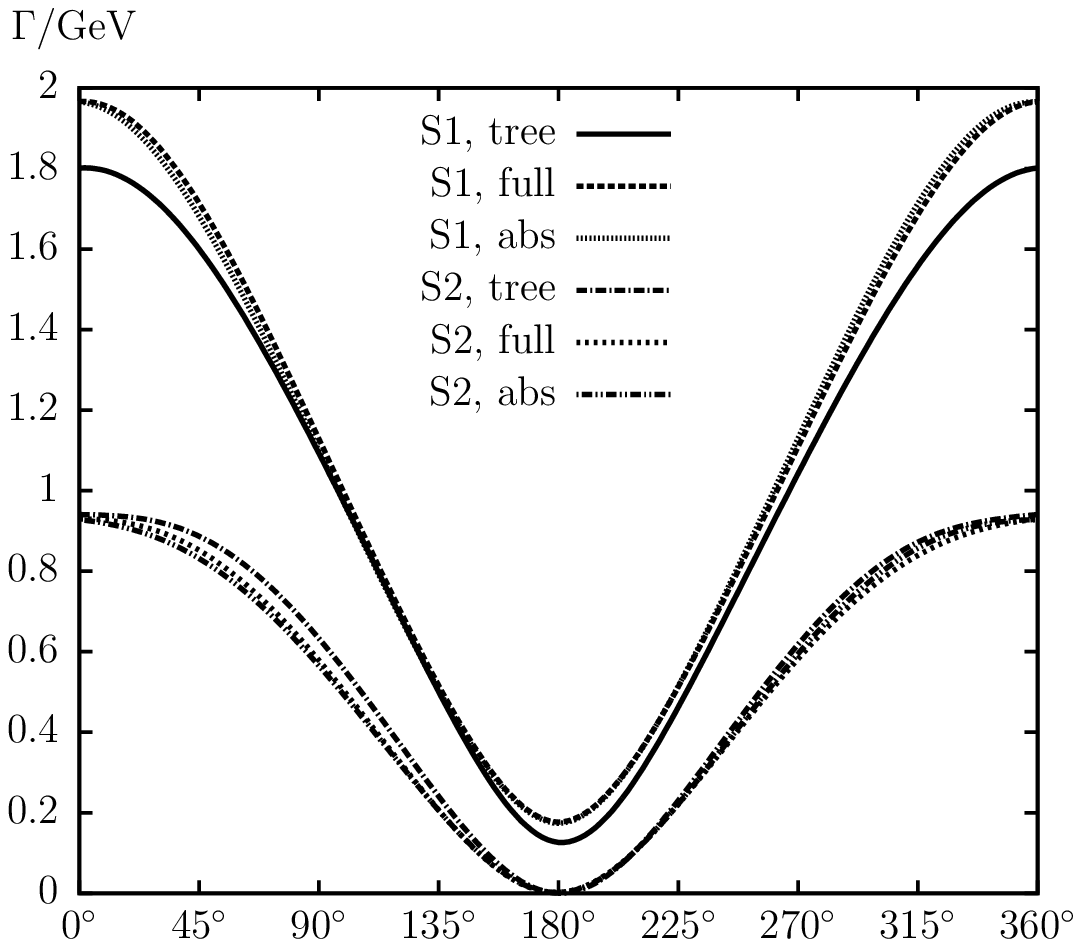}
\hspace{-4mm}
\includegraphics[width=0.49\textwidth,height=7.5cm]{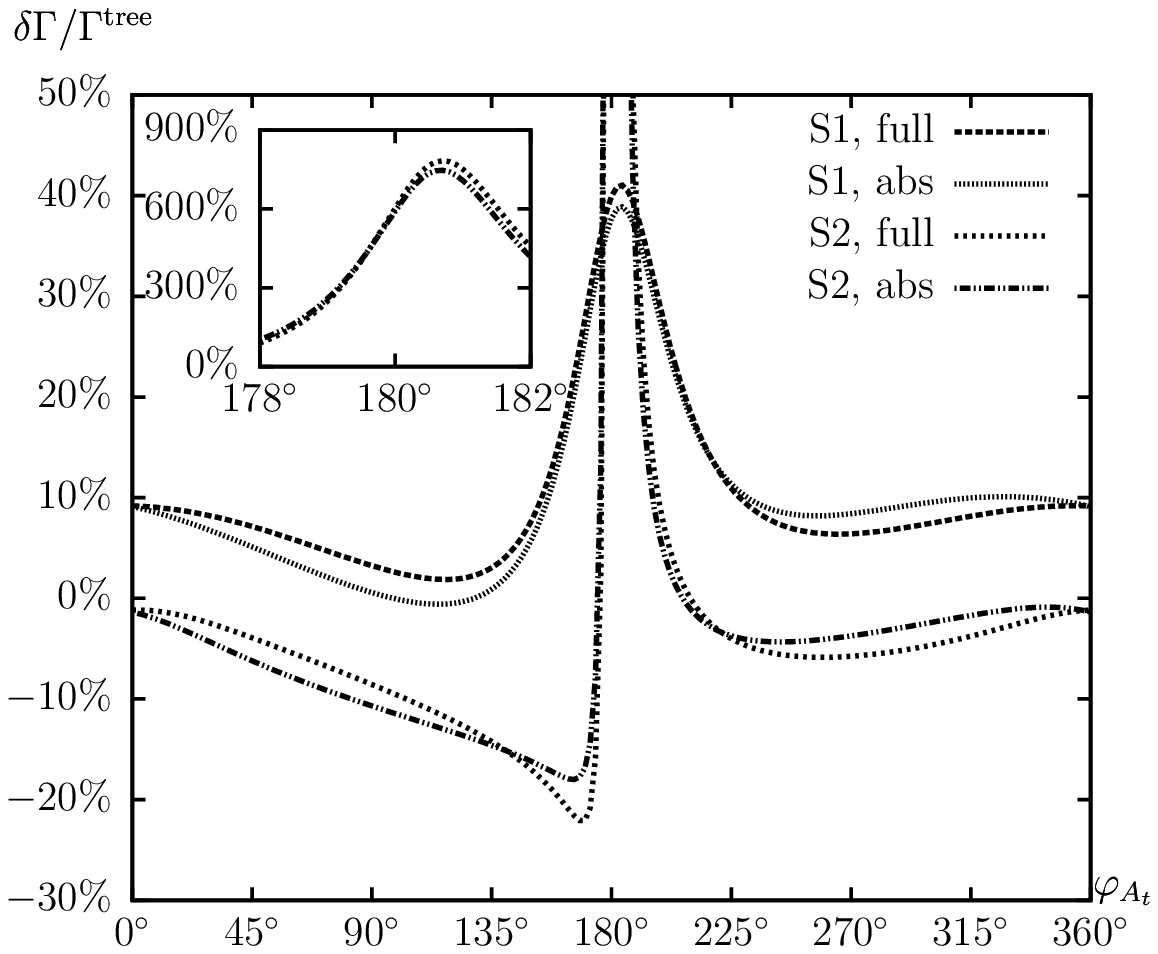}
\\[4em]
\includegraphics[width=0.49\textwidth,height=7.5cm]{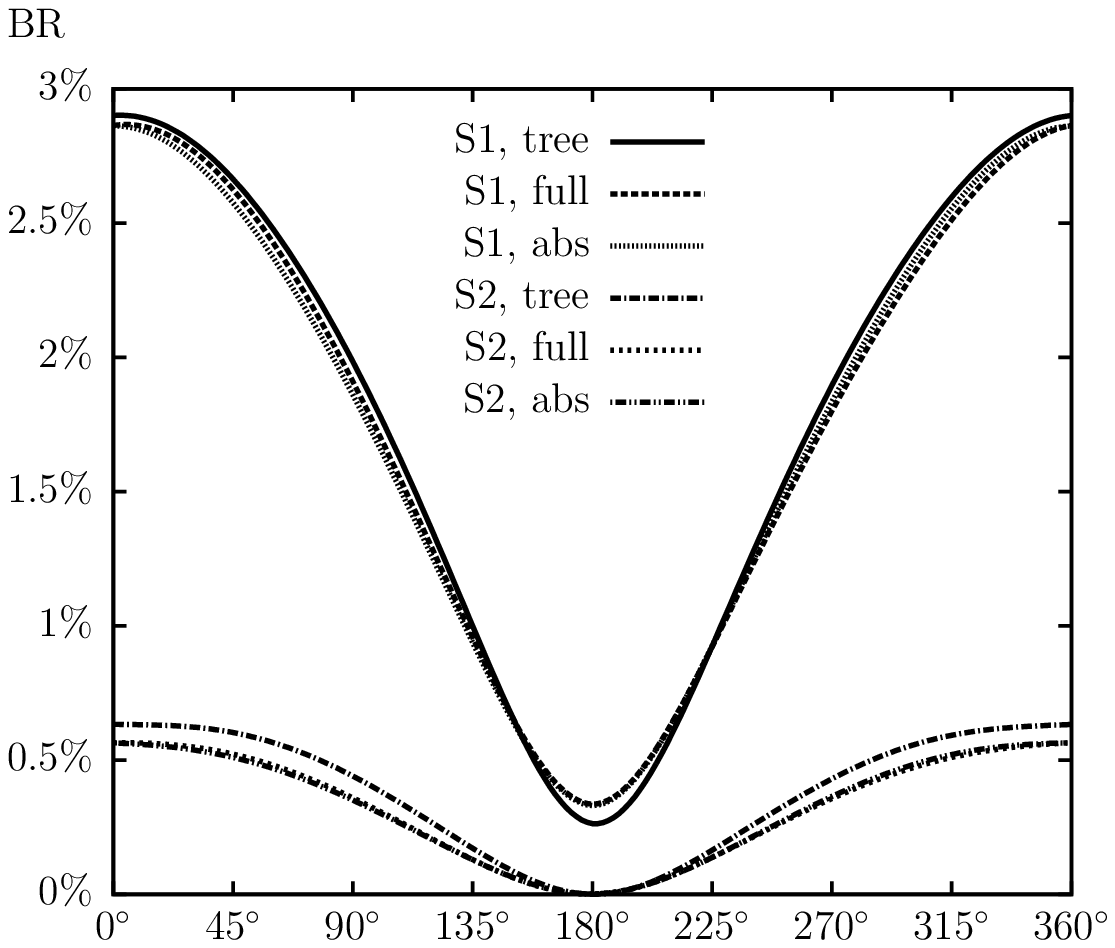}
\hspace{-4mm}
\includegraphics[width=0.49\textwidth,height=7.5cm]{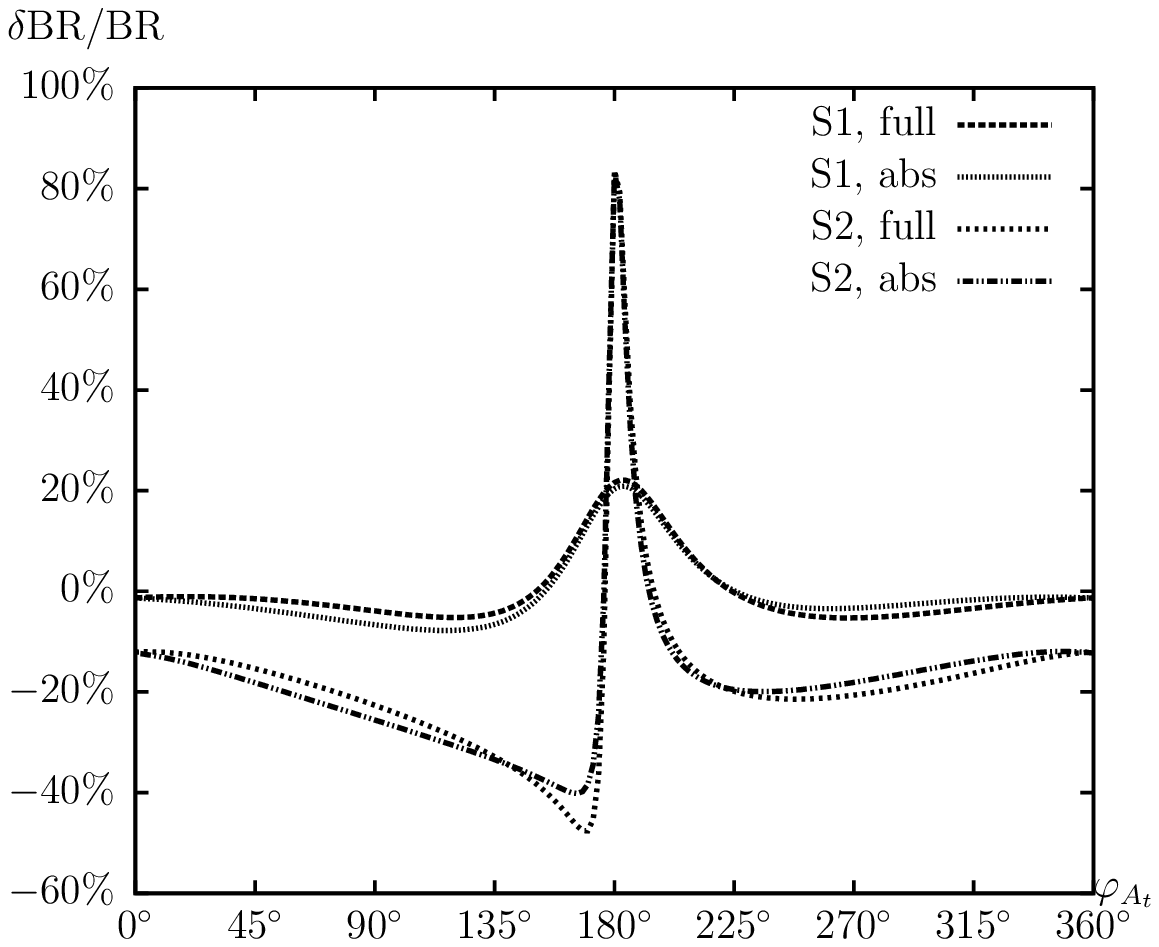}
\end{tabular}
\vspace{2em}
\caption{$\Ga(\decayA)$. 
  Tree-level (``tree'') and full one-loop (``full'') corrected 
  partial decay widths are shown. Also shown are the full one-loop
  corrected partial decay  
  widths including absorptive contributions (``abs''). 
  The parameters are chosen according to \SE\ and \SZ\ (see \refta{tab:para}), 
  with $\phiat$ varied.
  The upper left plot shows the partial decay width; 
  the upper right plot shows the corresponding  relative size of the corrections. 
  The lower left plot shows the BR; 
  the lower right plot shows the relative correction of the BR.
}
\label{fig:PhiAt.st2st1h3}
\end{center}
\end{figure}

As before we start with the decays to Higgs bosons, $\decayhn$ ($n = 1,2,3$)
shown in \reffis{fig:PhiAt.st2st1h1} -- \ref{fig:PhiAt.st2st1h3}.
The arrangement of the panels is the same as in the previous subsection.
In \reffi{fig:PhiAt.st2st1h1}, where the  partial decay width
$\Ga(\decayh)$ is given as a 
function of $\phiat$, one can see that the size of the corrections to
the partial decay width vary substantially with $\phiat$. The one-loop effects
range from $+21\ (+16)\%$ to $+6\ (+4)\%$ in \SE\ (\SZ). 
The effect of the absorptive parts of self-energy type corrections on
external legs  (called ``absorptive contributions'' from now on) are at
the few percent level. For $\phiat = 0, \pi, 2\pi$ these effects 
vanish (by construction).

It should be kept in mind that the parameters are chosen such that 
$e^+e^- \to \aStope\Stopz$ is kinematically possible 
at the ILC(1000) in 
\SE, where the knowledge of such a large variation can be very important. 
Also for $\decayH$, shown in \reffi{fig:PhiAt.st2st1h2}, the variation
with $\phiat$ is very large, ranging from $-6\%$ to $-24\%$,
again with a non-negligible shift from the absorptive contributions.
The wiggles in the size of the relative corrections to the partial decay width
is a result of small numerical variations and not visible in the upper
left panel showing the full decay width. 
These variations are enhanced due to the smallness of the tree-level 
partial width; see \refeq{Garel}. 
The results for $\decayA$ can be found in \reffi{fig:PhiAt.st2st1h3}.
Also here the size of the corrections shows a large variation with
$\phiat$, again with non-negligible absorptive contributions. 
Within \SZ\ for real and negative $\At$ the partial width becomes 
extremely small at tree-level, leading to (formally) very large 
relative one-loop corrections.
The one-loop effects on the branching ratios also vary strongly
with $\phiat$, following the same pattern as the partial decay widths. Effects
up to $\pm 8\%$ are reached for $\br(\decayh)$, while the other two
decay modes reach large corrections only where the BRs are relatively
small, $\lsim 1\%$. The one-loop corrections to $\Ga(\decayh)$,
however, can easily exceed the ILC precision.

\begin{figure}[htb!]
\begin{center}
\begin{tabular}{c}
\includegraphics[width=0.49\textwidth,height=7.5cm]{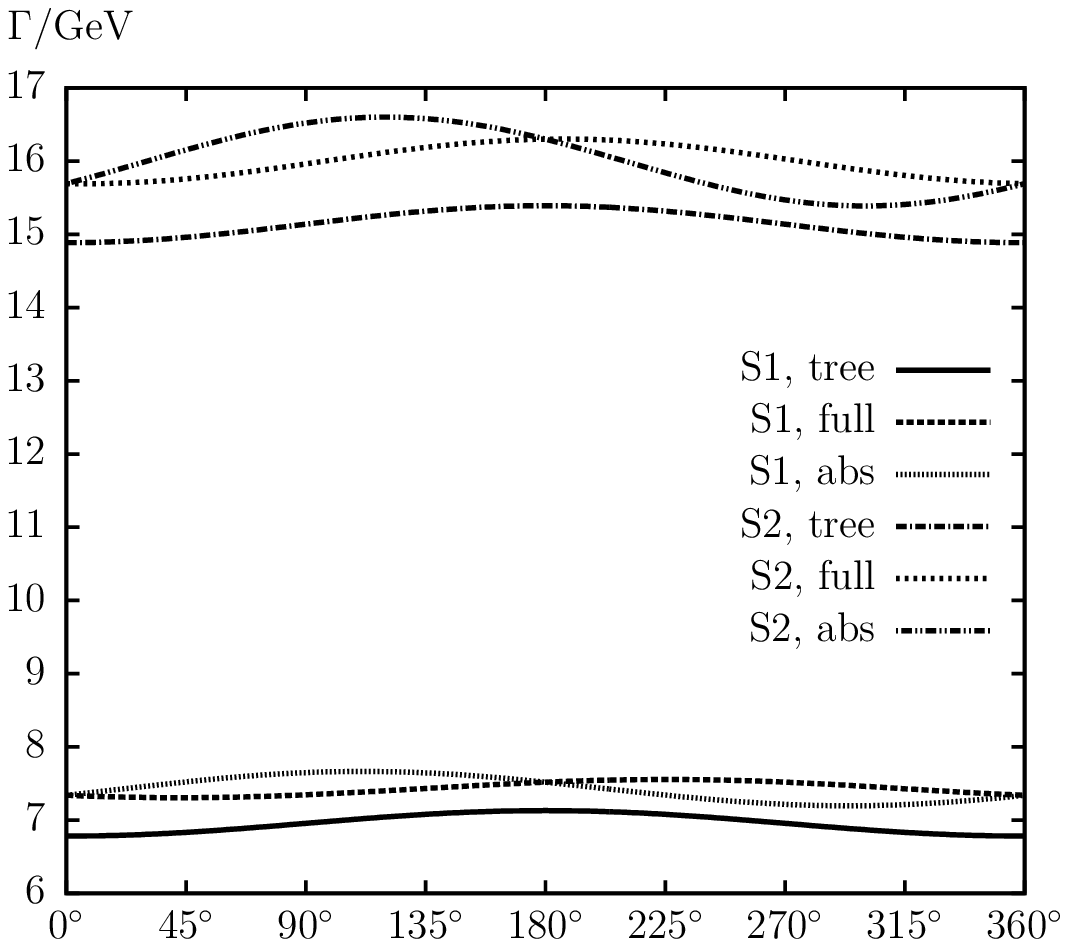}
\hspace{-4mm}
\includegraphics[width=0.49\textwidth,height=7.5cm]{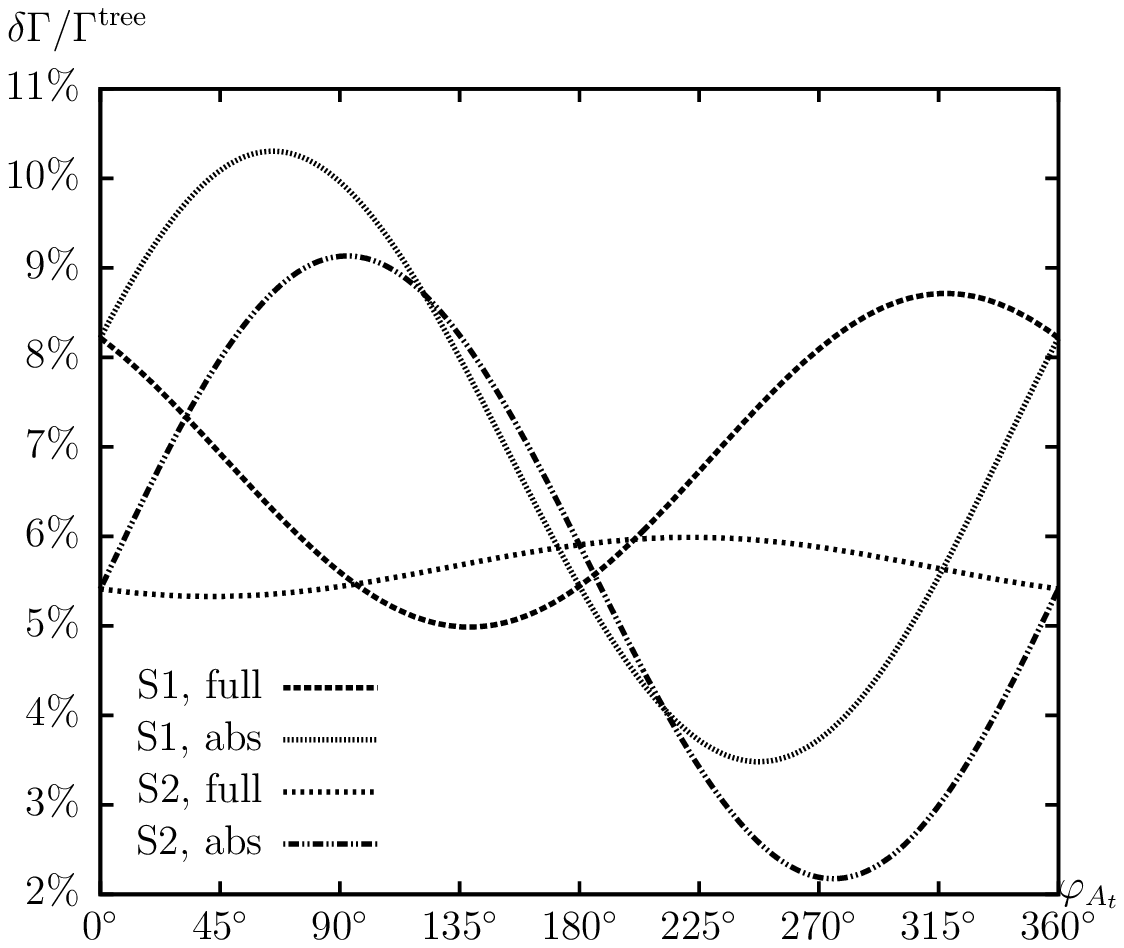}
\\[4em]
\includegraphics[width=0.49\textwidth,height=7.5cm]{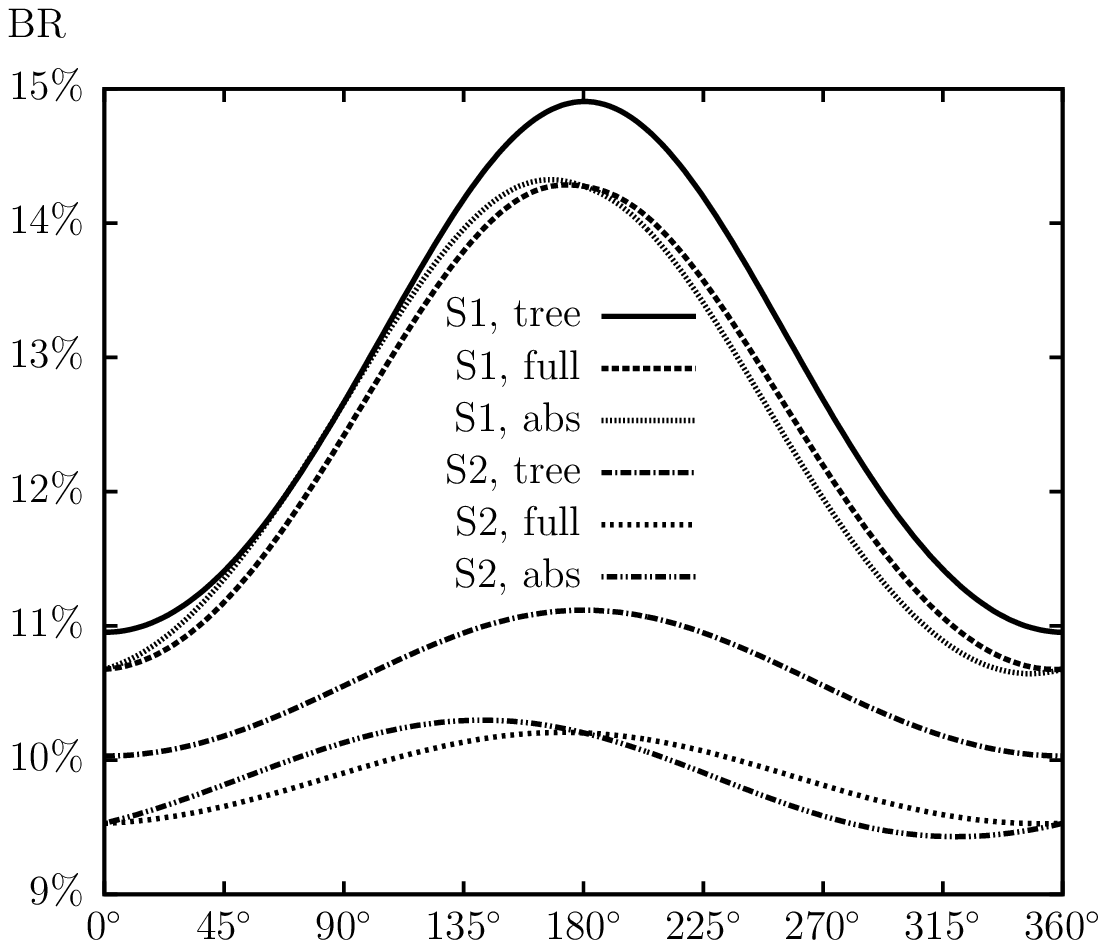}
\hspace{-4mm}
\includegraphics[width=0.49\textwidth,height=7.5cm]{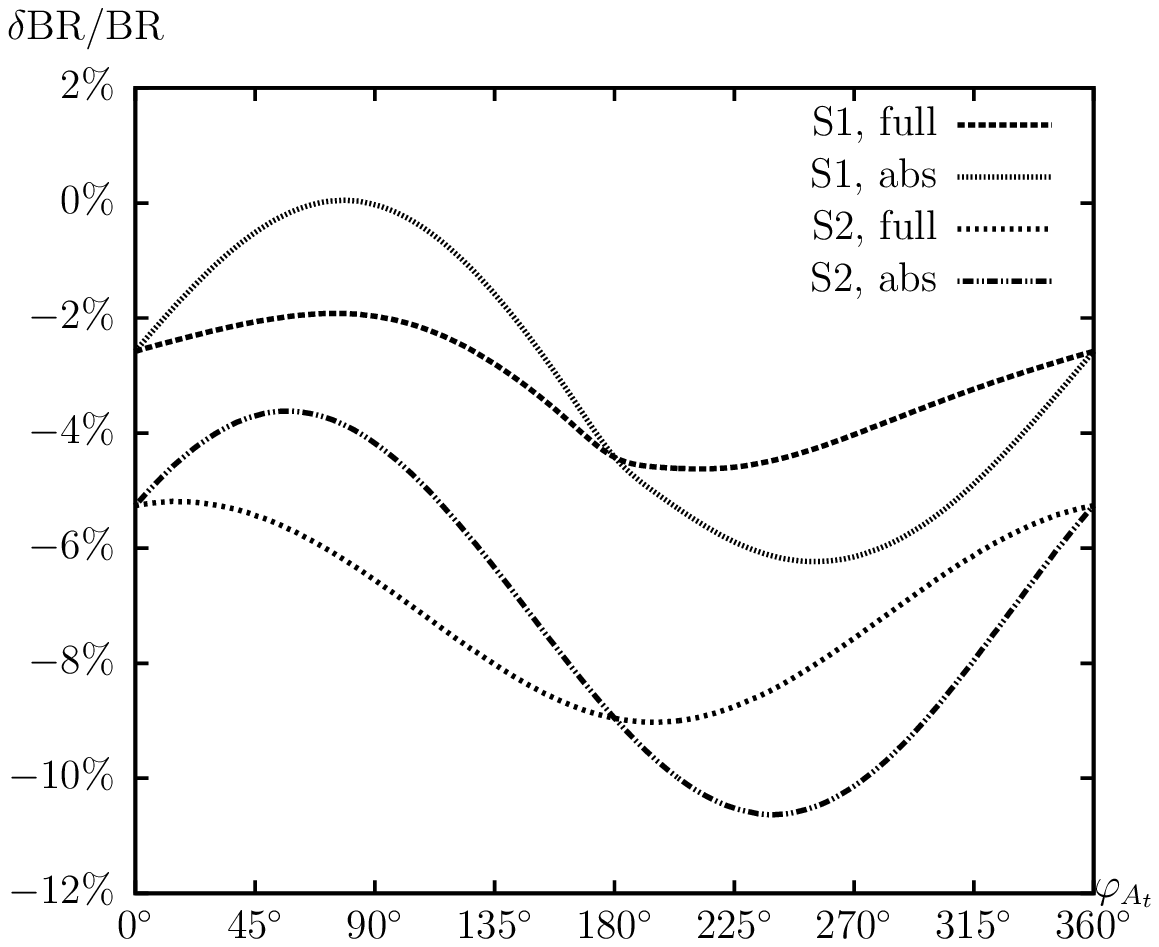}
\end{tabular}
\vspace{2em}
\caption{$\Ga(\decayZ)$. 
  Tree-level (``tree'') and full one-loop (``full'') corrected 
  partial decay widths are shown. Also shown are the full one-loop
  corrected partial decay  
  widths including absorptive contributions (``abs''). 
  The parameters are chosen according to \SE\ and \SZ\ (see \refta{tab:para}), 
  with $\phiat$ varied.
  The upper left plot shows the partial decay width; 
  the upper right plot shows the corresponding  relative size of the corrections. 
  The lower left plot shows the BR; 
  the lower right plot shows the relative correction of the BR.
}
\label{fig:PhiAt.st2st1Z}
\end{center}
\end{figure}

In \reffi{fig:PhiAt.st2st1Z} we present the phase dependence for the
decay mode $\decayZ$. While in \SE\ the effect of the one-loop corrections to 
$\Ga(\decayZ)$ varies from $\sim +8\%$ to $\sim +5\%$, within \SZ\ only a
very small variation can be observed.
These numbers change if the absorptive contributions are taken into
account. In both scenarios substantially larger variations are found.
Within \SE\ (\SZ) the branching ratio varies with $\phiat$ between 
$\sim 11\%$ and $\sim 15\%$ ($9.5\%$ and $11\%$). 
Again the variation of the relative correction of the BR increases visibly via
the inclusion of the absorptive contributions.
The relative corrections reach $-4\%$ ($-9\%$) and are relevant to match
the ILC precision in \SE.

\begin{figure}[htb!]
\begin{center}
\begin{tabular}{c}
\includegraphics[width=0.49\textwidth,height=7.5cm]{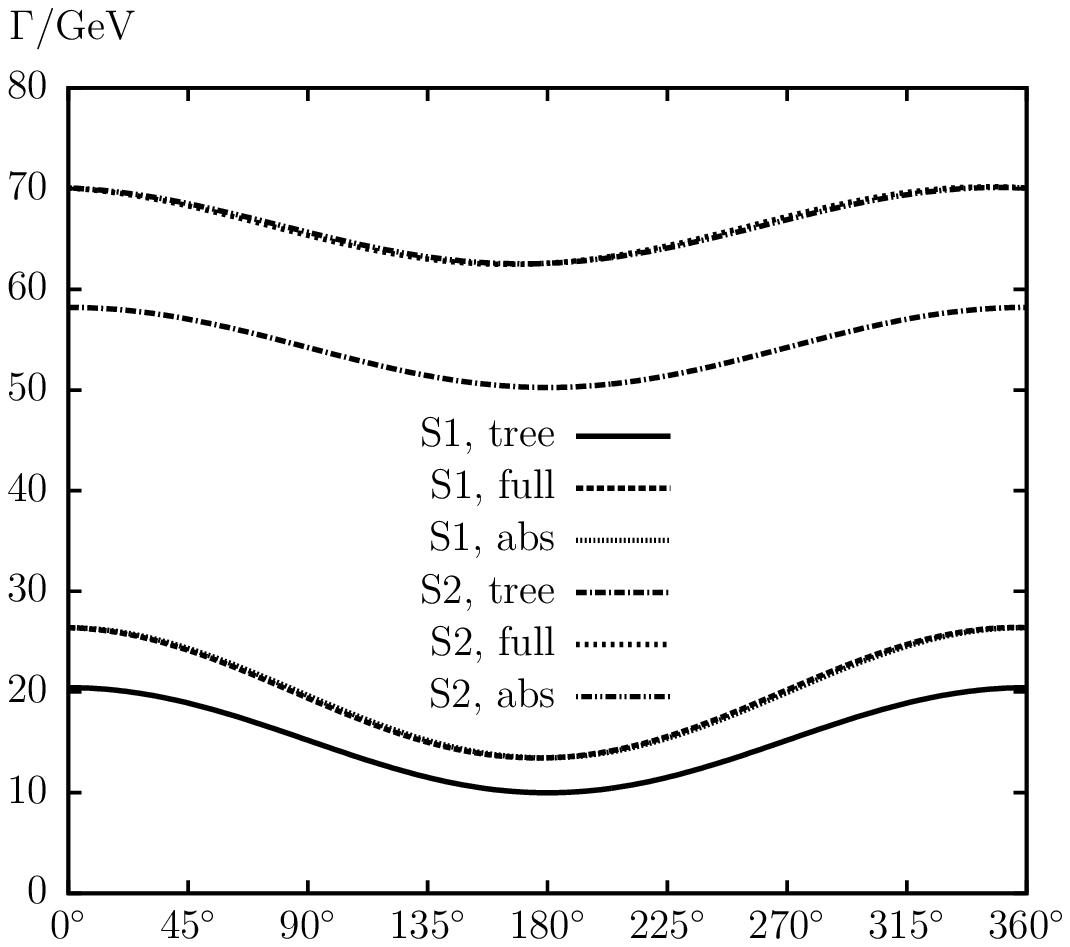}
\hspace{-4mm}
\includegraphics[width=0.49\textwidth,height=7.5cm]{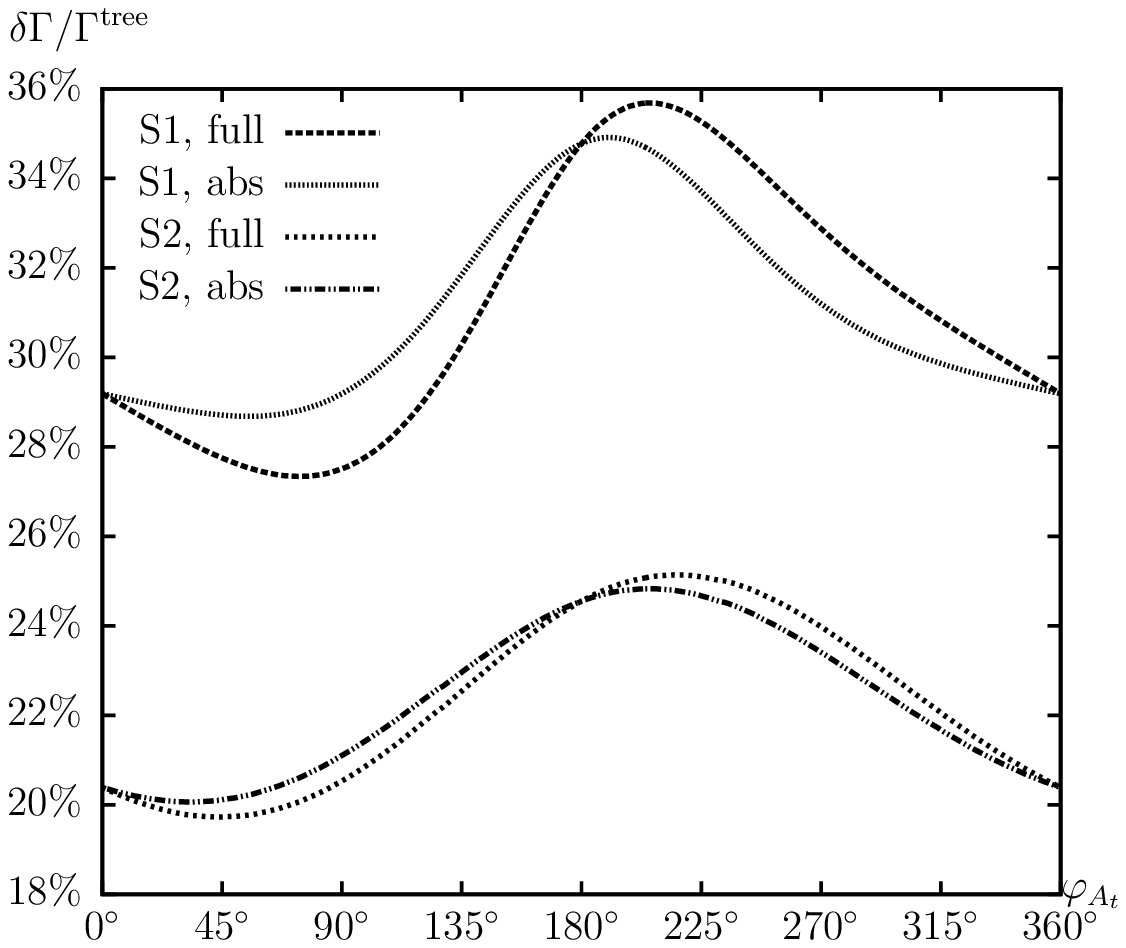}
\\[4em]
\includegraphics[width=0.49\textwidth,height=7.5cm]{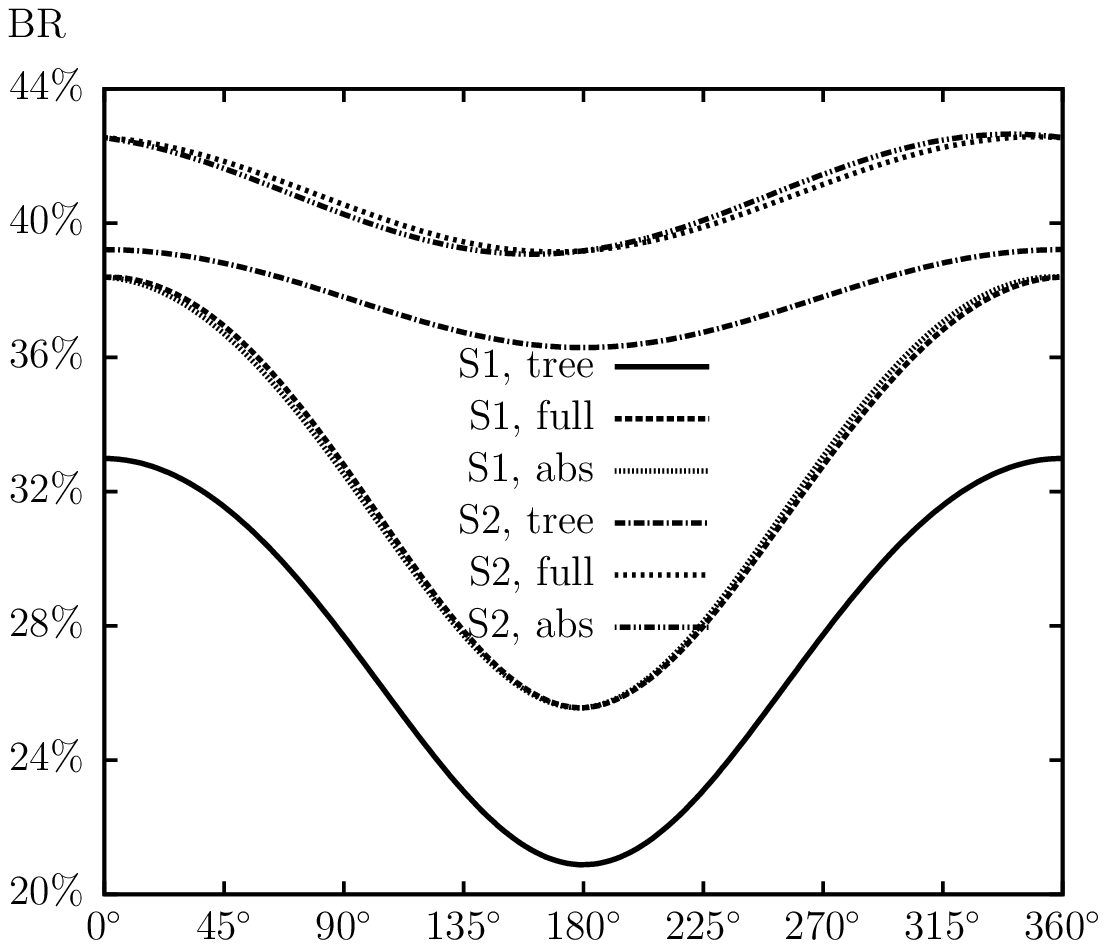}
\hspace{-4mm}
\includegraphics[width=0.49\textwidth,height=7.5cm]{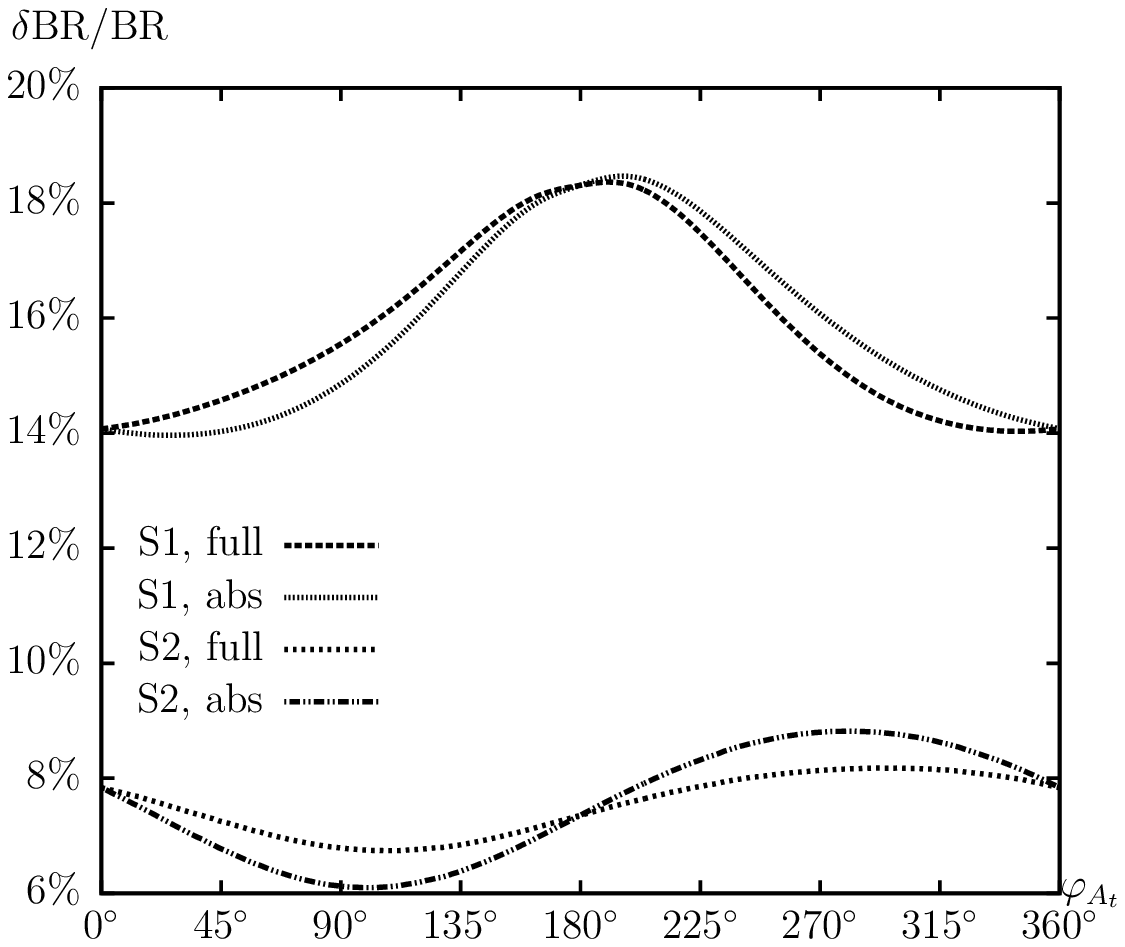}
\end{tabular}
\vspace{2em}
\caption{$\Ga(\decaygl)$.
  Tree-level (``tree'') and full one-loop (``full'') corrected 
  partial decay widths are shown. Also shown are the full one-loop
  corrected partial decay  
  widths including absorptive contributions (``abs''). 
  The parameters are chosen according to \SE\ and \SZ\ (see \refta{tab:para}), 
  with $\phiat$ varied.
  The upper left plot shows the partial decay width; 
  the upper right plot shows the corresponding relative size of the corrections. 
  The lower left plot shows the BR; 
  the lower right plot shows the relative correction of the BR.
}
\label{fig:PhiAt.st2tgl}
\end{center}
\end{figure}

Next we show the results for $\decaygl$ in \reffi{fig:PhiAt.st2tgl}. In
both numerical scenarios we find a substantial variation of the one-loop
effects with $\phiat$. The effects range from +28\% (+20\%) to +36\% (+24\%)
in \SE\ (\SZ), where the effect of the absorptive contributions remains 
relatively small.
The branching ratio varies strongly with $\phiat$ in \SE, ranging
from $\sim 38\%$ to $\sim 25\%$, while in \SZ\ it is larger 
and varies less, being around $\sim 41\%$. 
The one-loop corrections in \SE\ vary between $+14\%$ and $+18\%$ and 
are important for physics at the LHC and the ILC. 
Within \SZ\ they are found to be $\sim 8\%$.

\begin{figure}[htb!]
\begin{center}
\begin{tabular}{c}
\includegraphics[width=0.49\textwidth,height=7.5cm]{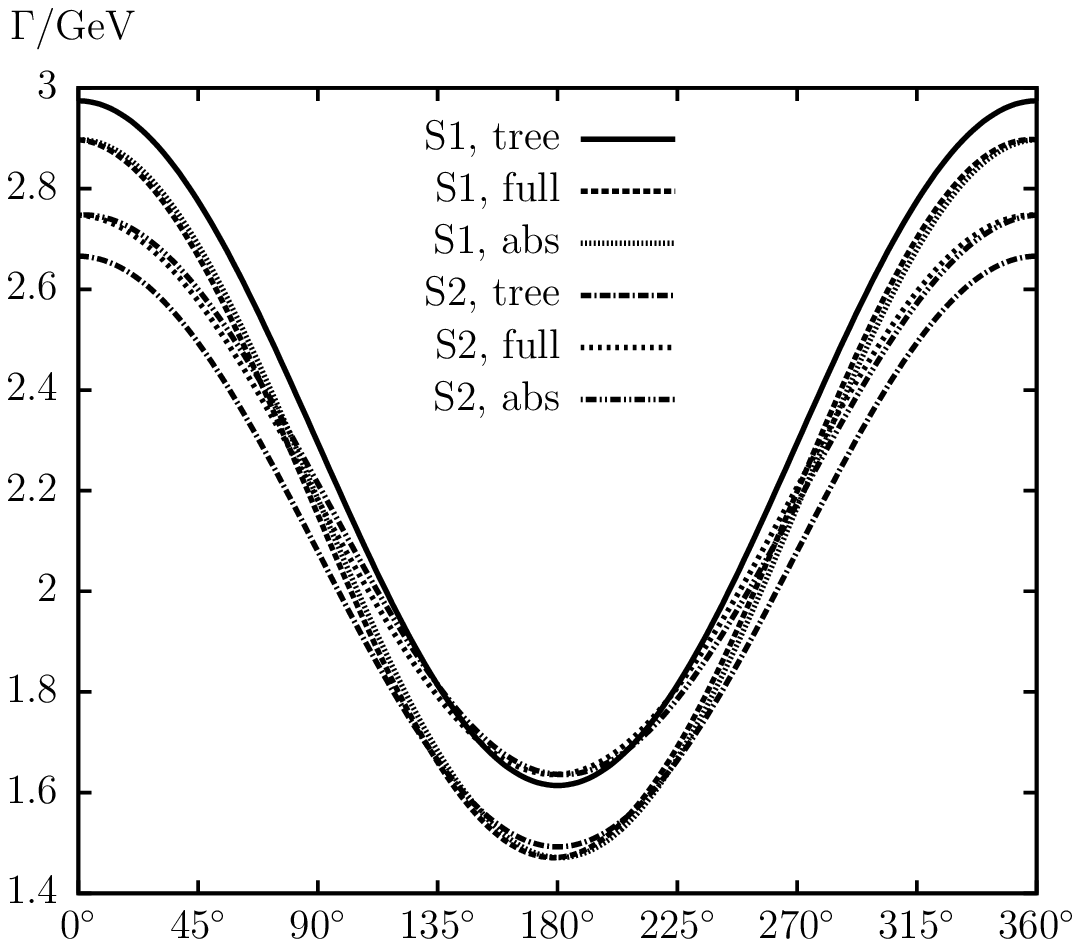}
\hspace{-4mm}
\includegraphics[width=0.49\textwidth,height=7.5cm]{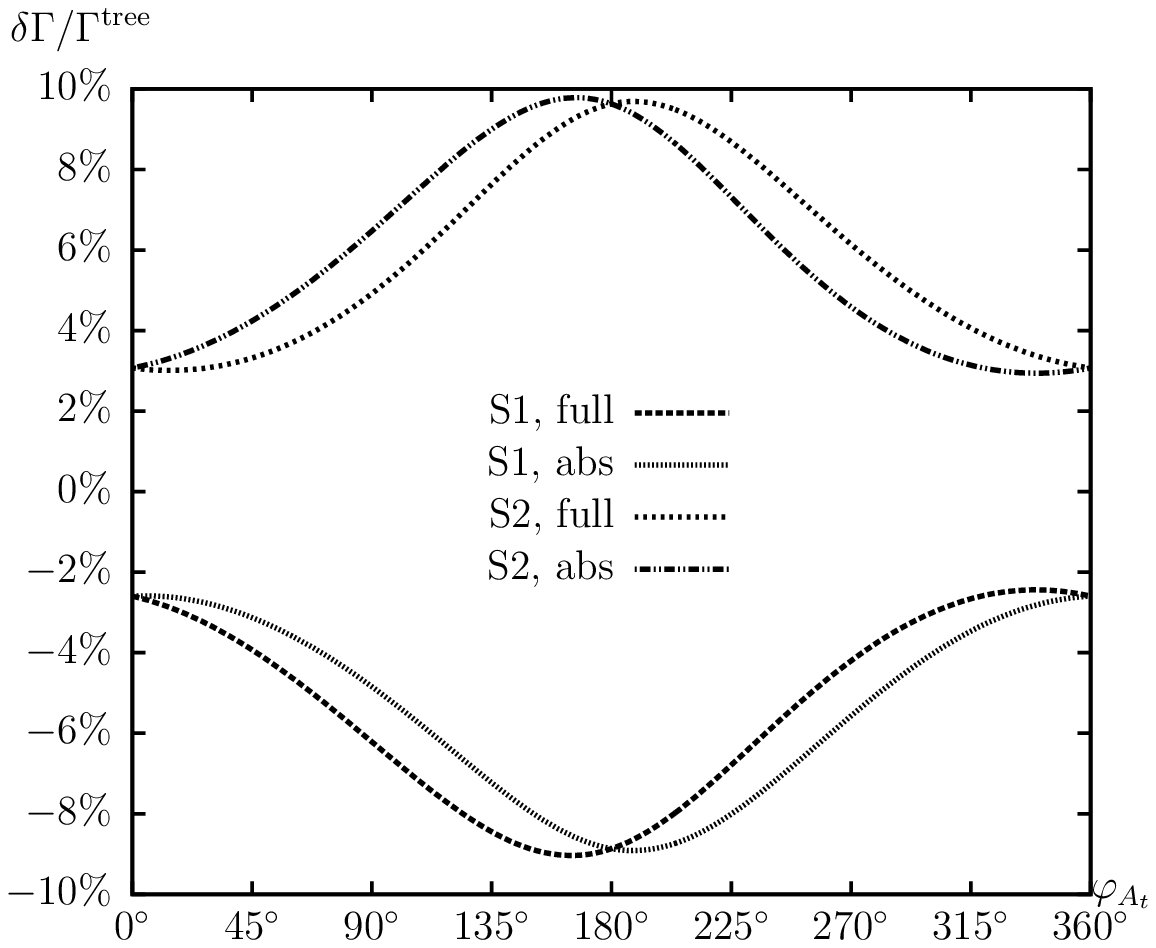}
\\[4em]
\includegraphics[width=0.49\textwidth,height=7.5cm]{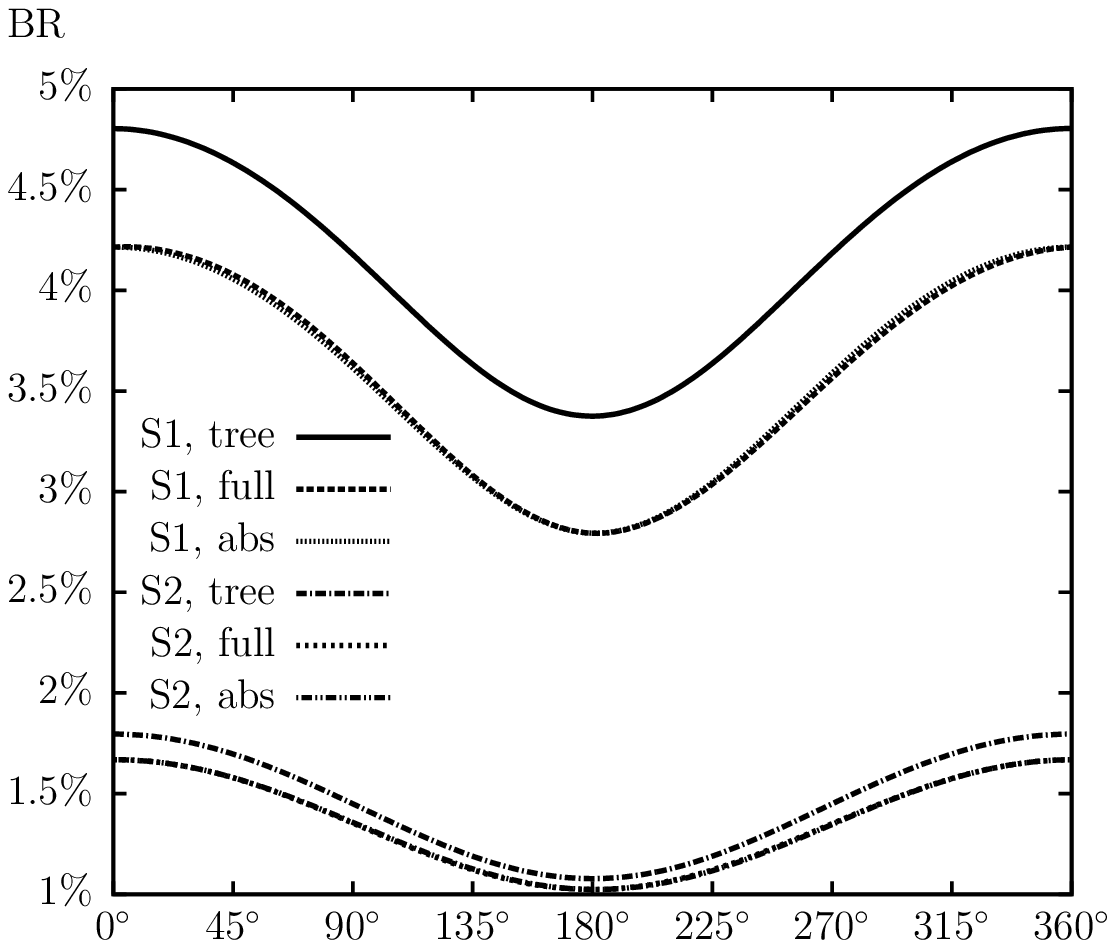}
\hspace{-4mm}
\includegraphics[width=0.49\textwidth,height=7.5cm]{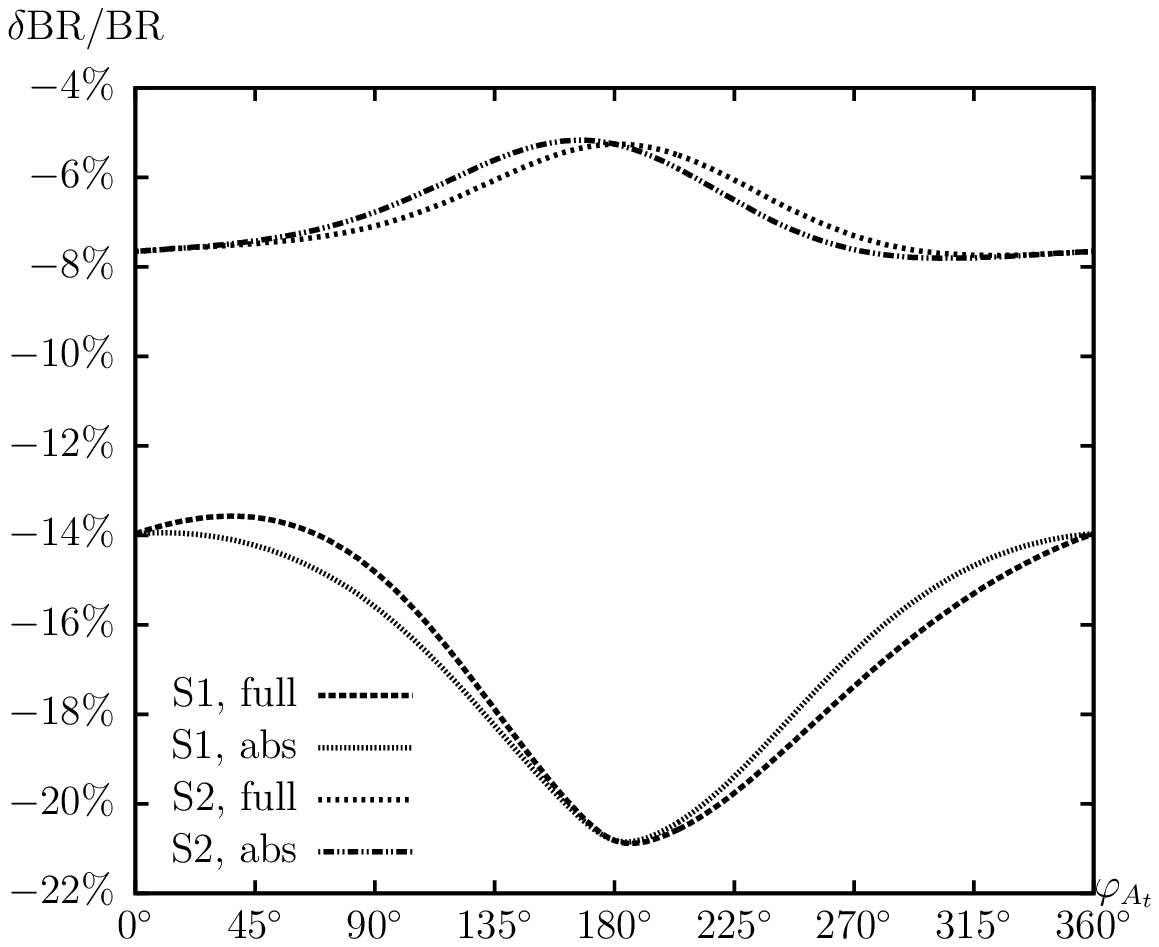}
\end{tabular}
\vspace{2em}
\caption{$\Ga(\decayNe)$. 
  Tree-level (``tree'') and full one-loop (``full'') corrected 
  partial decay widths are shown. Also shown are the full one-loop
  corrected partial decay  
  widths including absorptive contributions (``abs''). 
  The parameters are chosen according to \SE\ and \SZ\ (see \refta{tab:para}), 
  with $\phiat$ varied.
  The upper left plot shows the partial decay width;
  the upper right plot shows the corresponding  relative size of the corrections. 
  The lower left plot shows the BR; 
  the lower right plot shows the relative correction of the BR.
}
\label{fig:PhiAt.st2tneu1}
\end{center}
\end{figure}

\begin{figure}[htb!]
\begin{center}
\begin{tabular}{c}
\includegraphics[width=0.49\textwidth,height=7.5cm]{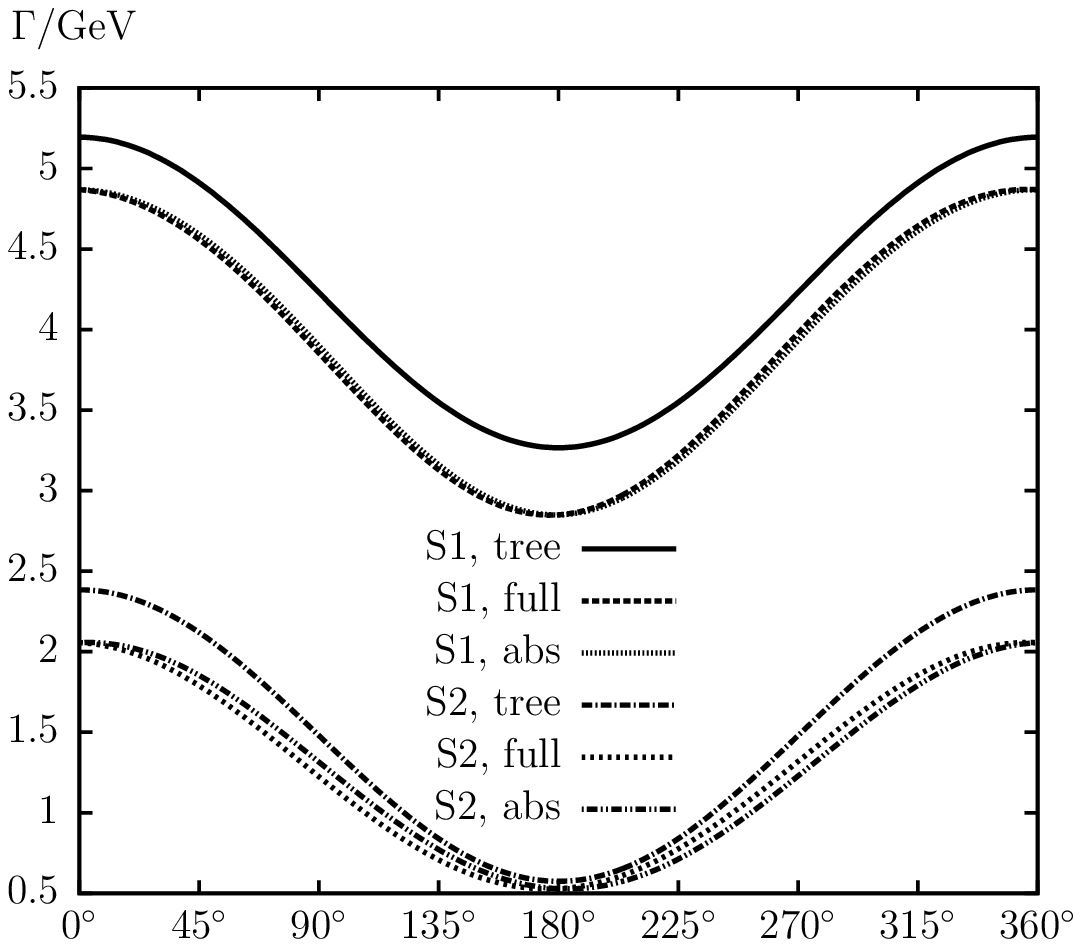}
\hspace{-4mm}
\includegraphics[width=0.49\textwidth,height=7.5cm]{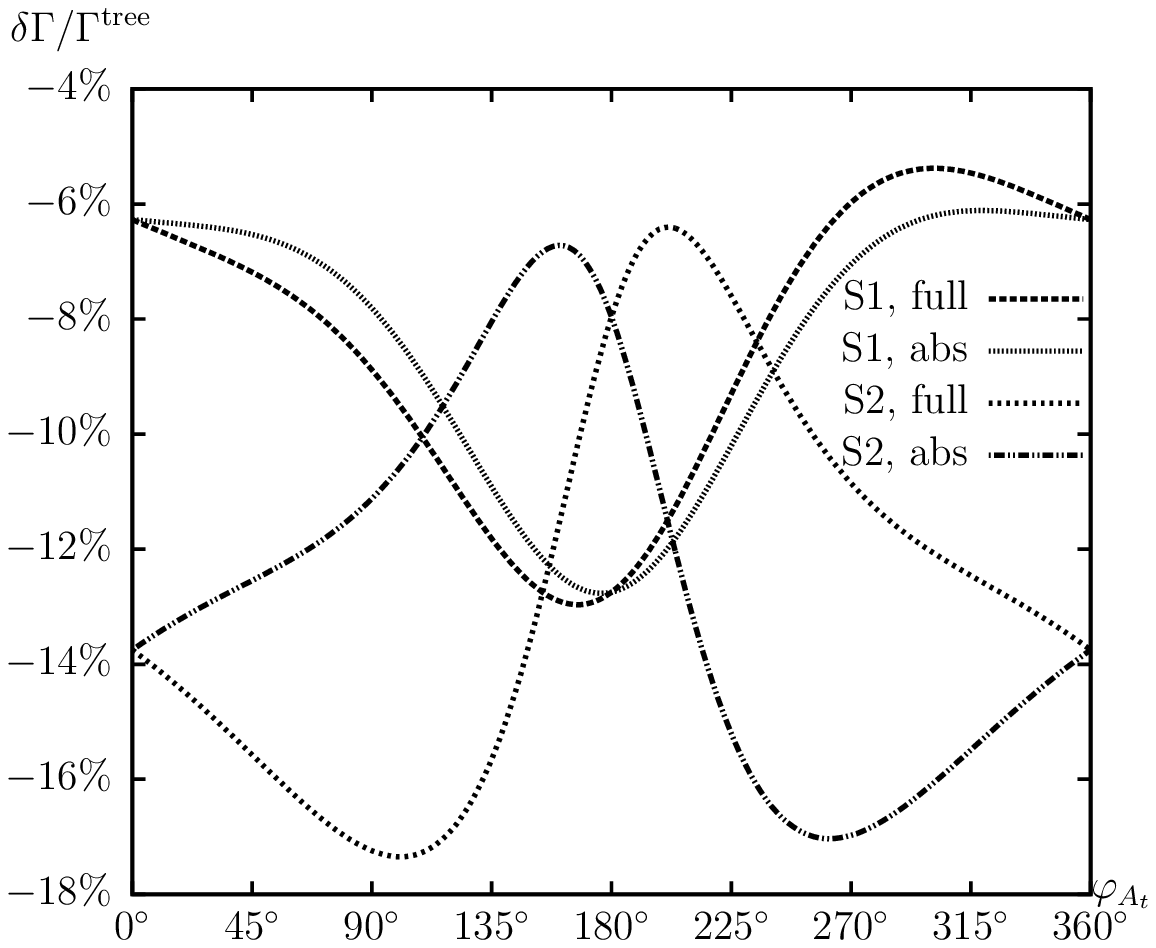}
\\[4em]
\includegraphics[width=0.49\textwidth,height=7.5cm]{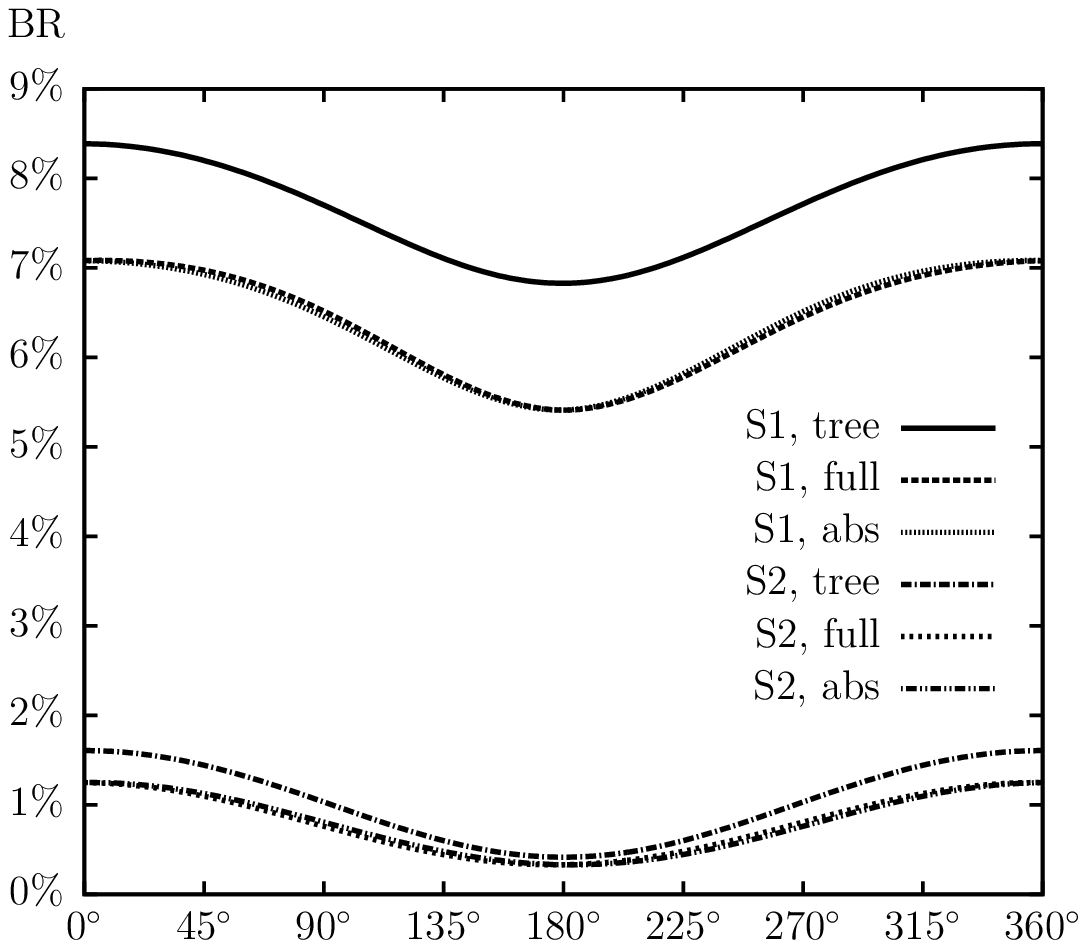}
\hspace{-4mm}
\includegraphics[width=0.49\textwidth,height=7.5cm]{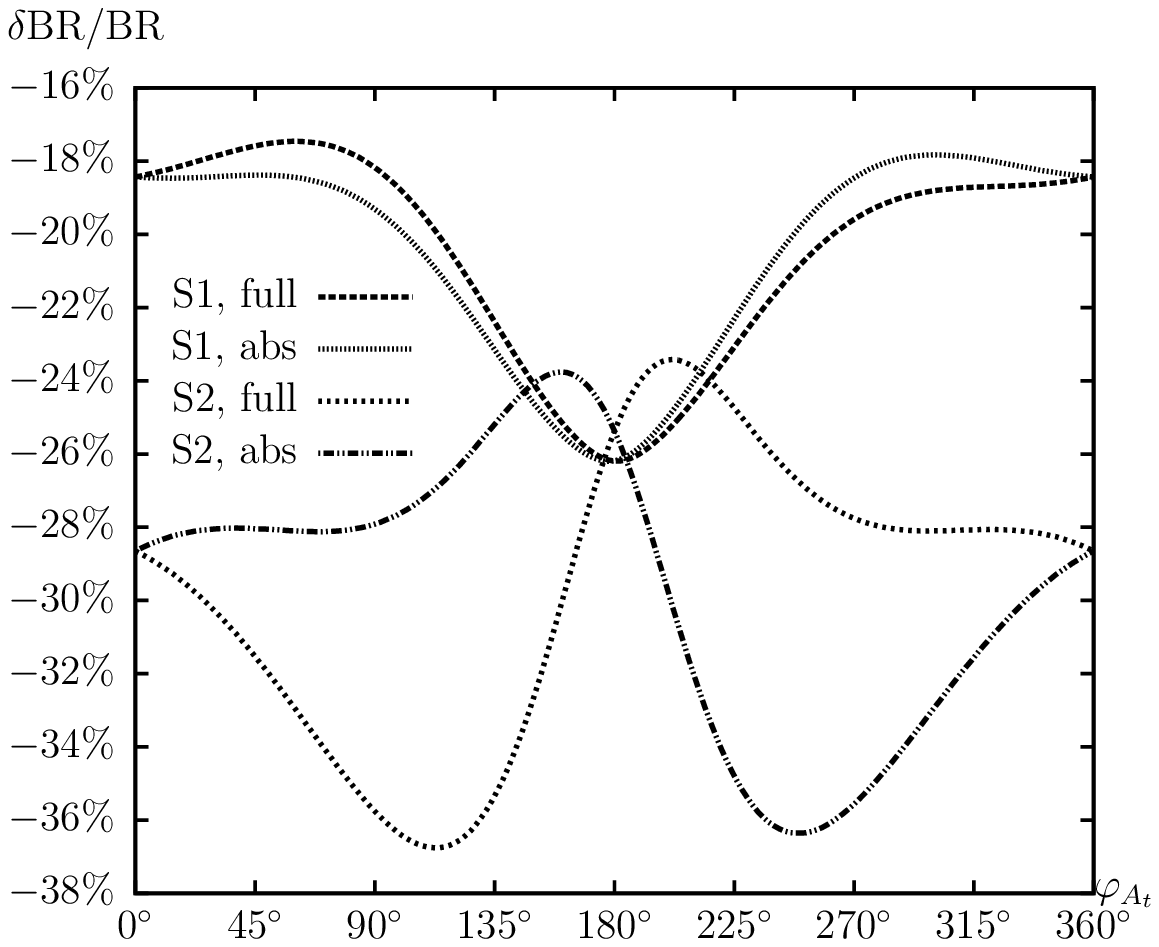}
\end{tabular}
\vspace{2em}
\caption{$\Ga(\decayNz)$.
  Tree-level (``tree'') and full one-loop (``full'') corrected 
  partial decay widths are shown. Also shown are the full one-loop
  corrected partial decay  
  widths including absorptive contributions (``abs''). 
  The parameters are chosen according to \SE\ and \SZ\ (see \refta{tab:para}), 
  with $\phiat$ varied.
  The upper left plot shows the partial decay width; 
  the upper right plot shows the corresponding  relative size of the corrections. 
  The lower left plot shows the BR; 
  the lower right plot shows the relative correction of the BR.
}
\label{fig:PhiAt.st2tneu2}
\end{center}
\end{figure}

\begin{figure}[htb!]
\begin{center}
\begin{tabular}{c}
\includegraphics[width=0.49\textwidth,height=7.5cm]{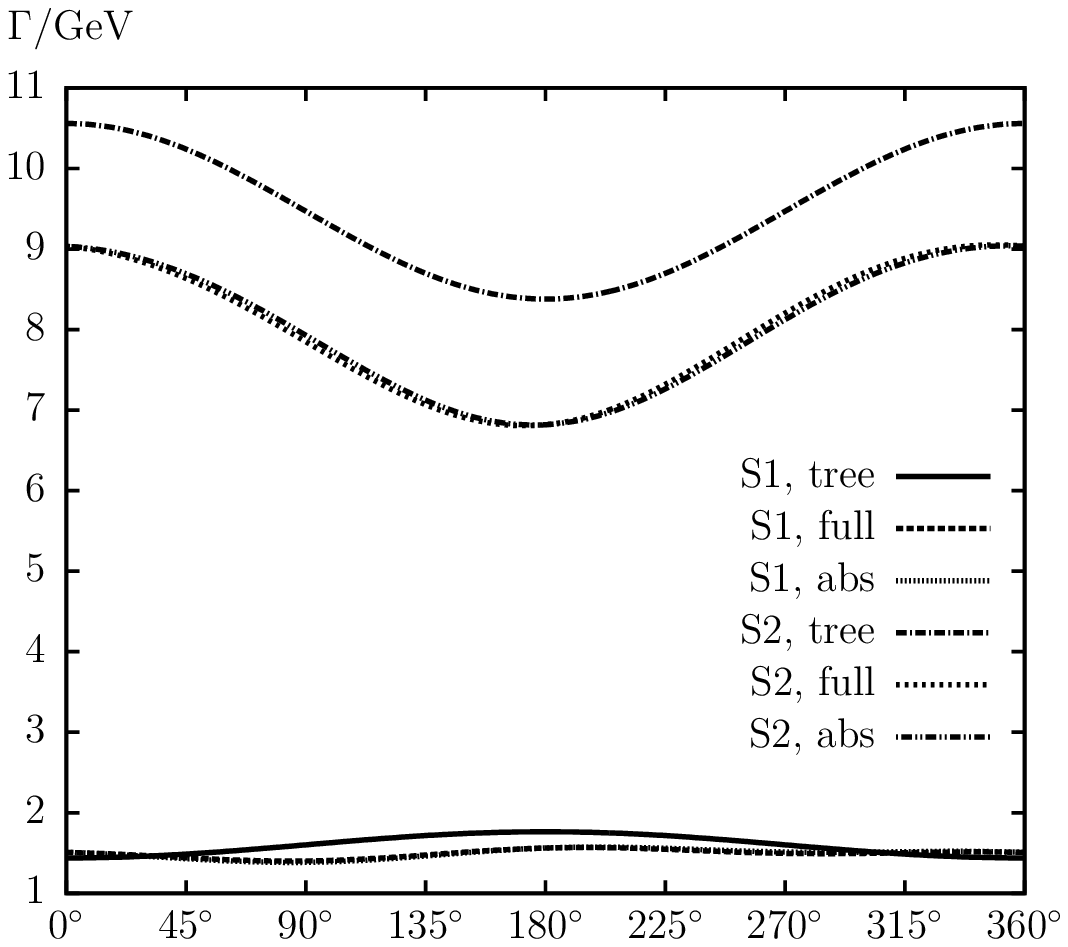}
\hspace{-4mm}
\includegraphics[width=0.49\textwidth,height=7.5cm]{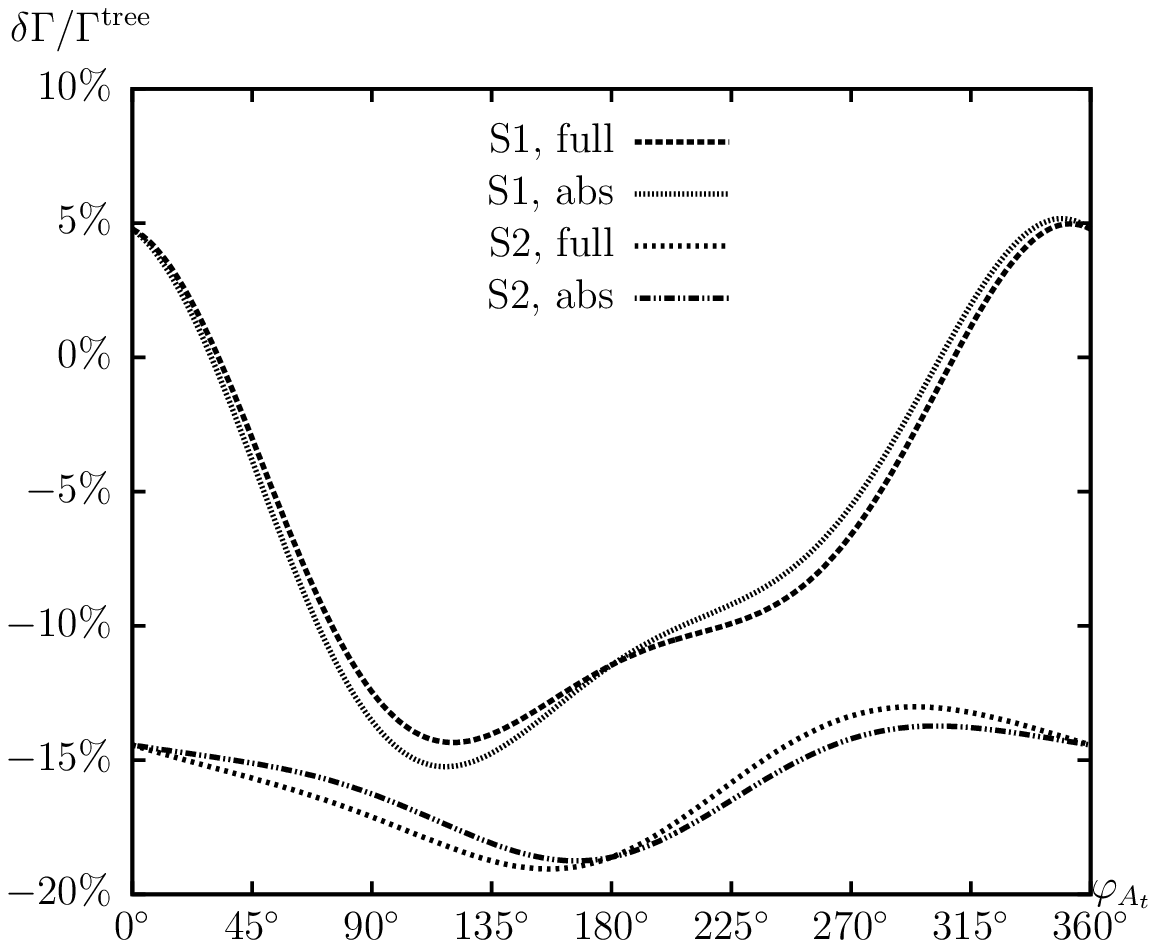}
\\[4em]
\includegraphics[width=0.49\textwidth,height=7.5cm]{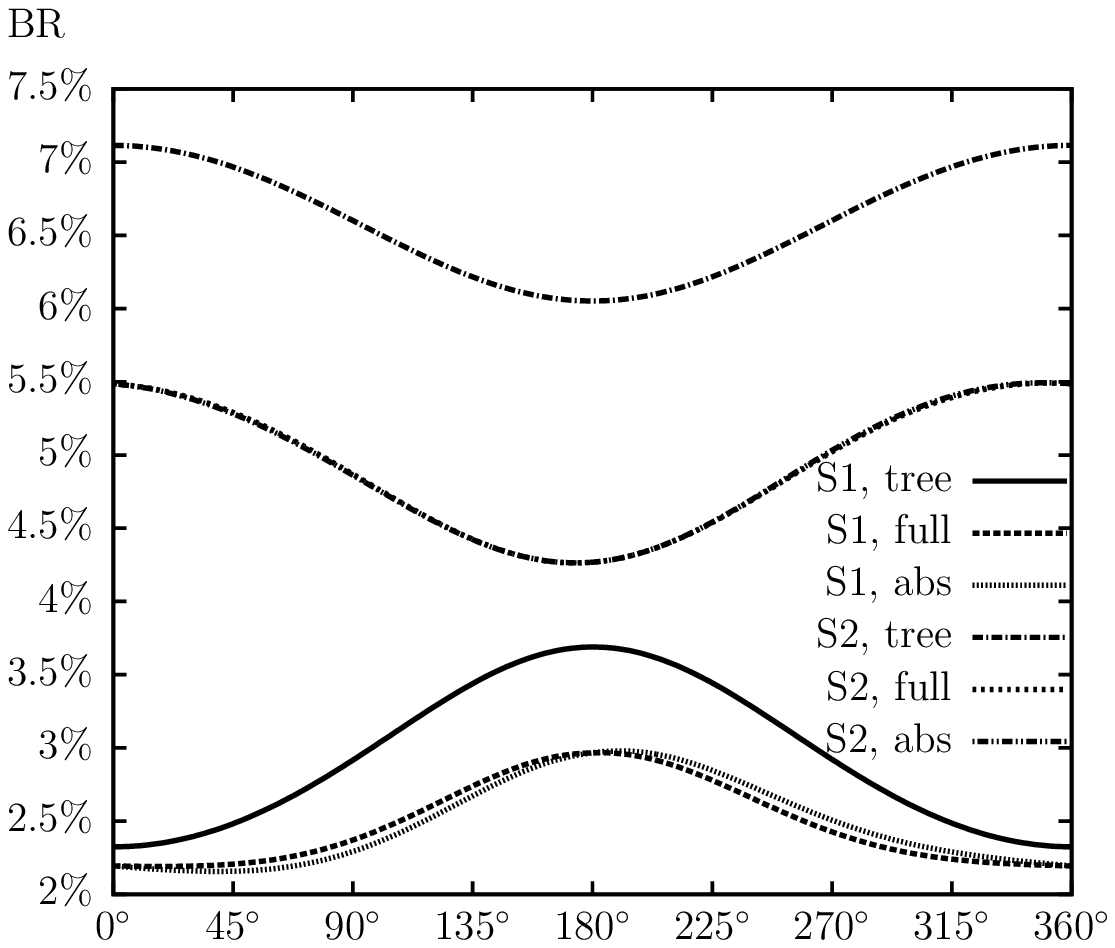}
\hspace{-4mm}
\includegraphics[width=0.49\textwidth,height=7.5cm]{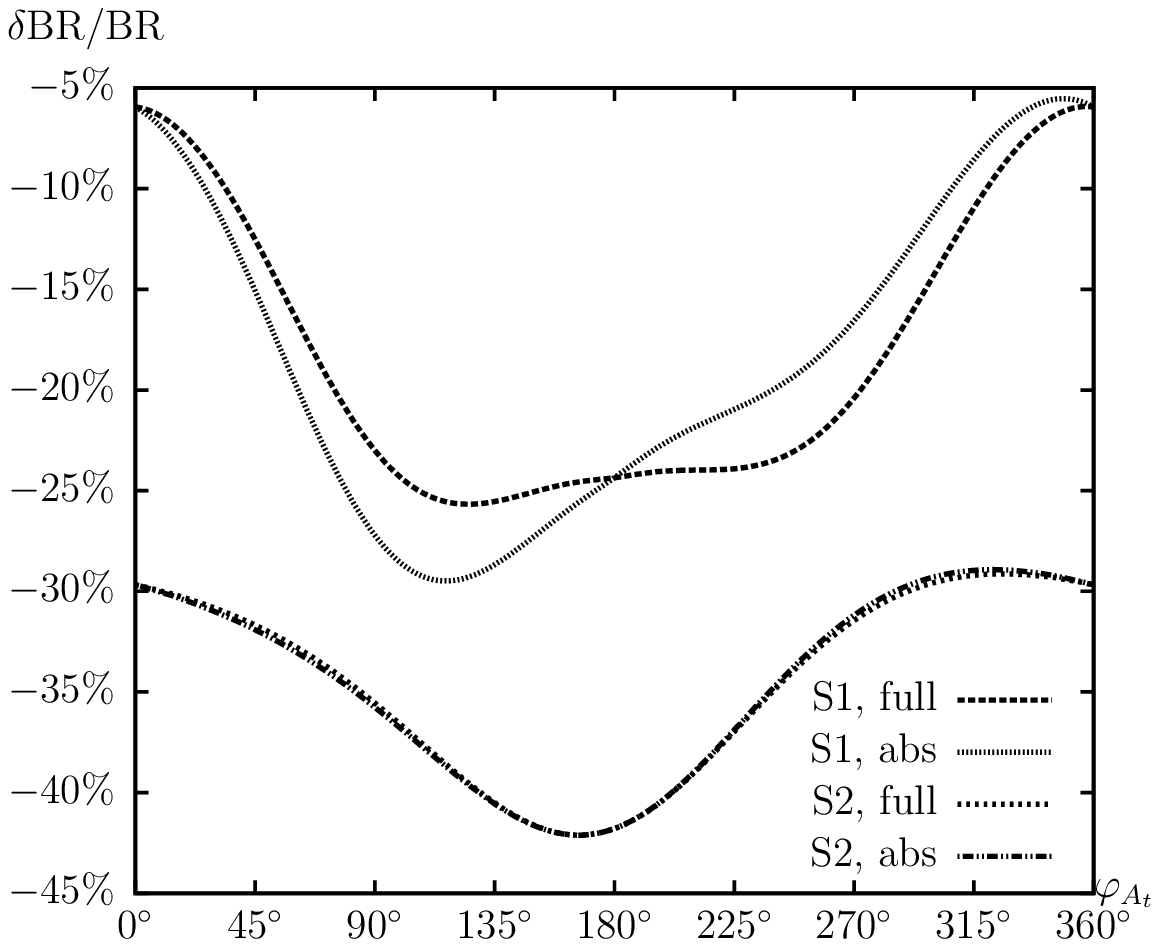}
\end{tabular}
\vspace{2em}
\caption{$\Ga(\decayNd)$.
  Tree-level (``tree'') and full one-loop (``full'') corrected 
  partial decay widths are shown. Also shown are the full one-loop
  corrected partial decay  
  widths including absorptive contributions (``abs''). 
  The parameters are chosen according to \SE\ and \SZ\ (see \refta{tab:para}), 
  with $\phiat$ varied.
  The upper left plot shows the partial decay width;
  the upper right plot shows the corresponding  relative size of the corrections. 
  The lower left plot shows the BR; 
  the lower right plot shows the relative correction of the BR.
}
\label{fig:PhiAt.st2tneu3}
\end{center}
\end{figure}

\begin{figure}[htb!]
\begin{center}
\begin{tabular}{c}
\includegraphics[width=0.49\textwidth,height=7.5cm]{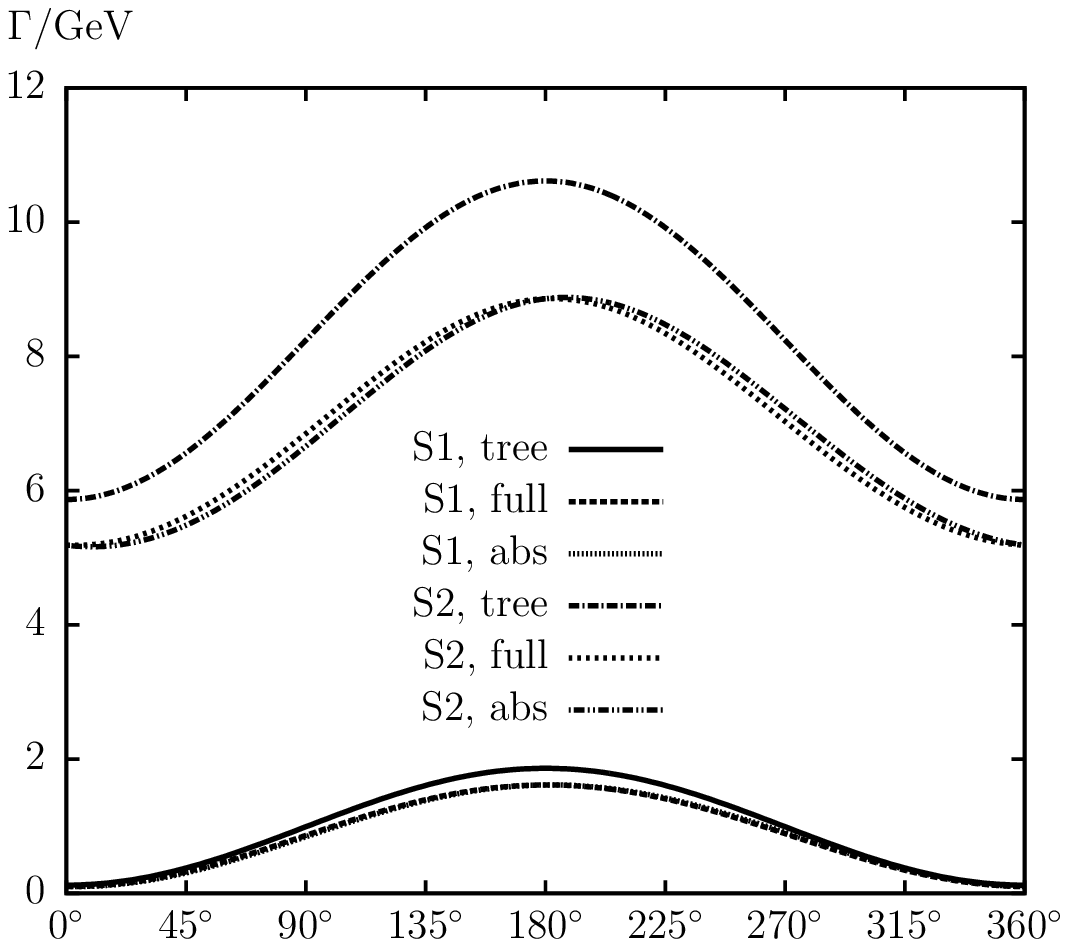}
\hspace{-4mm}
\includegraphics[width=0.49\textwidth,height=7.5cm]{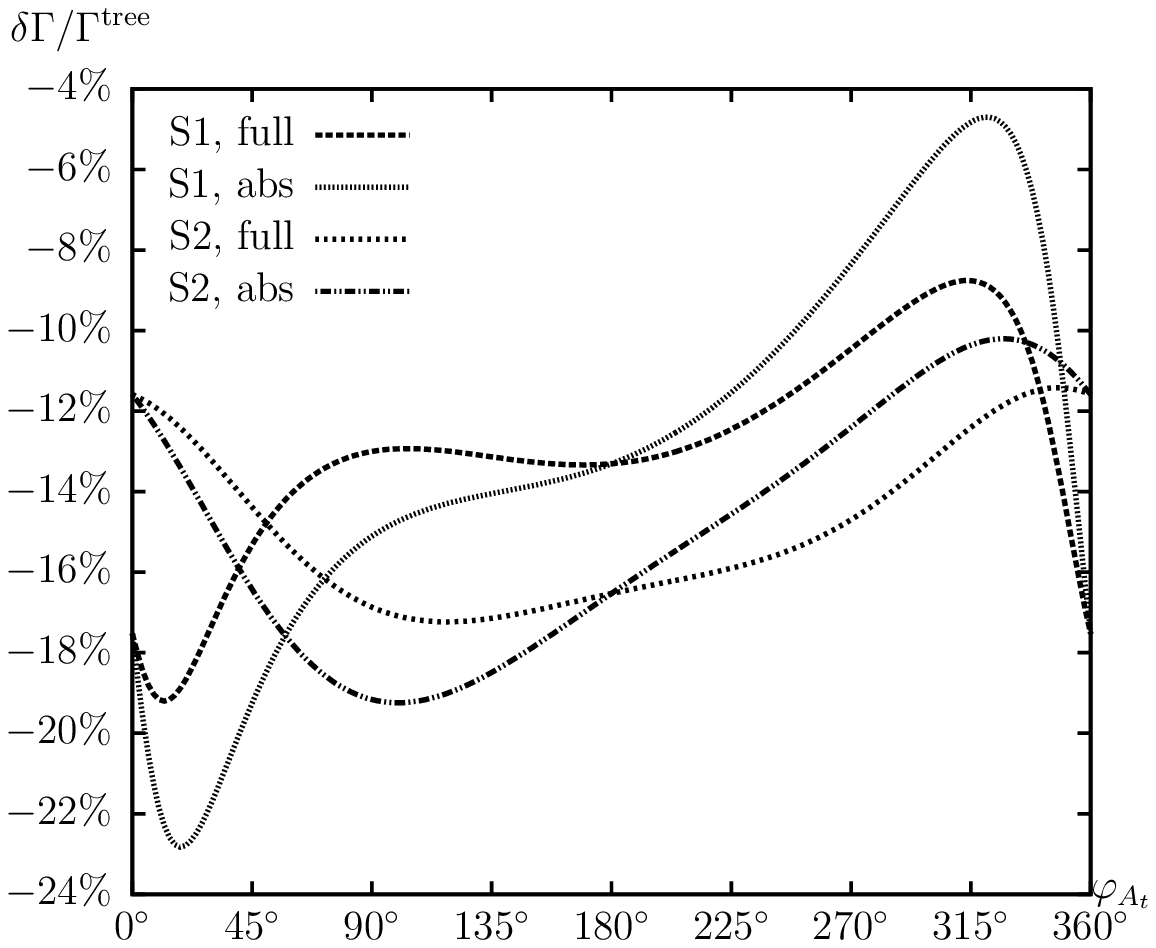}
\\[4em]
\includegraphics[width=0.49\textwidth,height=7.5cm]{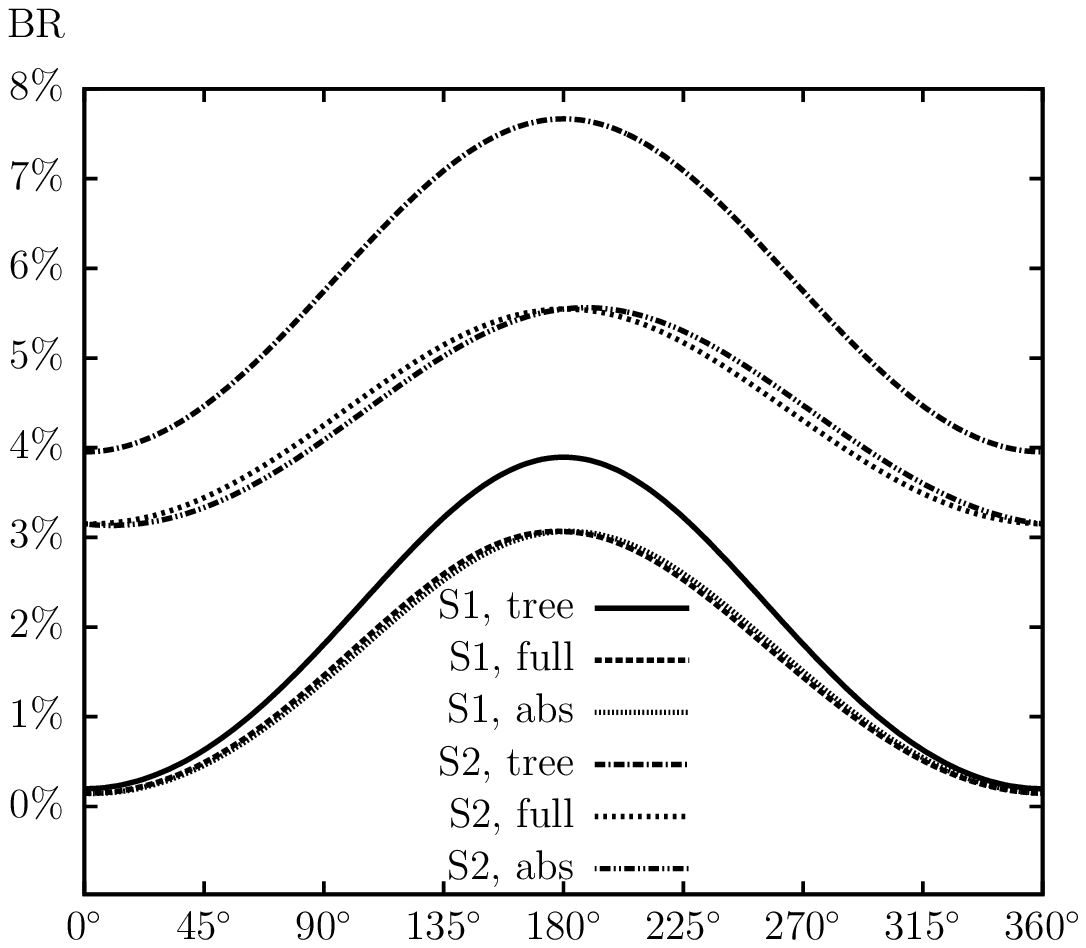}
\hspace{-4mm}
\includegraphics[width=0.49\textwidth,height=7.5cm]{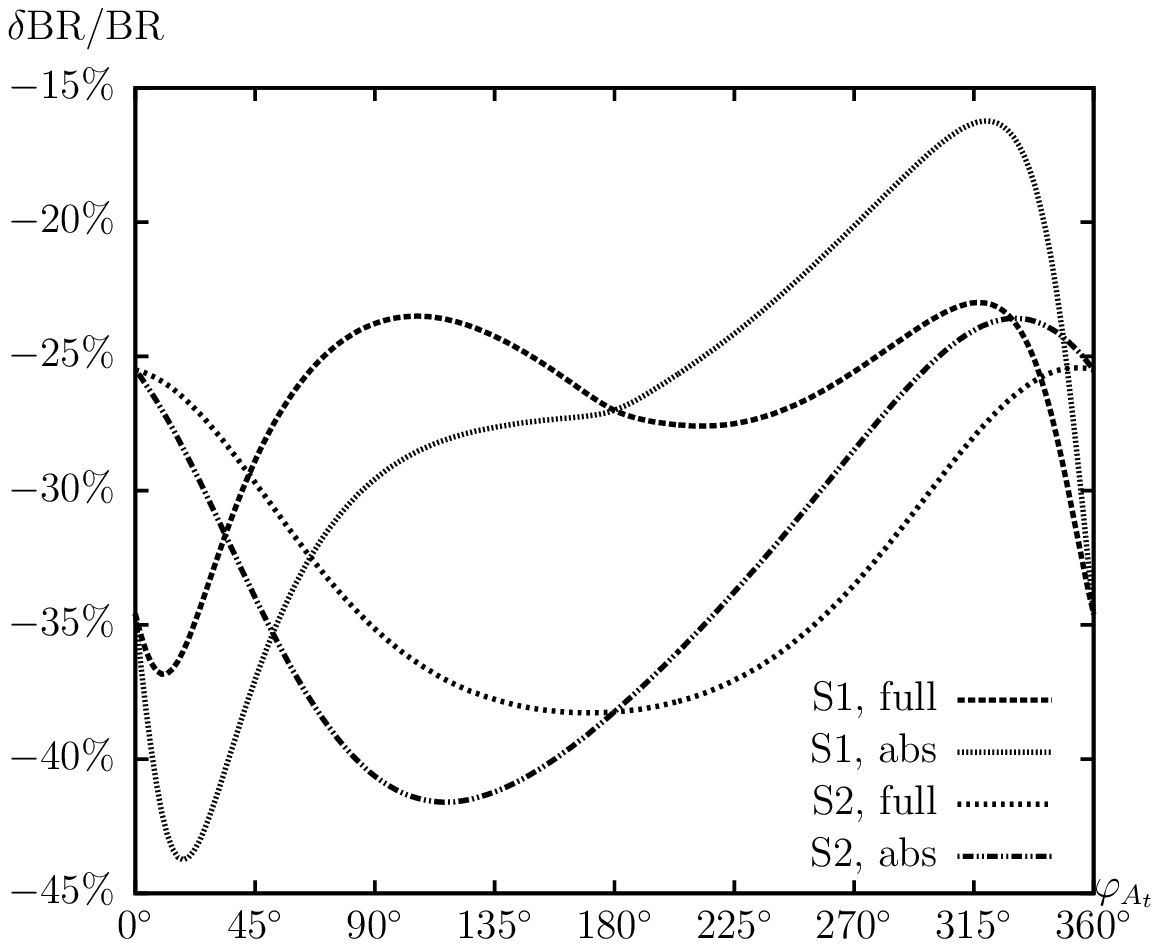}
\end{tabular}
\vspace{2em}
\caption{$\Ga(\decayNv)$.
  Tree-level (``tree'') and full one-loop (``full'') corrected 
  partial decay widths are shown. Also shown are the full one-loop
  corrected partial decay  
  widths including absorptive contributions (``abs''). 
  The parameters are chosen according to \SE\ and \SZ\ (see \refta{tab:para}), 
  with $\phiat$ varied.
  The upper left plot shows the partial decay width; 
  the upper right plot shows the corresponding relative size of the corrections. 
  The lower left plot shows the BR; 
  the lower right plot shows the relative correction of the BR.
}
\label{fig:PhiAt.st2tneu4}
\end{center}
\end{figure}

In \reffis{fig:PhiAt.st2tneu1} -- \ref{fig:PhiAt.st2tneu4} we present
the variation of $\Ga(\decayNk)$ ($k = 1,2,3,4$) as a function of
$\phiat$. 
As for the variation with $\mstz$ also here for $k = 1,2$
($k = 3,4$) larger values of the partial decay width are found  in
\SE\ (\SZ) with a similar size as before. 
The one-loop effects on $\Ga(\decayNe)$ for $\phiat = 0$ are relatively 
small, at the +3\% ($-3\%$) level in \SE\ (\SZ). 
The variation of $\phiat$, however, now yields one-loop corrections up to 
$\sim +10\%\ (-10\%)$ in the two scenarios, with a small shift induced 
by the absorptive contributions.
$\Ga(\decayNz)$ also exhibits a strong variation
with $\phiat$, ranging from $-12\%$ ($-18\%$) to $-6\%$ in \SE\ (\SZ).
The absorptive contributions in this case can change the result strongly, 
leading especially in \SZ\ to a substantially different shape.
For $\Ga(\decayNd)$ the variation in \SZ\ is small. Within \SE, however, the
effects for $\phiat = 0$ are at the $\sim +5\%$ level, while they reach 
nearly $-15\%$ for intermediate $\phiat$. 
The absorptive contributions are small. 
The last partial decay width of the four decay modes, $\Ga(\decayNv)$,
shows a large variation at the  one-loop level of nearly $-20\%$ in \SE,
where, however, the partial width itself is very small. 
For $\phiat \approx \pi$ a width of $\sim 2 \gev$ is reached with a
variation at the $-10\%$ level. Within \SZ\ $\Ga(\decayNv)$ varies around 
$\sim 7 \gev$ with a one-loop variation between $-12\%$ and $-17\%$. 
The absorptive contributions lead to a result smaller by a few percent.
Within \SE, i.e.\ with the ILC(1000) accessible parameter space, the
one-loop corrections reach $-20\%$ and more for $\br(\decayNk)$, $k = 1,2$, 
which can exceed the anticipated ILC precision. In general a strong
variation of the one-loop effects with $\phiat$ on the branching ratios
is found, where very large corrections are found in \SZ\ for $k = 3,4$,
where the one-loop contributions can change the BRs by up to $-40\%$.

\begin{figure}[htb!]
\begin{center}
\begin{tabular}{c}
\includegraphics[width=0.49\textwidth,height=7.5cm]{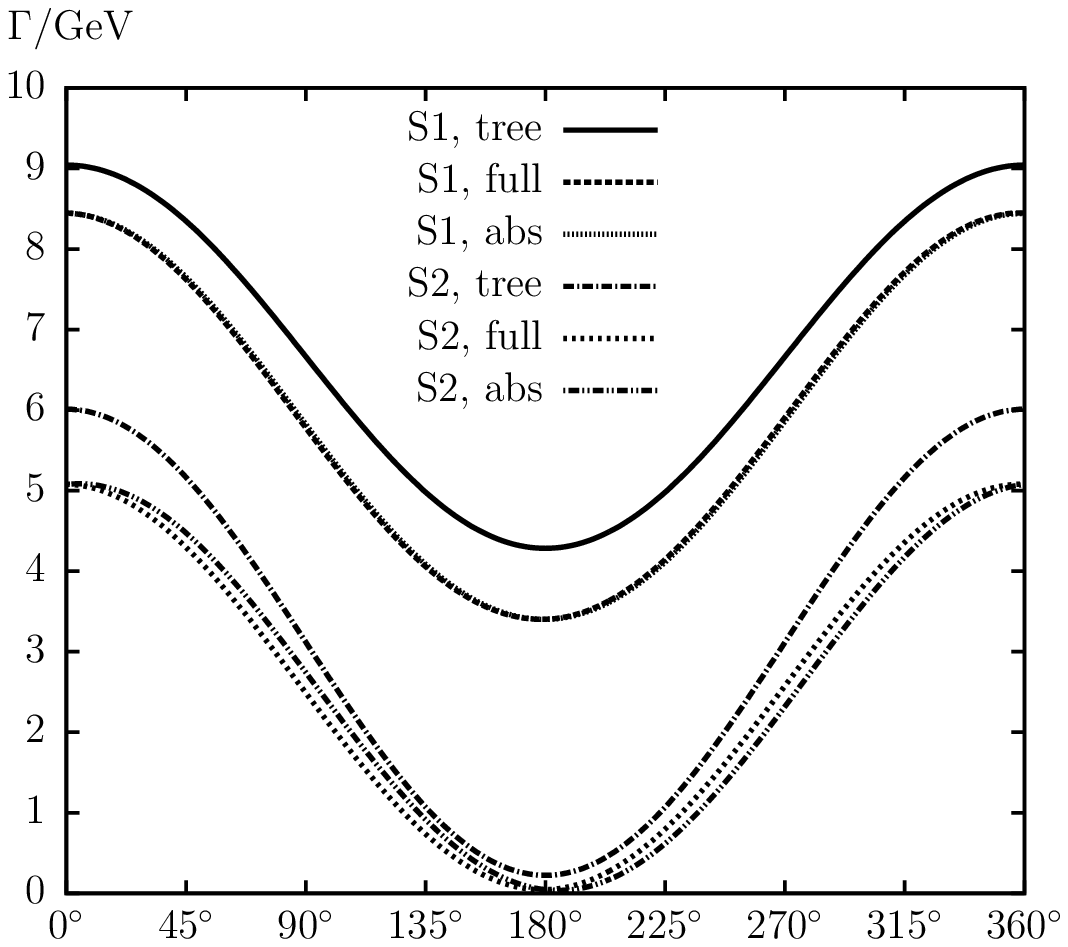}
\hspace{-4mm}
\includegraphics[width=0.49\textwidth,height=7.5cm]{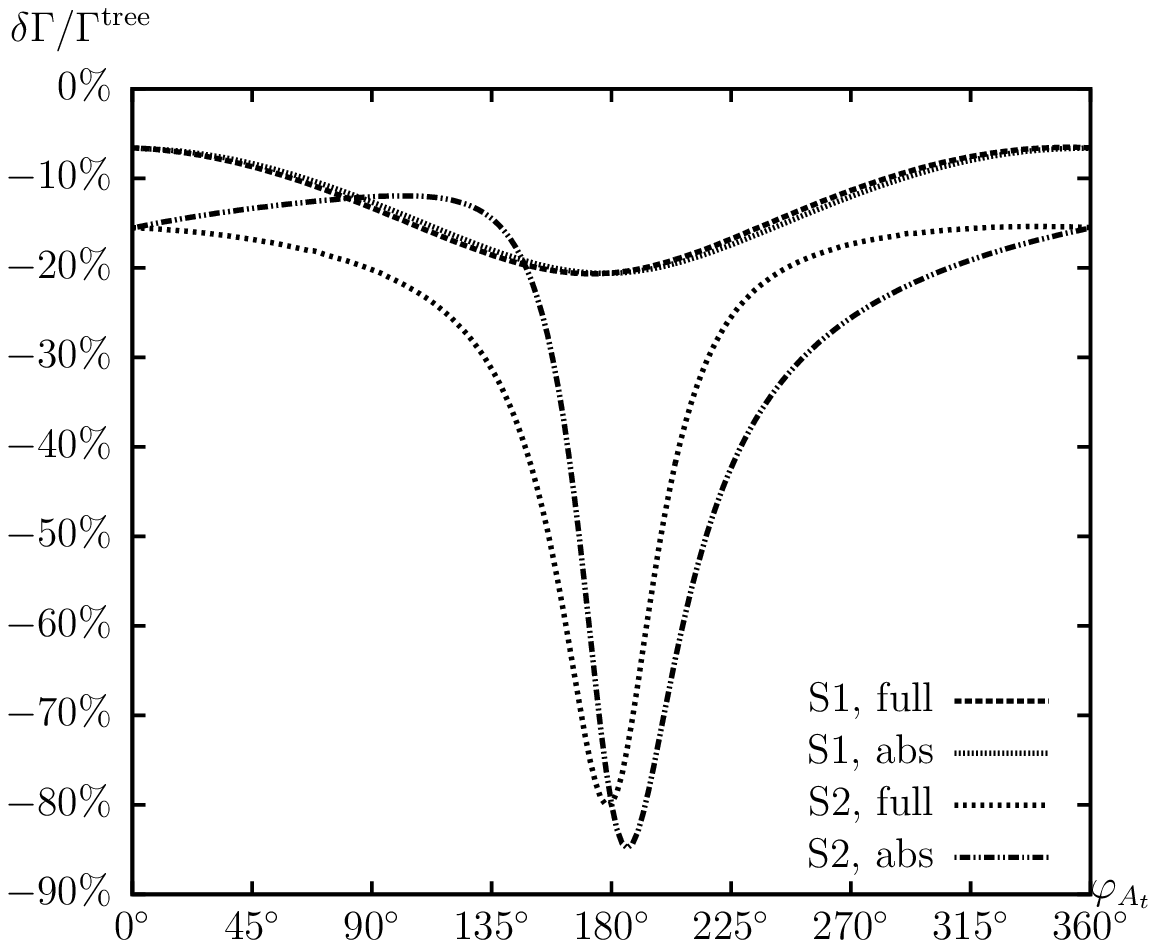}
\\[4em]
\includegraphics[width=0.49\textwidth,height=7.5cm]{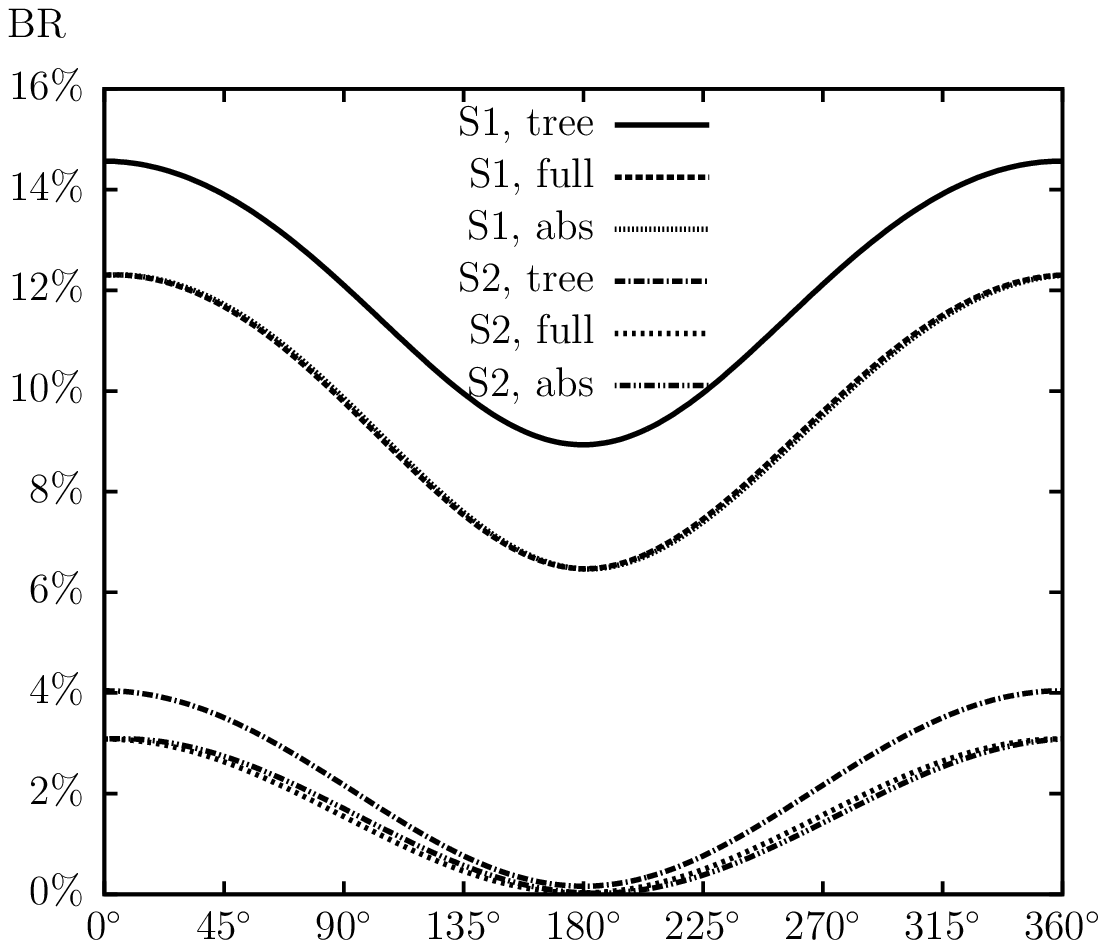}
\hspace{-4mm}
\includegraphics[width=0.49\textwidth,height=7.5cm]{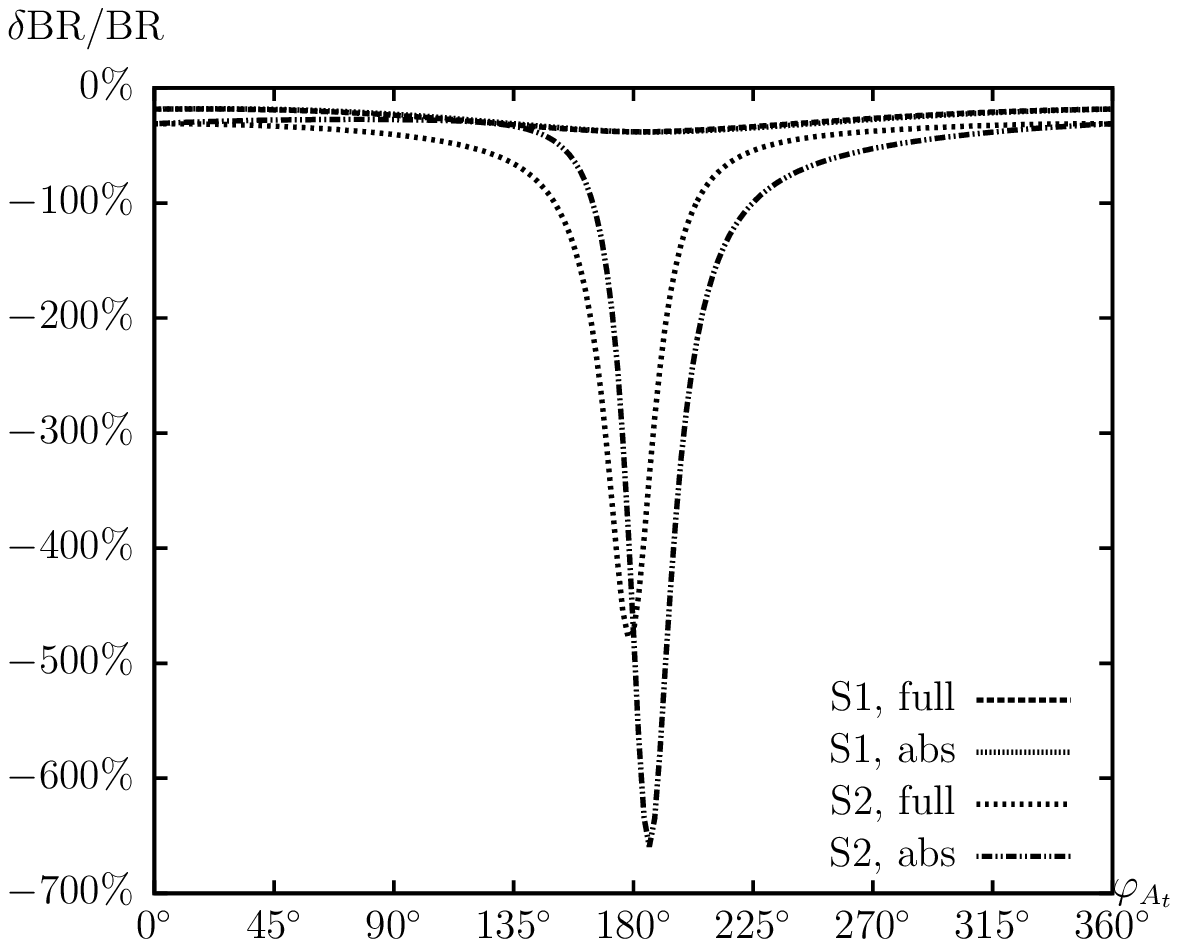}
\end{tabular}
\vspace{2em}
\caption{$\Ga(\decayCpe)$. 
  Tree-level (``tree'') and full one-loop (``full'') corrected 
  partial decay widths are shown. Also shown are the full one-loop
  corrected partial decay  
  widths including absorptive contributions (``abs''). 
  The parameters are chosen according to \SE\ and \SZ\ (see \refta{tab:para}), 
  with $\phiat$ varied.
  The upper left plot shows the partial decay width; 
  the upper right plot shows the corresponding  relative size of the corrections. 
  The lower left plot shows the BR; 
  the lower right plot shows the relative correction of the BR.
}
\label{fig:PhiAt.st2bcha1}
\end{center}
\end{figure}

\begin{figure}[htb!]
\begin{center}
\begin{tabular}{c}
\includegraphics[width=0.49\textwidth,height=7.5cm]{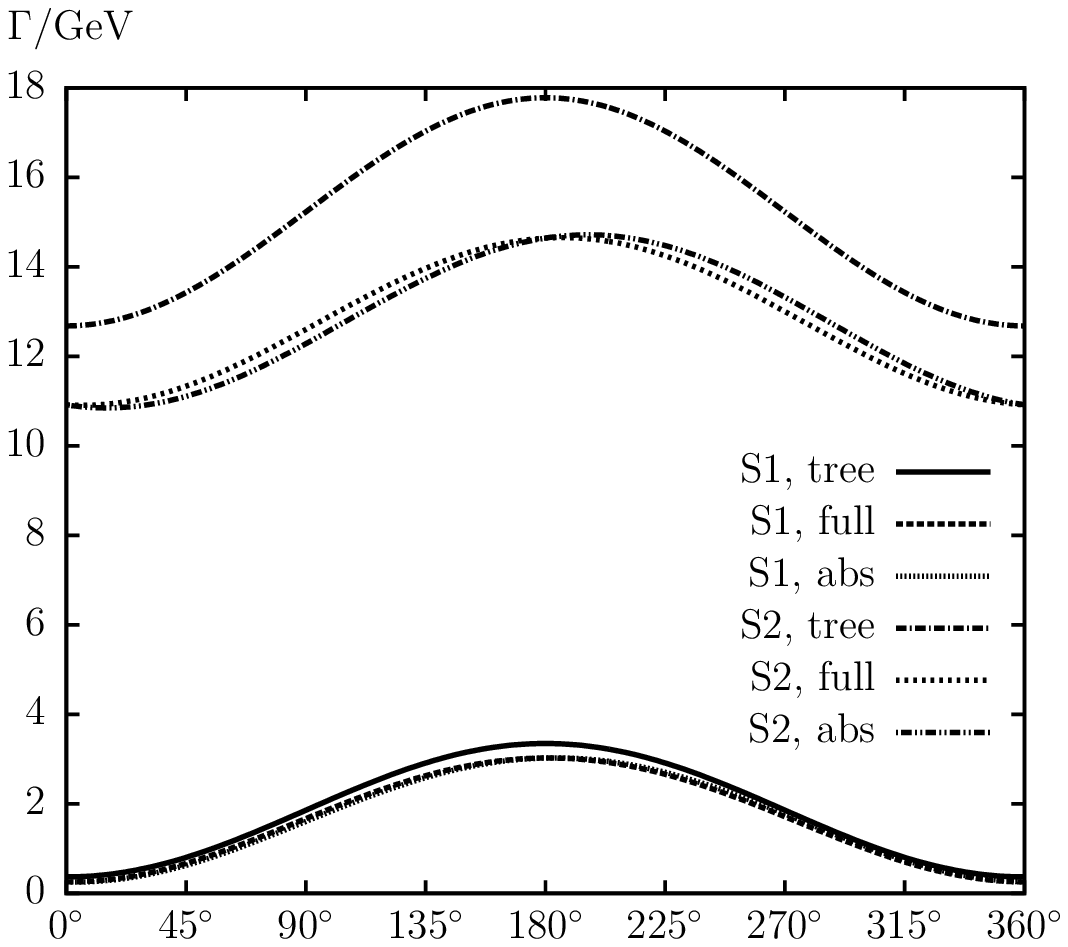}
\hspace{-4mm}
\includegraphics[width=0.49\textwidth,height=7.5cm]{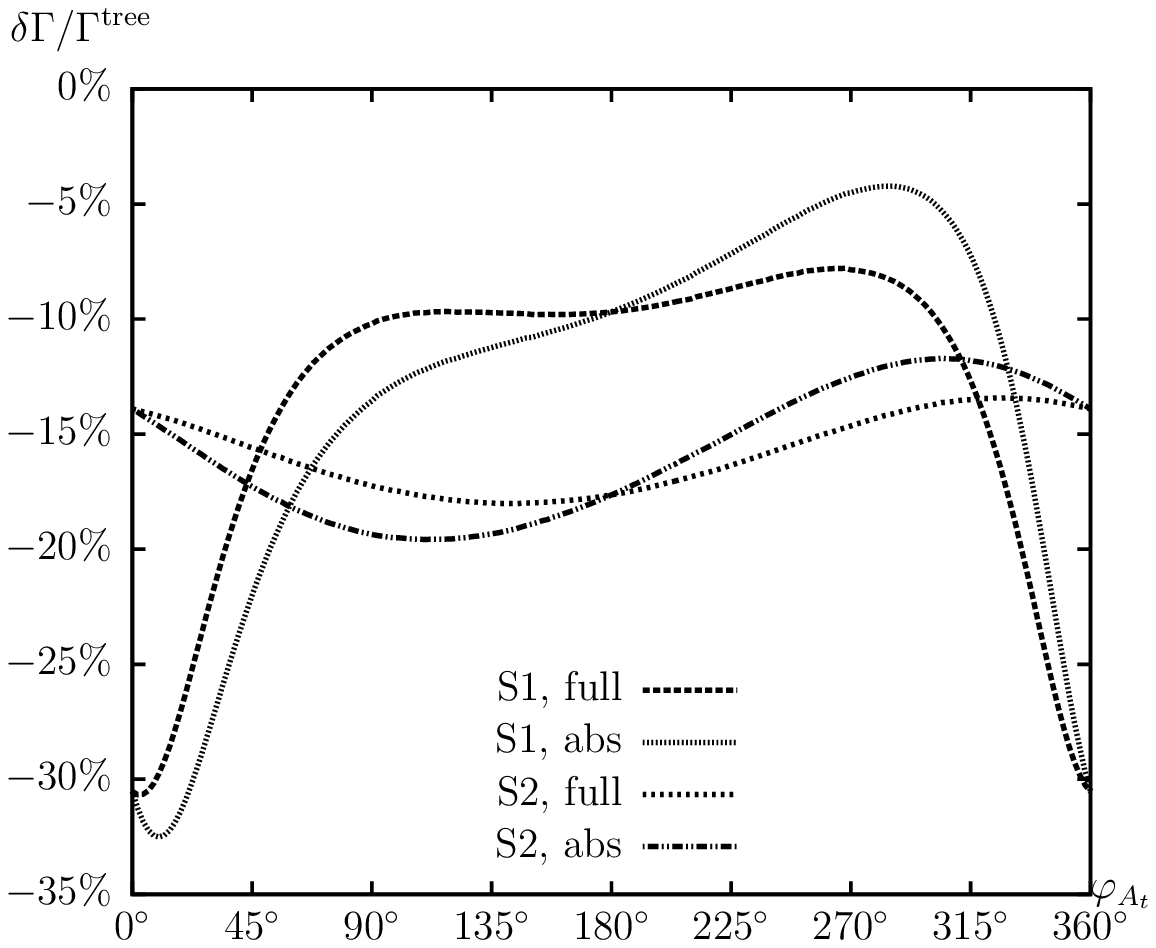}
\\[4em]
\includegraphics[width=0.49\textwidth,height=7.5cm]{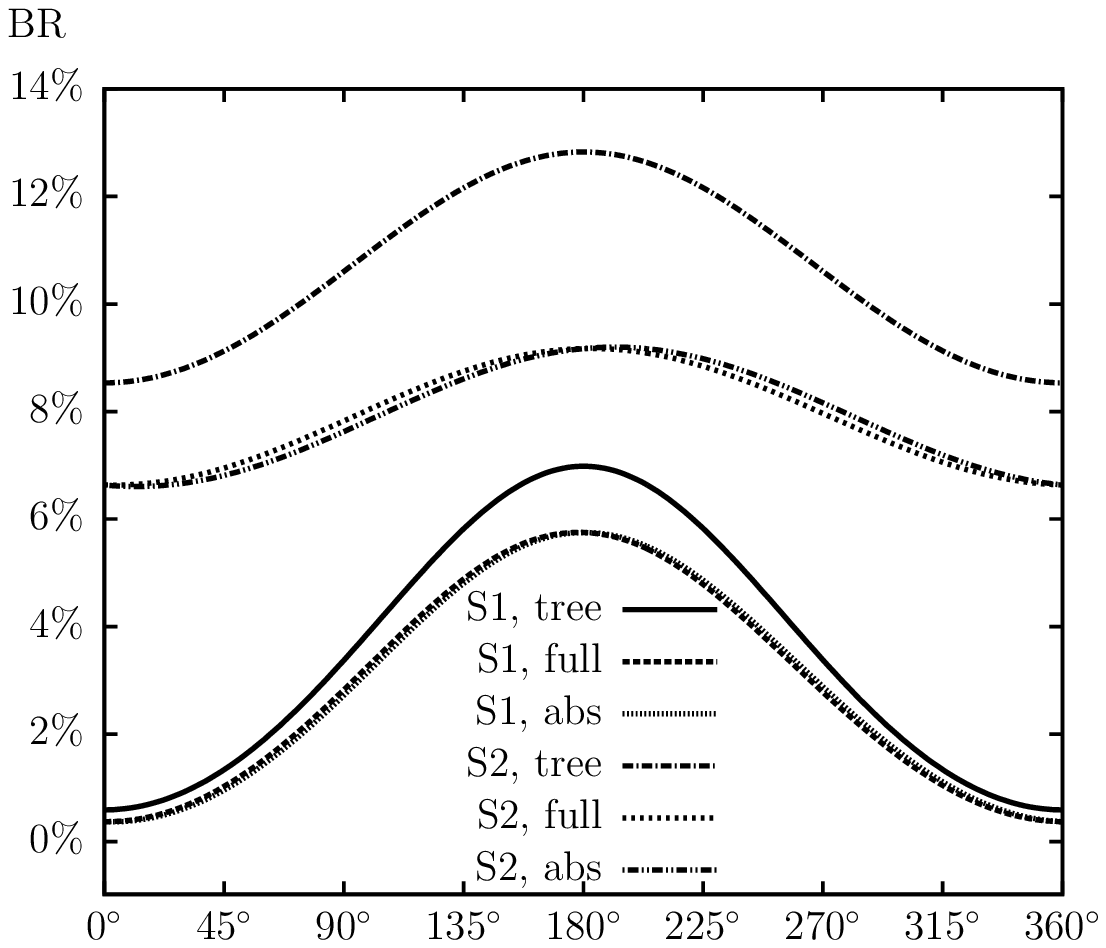}
\hspace{-4mm}
\includegraphics[width=0.49\textwidth,height=7.5cm]{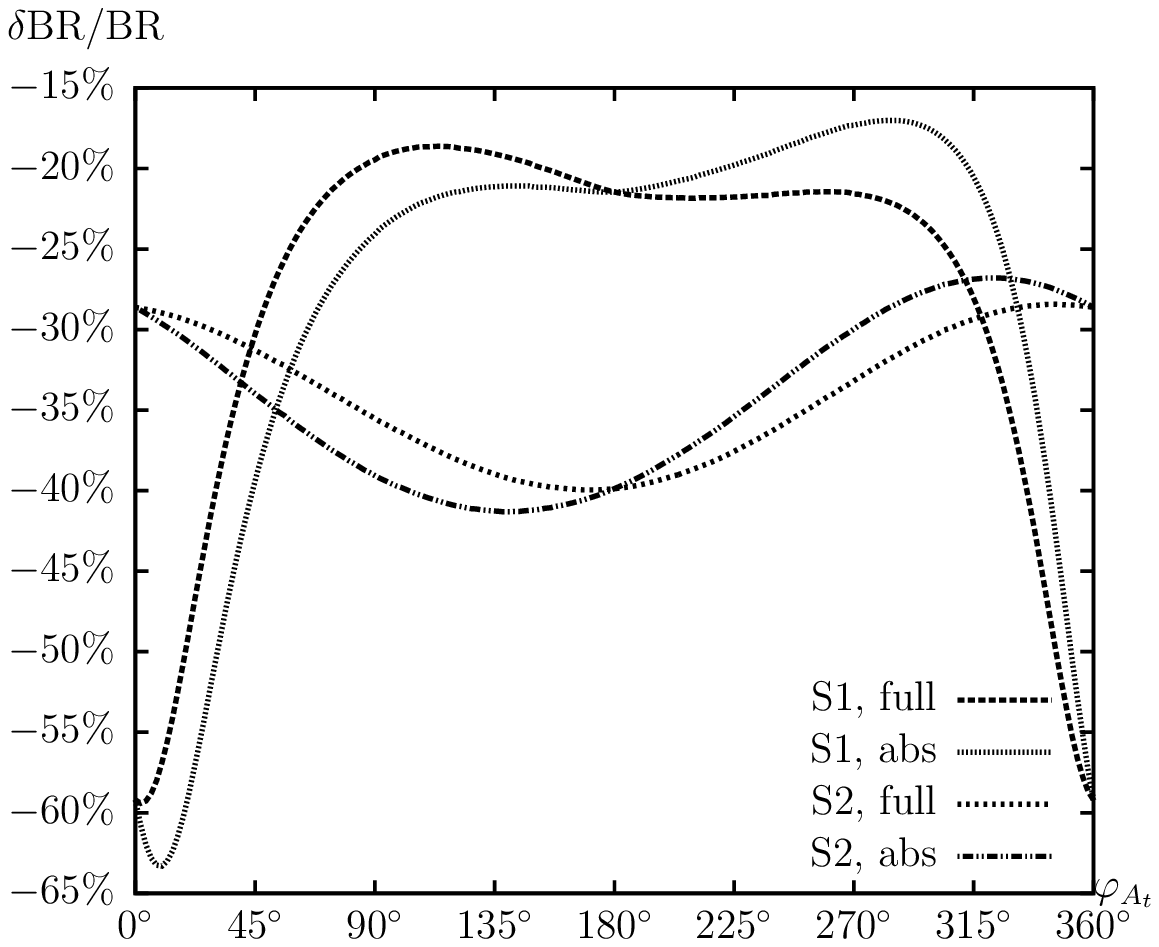}
\end{tabular}
\vspace{2em}
\caption{$\Ga(\decayCpz)$.
  Tree-level (``tree'') and full one-loop (``full'') corrected 
  partial decay widths are shown. Also shown are the full one-loop
  corrected partial decay  
  widths including absorptive contributions (``abs''). 
  The parameters are chosen according to \SE\ and \SZ\ (see \refta{tab:para}), 
  with $\phiat$ varied.
  The upper left plot shows the partial decay width; 
  the upper right plot shows the corresponding  relative size of the corrections. 
  The lower left plot shows the BR; 
  the lower right plot shows the relative correction of the BR.
}
\label{fig:PhiAt.st2bcha2}
\end{center}
\end{figure}

The results for $\Ga(\decayCpj)$ ($j = 1,2$) are shown in
\reffis{fig:PhiAt.st2bcha1}, \ref{fig:PhiAt.st2bcha2}. For
$\Ga(\decayCpe)$ the decay width changes substantially with $\phiat$.
The relative corrections are mostly between $-5\%$ and $-20\%$, except in
\SZ\ for $\phiat \sim \pi$, where $\Ga(\decayCpe)$ becomes very small.
The absorptive contributions lead to a visible shift in the
relative one-loop corrections in \SZ, where the largest effects are
found again where $\Ga(\decayCpe)$ is small.
For $\Ga(\decayCpz)$ we find a similar size of the corrections. 
Larger relative corrections of up to $-32\%$ are reached only in
\SE\ where the decay width itself becomes very small.
Within \SE\ the larger branching ratio values of $7\%$ - $12\%$ are found for 
$\decayCpe$.  Here the relative corrections are between $-18\%$ and
$-30\%$, with some variation induced by the absorptive
contributions, which can be relevant for the LHC and the ILC.
For $\decayCpz$ the larger BR is found in \SZ, where values around $8\%$
are found. The one-loop effects nearly reach $-40\%$, which can be
relevant for the LHC. 
Finally it should be noted that the apparently very 
large corrections on $\br(\decayCpe)$ in \SZ\
(see the lower right plot in \reffi{fig:PhiAt.st2bcha1})
do {\em not} correspond to a negative BR. 
At $\phiat \sim \pi$ the loop corrections are negative and comparably 
to the (very small) tree-level width, leading to 
$\br^{\rm full} \ll \br^{\rm tree}$ in \refeq{brrel}.
The effect of these relatively large loop corrections around
$\phiat \sim \pi$ can be sizably lowered by including 
higher-order corrections as, e.g., $|{\cal M}_{\rm loop}|^2$.

\begin{figure}[htb!]
\begin{center}
\begin{tabular}{c}
\includegraphics[width=0.49\textwidth,height=7.5cm]{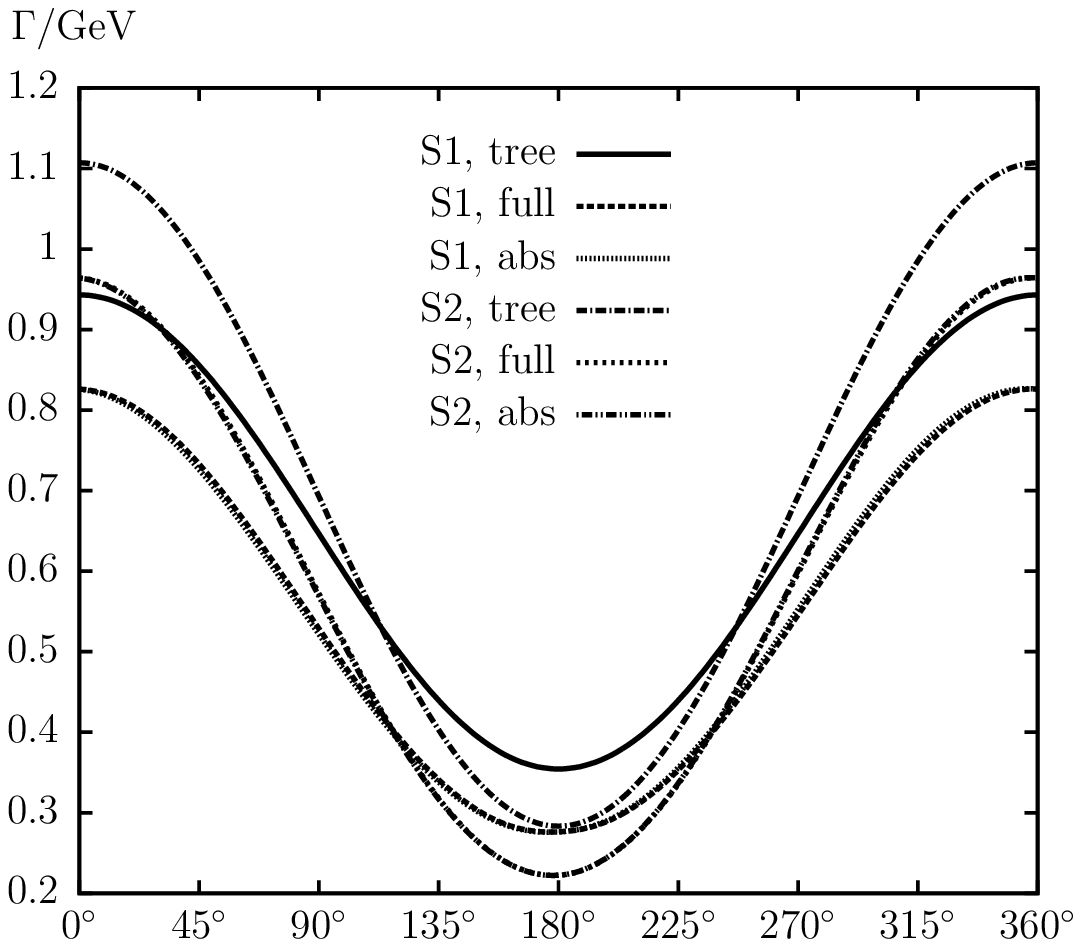}
\hspace{-4mm}
\includegraphics[width=0.49\textwidth,height=7.5cm]{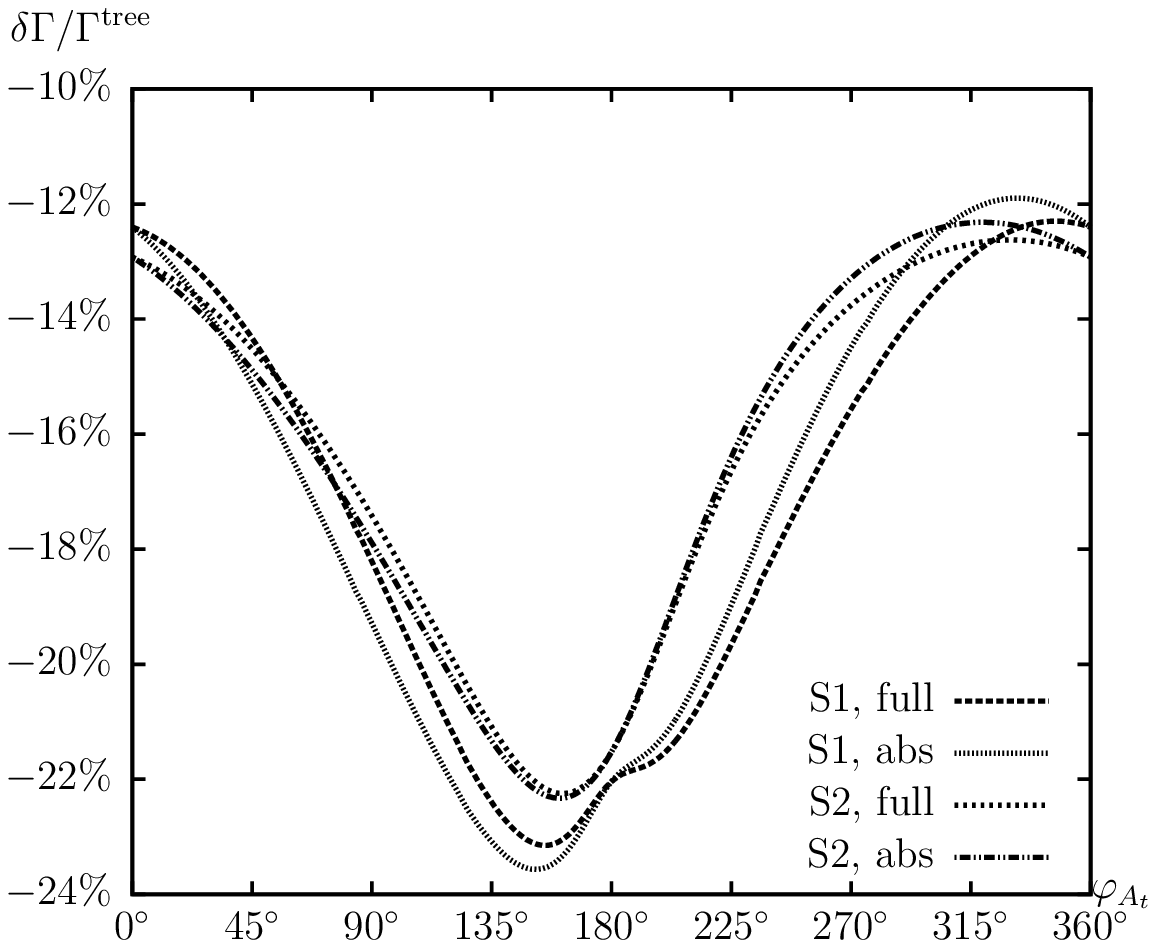}
\\[4em]
\includegraphics[width=0.49\textwidth,height=7.5cm]{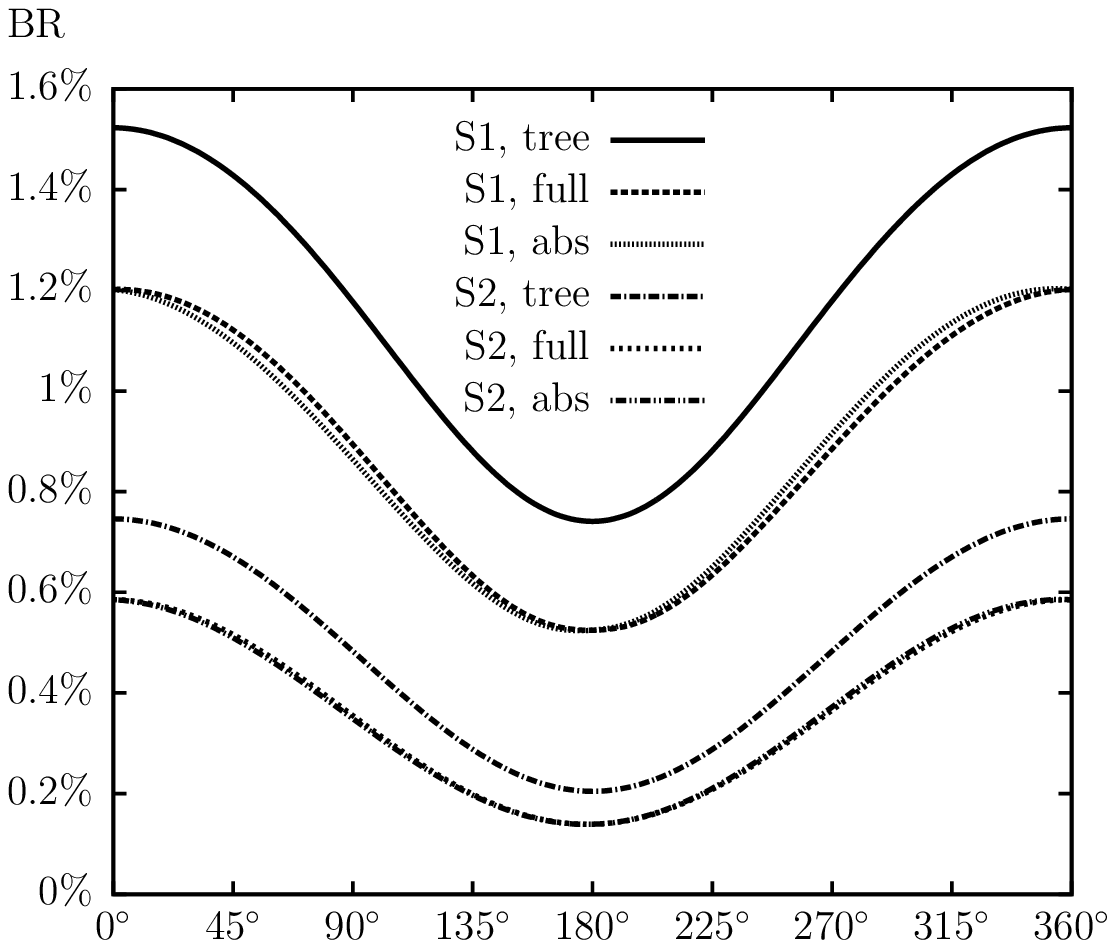}
\hspace{-4mm}
\includegraphics[width=0.49\textwidth,height=7.5cm]{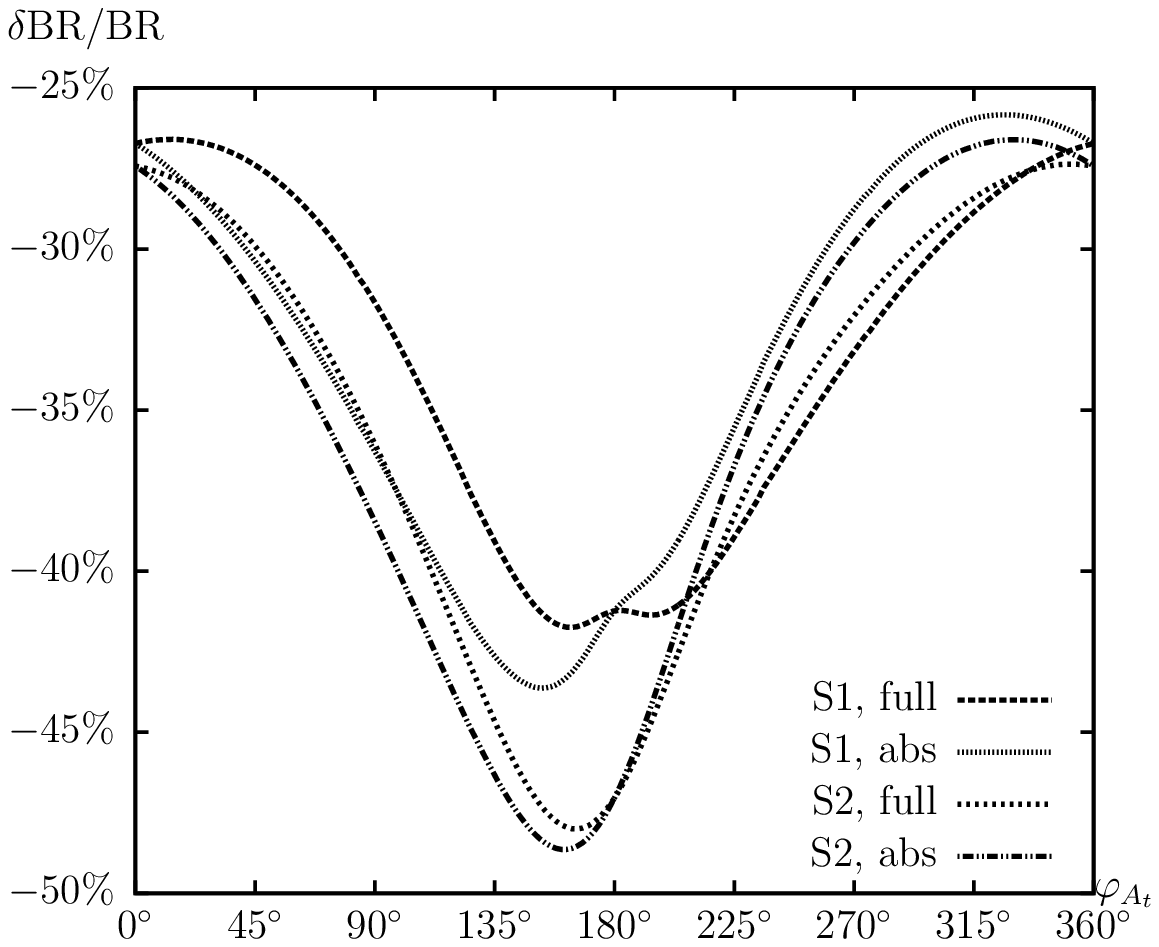}
\end{tabular}
\vspace{2em}
\caption{$\Ga(\decaySbeH)$. 
  Tree-level (``tree'') and full one-loop (``full'') corrected 
  partial decay widths are shown. Also shown are the full one-loop
  corrected partial decay  
  widths including absorptive contributions (``abs''). 
  The parameters are chosen according to \SE\ and \SZ\ (see \refta{tab:para}), 
  with $\phiat$ varied.
  The upper left plot shows the partial decay width;
  the upper right plot shows the corresponding  relative size of the corrections. 
  The lower left plot shows the BR; 
  the lower right plot shows the relative correction of the BR.
}
\label{fig:PhiAt.st2sb1H}
\end{center}
\end{figure}

\begin{figure}[htb!]
\begin{center}
\begin{tabular}{c}
\includegraphics[width=0.49\textwidth,height=7.5cm]{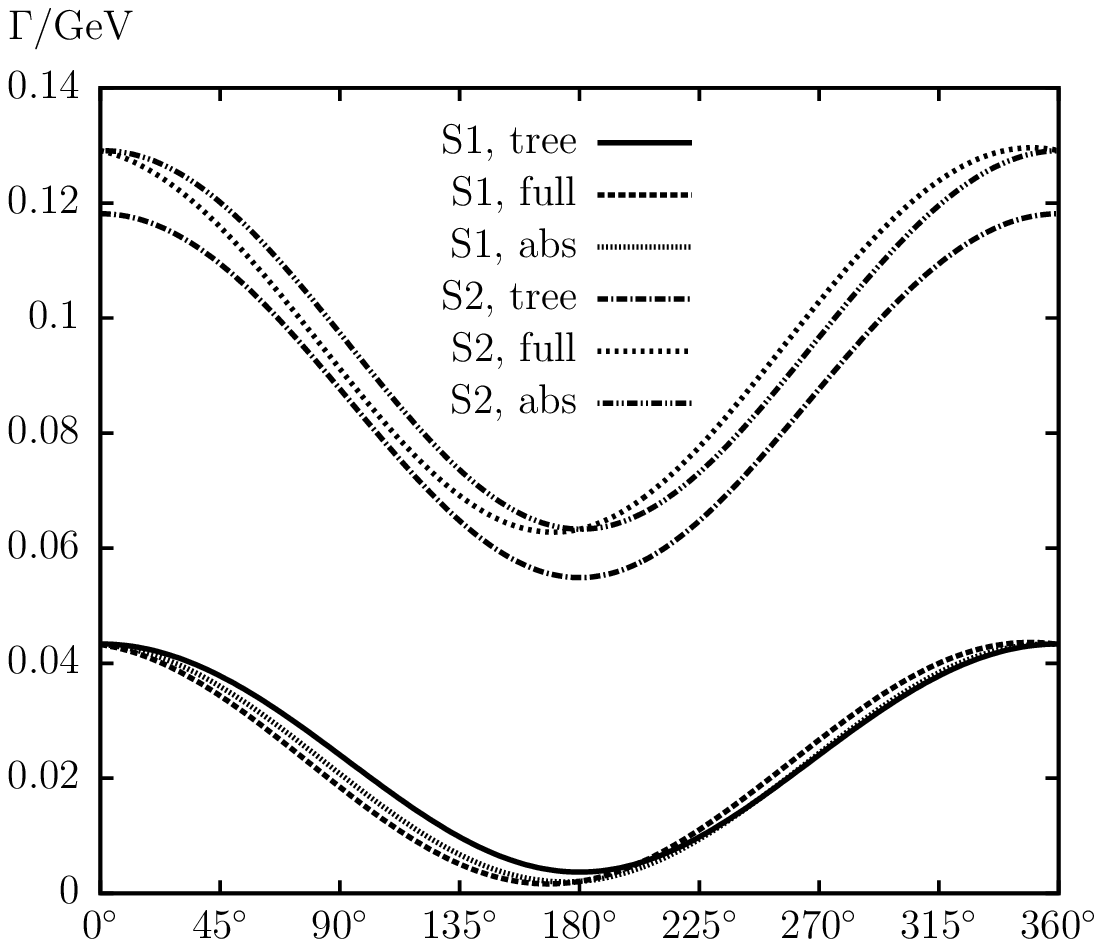}
\hspace{-4mm}
\includegraphics[width=0.49\textwidth,height=7.5cm]{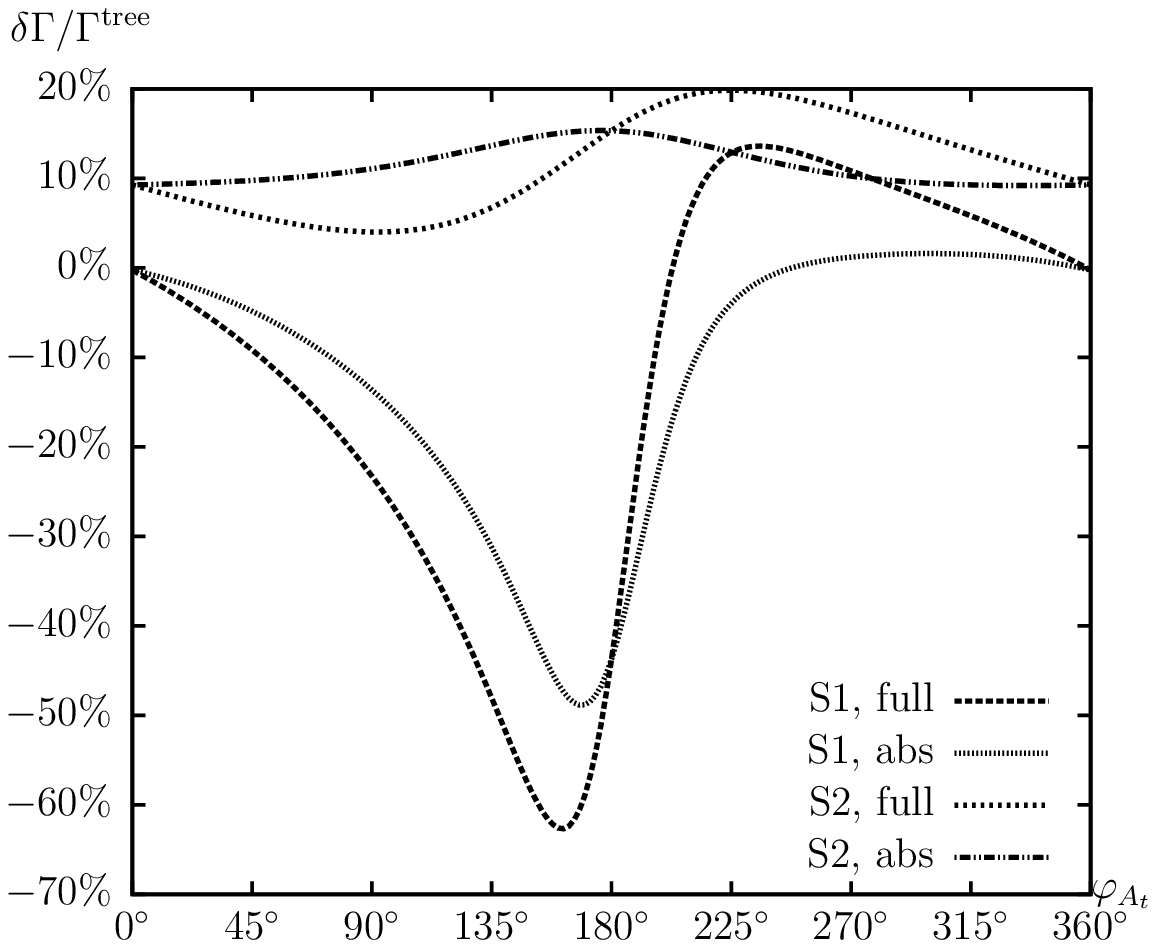}
\\[4em]
\includegraphics[width=0.49\textwidth,height=7.5cm]{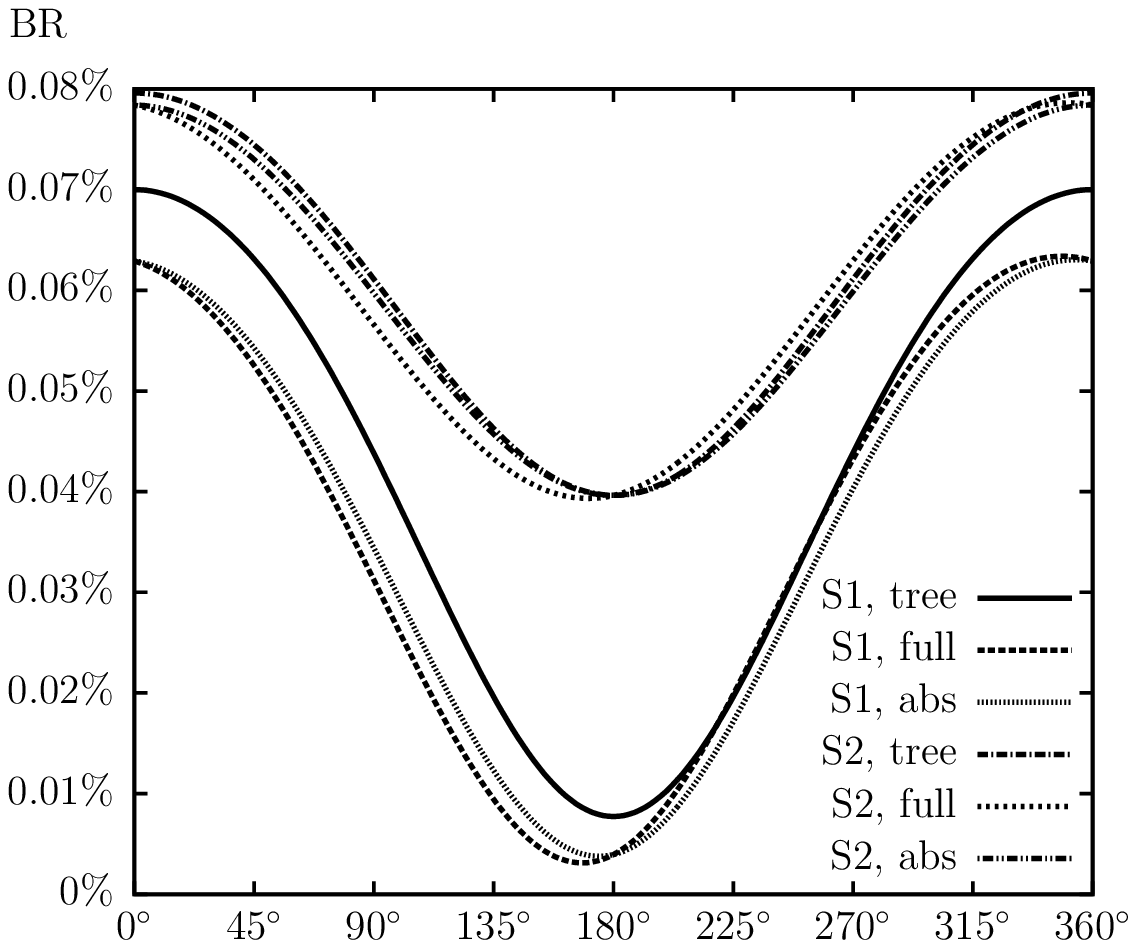}
\hspace{-4mm}
\includegraphics[width=0.49\textwidth,height=7.5cm]{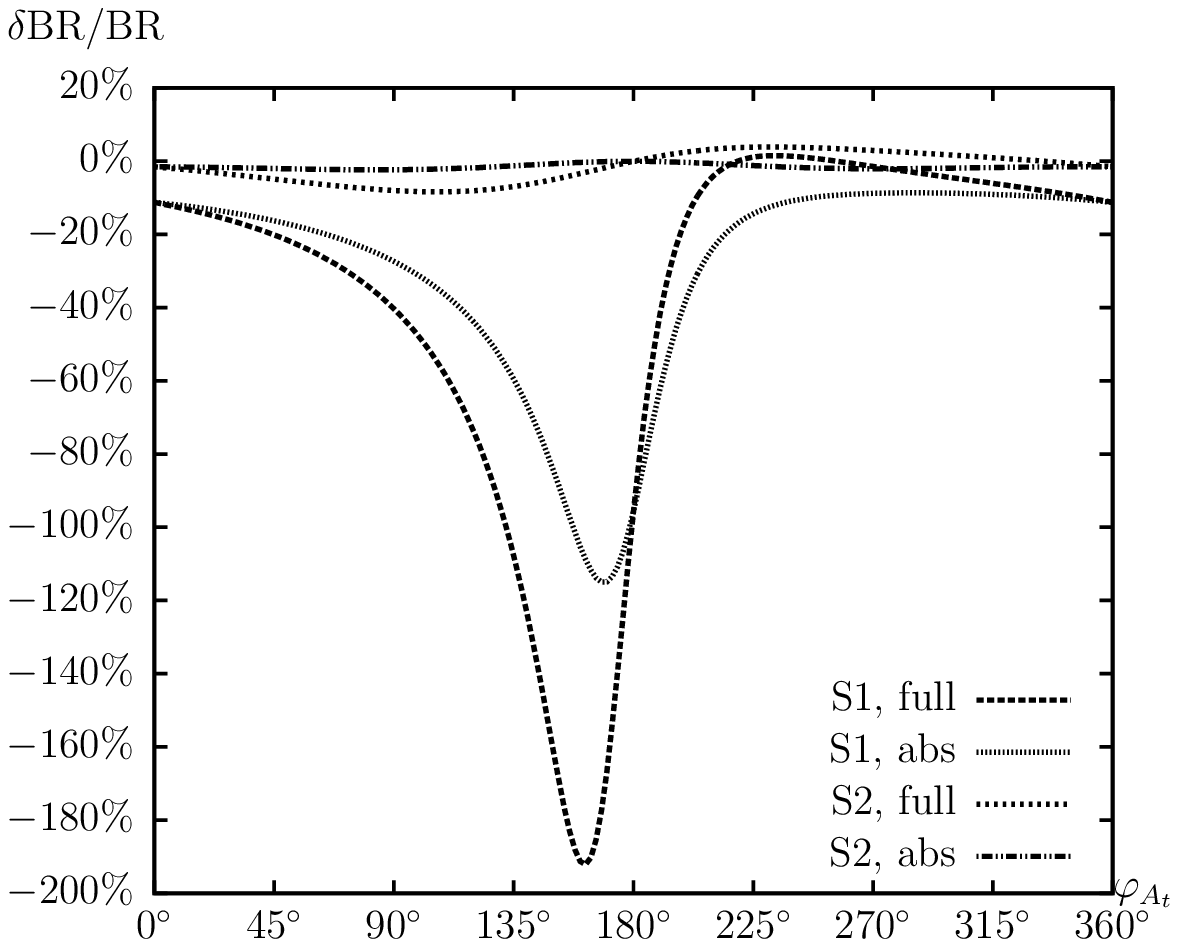}
\end{tabular}
\vspace{2em}
\caption{$\Ga(\decaySbzH)$. 
  Tree-level (``tree'') and full one-loop (``full'') corrected 
  partial decay widths are shown. Also shown are the full one-loop
  corrected partial decay  
  widths including absorptive contributions (``abs''). 
  The parameters are chosen according to \SE\ and \SZ\ (see \refta{tab:para}), 
  with $\phiat$ varied.
  The upper left plot shows the partial decay width;
  the upper right plot shows the corresponding relative size of the corrections. 
  The lower left plot shows the BR; 
  the lower right plot shows the relative correction of the BR.
}
\label{fig:PhiAt.st2sb2H}
\end{center}
\end{figure}

Now we turn to the decay modes involving scalar bottom quarks, which
have also been analyzed in \citere{SbotRen}. 
In \reffis{fig:PhiAt.st2sb1H}, \ref{fig:PhiAt.st2sb2H} the results 
for $\Ga(\decaySbiH)$ ($i = 1,2$) are presented. 
While we find $\Ga(\decaySbeH)$ at the $\sim 1 \gev$~level, 
$\Ga(\decaySbzH)$ is only around the $0.1 \gev$~level. The relative
variation of $\Ga(\decaySbeH)$ ranges from $\sim -13\%$ for large values
of the width to $\sim -23\%$ for small values, with some variations
induced by the absorptive contributions. For $\Ga(\decaySbzH)$ in \SE\ 
the relative variation can become very large, $-60\%$ (with a clearly 
visible shift from the absorptive contributions) but the partial decay 
width is negligibly small. 
Within \SZ, where $\Ga(\decaySbzH) \sim 0.1 \gev$
is realized the relative variation is at the 10\%~level.
As for the variation with $\mstz$ we find only small values for the
branching ratios at the $1\%$ ($0.1\%$)~level for the decay to the
lighter (heavier) sbottom. The one-loop effects on the BRs are only
important if other channels are kinematically suppressed. In this case
the effects can be of the same order as for the partial decay widths itself.
Again, the apparent very large effect on $\br(\decaySbzH)$ in \SE\ 
still corresponds to a positive BR; see \refeq{brrel}.

\begin{figure}[htb!]
\begin{center}
\begin{tabular}{c}
\includegraphics[width=0.49\textwidth,height=7.5cm]{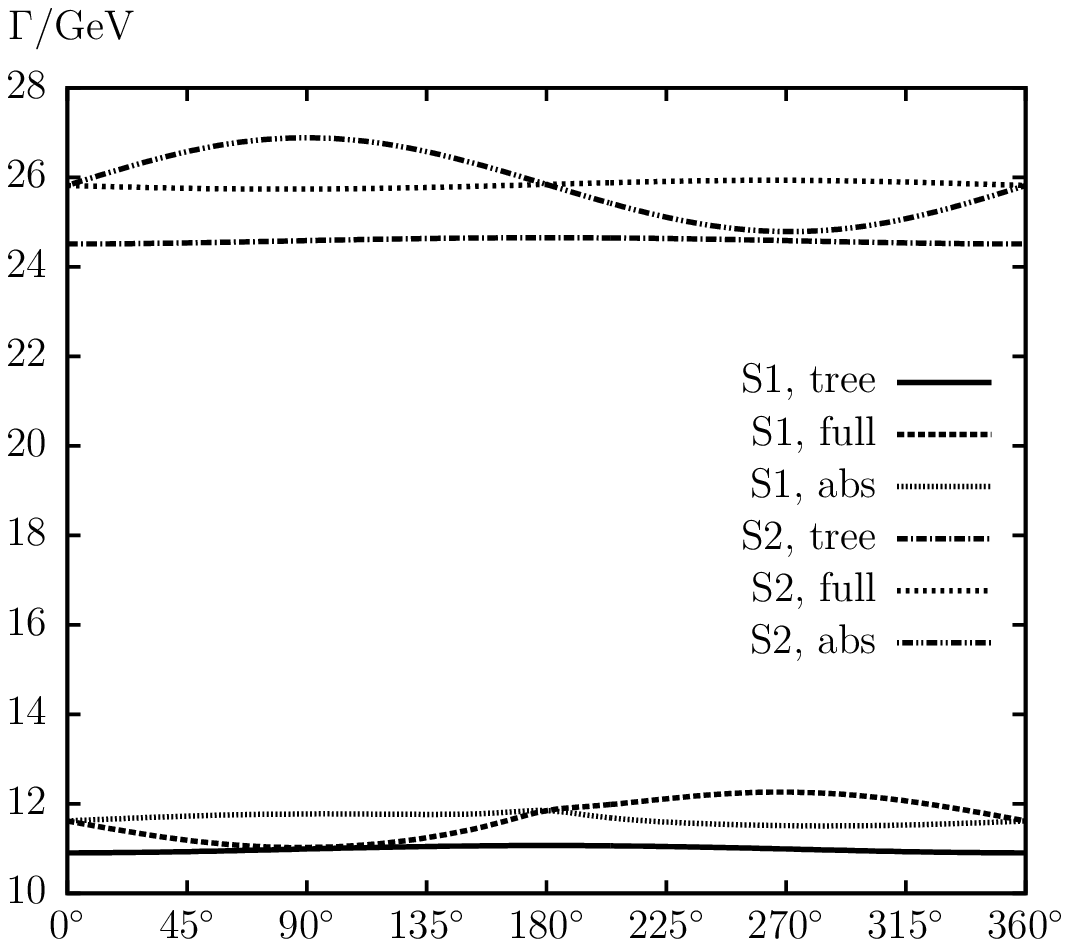}
\hspace{-4mm}
\includegraphics[width=0.49\textwidth,height=7.5cm]{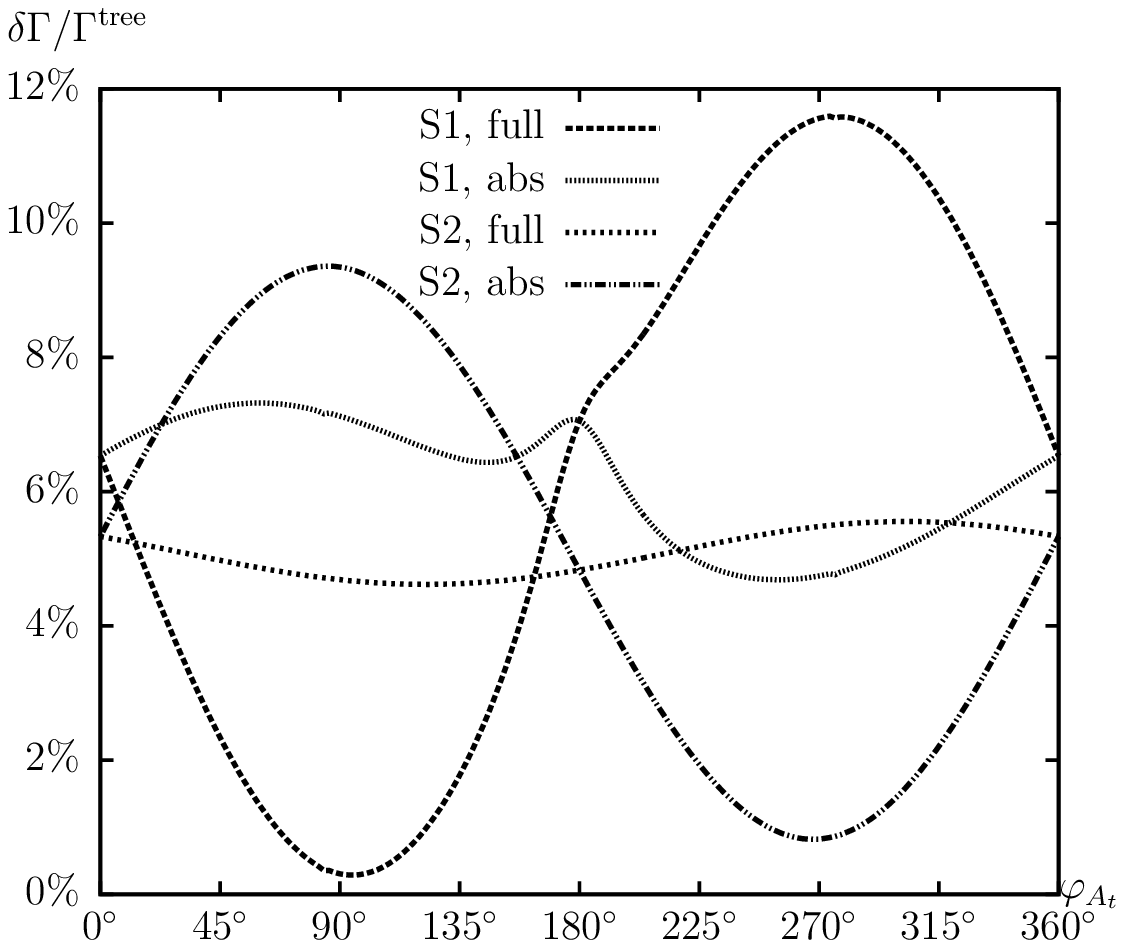}
\\[4em]
\includegraphics[width=0.49\textwidth,height=7.5cm]{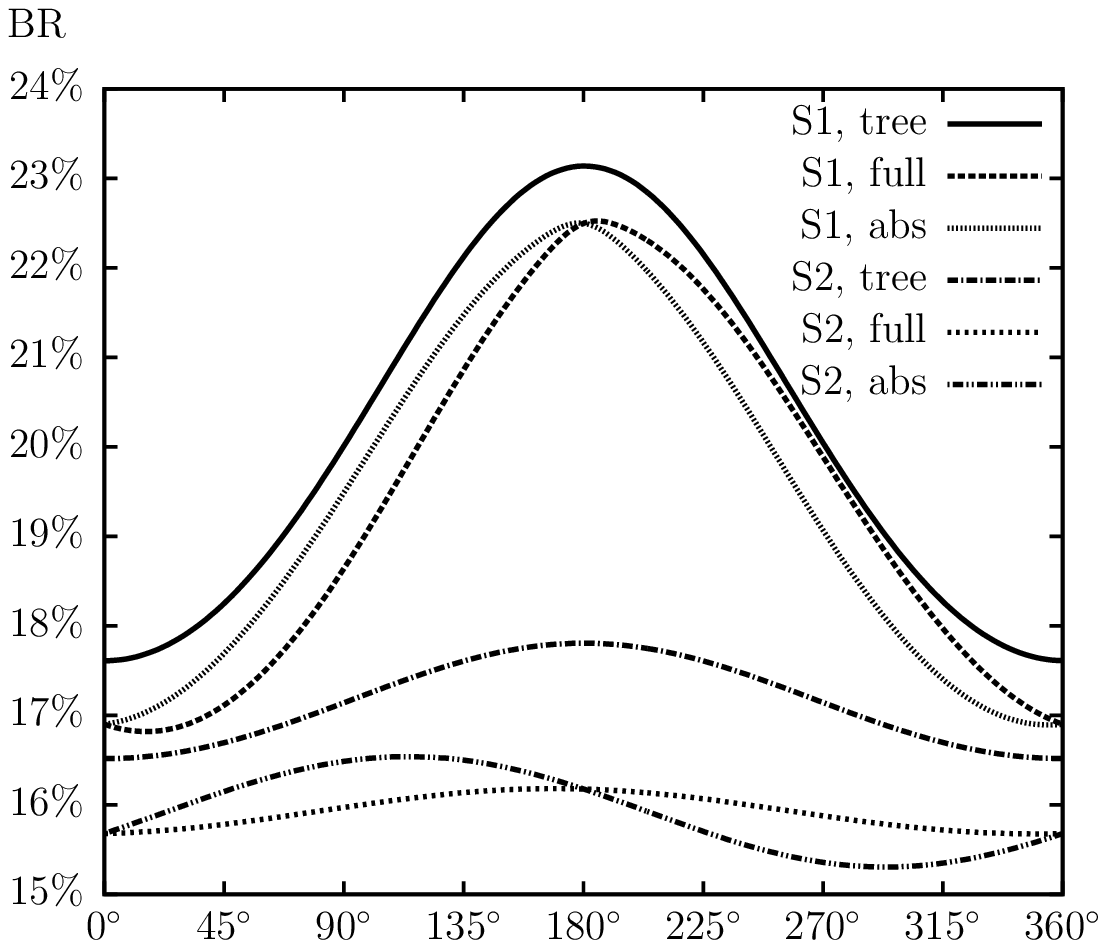}
\hspace{-4mm}
\includegraphics[width=0.49\textwidth,height=7.5cm]{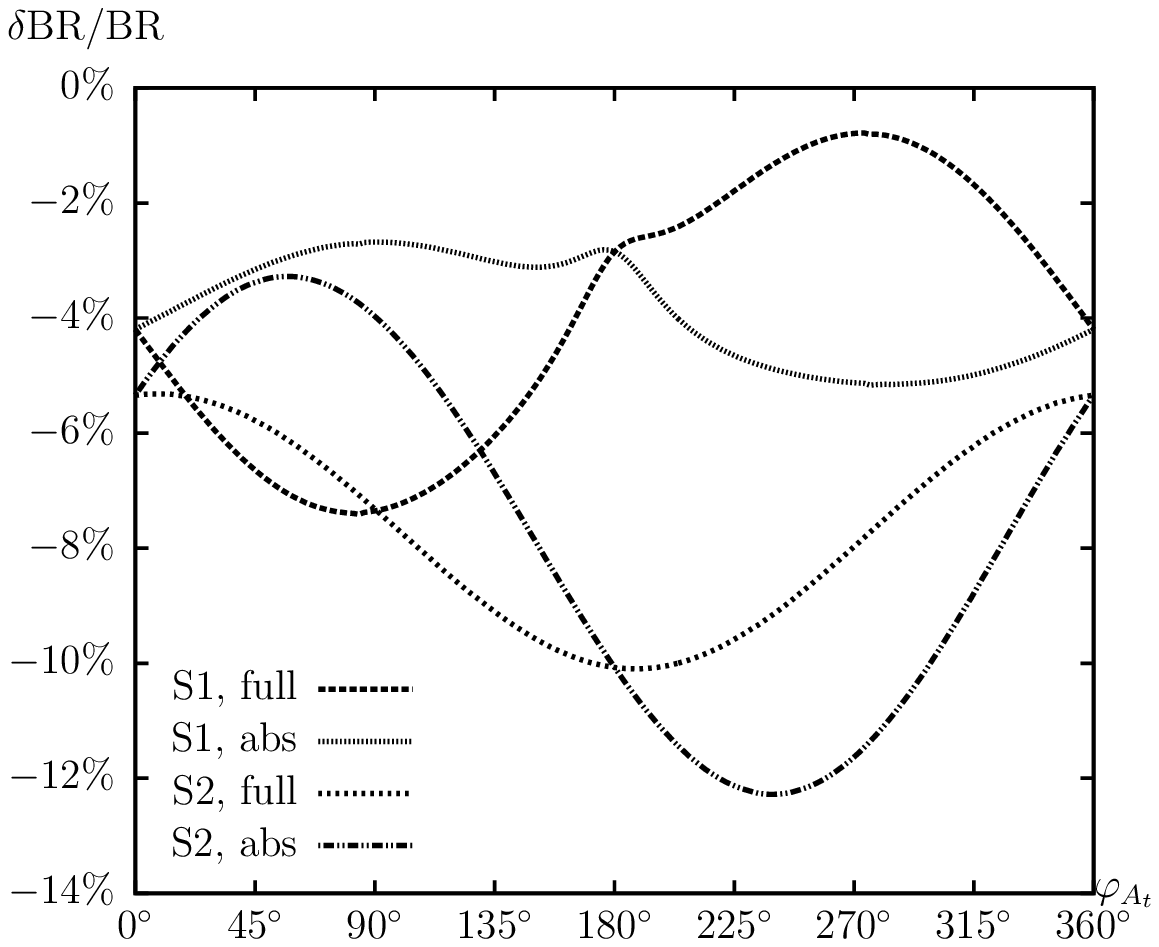}
\end{tabular}
\vspace{2em}
\caption{$\Ga(\decaySbeW)$. 
  Tree-level (``tree'') and full one-loop (``full'') corrected 
  partial decay widths are shown. Also shown are the full one-loop
  corrected partial decay  
  widths including absorptive contributions (``abs''). 
  The parameters are chosen according to \SE\ and \SZ\ (see \refta{tab:para}), 
  with $\phiat$ varied.
  The upper left plot shows the partial decay width; 
  the upper right plot shows the corresponding relative size of the corrections. 
  The lower left plot shows the BR; 
  the lower right plot shows the relative correction of the BR. 
}
\label{fig:PhiAt.st2sb1W}
\end{center}
\end{figure}

\begin{figure}[htb!]
\begin{center}
\begin{tabular}{c}
\includegraphics[width=0.49\textwidth,height=7.5cm]{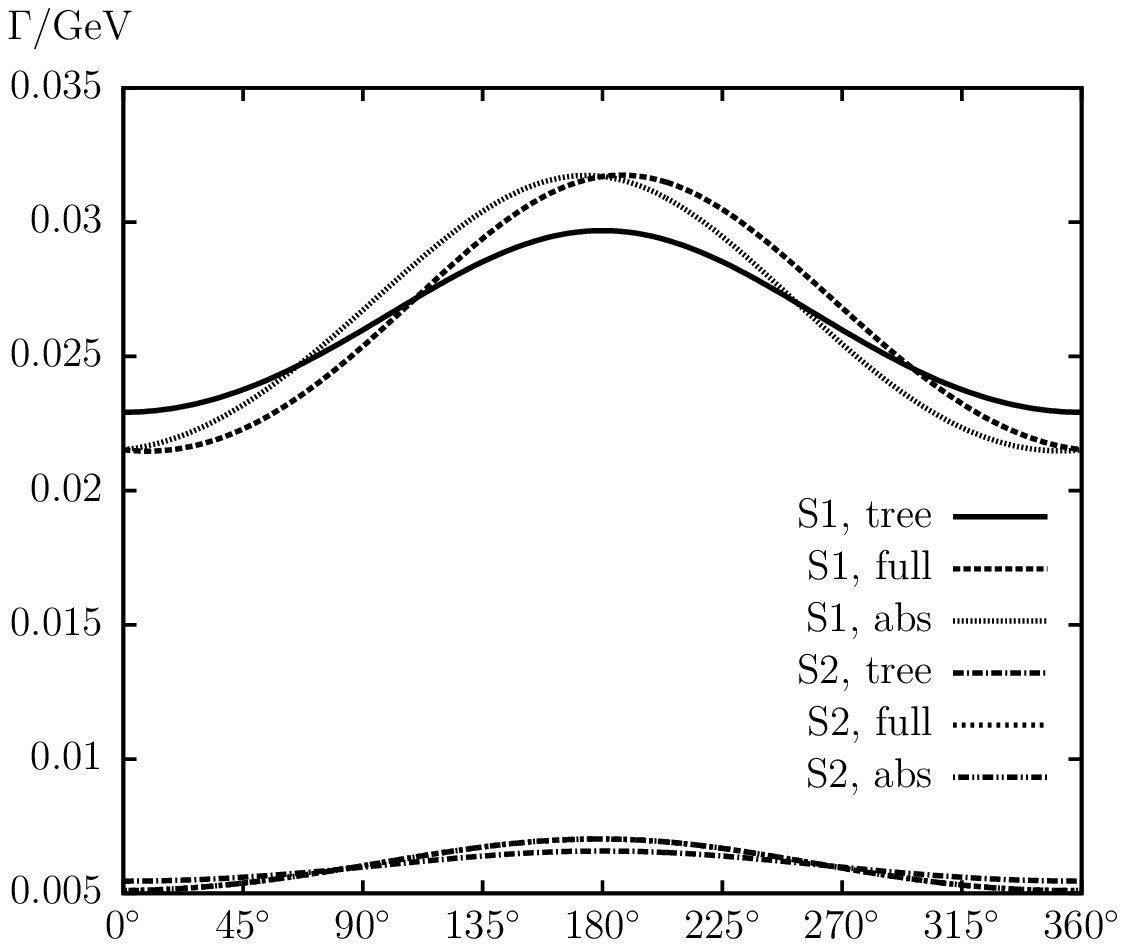}
\hspace{-4mm}
\includegraphics[width=0.49\textwidth,height=7.5cm]{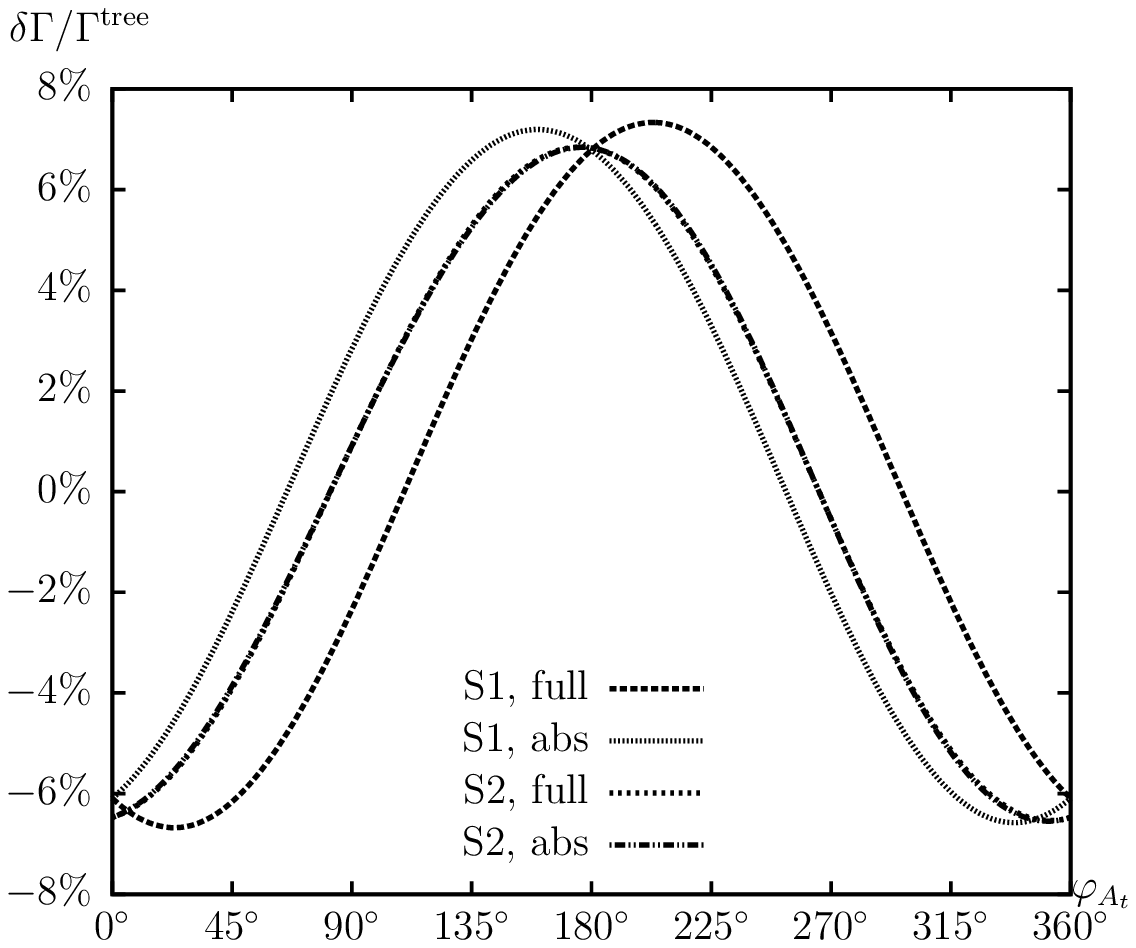}
\\[4em]
\includegraphics[width=0.49\textwidth,height=7.5cm]{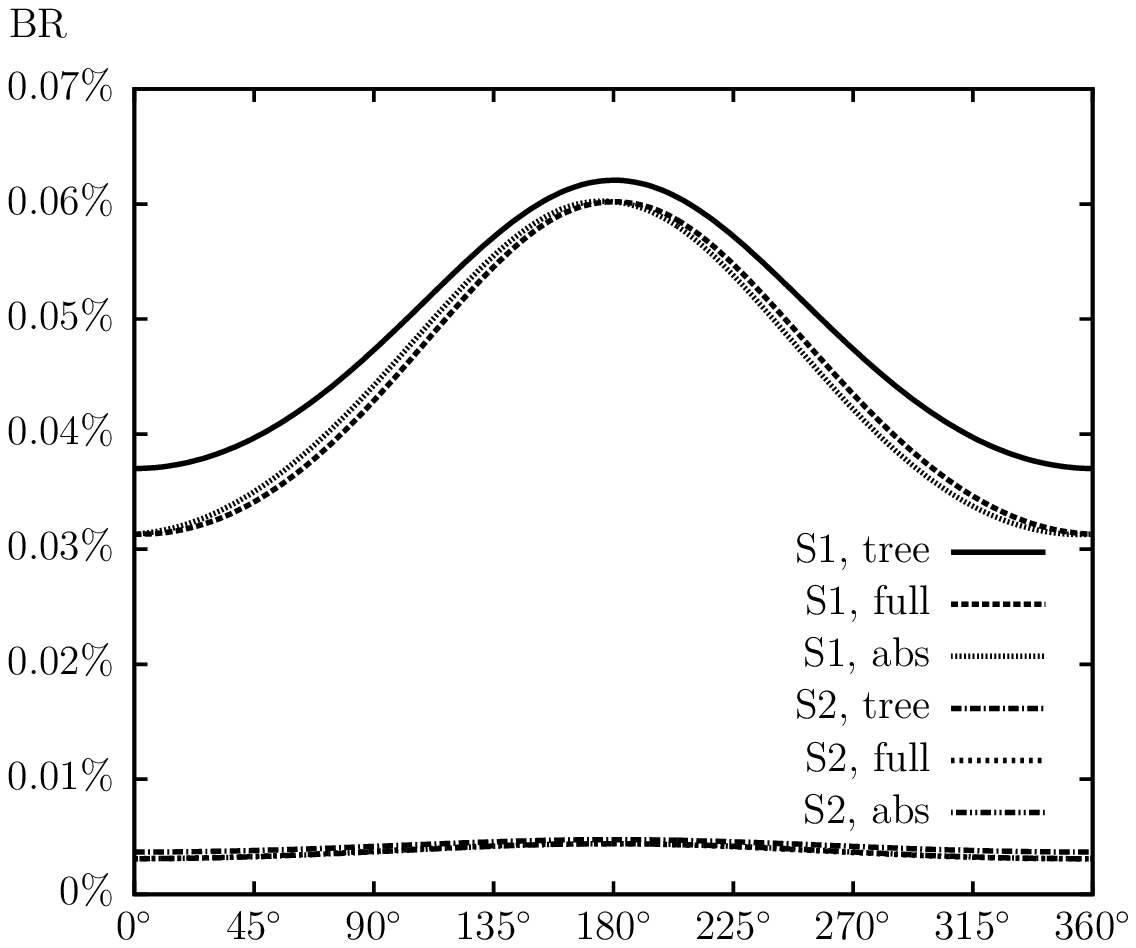}
\hspace{-4mm}
\includegraphics[width=0.49\textwidth,height=7.5cm]{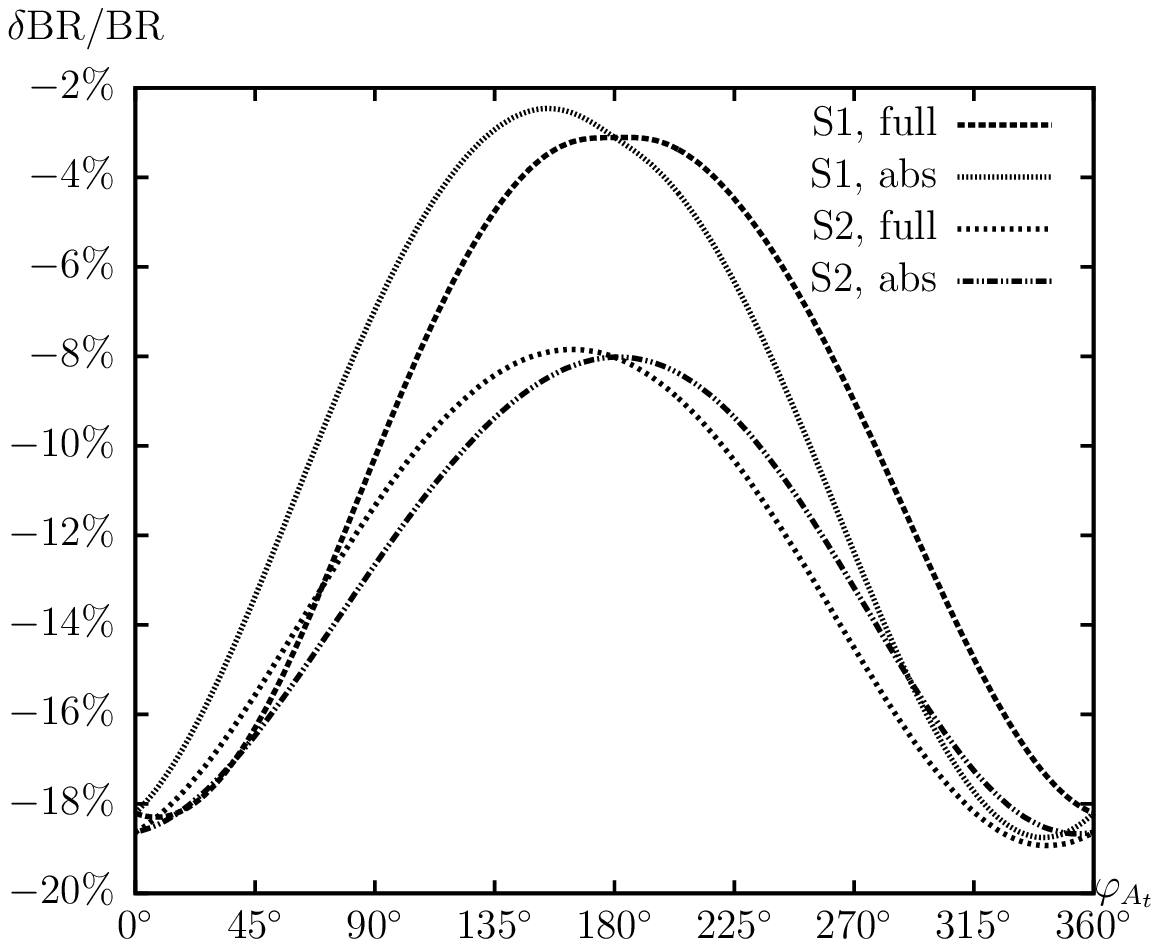}
\end{tabular}
\vspace{2em}
\caption{$\Ga(\decaySbzW)$. 
  Tree-level (``tree'') and full one-loop (``full'') corrected 
  partial decay widths are shown. Also shown are the full one-loop
  corrected partial decay  
  widths including absorptive contributions (``abs''). 
  The parameters are chosen according to \SE\ and \SZ\ (see \refta{tab:para}), 
  with $\phiat$ varied.
  The upper left plot shows the partial decay width; 
  the upper right plot shows the corresponding  relative size of the
  corrections.  
  The lower left plot shows the BR; 
  the lower right plot shows the relative correction of the BR. 
}
\label{fig:PhiAt.st2sb2W}
\end{center}
\end{figure}

The other decay modes involving scalar bottom quarks, 
$\decaySbiW$ ($i = 1,2$) are analyzed in \reffis{fig:PhiAt.st2sb1W}, 
\ref{fig:PhiAt.st2sb2W}. 
As in the analysis with $\mstz$ varied, we find $\Ga(\decaySbeW)$ at the 
$11\ (25) \gev$~level in \SE\ (\SZ). 
The relative correction without absorptive contributions changes 
sizably in \SE, ranging between 0\% and $+12\%$.
Taking into account the absorptive contributions this strongly reduces to 
5\% and 7\%. 
Within \SZ\ the corrections without absorptive contributions are around
$+5\%$ for all $\phiat$, but the absorptive contributions have the
opposite effect of strongly enhancing the variation.
$\Ga(\decaySbzW)$, again as in \refse{sec:full1L}, is very small and
stays below $\sim 0.03 \gev$. The variation of this negligibly small
partial decay width is found to be between $-6\%$ and +6\%; the shift
from the absorptive contributions remains relatively small.
Consequently, a relevant branching ratio is found only for
$\decaySbeW$, where values around $\sim 20\%$ ($\sim 16\%$) 
are found in \SE\ (\SZ). 
The relative effects of the one-loop corrections can reach 
$-7\%$ ($-10\%$), which is potentially important for physics at the 
ILC and the LHC.

\newpage

\newpage

\newpage

\newpage

\newpage

\newpage

\newpage

\newpage

\newpage

\newpage

\newpage

\newpage

\newpage

\newpage

\newpage

\subsection{The total decay width}
\label{sec:totdecay}

\begin{figure}[htb!]
\begin{center}
\begin{tabular}{c}
\includegraphics[width=0.49\textwidth,height=7.5cm]{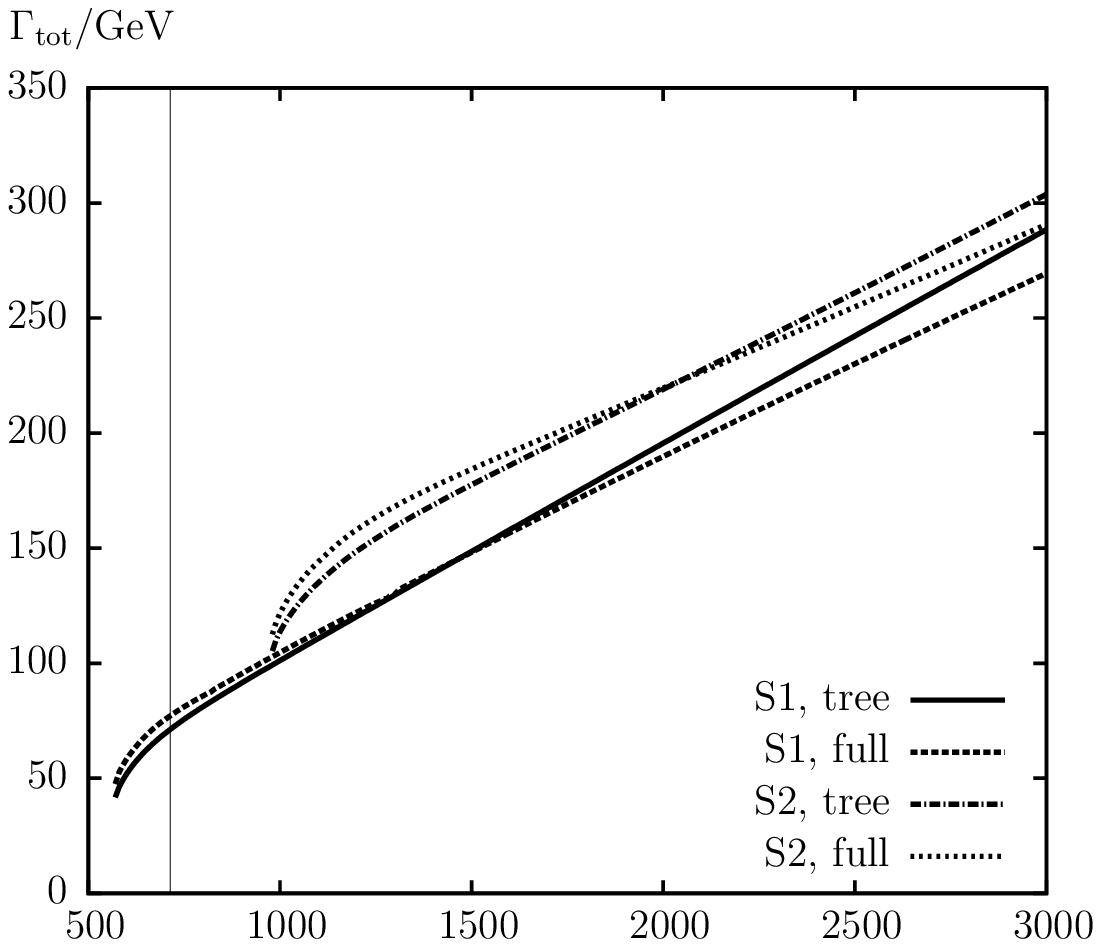}
\hspace{-4mm}
\includegraphics[width=0.49\textwidth,height=7.5cm]{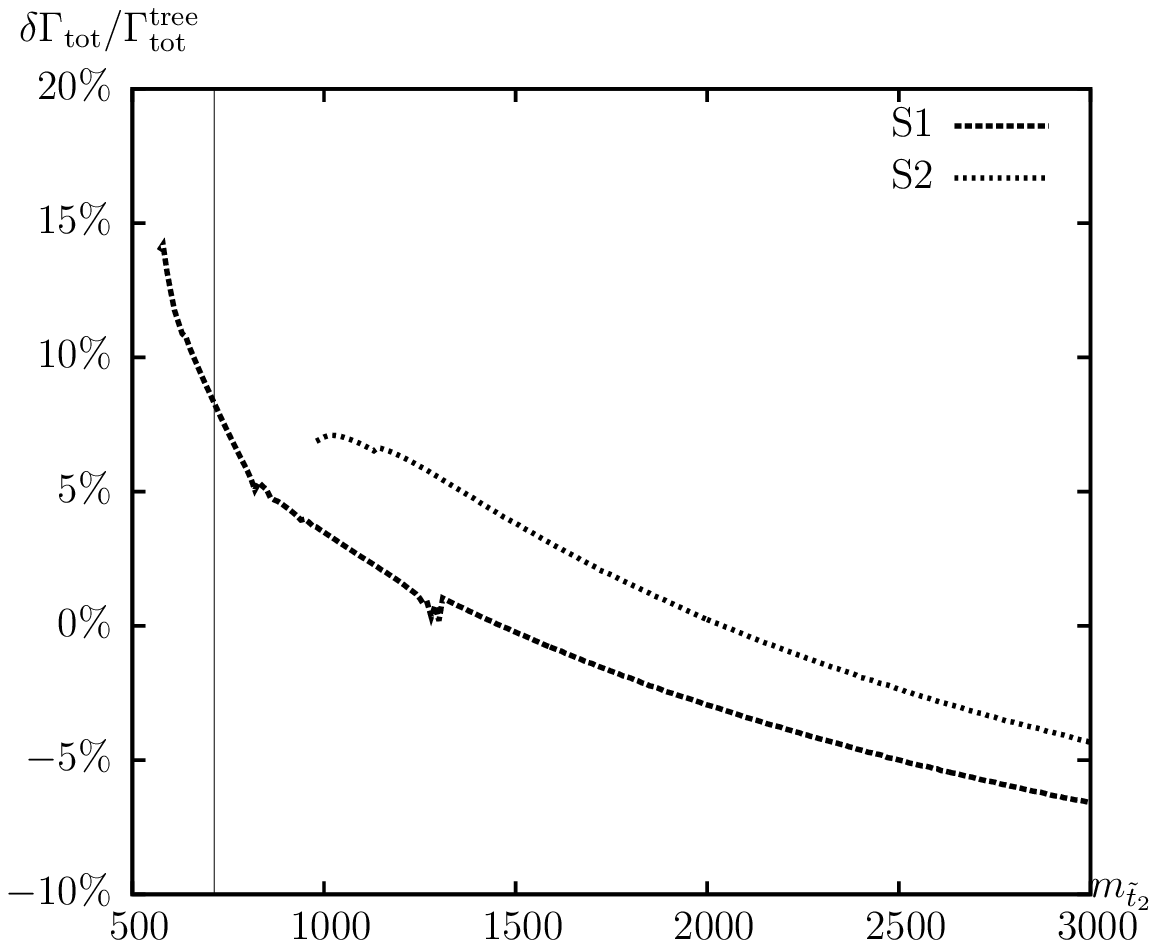}
\\[4em]
\includegraphics[width=0.49\textwidth,height=7.5cm]{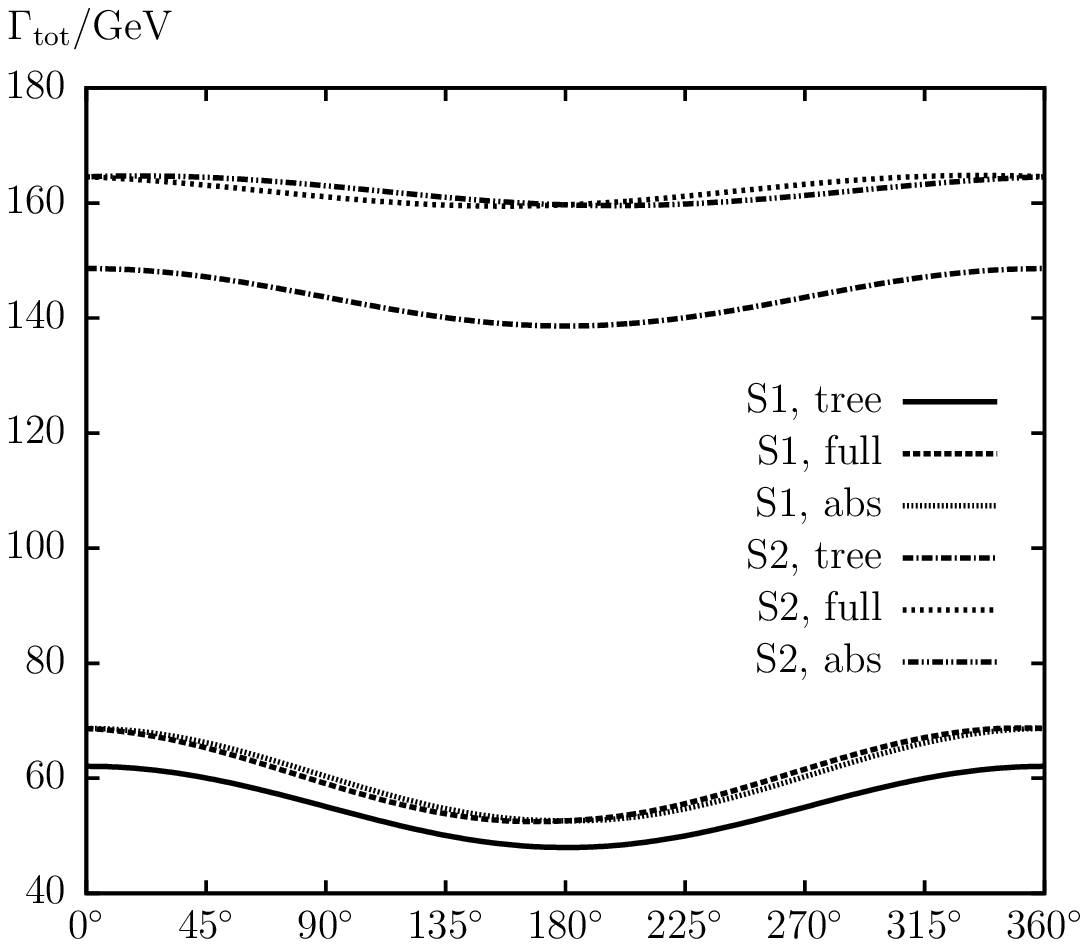}
\hspace{-4mm}
\includegraphics[width=0.49\textwidth,height=7.5cm]{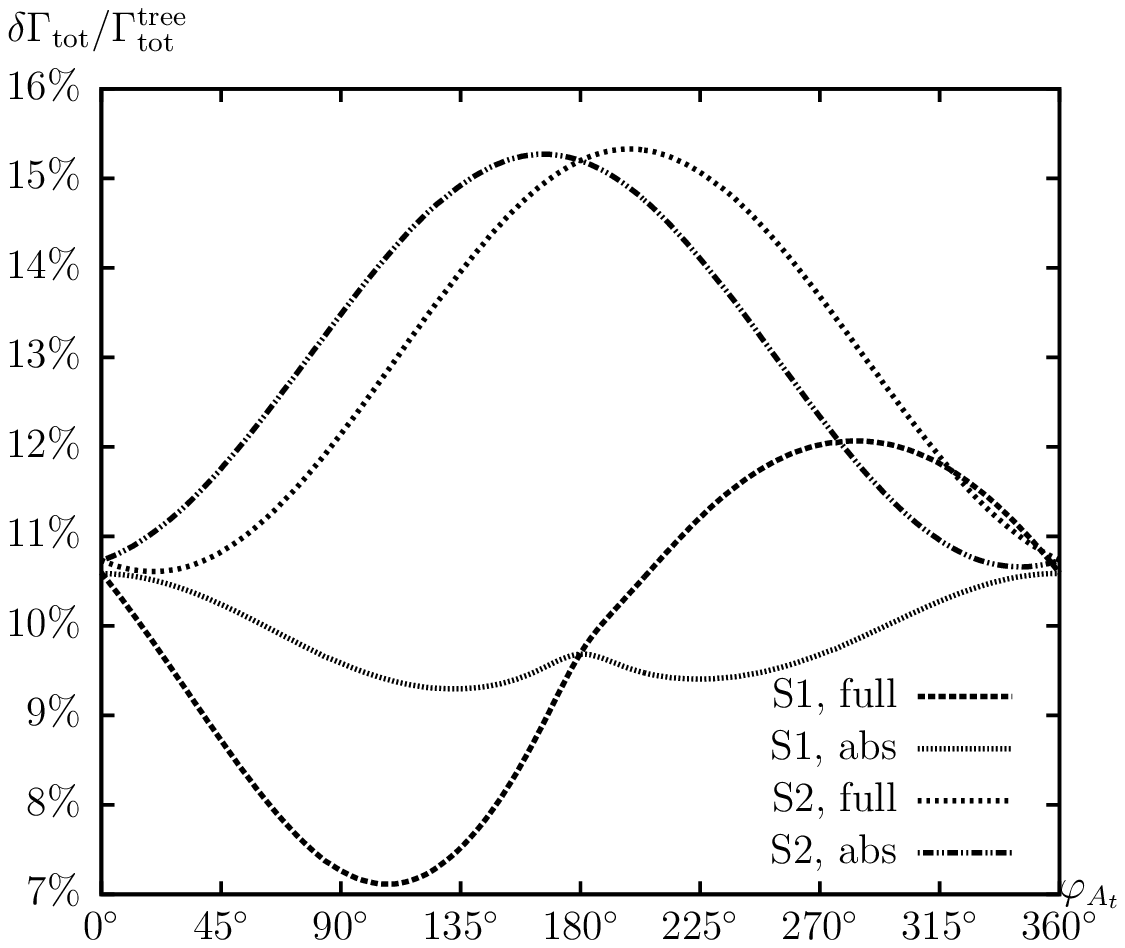}
\end{tabular}
\vspace{2em}
\caption{$\Ga_{\rm tot}$.
  The tree-level (``tree'') and full one-loop (``full'') corrected 
  total decay widths shown with the parameters chosen according to 
  \SE\ and \SZ\ (see \refta{tab:para}).
  The upper left plot shows the total decay width; 
  the upper right plot shows the corresponding relative size of the total 
  corrections, with $\mstz$ varied.  
  The lower plots show the same but with $\phiat$ varied. 
  Also shown is the full one-loop corrected total decay width including 
  absorptive contributions (``abs''). 
}
\label{fig:GammaTot}
\end{center}
\end{figure}

Finally we show the results for the total decay width of $\Stopz$. In
\reffi{fig:GammaTot} the upper panels show the absolute and relative
variation with $\mstz$. The lower panels depict the result for varying
$\phiat$. In \SE\ for small $\mstz$, $\mste + \mstz \le 1000 \gev$ the
size of the relative corrections of $\Ga_{\rm tot}$ ranges between
+15\% and +8\%. For larger 
$\mstz$ in the two numerical scenarios the variation ranges between 
$\sim +7\%$ down to $\sim -5\%$ for $\mstz = 3 \tev$. 
The variation with $\phiat$ is found to be large in both numerical
scenarios. Within \SE\ we find values of the relative correction between 
+13\% and +7\%, decreasing to a range of +9.5\% and +11\% 
once the absorptive contributions are taken into account. 
For \SZ\ the absolute values as well as the relative correction of 
$\Ga_{\rm tot}$, are larger than in \SE. The size of the relative
corrections ranges 
between +11\% and +15.5\%, where the absorptive contributions do 
not change the overall size of the effects but only affect the dependence 
on~$\phiat$.

\section{Conclusions}
\label{sec:conclusions}

We evaluate all partial decay widths corresponding to a two-body decay
of the heavy scalar top quark in 
the Minimal Supersymmetric Standard Model with complex parameters (cMSSM). 
The decay modes are given in \refeqs{ststphi} -- (\ref{stbcha}).
The evaluation is based on a full
one-loop calculation of all decay channels, also including hard QED and
QCD radiation. Such a calculation is necessary to derive a reliable
prediction of any two-body decay branching ratio.
Three-body decay modes can become sizable only if all the two-body decay
channels are kinematically (nearly) closed and have thus been neglected
throughout the paper. 

We first reviewed the one-loop renormalization procedure of the cMSSM,
which is relevant for our calculation. 
This includes the $t/\Stop$ and $b/\Sbot$ sector (which has been
chosen according to the analysis in \citere{SbotRen}),
the gluino sector, and the strong coupling constant. We furthermore
reviewed the required renormalization of the Higgs and SM gauge boson
sector as well as the chargino and neutralino sector in the cMSSM. 

We have discussed the calculation of the one-loop diagrams, the
treatment of UV and IR divergences that are canceled by the inclusion
of (hard and soft) QCD and QED radiation. 
Our calculation set up can easily be extended to other two-body decay modes
in the cMSSM. In fact in order to test our method we checked the
finiteness of various other partial decay widths (considering
neutralino, chargino, and Higgs boson decays).

For the numerical analysis we have chosen two parameter sets that allow
simultaneously {\em all} two-body decay modes 
(but could potentially be in conflict with the most recent SUSY
search results from the LHC).
The masses of the scalar top quarks in these scenarios are $260$
and $650 \gev$, and $720$ and $1200 \gev$ for the lighter and the
heavier stop, respectively. 
Consequently, both scenarios result in copious scalar
top quark production at the LHC. A decay of the heavy stop to a lighter
stop (or sbottom) and a neutral (or charged) Higgs boson can 
serve as a source of Higgs bosons
at the LHC, thus a precise knowledge of stop branching ratios is required.
The first scenario also allows $\aStope\Stopz$
production at the ILC(1000), where statistically dominated experimental
measurements of the heavy stop branching ratios will be possible. 
Depending on the integrated luminosity a precision at the few percent
level seems to be achievable.

In our numerical analysis we have shown results for varying $\mstz$ and
$\phiat$, the phase of the trilinear coupling~$\At$. In the results with
varied $\mstz$ only the lighter values allow $\aStope\Stopz$ 
production at the ILC(1000), whereas the results with varied $\phiat$ have
sufficiently light scalar top quarks to permit 
$e^+e^- \to \aStope\Stopz$. In the two
numerical scenarios we compared the tree-level partial widths with the
one-loop corrected  partial decay widths. In the analysis with $\phiat$
varied we showed explicitly the effect of the absorptive parts of
self-energy type corrections on external legs.
We also analyzed the relative change of the partial decay widths
to demonstrate the size of the loop corrections on each individual
channel. In order to see the effect on the experimentally accessible
quantities we also show the various branching ratios at tree-level (all
channels are evaluated at tree-level) and at the one-loop level (with
all channels evaluated including the full one-loop
contributions). Furthermore we presented the relative change of the BRs
that can directly be compared with the anticipated experimental
accuracy.

We found sizable, roughly \order{10\%}, corrections in all the
channels. For some parts of the parameter space (not only close to
thresholds) also larger corrections up to $30\%$~or~$40\%$ have been
observed. This applies especially to the $\br(\Stopz \to \Stope h_n)$
with $n = 1,2,3$. The size of the full one-loop corrections to the partial decay
widths and the branching ratios also depends strongly on $\phiat$. The
one-loop contributions, again being roughly of \order{10\%}, often vary
by a factor of $2-3$ as a function of $\phiat$. In some cases the
absorptive contributions can change the result visibly.
All results are given in detail in \refses{sec:full1L}
and~\ref{sec:full1Lphiat}.  

The numerical results we have shown are, of course, dependent on the choice 
of the SUSY parameters. Nevertheless, they give an idea of the relevance
of the full one-loop corrections. 
The largest partial decay width, if kinematically allowed, is
$\Ga(\decaygl)$ in our scenarios,  
dominating the total decay width, $\Ga_{\rm tot}$, and thus the various 
branching ratios. For other choices of $\mgl$ with $\mgl > \mstz$ the 
corrections to the partial decay widths would stay the same, but the
branching ratios would look very different. 
Decay channels (and their respective one-loop corrections) that may look 
unobservable due to the smallness of their BR in our numerical examples
could become important if other channels are kinematically forbidden.

Following our analysis it is evident that the full one-loop corrections
are mandatory for a precise prediction of the various branching ratios.
The results for the scalar top quark decays will be implemented into the
Fortran code {\tt FeynHiggs}.


\subsection*{Acknowledgements}

We thank  
P.~Bechtle, 
K.~Desch,
S.~Dittmaier,
A.~Fowler,
J.~Guasch, 
H.~Haber,
T.~Hahn, 
W.~Hollik,
L.~Mihaila,  
E.~Mirabella,
F.~von~der~Pahlen, 
T.~Plehn, 
D.~St\"ockinger,
and
G.~Weiglein
for helpful discussions and M.~Spira for critical comments.
The work of S.H.\ was supported in part by CICYT 
(Grant No.\ FPA 2007--66387), 
in part by CICYT (Grant No.\ FPA 2010--22163-C02-01), 
and by the Spanish MICINN's Consolider-Ingenio 2010 Program under 
Grant MultiDark No.\ CSD2009-00064. 
This work was supported in part by the European Community's 
Marie-Curie Research Training Network under Contract No.\
MRTN-CT-2006-035505
``Tools and Precision Calculations for Physics Discoveries at Colliders,''
Part of the work was performed while H.R. was at 
Albert-Ludwigs-Universit\"at Freiburg, Freiburg, Germany.


\newpage

\begin{appendix}

\section*{Appendix: Absorptive parts from self-energy type contributions}

As indicated in the main text, contributions to the partial
decay widths can arise from the product of the imaginary parts of the
loop functions (absorptive parts) of the self-energy type
contributions in the external legs and the imaginary parts of
complex couplings entering the decay vertex or the self-energies. 
In our calculation these corrections are taken into account via 
wave function correction factors $\de{\hat Z}$ (which should not be 
confused with the field renormalization constants $\dZ{}$, which have 
been introduced via the multiplicative renormalization procedure). 
For the off-diagonal wave function correction factors, this procedure has 
been checked against explicitly including the (renormalized) self-energy 
type corrections of the external legs, and full agreement has been
found. The corrections from the absorptive parts can be sizable. 

It is possible to combine the wave function correction factors 
with the field renormalization constants in a single $Z$~factor,
$\mathcal Z$; see e.g.~\citere{imim} and references therein. 

However, if the external particles were stable the wave function 
corrections would be fully taken into account via the field renormalization
constants. The $Z$~factors listed in Sec.~\ref{sec:renorm} also ensure 
that the external (stable) particle does not mix with other fields, 
which is one of the on-shell properties. 
In our scenarios, this is true e.g.\ the lightest neutralino.

In the case of quasi stable particles%
\footnote{
  Which means that considering them as external particles is an 
  approximation, which is justified because in our decays the 
  contributions from the (additional) diagonal $\de{\hat Z}$ are 
  numerically rather negligible.
}%
, additional contributions to the mixing can occur so that the field 
renormalization constants only partly ensure no mixing (for a more 
detailed explanation; see the subsection scalar quarks below). Extra
diagonal contributions can also be taken into account via
$\hat{Z}$~factors.%
\footnote{
  There is still an ongoing discussion whether the diagonal field
  renormalization constants take into account all the contributions 
  needed to ensure the on-shell properties of the external particles or 
  whether an extra wave function correction factor $\de{\hat Z}$ is needed.
}
Here we briefly list all the resulting constants. 


\subsection*{Scalar quarks}

\begin{itemize}
\item[(i)]
For an on-shell particle state, no mixing with another state should
occur, corresponding to
\begin{align}
\hSi_{\sq_{12}}(\msqe^2) = 0~, \qquad \hSi_{\sq_{12}}(\msqz^2) = 0~,\qquad
\hSi_{\sq_{21}}(\msqe^2) = 0~, \qquad \hSi_{\sq_{21}}(\msqz^2) = 0~,
\end{align}
in the case of squarks. Partly, this is already fulfilled due to our
renormalization conditions \refeq{residuumSqOS} in \refse{sec:topbottom}. 
As the considered external particles are quasi stable, in spite of the 
renormalization conditions above, there remains a contribution of the 
imaginary parts of the loop functions. 
This contribution is taken into account via wave function correction 
factors $\de\hat{Z}$, which are different for incoming squarks/outgoing 
antisquarks (unbarred) and outgoing squarks/incoming antisquarks (barred),
\begin{align}
\bigl[\de\hat{\matr Z}_{\sq}\bigr]_{12} &=
+ 2 i \frac{\wtim\Si_{\sq_{12}}(\msqz^2)}{(\msqe^2 - \msqz^2)}~, \quad\
\bigl[\de\hat{\matr Z}_{\sq}\bigr]_{21} =  
- 2 i \frac{\wtim\Si_{\sq_{21}}(\msqe^2)}{(\msqe^2 - \msqz^2)}~,  \\ 
\bigl[\de\bar{\hat{\matr Z}}_{\sq}\bigr]_{12} &= 
+ 2 i \frac{\wtim\Si_{\sq_{21}}(\msqz^2)}{(\msqe^2 - \msqz^2)}~,  \quad\
\bigl[\de\bar{\hat{\matr Z}}_{\sq}\bigr]_{21} = 
- 2 i \frac{\wtim\Si_{\sq_{12}}(\msqe^2)}{(\msqe^2 - \msqz^2)}~,
\end{align}
where $\wtim$ takes only the imaginary part of the loop functions.
Compact expressions for practical numerical 
calculations are obtained via the combined $Z$~factors $\de \cZ$,
\begin{align}
\bigl[\dcZ{\sq}\bigr]_{12} &= 
\bigl[\dZm{\sq} + \de\hat{\matr Z}_{\sq}\bigr]_{12} = 
+ 2 \frac{\Si_{\sq_{12}}(\msqz^2) - \de Y_q}{(\msqe^2 - \msqz^2)}~,\\
\bigl[\dcZ{\sq}\bigr]_{21} &= 
\bigl[\dZm{\sq} + \de\hat{\matr Z}_{\sq}\bigr]_{21} = 
- 2 \frac{\Si_{\sq_{21}}(\msqe^2) - \de Y_q^*}{(\msqe^2 - \msqz^2)}~, \\ 
\bigl[\de\bar{\cZ}_{\sq}\bigr]_{12} &= 
\bigl[\dZm{\sq}^* + \de\bar{\hat{\matr Z}}_{\sq}\bigr]_{12} =
+ 2 \frac{\Si_{\sq_{21}}(\msqz^2) - \de Y_q^*}{(\msqe^2 - \msqz^2)}~,  \\ 
\bigl[\de\bar{\cZ}_{\sq}\bigr]_{21} &= 
\bigl[\dZm{\sq}^* + \de\bar{\hat{\matr Z}}_{\sq}\bigr]_{21} = 
- 2 \frac{\Si_{\sq_{12}}(\msqe^2) - \de Y_q}{(\msqe^2 - \msqz^2)}~.
\end{align}
\item[(ii)]
The diagonal contributions result in the following combined $Z$~factors:
\begin{align}
\bigl[\de{\cZ}_{\sq}\bigr]_{ii} &= 
     \bigl[\dZm{\sq} + \de{\hat{\matr Z}}_{\sq}\bigr]_{ii} 
 = - \KKL \wtre \Si_{\sq_{ii}}'(\msqi^2)
      + i \wtim \Si_{\sq_{ii}}'(\msqi^2) \KKR
= - \Si_{\sq_{ii}}'(\msqi^2)~, \\
\bigl[\de{\bar{\cZ}}_{\sq}\bigr]_{ii} &=
     \bigl[\dZm{\sq}^* + \de{\bar{\hat{\matr Z}}}_{\sq}\bigr]_{ii}
 = - \KKL \wtre \Si_{\sq_{ii}}'(\msqi^2) 
      + i \wtim \Si_{\sq_{ii}}'(\msqi^2) \KKR
= \bigl[\de{\cZ}_{\sq}\bigr]_{ii}~.
\end{align}
\end{itemize}
It should be noted that $\wtre\Si_{ij}(p^2) = (\wtre\Si_{ji}(p^2))^*$ holds
due to ${\cal CPT}$ invariance and the squark field renormalization 
constants obey $[\wtre\,\de\bar{\cZ}_{\sq}]_{ij} = 
[\wtre\,\de{\cZ}_{\sq}]_{ij}^* = [\dZm{\sq}]_{ij}^*~$,
which is exactly the case without absorptive contributions as described 
in \refse{sec:topbottom}.%

In the following we will only give the $\cZ$~factors that combine the
renormalization factors and the additional wave function correction
factors. The derivation is analogous to the one performed in the squark
sector.


\subsection*{Quarks}

The new (diagonal) combined factors $\cZ_{q}$, 
taking into account the absorptive part of the self-energy type 
contribution on an external quark leg  are different for incoming 
quarks/outgoing antiquarks (unbarred) and 
outgoing quarks/incoming antiquarks (barred),
\begin{align}
\dcZ{q}^{L/R} &= - \Big[ \Si_q^{L/R} (m_q^2)
    + m_q^2 \KL {{\Si}_q^{L}}'(m_q^2) + {{\Si}_q^{R}}'(m_q^2) \KR
    + m_q \KL {{\Si}_q^{SL}}'(m_q^2) + {{\Si}_q^{SR}}'(m_q^2) \KR
                  \Big] \non \\
&\qquad \pm \frac{1}{2\, \mq} 
        \KKL {\Si}_q^{SL}(\mq^2) - {\Si}_q^{SR}(\mq^2) \KKR~, \\
\dbcZ{q}^{L/R} &= - \Big[ \Si_q^{L/R} (m_q^2)
    + m_q^2 \KL {{\Si}_q^{L}}'(m_q^2) + {{\Si}_q^{R}}'(m_q^2) \KR
    + m_q \KL {{\Si}_q^{SL}}'(m_q^2) + {{\Si}_q^{SR}}'(m_q^2) \KR
                  \Big] \non \\
&\qquad \mp \frac{1}{2\, \mq} 
        \KKL {\Si}_q^{SL}(\mq^2) - {\Si}_q^{SR}(\mq^2) \KKR~.
\end{align}
The diagonal quark field renormalization constants obey
$\wtre\,\dbcZ{q}^{L/R} = [\dZ{q}^{L/R}]^*$,
which is exactly the case without absorptive contributions as 
described in \refse{sec:topbottom}.

There are no additional off-diagonal terms to the absorptive contributions 
because the CKM matrix has been set to unity.


\subsection*{Gluinos}

The new combined factors $\cZ_{\gl}$, taking into account the
absorptive part of the self-energy type contribution on the external
gluino leg are unbarred (barred) for an incoming (outgoing) gluino,
\begin{align}
\dcZ{\gl}^{L/R} &= - \Big[ \Si_{\gl}^{L/R}(\mgl^2)
    + \mgl^2 \KL \Si_{\gl}^{L'}(\mgl^2) + \Si_{\gl}^{R'}(\mgl^2) \KR
    + \mgl \KL \Si_{\gl}^{SL'}(\mgl^2) + \Si_{\gl}^{SR'}(\mgl^2) \KR
                   \Big] \non \\
&\qquad \pm \ed{2 \mgl} \KKL 
            \Si_{\gl}^{SL}(\mgl^2) - \Si_{\gl}^{SR}(\mgl^2) \KKR~, \\
\dbcZ{\gl}^{L/R} &= \dcZ{\gl}^{R/L}~.
\end{align}
The last formula holds due to the Majorana character of the gluino and
the $\cZ_{\gl}$ factors obey
$\wtre\,\dbcZ{\gl}^{L/R} = \wtre\,\dcZ{\gl}^{R/L} = \dZ{\gl}^*/\dZ{\gl}$, 
which is exactly the case without absorptive contributions as 
described in \refse{sec:gluino}.


\subsection*{Higgs bosons}

Finite contributions from the neutral Higgs wave function correction
factors are taken into account via the $\matr{Z}$~matrix; see
\refeq{eq:zfactors123}, which is a complex quantity. The application
of the $\matr{Z}$~matrix at the amplitude level automatically takes
any absorptive contribution into account.

For the charged Higgs bosons, the new combined factors $\cZ_{H^-H^+}$ 
(unbarred (barred) for an incoming (outgoing) Higgs) read
\begin{align}
\dcZ{H^-H^+} &= - \Sip_{H^-H^+}(\MHp^2)~, \\
\dbcZ{H^-H^+} &= \dcZ{H^-H^+}
\end{align}
instead of \refeq{dZHpHm} in addition with \refeq{dhZHpHm}.


\subsection*{Vector bosons}

For the vector bosons, the new combined factors $\cZ_{\{WW,ZZ\}}$ are
\begin{align}
\dcZ{ZZ} &= -\Sigma_{ZZ}^{\trans\prime}(\MZ^2)~,
& \dcZ{WW} &= -\Sigma_{WW}^{\trans\prime}(\MW^2)~, \\
\dbcZ{ZZ} &= \dcZ{ZZ}~, & \dbcZ{WW} &= \dcZ{WW}~.
\end{align}
However, we found that the additional corrections from vector boson 
self-energies due to the imaginary parts do not give a contribution 
(because in this paper all SUSY masses are larger than $\MZ$), 
and hence no change in the $Z$~factors is required.


\subsection*{Charginos and neutralinos}

More details to the new combined factors $\cZ_{\chapm{}}$ and $\cZ_{\neu{}}$  
(taking into account the absorptive part of the self-energy type 
contributions on the external legs) can be found in \citere{dissAF}.
In our notation they read (unbarred for an incoming neutralino or a
negative chargino, barred for an outgoing neutralino or negative chargino) 
\begin{align}
\KKL \dcZ{\chapm{}}^{L/R} \KKR_{ii} &=
        - \Big[ \Si_{\chapm{}}^{L/R}(\mcha{i}^2) \\
&\qquad + \mcha{i}^2 \KL \Si_{\chapm{}}^{L'}(\mcha{i}^2)
                       + \Si_{\chapm{}}^{R'}(\mcha{i}^2) \KR
        + \mcha{i} \KL \Si_{\chapm{}}^{SL'}(\mcha{i}^2)
                    +  \Si_{\chapm{}}^{SR'}(\mcha{i}^2) \KR \Big]_{ii}~\non \\
&\qquad \pm \ed{2 \mcha{i}} \KKL 
                      \Si_{\chapm{}}^{SL}(\mcha{i}^2)
                    - \Si_{\chapm{}}^{SR}(\mcha{i}^2)
                    - \de\matr{M}_{\cham{}}
                    + \de\matr{M}_{\cham{}}^{*} \KKR_{ii}~, \non \\
\KKL \dcZ{\chapm{}}^{L/R} \KKR_{ij} &= \frac{2}{\mcha{i}^2 - \mcha{j}^2} 
        \Big[ \mcha{j}^2 \Si_{\chapm{}}^{L/R}(\mcha{j}^2) 
             +\mcha{i} \mcha{j} \Si_{\chapm{}}^{R/L}(\mcha{j}^2) \\
&\qquad + \mcha{i} \Si_{\chapm{}}^{SL/SR}(\mcha{j}^2)
        + \mcha{j} \Si_{\chapm{}}^{SR/SL}(\mcha{j}^2)
        - \mcha{i/j} \de\matr{M}_{\cham{}}
        - \mcha{j/i} \de\matr{M}_{\cham{}}^\dagger \Big]_{ij}~, \non \\
\KKL \dcZ{\neu{}}^{L/R} \KKR_{kk} &= 
  - \Big[ \Si_{\neu{}}^{L/R}(\mneu{k}^2) \\
&\qquad     + \mneu{k}^2 \KL \Si_{\neu{}}^{L'}(\mneu{k}^2)
                            +\Si_{\neu{}}^{R'}(\mneu{k}^2) \KR
            + \mneu{k} \KL \Si_{\neu{}}^{SL'}(\mneu{k}^2)
                           +\Si_{\neu{}}^{SR'}(\mneu{k}^2) \KR
       \Big]_{kk} \non \\
&\qquad \pm \ed{2 \mneu{k}} \KKL \Si_{\neu{}}^{SL}(\mneu{k}^2)
            - \Si_{\neu{}}^{SR}(\mneu{k}^2)
            - \de\matr{M}_{\neu{}} 
            + \de\matr{M}_{\neu{}}^{*} \KKR_{kk}~, \non \\
\KKL \dcZ{\neu{}}^{L/R} \KKR_{kl} &= \frac{2}{\mneu{k}^2 - \mneu{l}^2}
  \Big[ \mneu{l}^2 \Si_{\neu{}}^{L/R}(\mneu{l}^2) 
            + \mneu{k}\mneu{l} \Si_{\neu{}}^{R/L}(\mneu{l}^2) \\
&\qquad + \mneu{k} \Si_{\neu{}}^{SL/SR}(\mneu{l}^2)
        + \mneu{l} \Si_{\neu{}}^{SR/SL}(\mneu{l}^2)
        - \mneu{k/l} \de\matr{M}_{\neu{}} 
        - \mneu{l/k} \de\matr{M}_{\neu{}}^{\dagger} \Big]_{kl}~, \non
\end{align}
\begin{align}
\KKL \dbcZ{\chapm{}}^{L/R} \KKR_{ii} &=
        - \Big[ \Si_{\chapm{}}^{L/R}(\mcha{i}^2) \\
&\qquad + \mcha{i}^2 \KL \Si_{\chapm{}}^{L'}(\mcha{i}^2)
                       + \Si_{\chapm{}}^{R'}(\mcha{i}^2) \KR
        + \mcha{i} \KL \Si_{\chapm{}}^{SL'}(\mcha{i}^2)
                    +  \Si_{\chapm{}}^{SR'}(\mcha{i}^2) \KR \Big]_{ii}~\non \\
&\qquad \mp \ed{2 \mcha{i}} \KKL 
                      \Si_{\chapm{}}^{SL}(\mcha{i}^2)
                    - \Si_{\chapm{}}^{SR}(\mcha{i}^2)
                      - \de\matr{M}_{\cham{}}
                      + \de\matr{M}_{\cham{}}^{*} \KKR_{ii}~, \non \\
\KKL \dbcZ{\chapm{}}^{L/R} \KKR_{ij} &= \frac{2}{\mcha{j}^2 - \mcha{i}^2} 
        \Big[ \mcha{i}^2 \Si_{\chapm{}}^{L/R}(\mcha{i}^2) 
             +\mcha{i} \mcha{j} \Si_{\chapm{}}^{R/L}(\mcha{i}^2) \\
&\qquad + \mcha{i} \Si_{\chapm{}}^{SL/SR}(\mcha{i}^2)
       + \mcha{j} \Si_{\chapm{}}^{SR/SL}(\mcha{i}^2)
       - \mcha{i/j} \de\matr{M}_{\cham{}}
       - \mcha{j/i} \de\matr{M}_{\cham{}}^\dagger \Big]_{ij}~, \non \\
\label{diagneu}
\KKL \dbcZ{\neu{}}^{L/R} \KKR_{kk} &= 
 \KKL \dcZ{\neu{}}^{R/L} \KKR_{kk}~,  \\
\label{offdiagneu}
\KKL \dbcZ{\neu{}}^{L/R} \KKR_{kl} &= 
 \KKL \dcZ{\neu{}}^{R/L} \KKR_{lk}~. 
\end{align}
The chargino/neutralino $\cZ$ factors obey
$\wtre\,\dbcZ{\tilde{\chi}}^{L/R} = 
[\wtre\,\dcZ{\tilde{\chi}}^{L/R}]^\dagger =
[\dZm{\tilde{\chi}}^{L/R}]^\dagger$, 
which is exactly the case without absorptive contributions as 
described in \refse{sec:chaneu}, or in other words  
$\dbcZ{\tilde{\chi}}^{L/R} = 
[\dZm{\tilde{\chi}}^{L/R}]^\dagger + 
[\de\bar{\hat{\matr Z}}^{L/R}_{\tilde{\chi}}]$.
The Eqs.~\eqref{diagneu} and \eqref{offdiagneu} hold due to the Majorana 
character of the neutralinos.

\end{appendix}



\begin{thebibliography}{99}
\bibitem{feynarts-mf} The couplings can be found in the files
                      {\tt MSSM.ps.gz}, {\tt MSSMQCD.ps.gz} and 
                      {\tt HMix.ps.gz} as part of the
                      \fa~package~\cite{feynarts}.
\end{thebibliography}
\end{document}